\documentclass[a4paper,11pt]{article}
\pdfoutput=1 % if your are submitting a pdflatex (i.e. if you have
\usepackage{jcappub} % for details on the use of the package, please
\usepackage[T1]{fontenc} % if needed

\usepackage{bm}
\usepackage{url}
\usepackage{amsmath}
\usepackage{amsfonts}
\usepackage{morefloats}

\newcommand{\SolarProp}{{\sc HelioProp}}

\newcommand{\GeV}{{\rm GeV}}

\newcommand{\km}{{\rm km}}

\newcommand{\s}{{\rm s}}

\newcommand{\AU}{{\rm AU}}
\newcommand{\nT}{{\rm nT}}

\newcommand{\dbar}{{\bar d}}
\newcommand{\pbar}{{\bar p}}
\newcommand{\nbar}{{\bar n}}
\newcommand{\sqrts}{\sqrt {s}}
\newcommand{\sigmav}{\langle \sigma v \rangle}
\newcommand{\mdm}{m_{DM}}
\newcommand{\uubar}{u \bar u}
\newcommand{\bbbar}{b \bar b}
\newcommand{\ww}{W^+W^-}
\newcommand{\thermal}{$2.3 \times 10^{-26}$ cm$^3$ s$^{-1}$}
\newcommand{\cms}{cm$^3$s$^{-1}$}
\newcommand{\mcp}{MC($\Delta p$)}
\newcommand{\mcpr}{MC($\Delta p + \Delta r$)}

\title{Dark matter searches with cosmic antideuterons: status and perspectives}

\author[a,b]{N. Fornengo}
\author[c,d]{L. Maccione}
\author[a,b]{A. Vittino}

\affiliation[a]{Department of Physics, University of Torino \\ via P. Giuria 1, 10125 Torino, Italy}
\affiliation[b]{Istituto Nazionale di Fisica Nucleare \\ via P. Giuria 1, 10125 Torino, Italy}
\affiliation[c]{Ludwig-Maximilians-Universit\"{a}t, Theresienstra{\ss}e 37, D-80333 M\"{u}nchen, Germany}
\affiliation[d]{Max-Planck-Institut f\"{u}r Physik (Werner Heisenberg Institut), F\"{o}hringer Ring 6, D-80805 M\"{u}nchen, Germany}

\emailAdd{fornengo@to.infn.it}
\emailAdd{luca.maccione@lmu.de}
\emailAdd{vittino@to.infn.it}

\abstract{
The search for antideuterons in cosmic rays has been proposed as a promising channel for dark matter indirect detection, especially for dark matter particles with a low or intermediate mass. With 
the current operational phase of the AMS-02 experiment and the ongoing development of a future dedicated experiment, the General Antiparticle Spectrometer (GAPS), there are exciting prospects 
for a dark matter detection in the near future. 
In this paper we develop a detailed and complete re-analysis of the cosmic-ray antideuteron signal, by discussing the main relevant issues related to antideuteron production and  propagation through the interstellar medium and the heliosphere. 
In particular, we first critically revisit the coalescence mechanism for antideuteron production in dark matter annihilation processes. 
Then, since antideuteron searches have their best prospects of detection at low kinetic energies
where the effect of the solar wind and magnetic field are most relevant, we address the impact of solar modulation modeling on the antideuteron flux at the Earth by developing a
full numerical 4D solution of cosmic rays transport in the heliosphere.
We finally use these improved predictions to provide updated estimates of the reaching capabilities for AMS-02 and GAPS, compatible with the current constraints imposed by the antiprotons measurements of PAMELA. After the
antiproton bound is applied, prospects of detection of up to about 15 events  in GAPS LDB+
and AMS-02 missions are found, depending on the dark matter mass, annihilation rate and
production channel from one side, and on the coalescence process, galactic and solar
transport parameters on the other.}

\begin{document}
\maketitle
\flushbottom

\section{Introduction}
\label{sec:introduction}

Antideuterons have been proposed in Ref. \cite{DFS} as a very promising channel for dark matter (DM) indirect detection. The most relevant energy window for DM searches is below few GeV/n \cite{DFS,DFM}, where the secondary astrophysical background is significantly suppressed,
as a consequence of simple kinematical reasons pertaining the  secondary process
$p_{\rm CR} + p_{\rm ISM} \rightarrow \dbar + X$, where a cosmic-rays proton (or heavier nucleus) impinges on a proton (or heavier nucleus) of the interstellar medium. DM particles annihilate (or decay) almost at rest: therefore very low energy $\dbar$ may be produced. The local interstellar (LIS) $\dbar$ fluxes produced by DM annihilation may be significantly larger than the expected background \cite{DFS,DFM} for $\dbar$ kinetic energies below 1--3 GeV/n, even orders of magnitude larger. Transport inside the heliosphere softens this difference \cite{DFS,DFM}, but still allows the low-energy $\dbar$
to exceed the secondary background to a level that makes $\dbar$ a potential tool of discovery in
large portions of the DM parameter space (namely its mass $\mdm$ and annihilation cross section
$\sigmav$, or alternatively decay rate in case of decaying DM). Following the original proposal
in Ref. \cite{DFS}, an updated analysis was performed in Ref. \cite{DFM}, where uncertainties
in both the secondary production and the DM $\dbar$ signal were quantified. These uncertainties
are large, mostly due to the propagation inside the galactic environment, and (for the secondary
production) nuclear physics uncertainties in the production mechanism.

More recent analyses have shown the impact of different modeling for the antideuteron
formation mechanism, typically described in terms of a coalescence mechanism. This
may have a significant impact, especially for the DM signal. While
Refs. \cite{DFS,DFM} adopted a conservative approach, where the $\dbar$ formation
from a produced $\pbar\nbar$ pair occurs in a purely uncorrelated, isotropic way, Refs. \cite{strumia,ibarra} attempted a Monte Carlo (MC) modeling of the coalescence process, which allows
to take into account correlations in the formation of the $\pbar\nbar$, which then reflect into
a significant increase in the predicted $\dbar$ spectra from DM annihilation, especially
for heavy DM and at large $\dbar$ energies. Ref. \cite{Ibarra:2013qt}, instead,
addresses the same issue (correlated production of $\pbar$ and $\nbar$) in the process
$p_{\rm CR} + p_{\rm ISM} \rightarrow \dbar + X$, which is responsible for the
secondary background component: a decrease of about a factor of 2 in the prediction of the
secondary background is found to be a possibility, although relevant uncertainties remain
for this hadronic process (including the ability of the adopted MC to properly treat this process at the relatively low center-of-mass energies pertaining the $\dbar$ production process in the Galaxy).
In addition to the uncertainties arising from the coalescence modeling, the dependence of the
$\dbar$ spectra on the hadronization model itself has been addressed in Ref. \cite{Dal:2012my} and
shown to have an impact of a factor of 2-4. Prospects for $\dbar$ detection have then
been discussed in Refs \cite{Profumo,Randall,ibarra}, under different assumptions on the expected detector
sensitivities. Prospects for decaying DM \cite{Ibarra:2009tn} and $\dbar$ relevance for  heavy DM \cite{Brauninger:2009pe} have been discussed.
Antideuterons have also been discussed in the context of specific particle physics models,
where the DM is represented by neutralinos \cite{DFS,DFM,Profumo:2004ty,Bottino:2007qg,Bottino:2008mf}, sneutrinos \cite{Arina:2007tm} and Kaluza-Klein DM \cite{Profumo}.

The predicted  $\dbar$ fluxes are quite small, both for the DM signal \cite{DFS,DFM} (see also Refs.\cite{strumia,ibarra}) and for the secondary component \cite{DFS,DFM} (see also Refs. \cite{duperray1,duperray2,Ibarra:2013qt}). The current most constraining
experimental bound comes from BESS: \cite{Fuke:2005it} the 95\% C.L. upper bound on the
$\dbar$ flux is at the level
of  $1.9\times 10^{-4}$ (m$^2$ s sr GeV/n)$^{-1}$ in the energy interval $(0.17 \div 1.15)$ GeV/n.
This limit is more than 3 orders of magnitude far from the expected secondary background. However, in the
next years two experimental Collaborations, which employ two different detection
techniques, will significantly improve the sensitivity to
cosmic antideuterons, and are expected to access the flux levels relevant for DM investigation.
The Alpha Magnetic Spectrometer (AMS) \cite{Battiston:2008zza,Incagli:2010zz,Bertucci:2011zz,Kounine:2012zz} is currently under operation onboard the International Space Station. Its design
flux-limit (corresponding to 1 detected event for the original experimental set-up and data taking
period)  points at the level of $4.5\times 10^{-7}$ (m$^2$ s sr GeV/n)$^{-1}$ in the energy interval $(0.2 \div 0.8)$ GeV/n \cite{giovacchini,choutko,Arruda:2007fy,giovacchinipriv}. 
The General AntiParticle Spectrometer (GAPS) \cite{Mori:2001dv,Hailey:2009zz,Hailey:2013gwa} is specifically designed to
access very low $\dbar$ energies. The current proposed layout  is envisaged to operate in high atmosphere with balloon flights.
GAPS has already successfully performed a prototype flight (pGAPS) 
from the JAXA/ISAS balloon facility in Hokkaido, Japan in 2012 \cite{Mognet:2013dxa,Fuke:2013lca} and it is now in its development phase, in preparation of a science flight from Antarctica in 2017-2018.
Its current best expected-sensitivity can be as low as $2.8\times 10^{-7}$ (m$^2$ s sr GeV/n)$^{-1}$ in the energy interval $(0.1 \div 0.25)$ GeV/n \cite{haileypriv}, with an extended
long-duration balloon flight of 210-day (GAPS LBD+).

It is therefore timely to re-analyze the theoretical predictions for the DM antideuteron flux, in order
to quantify more precisely the size of the signal, the impact of the sources of uncertainties, and
the potential reaching capabilities for different DM models.
In this paper we therefore perform a full reanalysis of our previous studies \cite{DFS,DFM}, in the light 
also of the recent theoretical developments discussed above, and in the light and in preparation
of the experimental data which are expected in the next years from AMS-02 and GAPS.
We concentrate here on a detailed investigation of the DM signal.
For the secondary background-component we use, as a reference model, the one derived in our
previous analyses \cite{DFS,DFM}. 

In this new study,
we first critically discuss and attempt a  re-assessment of the coalescence process, by employing MC techniques like those discussed in Refs. \cite{strumia,ibarra}, with a careful and detailed modeling of the $\dbar$ formation
process. In particular, we 
discuss some fine details that occur for various DM annihilation channels, relevant especially for
light DM producing $\dbar$ through a heavy--quark channel, especially $\bar b b$ (a situation of interest in many specific realizations of New Physics models, where the DM annihilation process proceeds mostly through a coupling to scalars, like e.g. in supersymmetry for neutralino or sneutrino DM, where the annihilation process is mediated by
higgs exchange in a large portion of the parameter space; and of interest also in connection to the DM direct detection results, which have an intriguing potential
indication for light DM). 

The second relevant point of the new analysis is the improvement
of the modeling of the $\dbar$ transport inside the heliosphere: the DM signal is expected to emerge with more distinction at low $\dbar$ energies, which are also those energies where the impact
of solar modulation is the largest. By employing a novel approach of solution of the transport equation,
based on stochastic techniques, we discuss the impact of solar modulation modeling on the
$\dbar$ low-energy spectra and on the ability of cosmic antideuteron detectors (namely GAPS and
AMS-02) to discover a true DM signal. 

In all our analyses, we impose the current stringent bounds which
are coming from the observation of the antiproton component in cosmic rays, for which we
adopt the PAMELA observations \cite{Adriani:2010rc}. Since both $\bar p$
and $\dbar$ are produced by DM in the same hadronic processes, the amount of $\dbar$ from
DM production is correlated to the amount of produced $\bar p$.
The actual impact of the antiproton bounds is nevertheless dependent also on the impact
of solar modulation, which may act differently on cosmic rays of different type and which
depends on the actual period of data taking. While currently-available PAMELA 
antiproton data refer to a period of negative polarity of the solar cycle, the expected
data taking periods for both AMS-02 (ongoing) and GAPS (in the future years) occur
in a period of positive polarity. Detailed modeling of the transport inside the heliosphere
is therefore an important element to consider, and we will discuss the size and relevance
of it.
 Also gamma--rays
\cite{gammabound1,gammabound2,gammabound3,gammabound4,gammabound5,
gammabound6,gammabound7,gammabound8} and radio signals
\cite{radiobound1,radiobound2,radiobound3,radiobound4} (as well as
effects induced on the cosmic microwave background by particle injection
from dark matter annihilation in the early stages after recombination
\cite{cmbbound1,cmbbound2,cmbbound3,cmbbound4,cmbbound5}) are currently
providing stringent bounds on the properties (masses and annihilation
cross sections) of the DM particle. We choose not to adopt
those bounds in the present analysis, since these indirect DM
signals involve layers of astrophysical modeling which are in large part
different from the modeling required for the antideuteron signal.
Antiprotons, instead, are more directly correlated with antideuterons,
and therefore we consider them as a necessary bound to be applied.

The paper is organized as follows. In Section \ref{sec:coalescence} we discuss the process of antideuteron formation and 
we implement the coalescence models then used in the subsequent analysis. In Section \ref{sec:galaxy} we
briefly recall the main elements of $\dbar$ propagation in the Galaxy, while in Section \ref{sec:solarmod}
we introduce advanced modeling of cosmic rays transport in the heliosphere, based on stochastic
techniques. In Section \ref{sec:signals} we discuss in details the DM signals and the relevant sources of
uncertainties arising from the coalescence process, transport in the galaxy and solar modulation.
In Section \ref{sec:prospects} we then move to discuss the prospects of detection of a DM signal in AMS-02
and GAPS and correlate those prospects to the current bounds arising from the antiproton
flux measured by PAMELA. Finally, Section \ref{sec:conclusions} reports our conclusions.

\section{Process of antideuteron formation: the coalescence mechanism}
\label{sec:coalescence}

\subsection{General formalism and techniques}
\label{sec:coalescence-form}

Antideuteron production can be described by the so-called coalescence mechanism \cite{coalescence1,coalescence2}. In this approach,
an antiproton $\pbar$ and an antineutron $\nbar$ are first produced: then, if they happen to be in the
correct sector of their mutual phase space such that the formation of a bound state is possible,
they merge (coalesce) to form an antideuteron $\dbar$, with a momentum fixed by energy-momentum
conservation. Coalescence is therefore a very rare process and the expected number of antideuterons is  
consequently suppressed, when compared to the $\pbar$ and $\nbar$ production multiplicities.

The number of antideuterons produced in a process occurring at a center-of-mass
energy $\sqrts$ can be expressed as \cite{Chardonnet:1997dv,DFS,DFM}: 

\begin{equation}
dN_{\dbar} = \frac{1}{\sigma_{\rm tot}}d^3\sigma_{\dbar}(\sqrts, \vec{k}_{\dbar})
= F_{\dbar}(\sqrt{s}, \vec{k}_{\dbar}) \;d^3\vec{k}_{\bar{d}}
\label{eq:coa1}
\end{equation}
where $\sigma_{\rm tot}$ and $d^3\sigma_{\dbar}$  are the total and differential cross sections for 
antideuteron production in the process under study (i.e. DM annihilation, or cosmic-rays spallation on the interstellar medium (ISM), in the case of  the astrophysical background). The phase space $\dbar$ distribution $F_{\dbar}(\sqrt{s}, \vec{k}_{\dbar})$
depends on the momentum distribution 
$F_{(\pbar\nbar)}(\sqrt{s}, \vec{k}_{\pbar},\vec{k}_{\nbar}) $
of the $(\pbar,\nbar)$ pair produced in the physical process, and on the probability
${\cal C}(\sqrts, \vec{k}_{\pbar},\vec{k}_{\nbar} | \vec{k}_{\dbar})$
that a $\pbar$ with momentum $\vec{k}_{\pbar}$ and a $\nbar$ with momentum
$\vec{k}_{\nbar}$ merge to form a $\dbar$ with momentum $\vec{k}_{\dbar}$:
\begin{equation}
 F_{\dbar}(\sqrt{s}, \vec{k}_{\dbar}) = \int  F_{(\pbar\nbar)}(\sqrt{s}, \vec{k}_{\pbar},\vec{k}_{\nbar}) \; {\cal C}(\sqrts, \vec{k}_{\pbar},\vec{k}_{\nbar} | \vec{k}_{\dbar})\;
 d^3\vec{k}_{\nbar} \; d^3\vec{k}_{\nbar}
\label{eq:coa2}
\end{equation}
Momentum conservation allows us to factorize the coalescence function as follows:
\begin{equation}
{\cal C}(\sqrts, \vec{k}_{\pbar},\vec{k}_{\nbar} | \vec{k}_{\dbar})
= C(\vec{k}_{\pbar},\vec{k}_{\nbar})\; \delta^{(3)}(\vec{k}_{\dbar}-\vec{k}_{\pbar}-\vec{k}_{\nbar})
\label{eq:coa3}
\end{equation}
We have implicitly assumed here that $C(\vec{k}_{\pbar},\vec{k}_{\nbar})$ is independent
on the energy $\sqrts$ of the production process: however, in general grounds, the coalescence
probability may evolve with the production energy. We will come to this point later.

With the factorization of Eq. (\ref{eq:coa3}), the $\dbar$ phase space distribution simply becomes:
\begin{equation}
 F_{\dbar}(\sqrt{s}, \vec{k}_{\dbar}) = \int  F_{(\pbar\nbar)}(\sqrt{s}, \vec{k}_{\pbar},\vec{k}_{\nbar}) \; 
 C(\vec{k}_{\pbar},\vec{k}_{\nbar})\; \delta^{(3)}(\vec{k}_{\dbar}-\vec{k}_{\pbar}-\vec{k}_{\nbar})\;
 d^3\vec{k}_{\nbar} \; d^3\vec{k}_{\nbar}
\label{eq:coa4}
\end{equation}
Integration variables can be conveniently rotated to the total momentum 
$\vec{k}_{\pbar} + \vec{k}_{\nbar} = \vec{k}_{\dbar}$ and to the relative momentum
 $\vec{k}_{\pbar} - \vec{k}_{\nbar} = \vec{\Delta}$ of the pair:
\begin{equation}
 F_{\dbar}(\sqrt{s}, \vec{k}_{\dbar}) = \int  F_{(\pbar\nbar)}(\sqrt{s}, \vec{k}_{\pbar},\vec{k}_{\nbar}) \; 
 C(\vec{\Delta})\; \delta^{(3)}(\vec{k}_{\dbar}-\vec{k}_{\pbar}-\vec{k}_{\nbar})\;
\frac{1}{8} \; d^3\vec{\Delta} \; d^3\vec{k}_{\dbar}
\label{eq:coa5}
\end{equation}
This allows to easily perform the integral on the $\dbar$ momenta, thanks to the $\delta$-function of momentum conservation, while
the coalescence function $C$ depends only on the relative momentum $\vec{\Delta}$. 
 Notice that in our previous papers \cite{DFS,DFM} we used a different
convention on the relative momentum: $\vec{k}_{\pbar} - \vec{k}_{\nbar} = 2 \vec{\Delta}$.
This simply reflects in the numerical factor 1/8 in Eq. (\ref{eq:coa5}),  arising from the Jacobian $J$ of the change of variables
from $\{\vec{k}_{\pbar}, \vec{k}_{\nbar}\}$ to $\{\vec{k}_{\dbar}, \vec{\Delta} \}$:
$J=1/8$ for the current definition, while with the old definition we had $J=1$. While this has clearly
no effect on the determination of the $\dbar$ spectra, it will introduce a numerical difference in the 
specific values of the ``coalescence momentum'' $p_0$, which will be defined below.
The $\dbar$ phase space distribution therefore takes the form:
\begin{equation}
 F_{\dbar}(\sqrt{s}, \vec{k}_{\dbar}) =  \frac{1}{8} \int  F_{(\pbar\nbar)}(\sqrt{s}, 
\vec{k}_{\pbar} = \vec{k}_{\pbar} ^{\;*},
\vec{k}_{\nbar} =  \vec{k}_{\nbar}^{\;*}) \; 
 C(\vec{\Delta})\;  d^3\vec{\Delta}
\label{eq:coa6}
\end{equation}
where $\vec{k}_{\pbar} ^{\;*} = (\vec{k}_{\dbar} + \vec{\Delta} )/2$ and 
$\vec{k}_{\nbar} ^{\;*} = (\vec{k}_{\dbar} - \vec{\Delta} )/2$.

We can notice that Eq.(\ref{eq:coa6}) is expressed in terms of the $\dbar$ and $(\pbar,\nbar)$ phase spaces:
\begin{eqnarray}
F_{\dbar}(\sqrt{s}, \vec{k}_{\dbar}) &=&  \frac{dN_{\dbar}}{d^3 \vec{k}_{\dbar}} \label{eq:coa7a}\\
F_{(\pbar\nbar)}(\sqrt{s}, \vec{k}_{\pbar},\vec{k}_{\nbar}) &=&  
\frac{dN_{(\pbar\nbar)}}{d^3 \vec{k}_{\pbar} d^3 \vec{k}_{\nbar}}
\label{eq:coa7b}
\end{eqnarray}
which are not Lorentz invariant. On the other hand, the same expression in Eq. (\ref{eq:coa6})
can be cast in a Lorentz-invariant form by using the Lorentz-invariant phase space:
\begin{eqnarray}
F_{\dbar}(\sqrt{s}, \vec{k}_{\dbar}) &\;\;\rightarrow\;\;& \gamma_{\dbar} \; F_{\dbar}(\sqrt{s}, \vec{k}_{\dbar}) \\
F_{(\pbar\nbar)}(\sqrt{s}, \vec{k}_{\pbar},\vec{k}_{\nbar}) &\;\;\rightarrow\;\;&  
\gamma_{\pbar} \gamma_{\nbar} \; F_{(\pbar\nbar)}(\sqrt{s}, \vec{k}_{\pbar},\vec{k}_{\nbar})
\label{eq:coa8}
\end{eqnarray}
By directly applying Eqs. (\ref{eq:coa7a},\ref{eq:coa7b}) allows to rewrite
Eq. (\ref{eq:coa6}) as:
\begin{equation}
\gamma_{\dbar} \; F_{\dbar}(\sqrt{s}, \vec{k}_{\dbar}) =  
\frac{\gamma_\dbar}{\gamma_{\pbar} \gamma_{\nbar}}\;
\frac{1}{8}\; \int   d^3\vec{\Delta} \; C(\vec{\Delta})\;  
\gamma_{\pbar} \gamma_{\nbar} \; F_{(\pbar\nbar)}(\sqrt{s}, 
\vec{k}_{\pbar} = \vec{k}_{\pbar} ^{\;*},
\vec{k}_{\nbar} =  \vec{k}_{\nbar}^{\;*})
\label{eq:coa9}
\end{equation}
Since the coalescence function has effect only when the relative momentum of the
$(\pbar,\nbar)$ pair is very small, we can apply the following approximation:
\begin{equation}
\gamma_{\dbar} \; F_{\dbar}(\sqrt{s}, \vec{k}_{\dbar}) \simeq 
\frac{1}{8}\;
\left [
\frac{\gamma_\dbar}{\gamma_{\pbar} \gamma_{\nbar}}\;
 \int   d^3\vec{\Delta} \; C(\vec{\Delta})
 \right ]
 \;  
\gamma_{\pbar} \gamma_{\nbar} \; F_{(\pbar\nbar)}(\sqrt{s}, 
\vec{k}_{\pbar} = \vec{k}_{\pbar} ^{\;**},
\vec{k}_{\nbar} =  \vec{k}_{\nbar}^{\;**})
\label{eq:coa10}
\end{equation}
where now $\vec{k}_{\pbar} ^{\;**} = \vec{k}_{\dbar}/2$ and 
$\vec{k}_{\nbar} ^{\;**} = \vec{k}_{\dbar}/2$. This approximation
holds since the range of relative momenta where $C(\Delta)$ is
effective can be estimated as $|\vec{\Delta}| \ll p_{\rm cut}$, where
$p_{\rm cut}$ is expected to range from
$p_{\rm cut}\sim \sqrt{m_p B} \sim 46\, {\rm MeV}$ (as derived from
$\dbar$ binding energy \cite{DFS})
up to about 180-200 MeV (as derived from a Hulthen parameterization of the
$\dbar$ wave function \cite{DFS,braun}.
In both cases $p_{\rm cut} \ll k_{\dbar}\equiv |\vec{k}_{\dbar}| $) 
and this justifies the approximations which lead to the factorization of Eq. (\ref{eq:coa10}).
Eq. (\ref{eq:coa10}) has a convenient form, since it is expressed in terms of the invariant
phase spaces: this implies that the term inside the square brackets is also Lorentz invariant.
It is therefore convenient to determine it in the $\dbar$ rest frame, where
$\vec{k}_{\dbar} = 0$ and
$\vec{k}_{\pbar} = \vec{k}_{\pbar}^{\;*} = 0 = \vec{k}_{\nbar}^{\;*} = \vec{k}_{\nbar}$.
This implies that the Lorentz factors $\gamma_{\dbar}$,  $\gamma_{\pbar}$ and $\gamma_{\nbar}$ are approximately unity, and we get that the term in square brackets can be expressed as:
\begin{equation}
\frac{\gamma_\dbar}{\gamma_{\pbar} \gamma_{\nbar}}\;
 \int   d^3\vec{\Delta} \; C(\vec{\Delta})
 \;\;\; \xrightarrow{\mbox{$\dbar$ rest frame}} \;\;\; 
 \int d^3\vec{\Delta} \; C(\vec{\Delta})\; = V_{\rm coal} \; = \frac{4\pi p_0^3}{3}
\label{eq:coa11}
\end{equation}
Eq. (\ref{eq:coa11}) defines the ``coalescence momentum'' $p_0$ as an effective parameter which determines the size of the phase-space volume $V_{\rm coal}$ in which the $(\pbar,\nbar)$ pair
can merge. Notice that Eq. (\ref{eq:coa11}) defines $p_0$ in the $\dbar$ rest frame.
Due to Lorentz invariance,
we can cast this expression back into Eq. (\ref{eq:coa10}), and obtain our final expression for the
$\dbar$ distribution:
\begin{equation}
\gamma_{\dbar} \; \frac{dN_{\dbar}}{d^3 \vec{k}_{\dbar}} = 
\frac{1}{8}\;
\frac{4\pi p_0^3}{3}
 \;  
\gamma_{\pbar} \gamma_{\nbar} \; F_{(\pbar\nbar)}(\sqrt{s}, 
\vec{k}_{\pbar} = \vec{k}_{\dbar}/2,
\vec{k}_{\nbar} = \vec{k}_{\dbar}/2)
\label{eq:coa12}
\end{equation}
Due to the different definition of the relative momentum $\vec\Delta$ adopted
here as compared to our previous analyses \cite{DFS,DFM}, we have that:
\begin{equation}
p_0^{\rm current~definition} = 2 p_0^{\rm previous~definition}
\label{eq:coa13}
\end{equation}
Moreover, the definition of $V_{\rm coal}$ in Eq. (\ref{eq:coa11}) is equivalent
to assume, in the $\dbar$ rest frame, the following form for the coalescence function:
\begin{equation}
C(\vec{\Delta}) = \theta (\Delta^2 - p_0^2) \; .
\label{eq:coa14}
\end{equation}

The relevant information which is needed in order to determine the $\dbar$ energy distribution
is now the phase-space distribution $F_{(\pbar\nbar)}(\sqrt{s}, \vec{k}_{\pbar},\vec{k}_{\nbar})$
of the $(\pbar,\nbar)$ pair. We discuss three models, which will be analyzed in detail in the following
Sections.

\subsection{Analytical model: assumption of factorization with independent, isotropic $\pbar\nbar$ production}
\label{sec:oldmodel}

The simplest assumption is that the production of the $\pbar$ and of the $\nbar$ is totally
uncorrelated, with an isotropic distribution of both nucleons. This assumption allows to analytically
derive the energy distribution of the $\dbar$, and is the approximation which was adopted in
our previous analyses \cite{DFS,DFM}. Pure factorization implies:
\begin{equation}
F_{(\pbar\nbar)}(\sqrt{s}, \vec{k}_{\pbar},\vec{k}_{\nbar}) =
F_{\pbar}(\sqrt{s}, \vec{k}_{\pbar}) \; \times \;
F_{\nbar}(\sqrt{s}, \vec{k}_{\nbar})
\label{eq:coa15}
\end{equation}
As discussed in Ref. \cite{DFS}, at relatively low $\sqrts$ (as it occurs in the case of
secondary $\dbar$, where the bulk of the antideuteron production comes from $\sqrts \sim 10$ GeV,
a value about of the same order of magnitude as the $\dbar$ mass)
exact factorization should break, due to energy conservation, and some adjustment should be done. 
A viable correction to Eq.(\ref{eq:coa15}), used in Ref. \cite{DFS} is:
\begin{equation}
F_{(\pbar\nbar)}(\sqrt{s}, \vec{k}_{\pbar},\vec{k}_{\nbar}) =
\frac{1}{2} F_{\pbar}(\sqrt{s}, \vec{k}_{\pbar}) \; 
F_{\nbar}(\sqrt{s} - 2E_\pbar, \vec{k}_{\nbar}) +
\frac{1}{2} F_{\pbar}(\sqrt{s} - 2E_\nbar, \vec{k}_{\pbar}) \;
F_{\nbar}(\sqrt{s}, \vec{k}_{\nbar}) 
\label{eq:coa15b}
\end{equation}
where it is assumed that the center-of-mass energy available for the production of the second antinucleon
is reduced by twice the energy carried away by the first antinucleon. In the case of $\dbar$ arising from DM production, DM annihilation at rest implies that $\sqrts = 2 \mdm$: therefore,
as long as $2 \mdm \gg m_\dbar$, factorization as in Eq. (\ref{eq:coa15}) can be safely assumed to hold. For DM lighter than about a few GeV, a correction as in Eq. (\ref{eq:coa15b}) is more appropriate.
In the following, for the analytical model we will assume exact factorization to hold: we will refer
to this modeling as the ``old model'' \cite{DFS,DFM}. By simple algebra the expression for $\dbar$ spectra can be recast as \cite{DFS,DFM}:
\begin{equation}
\frac{dN_\dbar}{dT_{\dbar}}\;=\; \frac{p_0^3}{6 k_{\dbar}}\; \frac{m_\dbar}{m_\pbar m_\nbar} \;
\left. \frac{dN}{dT_{\pbar}}\right\vert^{**} \left. \frac{dN}{dT_{\nbar}}\right\vert^{**}
\label{eq:oldmodel}
\end{equation}
where $T_i = E_i - m_i$ is the kinetic energy of $i=\dbar,\pbar,\nbar$ and the $\vert^{**}$
notation recalls that the $\pbar$ and $\nbar$ spectra must be evaluated at $T_\pbar = T_\dbar/2$
and $T_\nbar = T_\dbar/2$, respectively (as dictated by Eq. (\ref{eq:coa12})).
In deriving Eq. (\ref{eq:oldmodel}), we have clearly assumed $m_\pbar = m_\nbar = m_\dbar/2$.

\subsection{Monte Carlo model: correlated $\pbar\nbar$ production}
\label{sec:MCmodel}

Depending on the microphysical details of the $\pbar$ and $\nbar$ production, correlations
or anticorrelations between the two antinucleons may be present. This may depend on the
specific production channel, as well as on the center-of-mass energy of the process. In order
to understand the kinematical details of the $\pbar$ and $\nbar$ production, a Monte Carlo (MC) approach
is appropriate. In this way, we can check the possibility for the $(\pbar$, $\nbar)$ pair to coalesce on 
an event-by-event basis, including their mutual correlations (or anticorrelations), whenever present. This approach has been
investigated in Refs.~\cite{strumia,ibarra} where it has been shown that enhancements (with respect to the analytical
isotropic model) of the $\dbar$ multiplicity in DM annihilation processes are present, especially
at large energies and for heavy DM. In the following we will investigate the MC approach in
detail, by elucidating some fine details. 

The key element that can be extracted from the MC is the $\pbar$ and $\nbar$ phase space
distribution $F^{\rm MC}_{(\pbar\nbar)}(\sqrt{s}, \vec{k}_{\pbar},\vec{k}_{\nbar})$: this
can be done without any assumption regarding factorization or isotropy.
The $\dbar$ distribution is then constructed by defining an appropriate procedure to merge the $\pbar$ and $\nbar$ 
to form a $\dbar$.
It is convenient to start back from Eq.~(\ref{eq:coa4}), together with the assumption that
the coalescence function depends only on the relative momentum of the pair:
$C(\vec{k}_{\pbar},\vec{k}_{\nbar}) = C(\vec\Delta)$. 
We then build the $\dbar$ distribution function as follows:
\begin{equation}
 F^{\rm MC}_{\dbar}(\sqrt{s}, \vec{k}_{\dbar}) = \int  F^{\rm MC}_{(\pbar\nbar)}(\sqrt{s}, \vec{k}_{\pbar},\vec{k}_{\nbar}) \; 
C(\Delta)\; \delta^3(\vec{k}_{\dbar}-\vec{k}_{\pbar}-\vec{k}_{\nbar})\;
 d^3\vec{k}_{\nbar} \; d^3\vec{k}_{\nbar}\;.
\label{eq:coa16}
\end{equation}
We sample the $(\pbar,\nbar)$ pair with the MC, and we consider a $\dbar$ with momentum
$\vec{k}_\dbar = \vec{k}_\pbar + \vec{k}_\nbar$ to be formed when a condition related to the coalescence process is met. We adopt two techniques:

\begin{enumerate}

\item
Model ``MC($\Delta p$)'': it takes the form of the coalescence function as in Eq.~(\ref{eq:coa14}).
Therefore the $\pbar$ and the $\nbar$ form a $\dbar$ when their relative momentum (in the rest frame
of the $\pbar\nbar$ pair)  is smaller than the ``coalescence momentum'' $p_0$. This condition is the same as the one used in the analytical model, although the value of $p_0$ in the two approaches does
not need to be the same (in fact it is not, as discussed below).

\item
Model ``MC($\Delta p + \Delta r$)'': it also takes the form of the coalescence function as in Eq.~(\ref{eq:coa14}),
but in addition it requires that the physical distance $\Delta r$ between the $\pbar$ and the $\nbar$
along their trajectories (in the $\dbar$ rest frame) verify the condition:
\begin{equation}
\Delta r \leq R_\star \;.
\label{eq:MC2}
\end{equation}
$R_\star$ needs to be of the order of the spatial extension of the $\dbar$ wave-function.
For definiteness, we use $R_\star = 2$ fm. Also in this case the model has only one free parameter,
i.e. $p_0$.  $R_\star$ can be made an additional free parameter, but its actual value only mildly
affects the $\dbar$ production results, unless $R_\star$ is made unreasonably large.
\end{enumerate}
Therefore, in both MC approaches we numerically sample $F^{\rm MC}_{(\pbar\nbar)}(\sqrt{s}
, \vec{k}_{\pbar},\vec{k}_{\nbar})$ (in the DM annihilation reference frame, which is where we need the $\dbar$ spectra); we then impose on an event-by-event basis the coalescence
condition(s) described above (in the $\dbar$ rest frame, which corresponds
to the $(\pbar,\nbar)$ pair center-of-mass frame) and we build the $\dbar$ distribution $F^{\rm MC}_{\dbar}(\sqrt{s}, \vec{k}_{\dbar})$. The MC $\dbar$ spectra are then simply:
\begin{equation}
\frac{dN_\dbar}{dT_\dbar} =  (4\pi E_\dbar\,  k_\dbar) \, F^{\rm MC}_{\dbar}(\sqrt{s}, \vec{k}_{\dbar}) 
\label{eq:MC3}
\end{equation}

The three coalescence models discussed above will be first tuned to experimental data (in order to fix the coalescence
momentum $p_0$) and then exploited to predict DM annihilation signals and discuss 
differences among them. We adopt the following standard approach:
the physical production process (in this case: DM annihilation) is assumed to produce in first instance quarks, gauge and higgs bosons (as well as leptons, which however are not relevant for $\pbar$ and $\nbar$ production); the monte carlo event generator PYTHIA 6.4.26 \cite{Pythia} is used to model the hadronization process and the ensuing hadron/meson decays; from phase space of the $\pbar\nbar$ pair produced by PYTHIA, we apply the coalescence models introduced above. In the case of the ``old model'' we just build the energy spectra of $\pbar$ and $\nbar$ and then coalesce them with Eq.~(\ref{eq:oldmodel}). In the case of the two MC models, we use the full phase space information arising from the MC and adopt the techniques which lead to Eq.~(\ref{eq:MC3}).

\subsection{Determining the coalescence momentum $p_0$}
\label{sec:coalescence-p0}

The coalescence momentum $p_0$ (or more generally the coalescence function $C(\vec\Delta)$)
should depend on the specific process
of $\dbar$ production. Hadronic-processes production, like $pp  \rightarrow \dbar$ (as occurs for the $\dbar$ produced by cosmic rays spallation, i.e.
for the background component, for which data from collisions at $\sqrts=53$ GeV measured at the CERN ISR \cite{ISR} are mostly relevant) may involve microphysical features (e.g.~color flows at
the partonic level) not shared by leptonic-processes
production, like $e^+e^- \rightarrow \dbar$ studied with the ALEPH detector at LEP \cite{Aleph}, or semileptonic-processes production, like $e^-p \rightarrow \dbar$ studied by the ZEUS Collaboration at DESY \cite{ZEUS}. Antideuterons have also been measured by ARGUS \cite{ARGUS1,ARGUS2} and CLEO \cite{CLEO} 
in the decay of the $\Upsilon$(1S) and $\Upsilon(2s)$ mesons: the physical details for $\dbar$ production in this
process (resonance decay) should be different from the production in diffusion or annihilation processes: effects related to the bottomonium wave function and/or strong dinamical effects
at the resonance formation/decay are likely to make $\dbar$ formation in this process quite
different from the coalescence situations discussed above. 
Bottomonium decay to $\dbar$ could instead have some impact for DM production when $\mdm$ is 
close enough to the bottom mass to
allow the formation of $b$-meson resonances, with $\dbar$ production from the resonance
and not in the continuum (a very special situation, which we will not investigate here).

Since we are interested in determining the coalescence momentum for $\dbar$ production from DM annihilation, we choose to tune our coalescence models on the process which, under a physical point of view, may be assumed to possess more similarities: i.e. $e^+e^- \rightarrow \dbar$ (absence of colored particles at the elementary level in the initial state). For this
process, a measurement of the $\dbar$ production rate $R_\dbar$ in $e^+e^-$ collisions at the $Z$ resonance has been provided by ALEPH \cite{Aleph}: $R_\dbar = (5.9\pm1.8\pm 0.5)\times10^{-6}$ antideuterons per $Z$ decay, for $\dbar$ with a momentum in the range $(0.62,1.03)$ GeV and a polar angle $\theta$ in the interval $|\cos\theta| < 0.95$. 

%%%
\begin{figure}[t]
\centering
\includegraphics[width=0.45\textwidth]{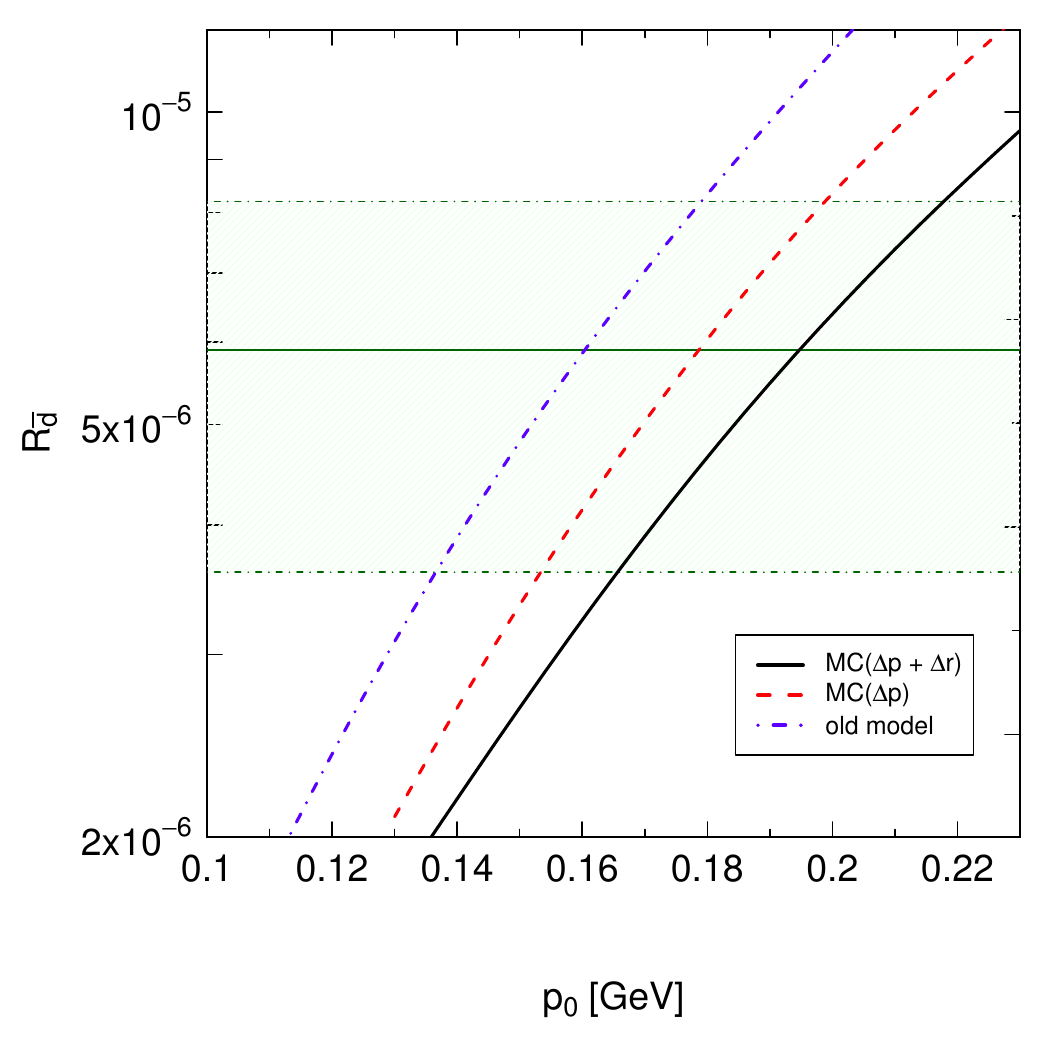} 
\caption{Production rate of antideuterons $\dbar$ in $e^+ e^-$ collisions at the $Z$--boson resonance, as a function of the coalescence momentum $p_0$, for the three coalescence mechanisms considered in the text: ``old model'' \cite{DFS,DFM} with uncorrelated $\bar n \bar p$ production (blue dot-dashed line); Monte Carlo model 
MC($\Delta p$) with the cut-off condition imposed on the relative momentum 
of the $\bar n \bar p$ pair (red dashed line); Monte Carlo model MC($\Delta p + \Delta r$) with the cut-off condition imposed both on the relative momentum and the physical distance of the $\bar n \bar p$ pair (black solid line). The horizontal solid line and the band show the ALEPH measurement \cite{Aleph}.
\label{fig:ALEPH}}
\end{figure}
%%%

We calculate $R_\dbar$ for the three coalescence models discussed above, as a function of the
coalescence momentum $p_0$. This is shown in Fig. \ref{fig:ALEPH}, together with
the ALEPH determination of $R_\dbar$. We notice that the three models predict different values of the coalescence momentum 
to reproduce the same $\dbar$ production rate at the $Z$ resonance. Assuming uncorrelated
production of $\pbar\nbar$ pairs (old model) requires a smaller extent of the coalescence
volume (smaller $p_0$) as compared to the MC models. By comparing
the MC models among themselves, the larger coalescence
volume required by \mcpr, as compared to \mcp, is instead related to
the fact that a fraction of the $\pbar\nbar$ pairs are not close enough in physical space
to merge. This issue will be
discussed in more details in Sec. \ref{sec:spectra} in connection to the DM $\dbar$ production. The best-fit values and $1\sigma$ intervals on $p_0$ are reported in Table \ref{tab:p0}.

Before moving to discuss $\dbar$ production from DM annihilation, we wish to comment that
the determination of the coalescence momentum from the $Z$ resonant decay in
the $e^+e^-$ process may not necessarily be directly applicable to any specific DM
annihilation channel. DM, in general, has branching ratios at the partonic level which are different from those of a $Z$ boson: differences in the $\pbar\nbar$ production from hadronization of heavy quarks as compared to light quarks could imply that the reconstructed $p_0$ from $Z$ decay may not directly apply to the case of a DM annihilating dominantly in
a specific channel (e.g., $b\bar b$ quarks). The flavour-blind $Z$ coupling to quarks somehow averages out  specific physical features occurring in the hadronization process of different quark flavors. We will see in Sec.~\ref{sec:spectra} that differences in the $\pbar\nbar$ production from light quarks and heavy quarks are actually present in the reconstructed MC distributions, both in energy-shape and in multiplicity. Nevertheless, the fact that these differences are not dramatic (although they have impact on DM searches, as we will discuss below) and the mere fact that it does not appear experimentally feasible to discern $\dbar$ production from specific partonic channels, allow us to conclude that the procedure to determine $p_0$ from the $e^+e^-$ collider data is fully justified.

\subsection{Remarks on the coalescence function}
\label{sec:coalescence-function}
%%%
\begin{table}[t]
\centering
\renewcommand{\arraystretch}{1.2}
    \begin{tabular}{|c|l|c|c|c|}
\hline
    $i$ & Coalescence model & $p_0$ (MeV) & $\Delta V_i/V_1$ & $\Delta V_i/V_2$ \\
\hline
    1 & ``old model'' \cite{DFS,DFM} &  $160 \pm 19$ & ---  & ---  \\
    2 & MC($\Delta p$)               &  $180 \pm 18$ & +42\% & ---  \\
    3 & MC($\Delta p + \Delta r$)    &  $195 \pm 22$ & +81\% & +27\% \\
\hline
    \end{tabular}
\caption{Coalescence momentum $p_0$ determined from the ALEPH data \cite{Aleph} for the three coalescence momentum discussed in the text: ``old model'' \cite{DFS,DFM} with uncorrelated $\bar n \bar p$ production; Monte Carlo model 
MC($\Delta p$) with the cut-off condition imposed on the relative momentum 
of the $\bar n \bar p$ pair; Monte Carlo model MC($\Delta p + \Delta r$) with the cut-off condition imposed both on the relative momentum and the physical distance of the $\bar n \bar p$ pair. The last two columns show relative differences of the coalescence volumes.}
\label{tab:p0}
\end{table}
%%%

The coalescence function $C(\vec\Delta)$ (or the coalescence momentum $p_0$) has been assumed 
above to be independent of the center-of-mass energy of the production process. This is approximately
verified in the energy interval (0.1 -- 300) GeV (incident energy on fixed target) for proton-nucleus
interactions (while some energy dependence is present in nucleus-nucleus collisions) \cite{duperray1,duperray2}. In the case of our reference-process for DM annihilation ($e^+e^-$ at colliders) for which only one data point is available from ALEPH (and which refers
to $\sqrts = m_Z$), we clearly do not have any experimental indication on possible energy dependence of the coalescence process. In absence of data which explicitly require energy dependence, we assume
$p_0$ constant in our current analysis. If and when additional data will become available, it will be possible to test whether energy-dependence is present and determine
the energy evolution of the coalescence momentum for each model introduced above, by adopting the same procedures of Sec.~\ref{sec:coalescence-p0} at different production energies.
Energy dependence on $p_0$ would impact on both the size and the spectral features of the predicted
DM signals.

The coalescence function $C(\vec\Delta)$ may be assumed to have a less sharp behavior,
in terms of the relative $\pbar\nbar$ momentum and physical distance, than the 
$\theta$-function shape we used in our modeling. A more physical assumption may be to adopt
information coming from the physics implied by the formation of the bound state, both
in momentum space and physical space. These refinements on the coalescence function
will not, however, introduce relevant changes in the results, given the resolution of the MC modeling and the current experimental data that can
be used to tune the models.

We finally wish to recall that a full understanding of the $\dbar$ production process is currently
not available, and different models have been discussed in the literature, like models based on
the fireball mechanism, which attempts to describe the $\dbar$ formation on the
basis of thermodynamical arguments \cite{Wes76,Bon78,Sie79,Scheibl:1998tk,Ioffe:2004rb}, or the diagrammatic model \cite{duperray1,duperray2,diagrammatic} which aims at understanding the microscopic fundamental details of the production process. 
These models have been developed mostly for $\dbar$ production in hadronic processes.

\subsection{Antideuteron injection spectra and comparison of the coalescence models}
\label{sec:spectra}

%%%
\begin{figure}[t]
\centering
\includegraphics[width=0.34\textwidth]{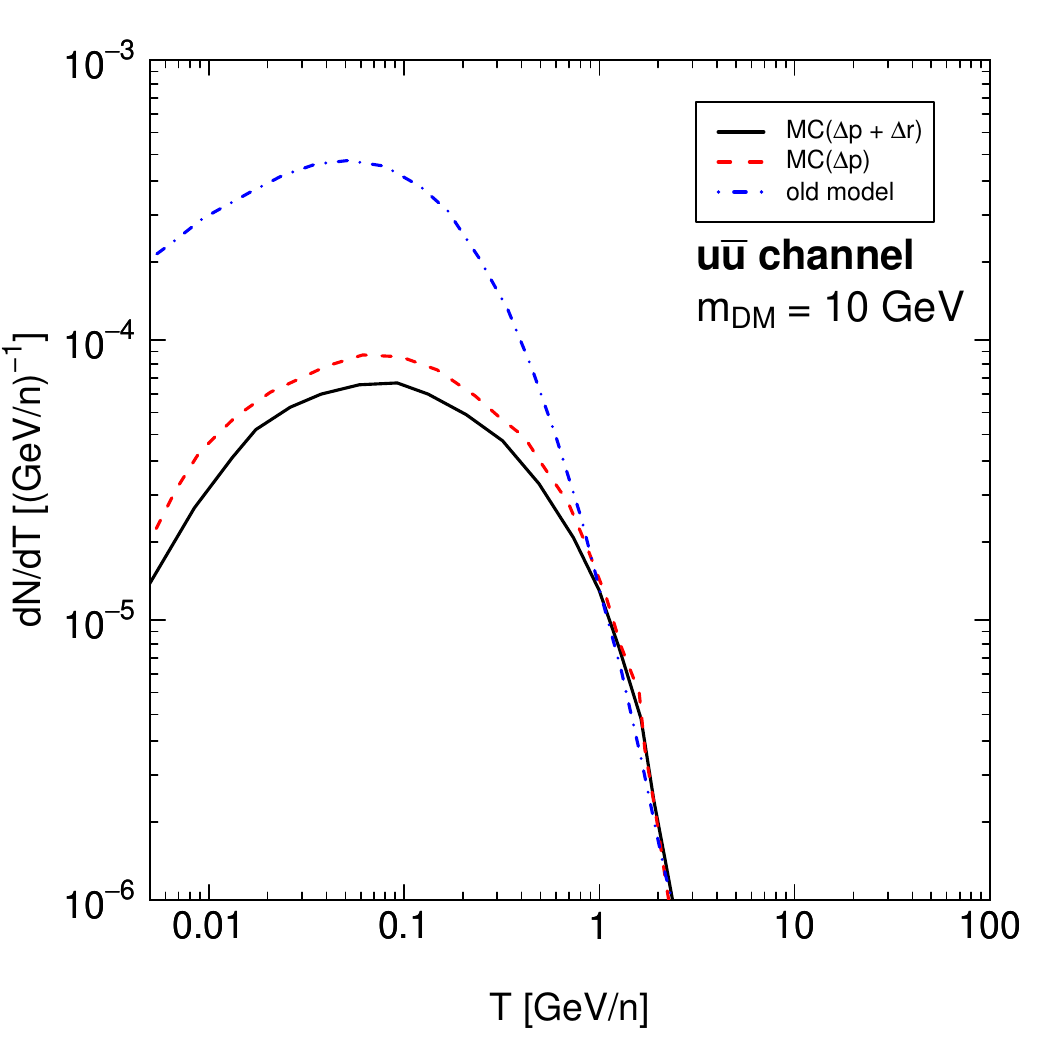}
\includegraphics[width=0.34\textwidth]{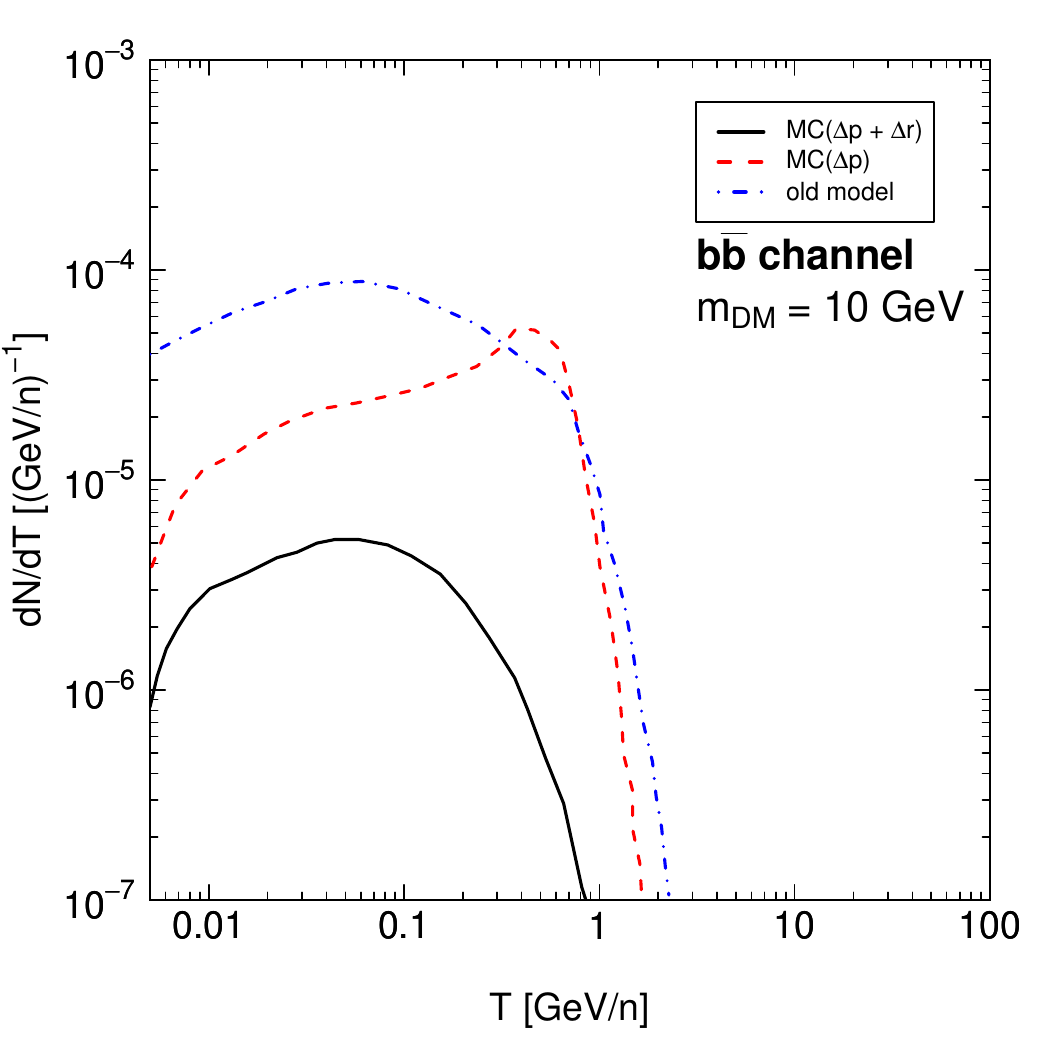}
\includegraphics[width=0.34\textwidth]{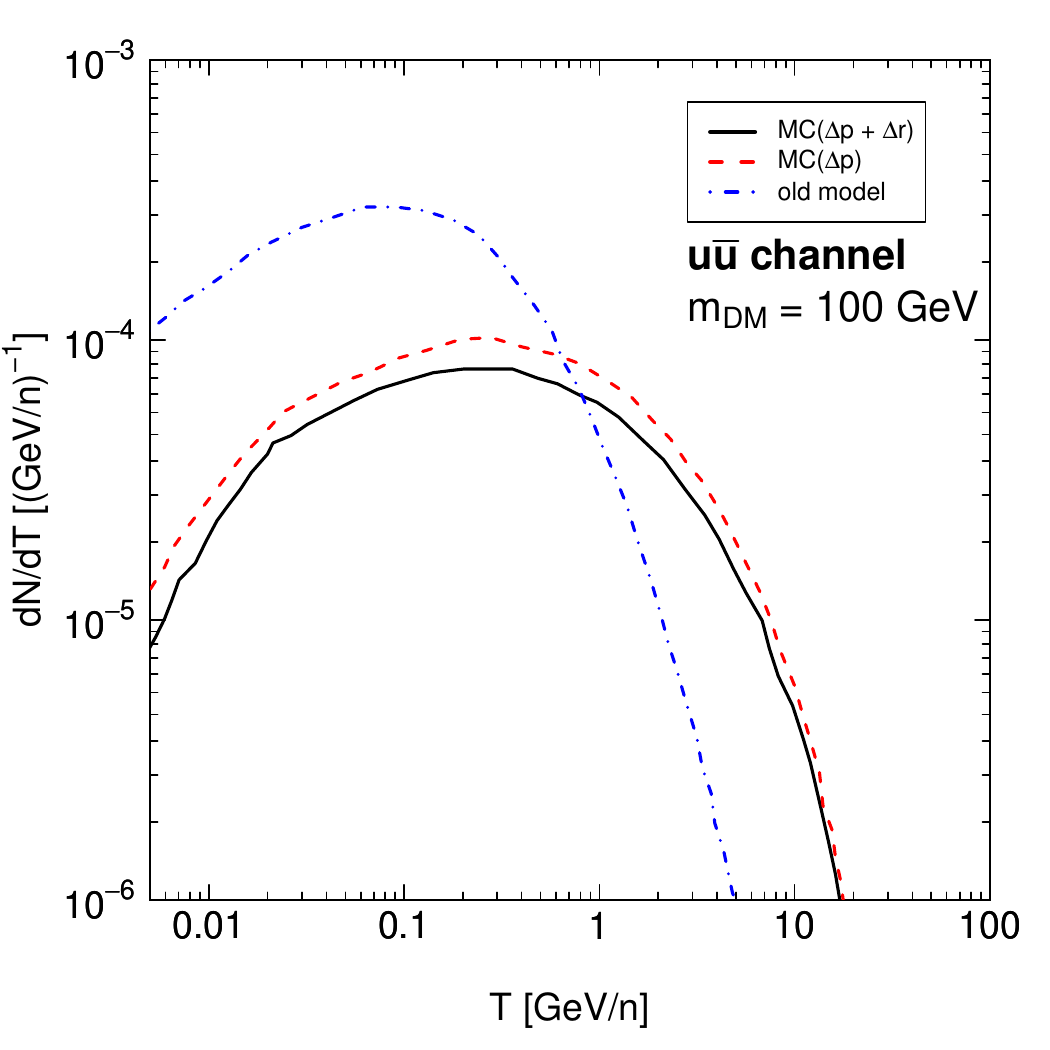}
\includegraphics[width=0.34\textwidth]{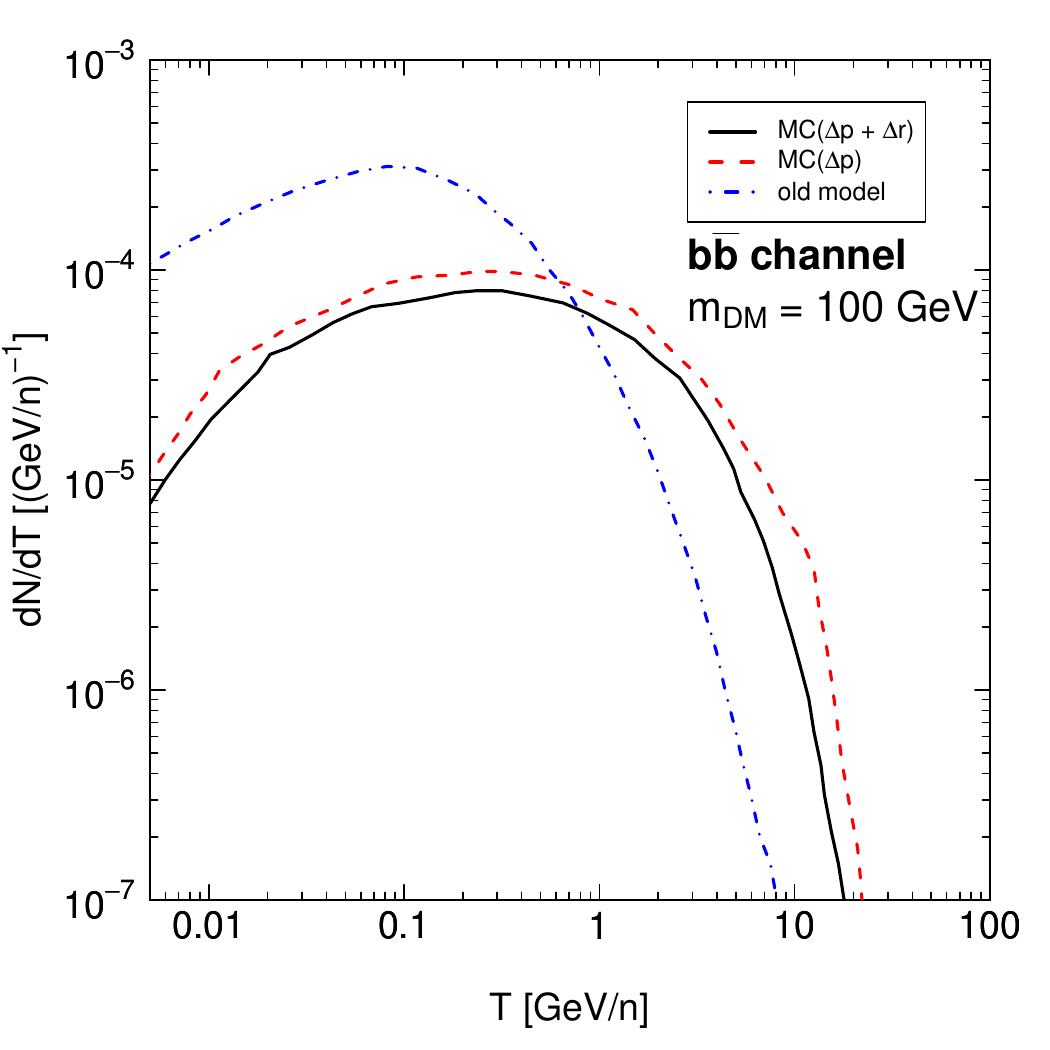}
\includegraphics[width=0.34\textwidth]{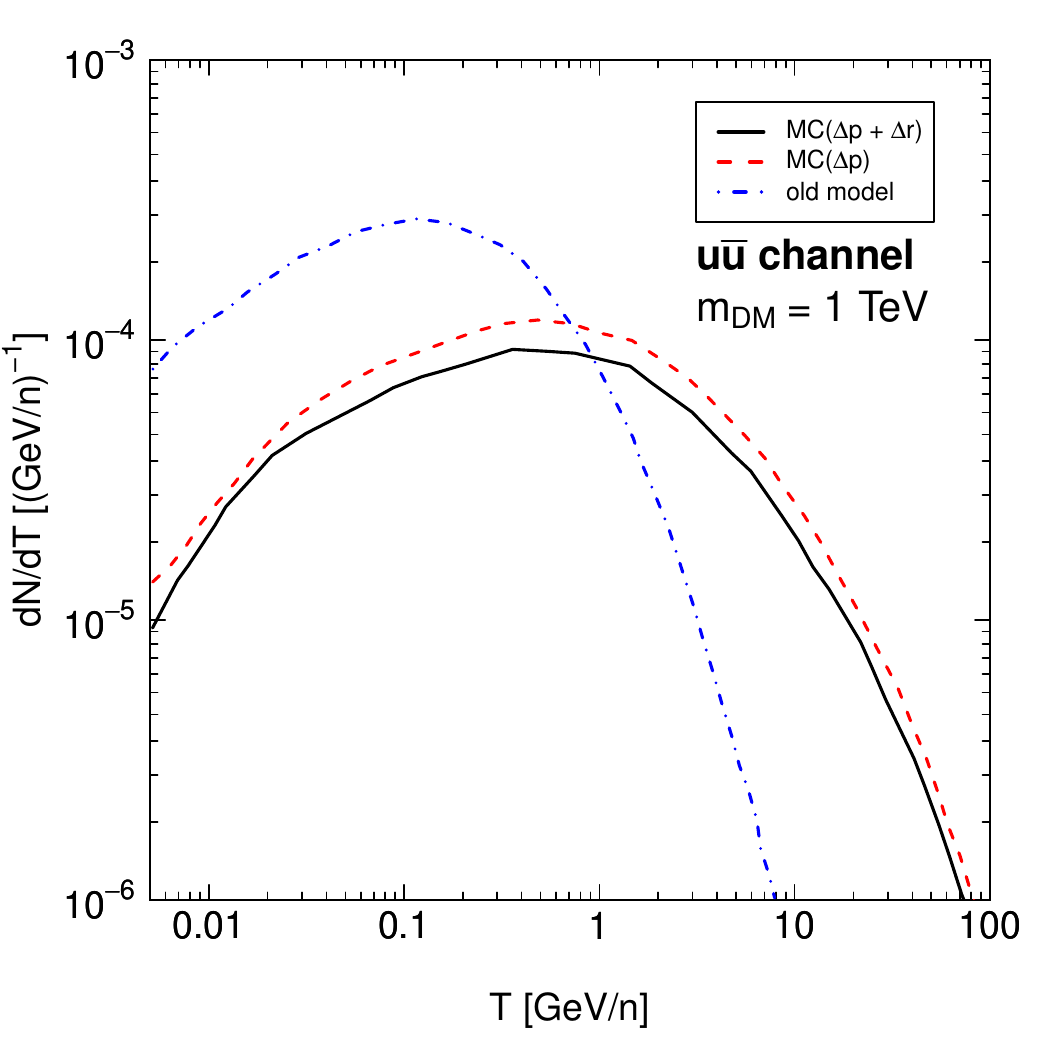}
\includegraphics[width=0.34\textwidth]{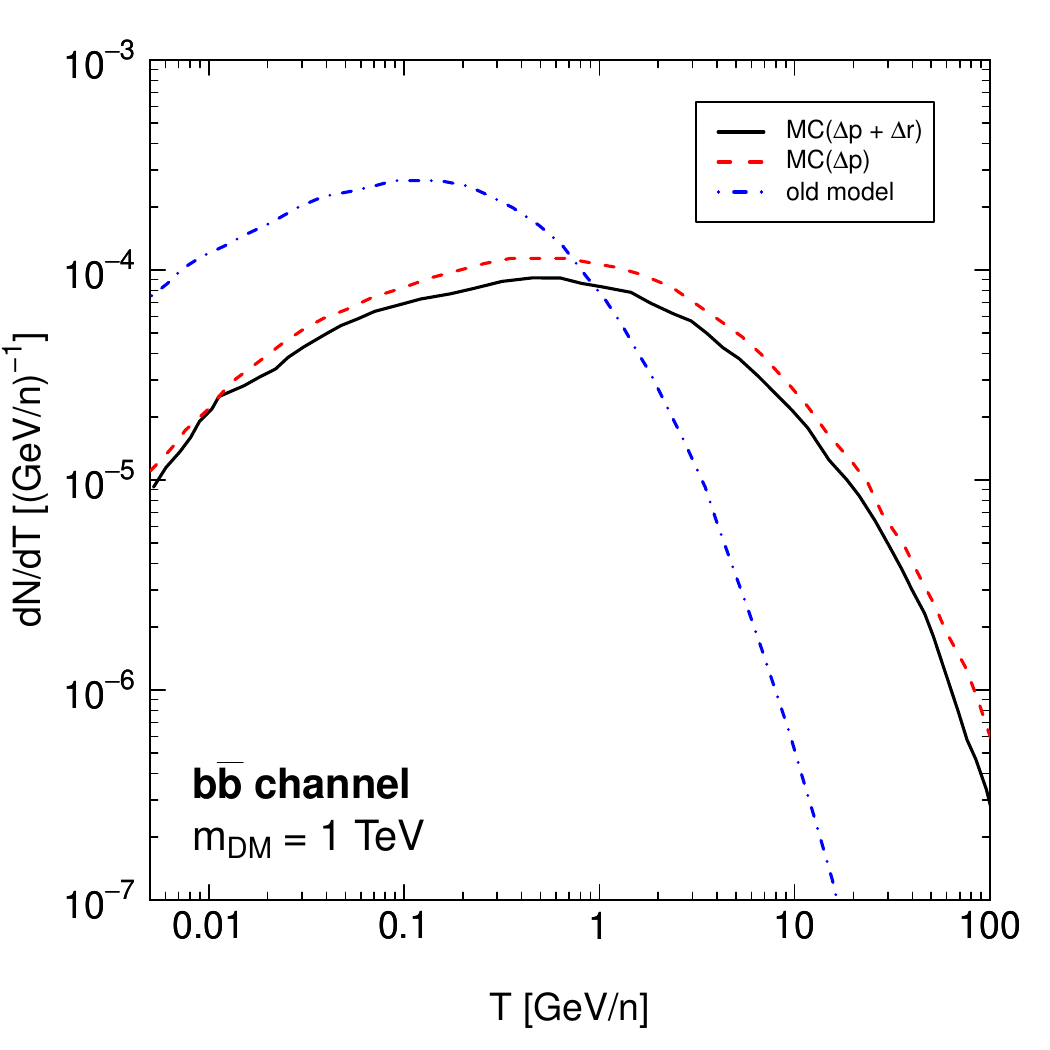}
\caption{$\bar{d}$ injection spectra for the $u\bar{u}$ (left column) and $b\bar{b}$ 
(right column) annihilation channels and for  $m_{DM}=10$ GeV (upper row),  $m_{DM}=100$ GeV (central row) and $m_{DM}=1$ TeV (lower row).
The different lines refer to the three different coalescence models discussed in the text:
old model with uncorrelated $\bar n \bar p$ production (blue dot-dashed line); MC model 
(MC($\Delta p$)) with the cut-off condition imposed on the relative momentum 
of the $\bar n \bar p$ pair (red dashed line); MC model (MC($\Delta p + \Delta r$)) with the cut-off condition imposed both on the relative momentum and on the physical distance of the $\bar n \bar p$ pair (black solid line).
\label{fig:spectra1}
}
\end{figure}
%%%

%%%
\begin{figure}[t]
\centering
\includegraphics[width=0.35\textwidth]{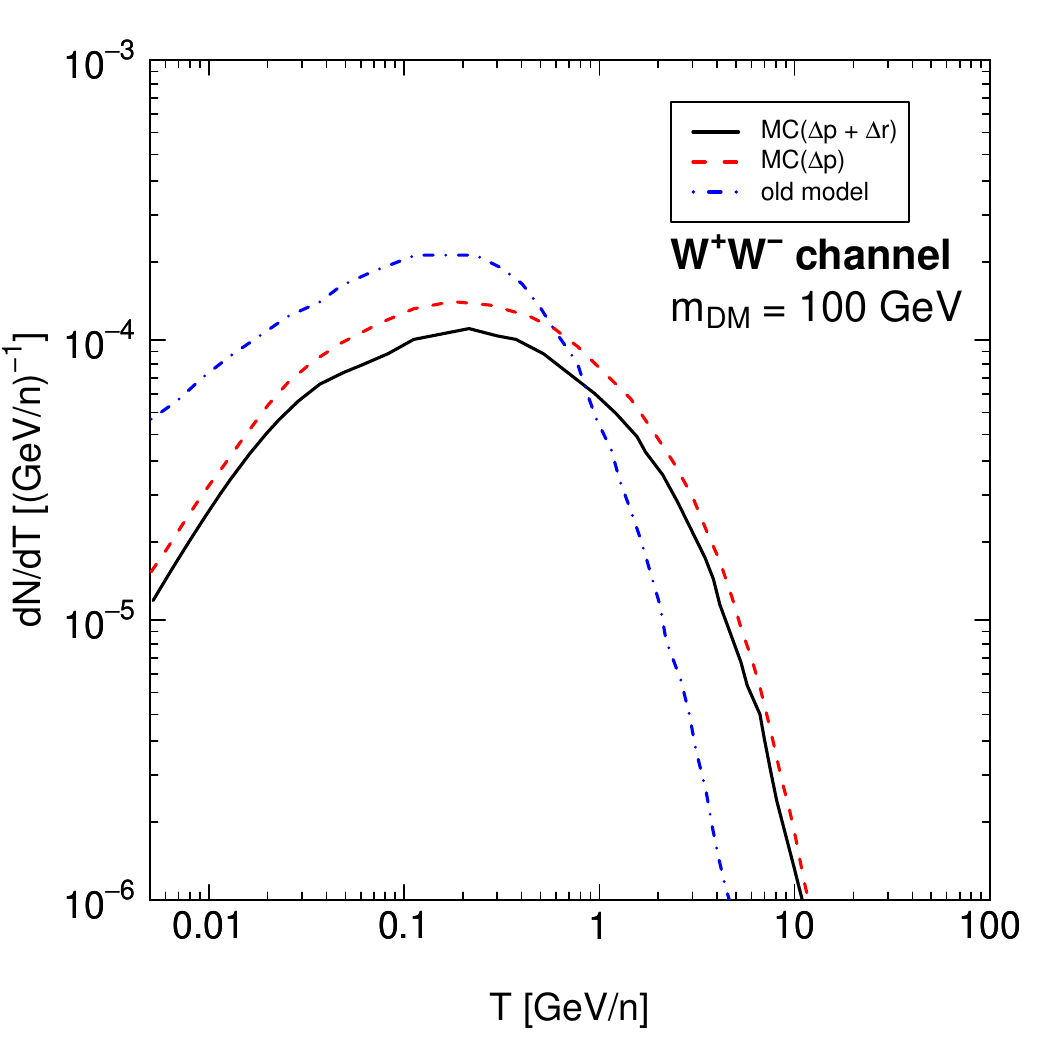}
\includegraphics[width=0.35\textwidth]{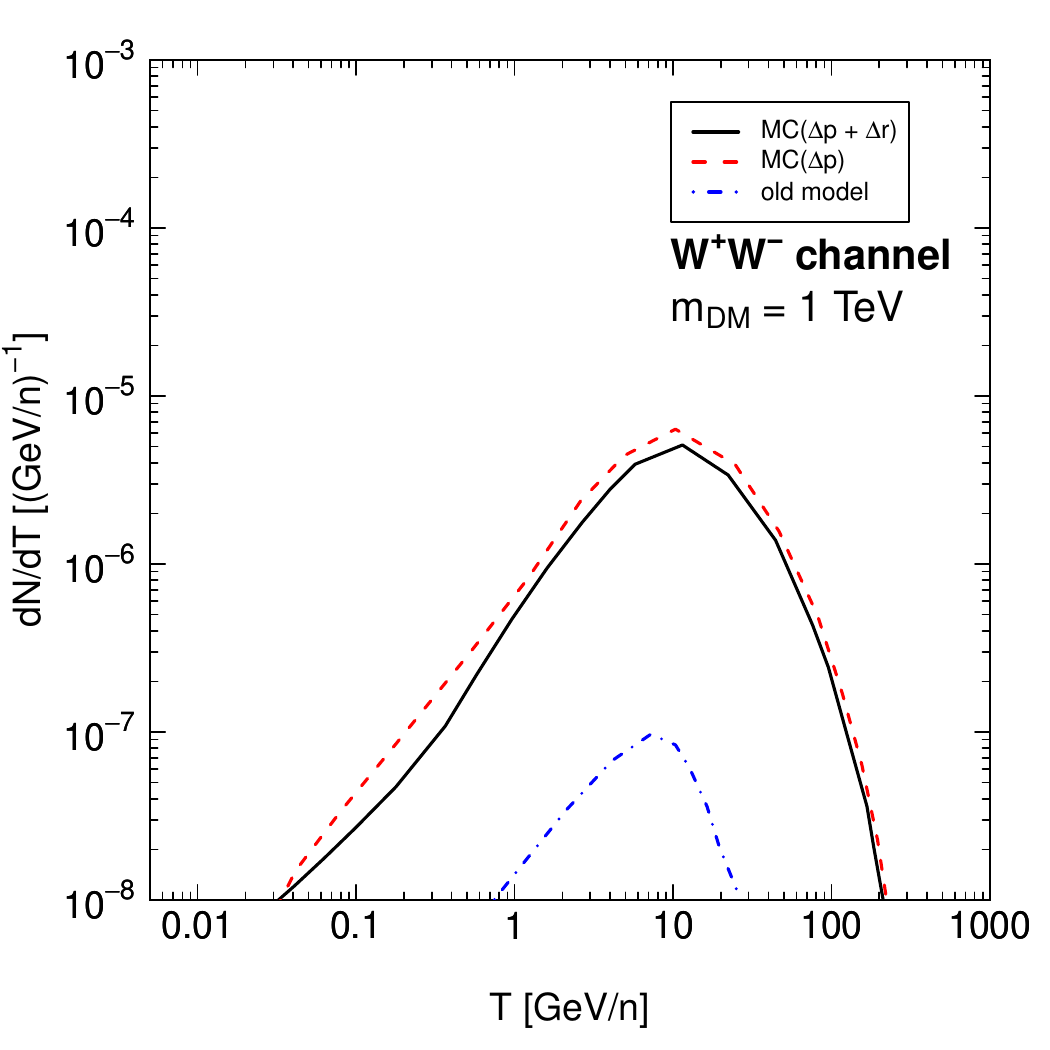}
\caption{$\bar{d}$ injection spectra for the $W^+W^-$ annihilation channel for $m_{DM}=100$
GeV (left) and $m_{DM}=1$ TeV (right). The different lines refer to the three different coalescence models discussed in the text:
old model with uncorrelated $\bar n \bar p$ production (blue dot-dashed line); MC model 
(MC($\Delta p$)) with the cut-off condition imposed on the relative momentum 
of the $\bar n \bar p$ pair (red dashed line); MC model (MC($\Delta p + \Delta r$)) with the cut-off condition imposed both on the relative momentum and the physical distance of the $\bar n \bar p$ pair (black solid line).
\label{fig:spectra2}
}
\end{figure}
%%%

We now move to the discussion of the $\dbar$ injection spectra from DM annihilation, obtained with the three coalescence models discussed in the previous Section. For definiteness, we
will concentrate our analysis on three representative annihilation channels: $\uubar$
(representative for light-quark production); $\bbbar$
(representative for heavy-quark production); $\ww$ (representative for gauge-boson production).

Figure \ref{fig:spectra1} shows the $\dbar$ energy spectra for the $\uubar$ and $\bbbar$
production channels, for three values of the DM mass, $\mdm=10,100,1000$ GeV (i.e. for three values of $\sqrts=2\mdm$ of the annihilation process, since annihilation occurs at rest) and for the three coalescence models under analysis. Figure \ref{fig:spectra2} shows the spectra for the $\ww$ channel and for $\mdm=100,1000$ GeV. We notice, as also reported in Refs.~\cite{strumia, ibarra}, that the spectra obtained with the old model and the MC models are significantly
different. 

In the case of annihilation into quarks, the model of uncorrelated $\pbar\nbar$ production tends to produce more $\dbar$ when the center-of-mass energy (in turn, the DM mass) is relatively small (first row in Fig.~\ref{fig:spectra1}), while when $\sqrts$ increases (second and
third row) the old model still produces more $\dbar$ at low kinetic energies $T$, but is strongly suppressed
at large kinetic energies, as compared to the MC model. This is due to the presence of
correlations in the $\pbar\nbar$ phase space, which become progressively more relevant
when the two quarks are produced at larger energies: in this case, the two back-to-back
emerging jets are more focused and the ensuing $\pbar\nbar$ possess, on average, closer
relative momenta; this becomes even more pronounced when $\pbar\nbar$ are more energetic, and therefore more $\dbar$ at large energies are produced.

%%%
\begin{figure}[t]
\centering
\includegraphics[width=0.30\textwidth]{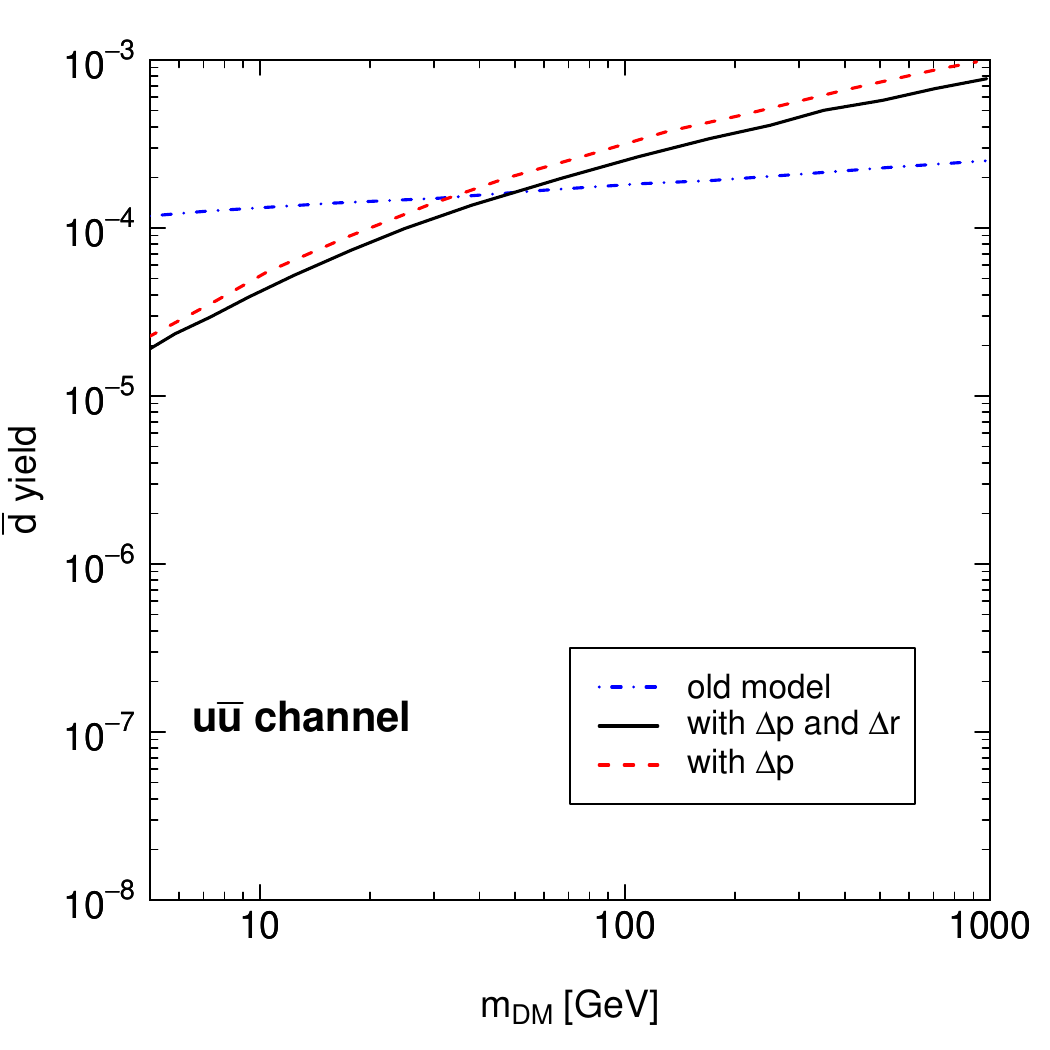}
\includegraphics[width=0.30\textwidth]{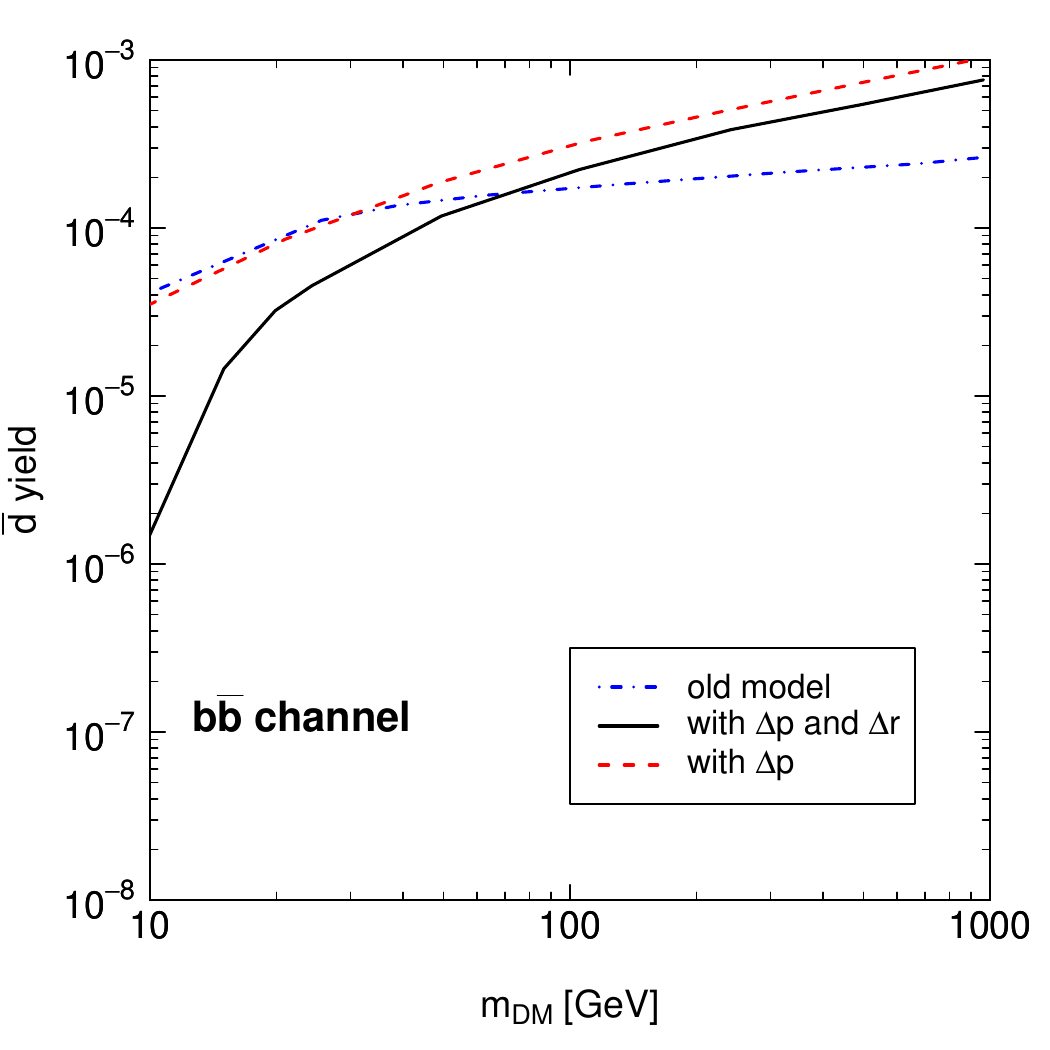}
\includegraphics[width=0.30\textwidth]{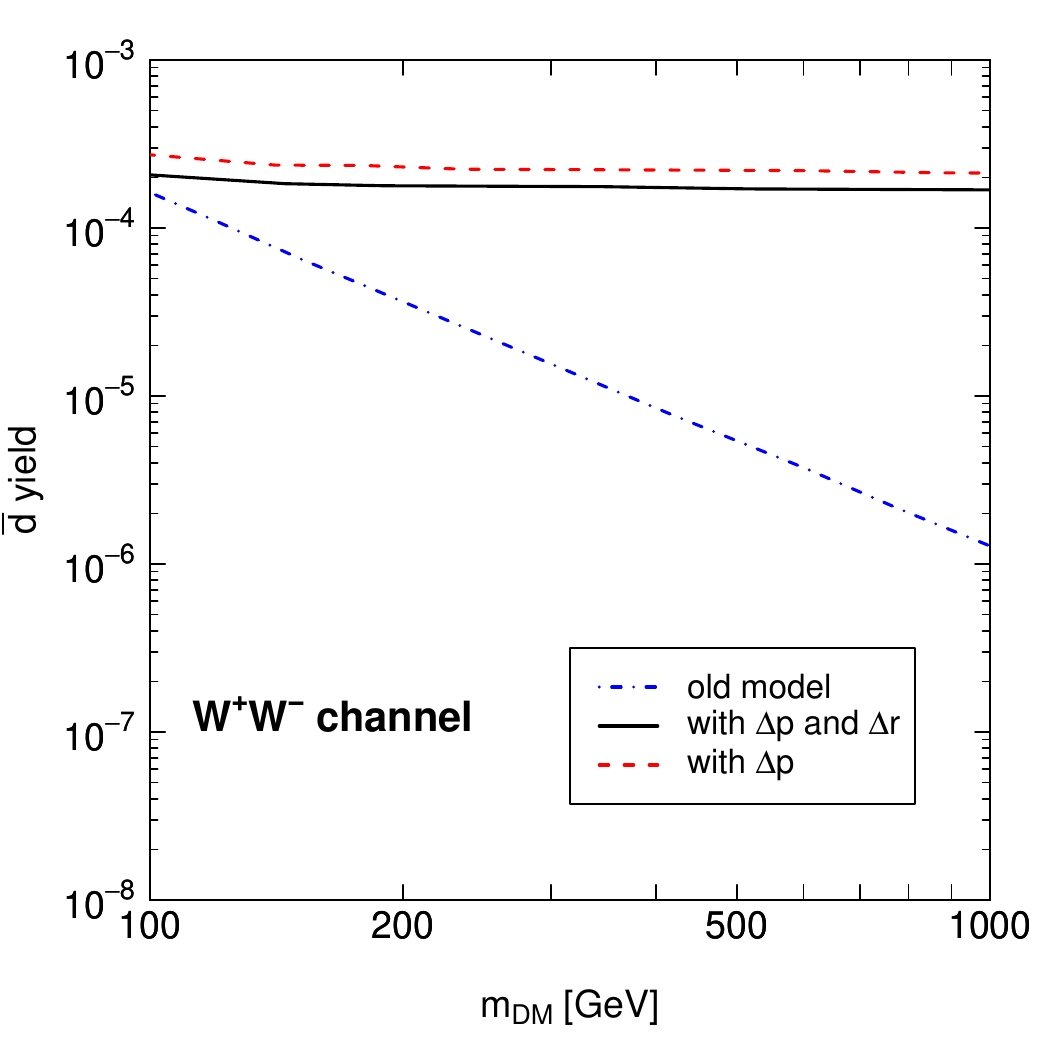}
\caption{Total $\bar{d}$ yield for single annihilation event, for the $u\bar{u}$ (left panel),
$b\bar{b}$ (central panel) and $W^+W^-$ (right panel) annihilation channels under the three different coalescence prescriptions: old model with uncorrelated $\bar n \bar p$ production (blue dot-dashed line); MC model 
(MC($\Delta p$)) with the cut-off condition imposed on the relative momentum 
of the $\bar n \bar p$ pair (red dashed line); MC model (MC($\Delta p + \Delta r$)) with the cut-off condition imposed both on the relative momentum and the physical distance of the $\bar n \bar p$ pair (black solid line).}
\label{fig:yield}
\end{figure}
%%%

The increase in $\dbar$ production for the MC models can be seen also in the $\dbar$ multiplicity distribution as a function
of the DM mass, reported in Fig.~\ref{fig:yield}. The left and central panels show that  for $\uubar$
and $\bbbar$ channels, the number of $\dbar$ produced in an annihilation event increases
much faster when the MC coalescence models are used. For the MC models, the number of
produced $\dbar$ by DM with a mass above 50-70 GeV is larger than in the case of the
uncorrelated old model. The difference can reach more than one order of magnitude for 
$\mdm > 1$ TeV. This implies that correlations in the $\pbar\nbar$ production are
expected to be present, especially for heavy DM \cite{strumia}, and therefore they have
to be considered in the determination of the DM $\dbar$ signal. 
For intermediate DM masses, the difference in the total number of $\dbar$ produced
is less pronounced, but still at the level of a factor of 1.5--2 for $\mdm \sim 100$ GeV.
Figure \ref{fig:yield} also shows that for low DM mass, a reduction in the $\dbar$ production
rate is present when (anti)correlations are taken into account. This is related to the fact,
manifest in Fig.~\ref{fig:spectra1}, that (anti)correlations deplete the number of low-kinetic energies 
$\dbar$ (roughly below 1 GeV for kinetic energy per nucleon): when the DM mass is small, 
only low-energy $\dbar$ can be produced, and therefore the total multiplicity is reduced.
For larger DM masses, highly boosted $\pbar\nbar$ are produced, with an enhanced
capability of merging (as discussed above), and larger $\dbar$ fluxes and multiplicities
occur.

When comparing the $\dbar$ production in the two MC models, we notice that for
the light-quarks case the effect of including also the distance in physical space
to determine coalescence has a reduction effect: spectra obtained in the \mcpr~model
are about 30--40\% smaller than those derived in the \mcp~model, with the reduction
factor almost independent of the kinetic energy and of the production energy (or DM mass). 
This is appreciable both in the spectra of Fig.~\ref{fig:spectra1} and in the multiplicities
shown in Fig.~\ref{fig:yield}. A quite different situation instead occurs for the $\bbbar$
channel: while for DM masses larger than about 30-40 GeV the \mcpr~model produces
a reduction of about 40--50\% (slightly larger than for the $\uubar$ case) in the
$\dbar$ spectrum, with a small dependence both on the $\dbar$ kinetic energy and
on the DM mass, for light DM the requirement imposed on the $\pbar\nbar$ distance
in physical space added in the \mcpr~model strongly suppresses $\dbar$ production.
This occurs for $\dbar$ kinetic energies close to their maximal values, and the impact
is a strong reduction (by almost an order of magnitude at $\mdm \sim 10$~GeV)
in total multiplicity.

The reason for this behavior resides in the fact that in the case of a $\bbbar$ production
heavy (anti)baryons are produced, which then subsequently decay down to $\pbar$
and $\nbar$: even though the $\pbar$ and $\nbar$ can possess small enough relative
momenta, nevertheless, due to delayed and uncorrelated production through these decays, 
the probability to find a $\pbar\nbar$ pair close enough in physical space to be able
to merge is very suppressed. This effect is more pronounced for small DM masses:
with a smaller production energy of the $\bbbar$ pair, a larger fraction of
heavy baryons with respect to direct production of nucleons occurs. 

%%%
\begin{figure}[t]
\centering
\includegraphics[width=0.35\textwidth]{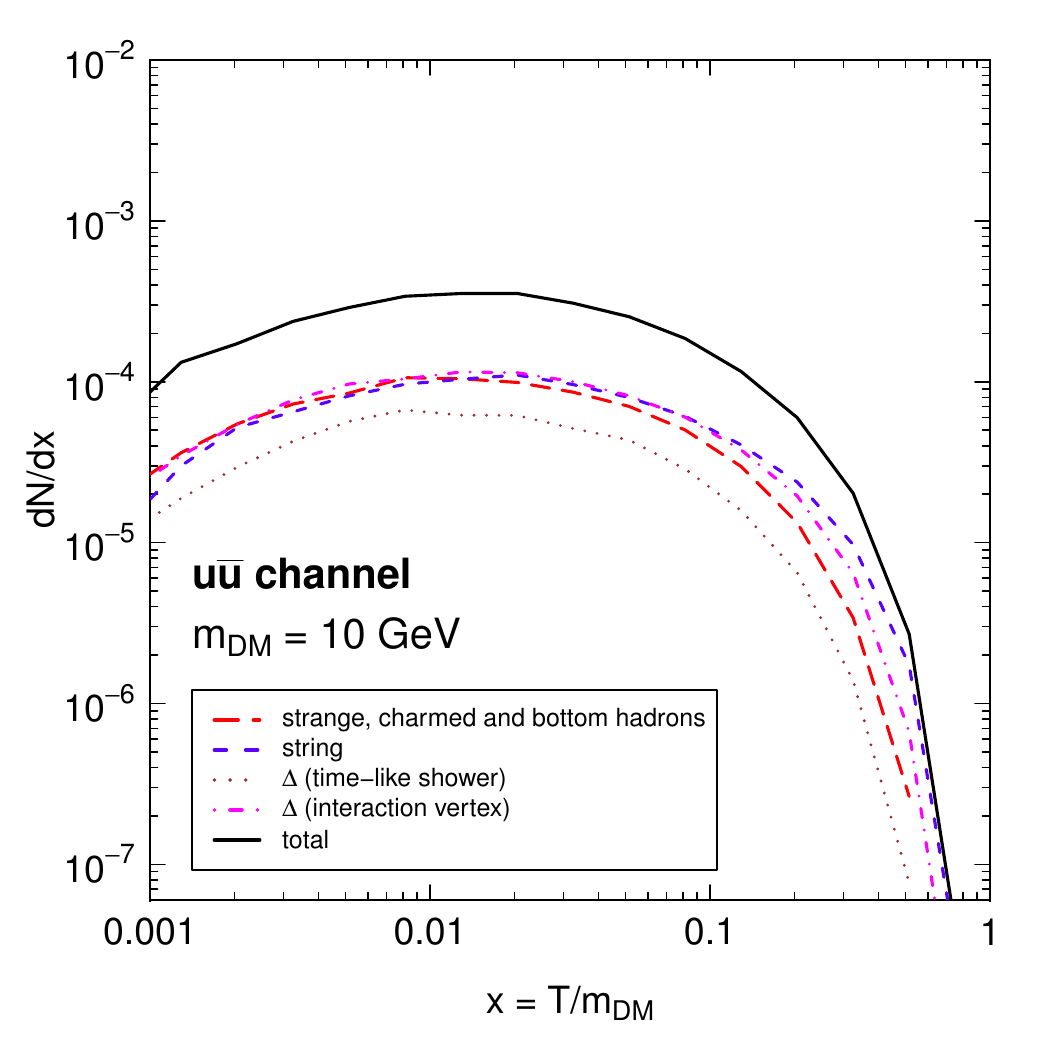}
\includegraphics[width=0.35\textwidth]{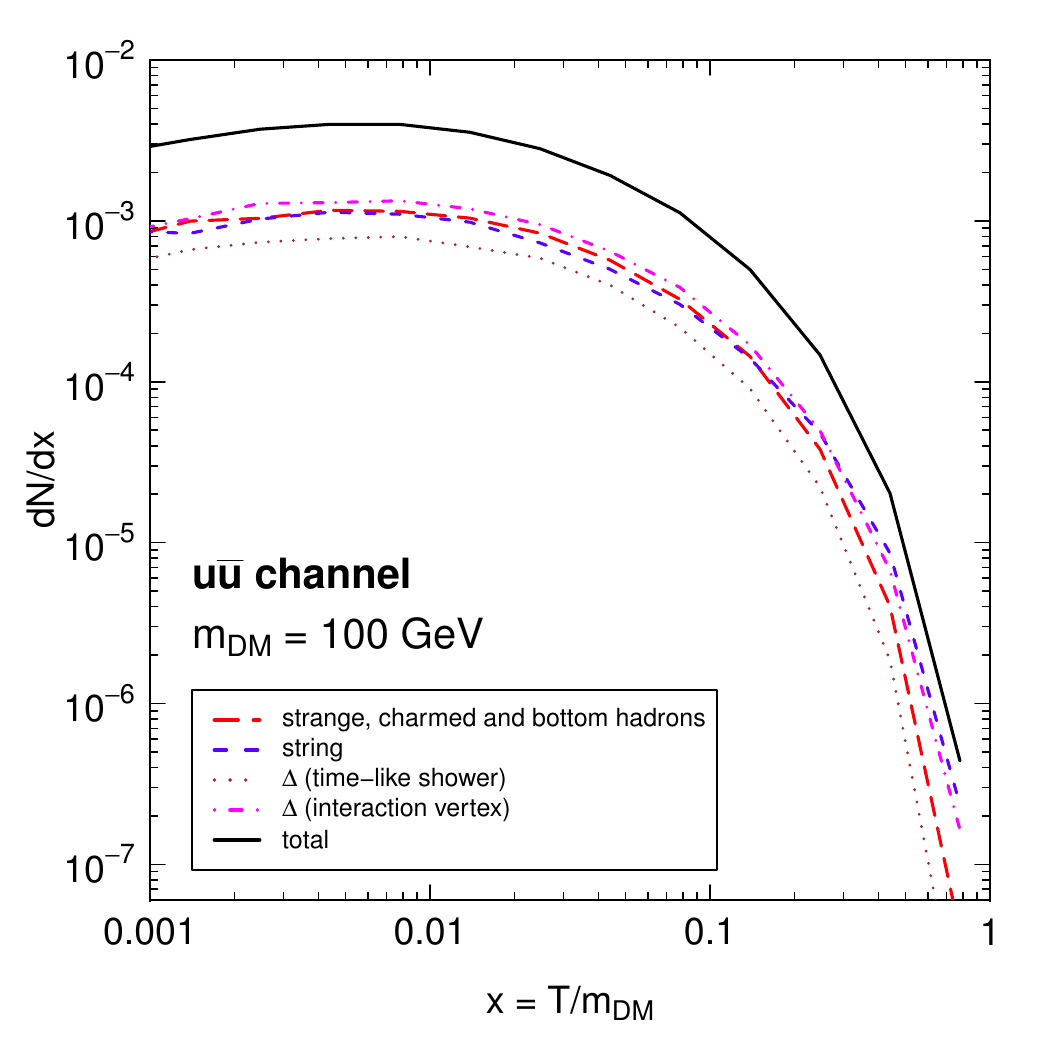}
\includegraphics[width=0.35\textwidth]{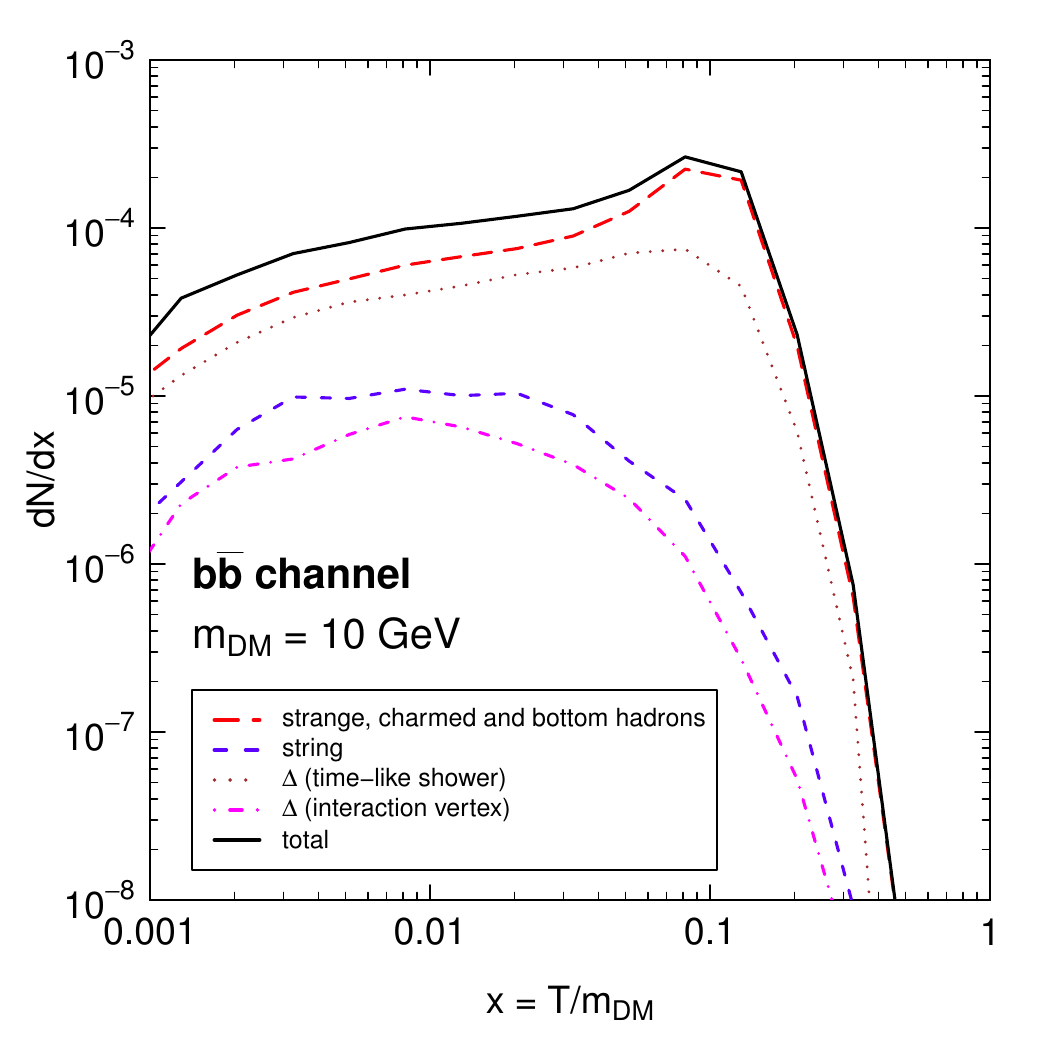}
\includegraphics[width=0.35\textwidth]{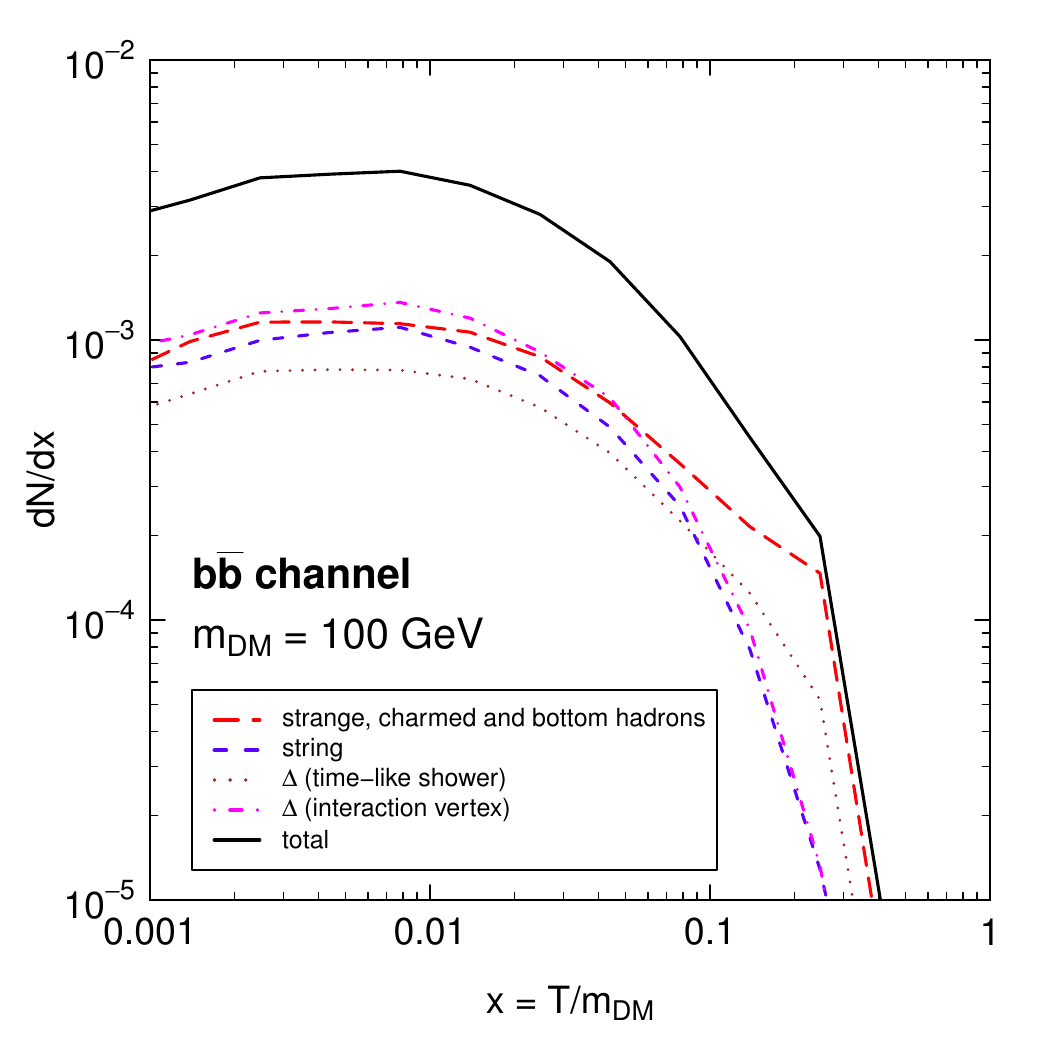}
\caption{Contributions to the $\bar{d}$ injection spectra in the \mcp~ model. The different lines refer to 
different origins of the produced $\bar{d}$: from strange, charm and bottom hadrons (long dashed);
from string hadronization (short dashed); from $\Delta$ resonance decay (dotted and dot-dashed).
Specifically, dotted lines refer to $\Delta$ resonances that
are generated in the time-like shower far from the dark matter annihilation vertex
(as the result of the decay of more massive hadrons), while dot-dashed
lines are associated to $\Delta$ resonances formed right at the
annihilation vertex.
The upper solid lines shows the total $\bar{d}$ specra.
\label{fig:spectra3}}
\end{figure}
%%%

%%%
\begin{figure}[t]
\centering
\includegraphics[width=0.33\textwidth]{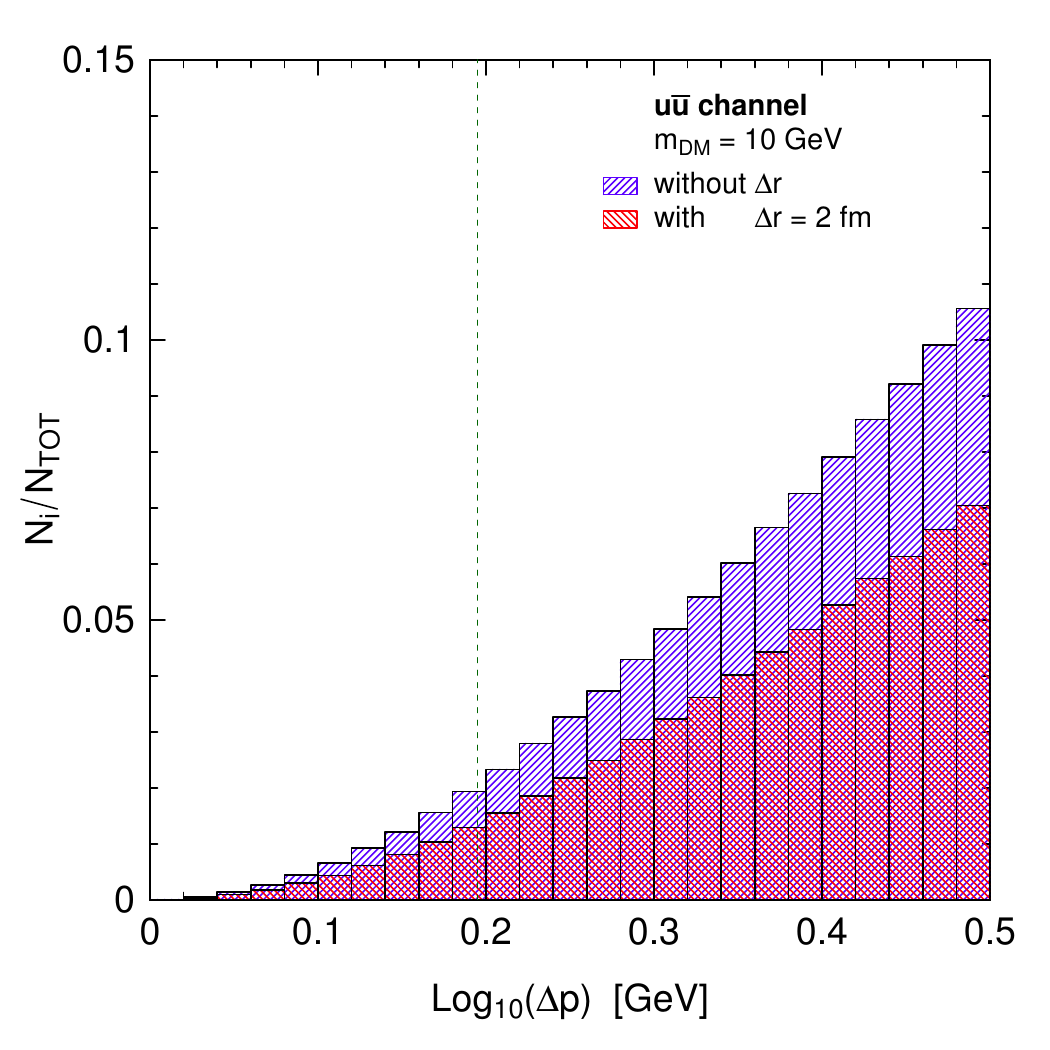}
\includegraphics[width=0.33\textwidth]{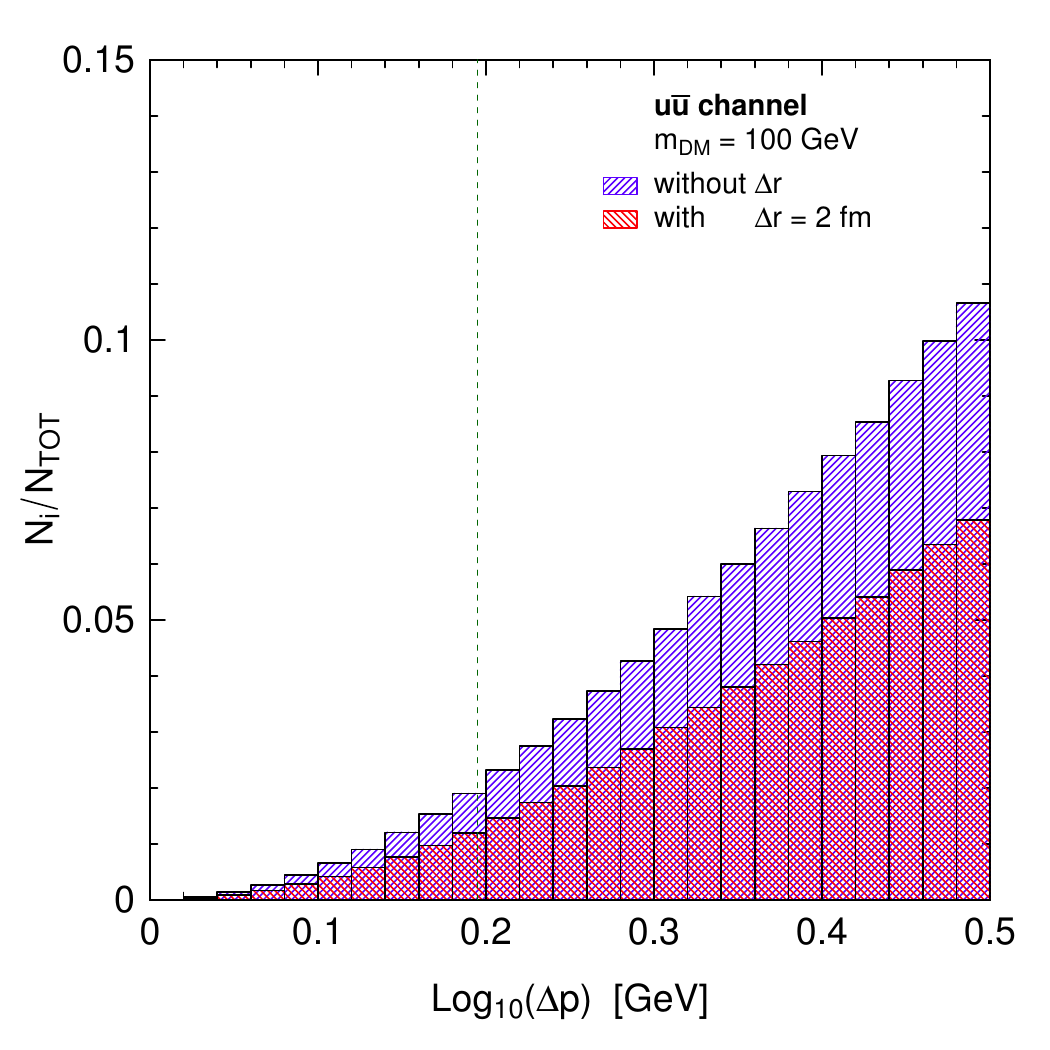}
\includegraphics[width=0.33\textwidth]{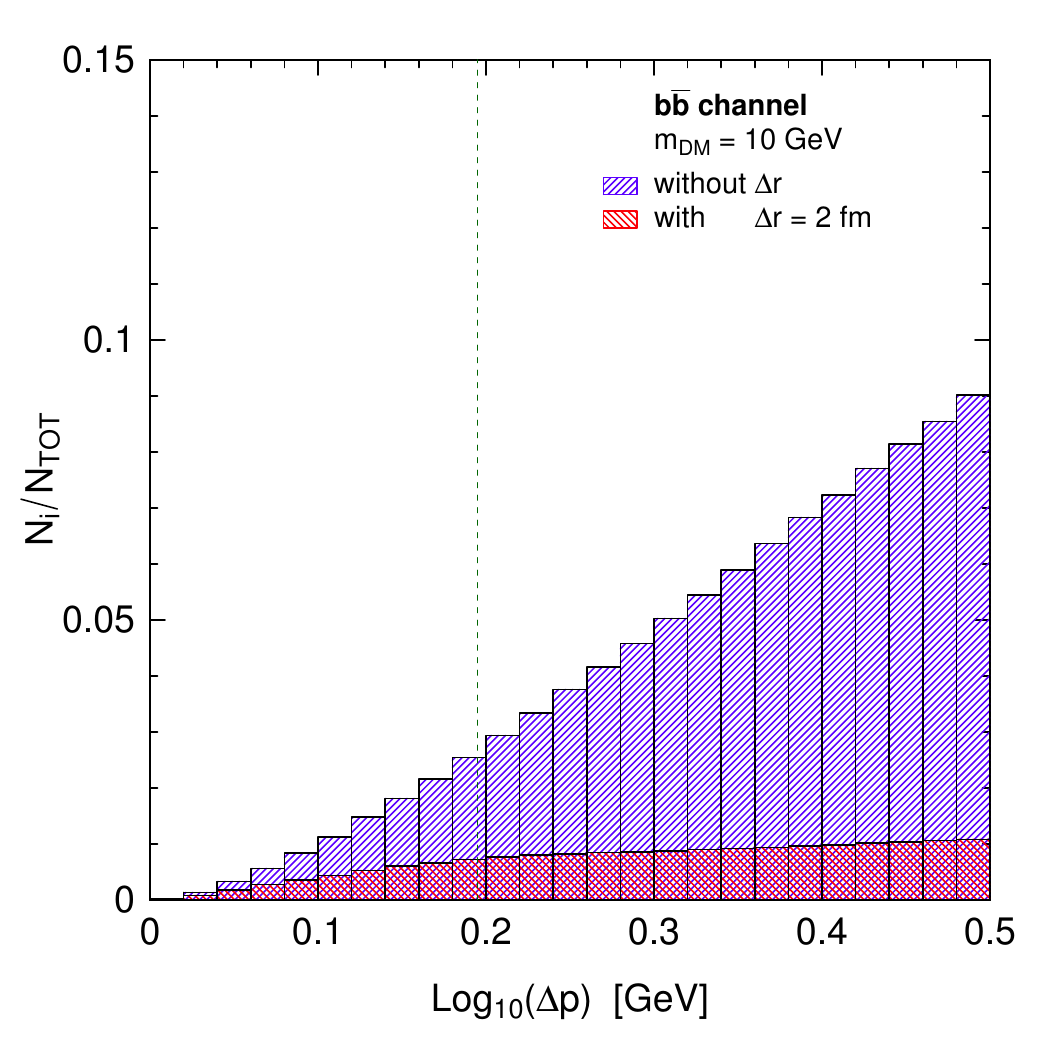}
\includegraphics[width=0.33\textwidth]{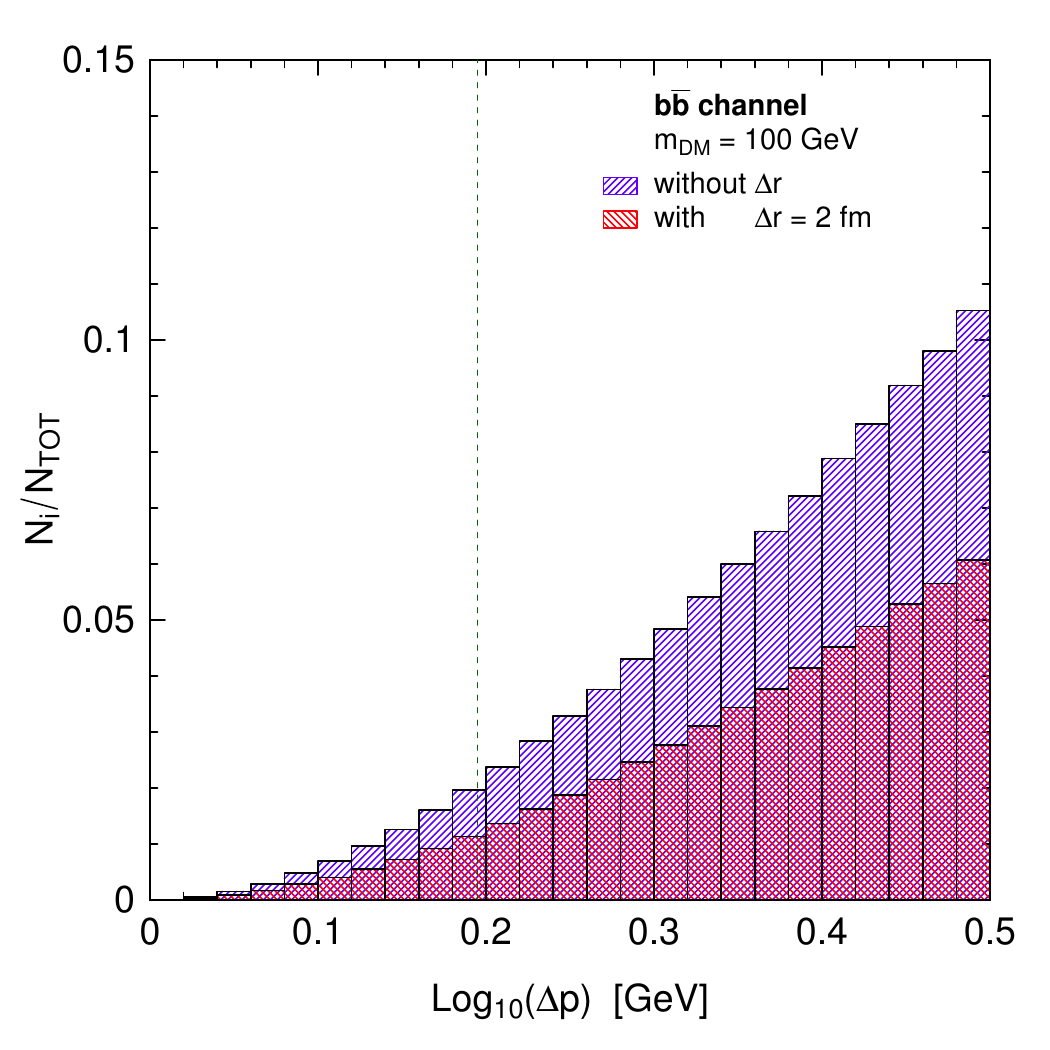}
\caption{Fractional number $N_i/N_{\rm TOT}$ of $\bar n \bar p$ pairs as a function of the relative momentum $\Delta p$ of the pair, occurring in dark matter annihilations into $u\bar u$ (upper row) and $b\bar b$ (lower row) final states. The left column stands for $m_{DM}=10$ GeV, the right column for $m_{DM}=100$ GeV.  $N_i$ refers for the number of $\bar n \bar p$ pairs in the corresponding
$i$-th bin in momentum, while $N_{\rm TOT}$ is the total number of $\bar n \bar p$ pairs for the
annihilation process. The (blue) 
higher histograms refer to the total fractional number of pairs for each case, while the (red) lower
histograms refer to the number of $\bar n \bar p$ pairs that, for each interval of $\Delta p$, also
possess a relative distance in physical space smaller than the $\dbar$ radius
(i.e. $\Delta r < 2$ fm), therefore showing the reduction in the coalescence process imposed
by this additional requirement. The vertical dotted line reports the value of the coalescence
momentum $p_0 = 195$ MeV: for $\Delta p < p_0$, the $\bar n$ and $\bar p$ of the pair
coalesce in the MC($\Delta p + \Delta r$) model, adopted in our analysis.
\label{fig:fractio1}
}
\end{figure} 
%%%

To illustrate
these points we show in Fig.~\ref{fig:spectra3} the contribution to the $\dbar$ spectra
arising from different sources of production of the $\pbar$ and $\nbar$. The upper row
refers to the $\uubar$ channel, while the lower row to the $\bbbar$ channel. Two
cases of production energies are plotted: $\mdm=10$~GeV and $\mdm=100$~GeV.
The upper solid lines show the total $\dbar$ spectra, while the other lines detail the
production channel of the $\pbar$ and $\nbar$: direct production from the hadronic ``string''
(dashed lines); production from $\Delta$ resonances generated in the time-like shower far from the dark matter annihilation vertex, as the result of the decay of more massive hadrons (dotted lines); 
production from $\Delta$ resonances formed right at the annihilation vertex (dot-dashed lines);
production from the decay of strange, charm and bottom hadrons (long-dashed lines).
Notice that Fig.~\ref{fig:spectra3} refers to the \mcp~model, i.e. only the condition $|\vec\Delta| < p_0$
is applied, and not the condition $\Delta r < 2$ fm. We can notice that in the case of $\uubar$
production, nucleons coming directly from the production vertex or from the string 
and nucleons produced in baryon decays are approximately the same amount, both
at small and large $\sqrts$. The situation is quite different for $\dbar$ production in the
$\bbbar$ channel, especially at low $\sqrts$: for this channel, when $\mdm=10$~GeV,
the dominant contribution to $\dbar$ in the \mcp~model arises from the decay of heavy
baryons, which are more easily produced in this case, since  the original parton 
is a heavy quark (a $b$ quark). The decay of heavy baryons produces a large
excess of $\pbar$ and $\nbar$ which then coalesce to form a $\dbar$. This process is
largely dominant for $\dbar$ with kinetic energy larger than about 0.1 times the 
total available energy (i.e.~the DM mass). Nucleons arising from the string are more likely 
to be produced close in physical space, and therefore when the coalescence condition  
$\Delta r < 2$ fm is further imposed, they are able to merge. On the contrary, nucleons
produced  by heavy baryons decay (or by $\Delta$ resonances produced themselves by
heavy baryon decay) are typically further apart, and cannot coalesce. This fact explains
why, in Fig.~\ref{fig:spectra1}, for the $\bbbar$ channel the spectra in the \mcpr~model
are significantly suppressed as compared to the \mcp~model, especially for light DM.
By comparing Fig.~\ref{fig:spectra1} and Fig.~\ref{fig:spectra3} we can notice that
approximately the total spectra in the \mcpr~model are originating from the 
partial spectra in Fig.~\ref{fig:spectra3} referring to the vertex and string nucleon production.

%%%
\begin{figure}[t]
\centering
\includegraphics[width=0.35\textwidth]{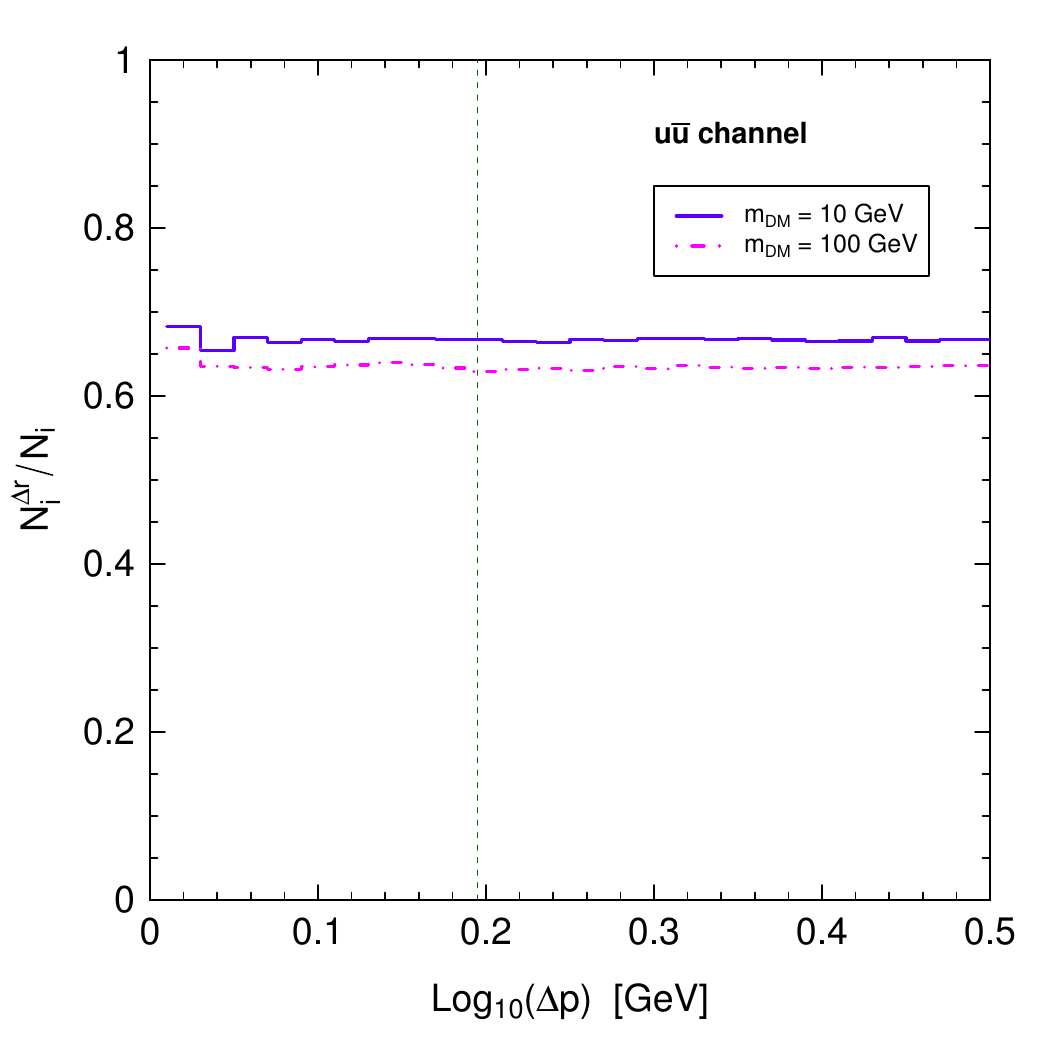}
\includegraphics[width=0.35\textwidth]{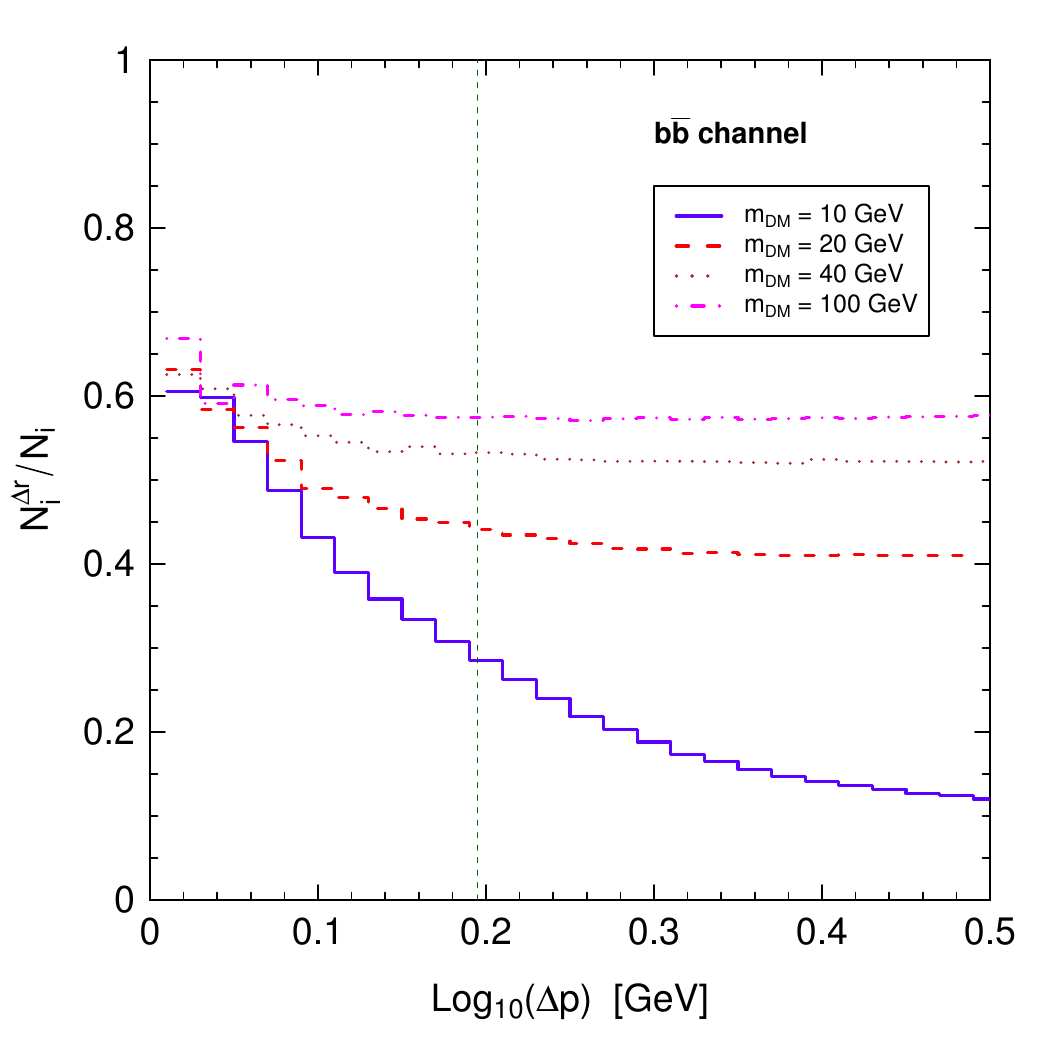}
\caption{The same as in figure \ref{fig:fractio1}, except that each momentum bin shows the
ratio between the two types of histograms reported in the panels of figure \ref{fig:fractio1}, 
i.e. the ratio between the number $N_i^{\Delta r}$ of
$\bar n \bar p$ pairs close in physical space ($\Delta r < 2$ fm)
in the $i$-th bin in momentum, and the total number $N_i$ of $\bar n \bar p$ pairs in the same
momentum bin.
The vertical dotted line reports the value of the coalescence
momentum $p_0 = 195$ MeV: for $\Delta p < p_0$, the $\bar n$ and $\bar p$ of the pair
coalesce in the MC($\Delta p + \Delta r$) model, adopted in our analysis.
\label{fig:fractio2}}
\end{figure} 
%%%

To further elucidate this point, in Fig.~\ref{fig:fractio1} we show the ratio between the
number $N_i$ of $\pbar\nbar$ in a given bin of relative-momentum difference $\Delta p$
and the total number $N_{\rm TOT}$ produced in the DM annihilation process. The upper
row reports the case of the $\uubar$ channel; the lower row stand for the $\bbbar$ channel.
The left column stands for $m_{DM}=10$~GeV, the right column for $m_{DM}=100$~GeV.  
The (blue) higher histograms show the fractional number $N_i/N_{\rm TOT}$ in each
 $\Delta p$ bin; the (red) lower histograms show the fractional number $N_i/N_{\rm TOT}$
 of $\pbar\nbar$ that possess a relative distance in physical space smaller than the $\dbar$ radius
(i.e. $\Delta r < 2$~fm). The vertical dotted line denotes the value of the coalescence
momentum $p_0 = 195$~MeV obtained in the \mcpr~model. The lower histograms clearly
show that the coalescence criterion $\Delta r < 2$ fm has in fact an impact: in the
$\uubar$ channel, a relatively small fraction of $\pbar\nbar$ pairs does not fulfill the
$\Delta r$ condition when  $\Delta p < p_0$. This occurs also for the $\bbbar$ channel,
when the DM mass is large. Instead, for light DM producing $\bbbar$, only a small
fx of $\pbar\nbar$ pairs happen to be close enough in physical space to merge.

%%%
\begin{figure}[t]
\centering
\includegraphics[width=0.34\textwidth]{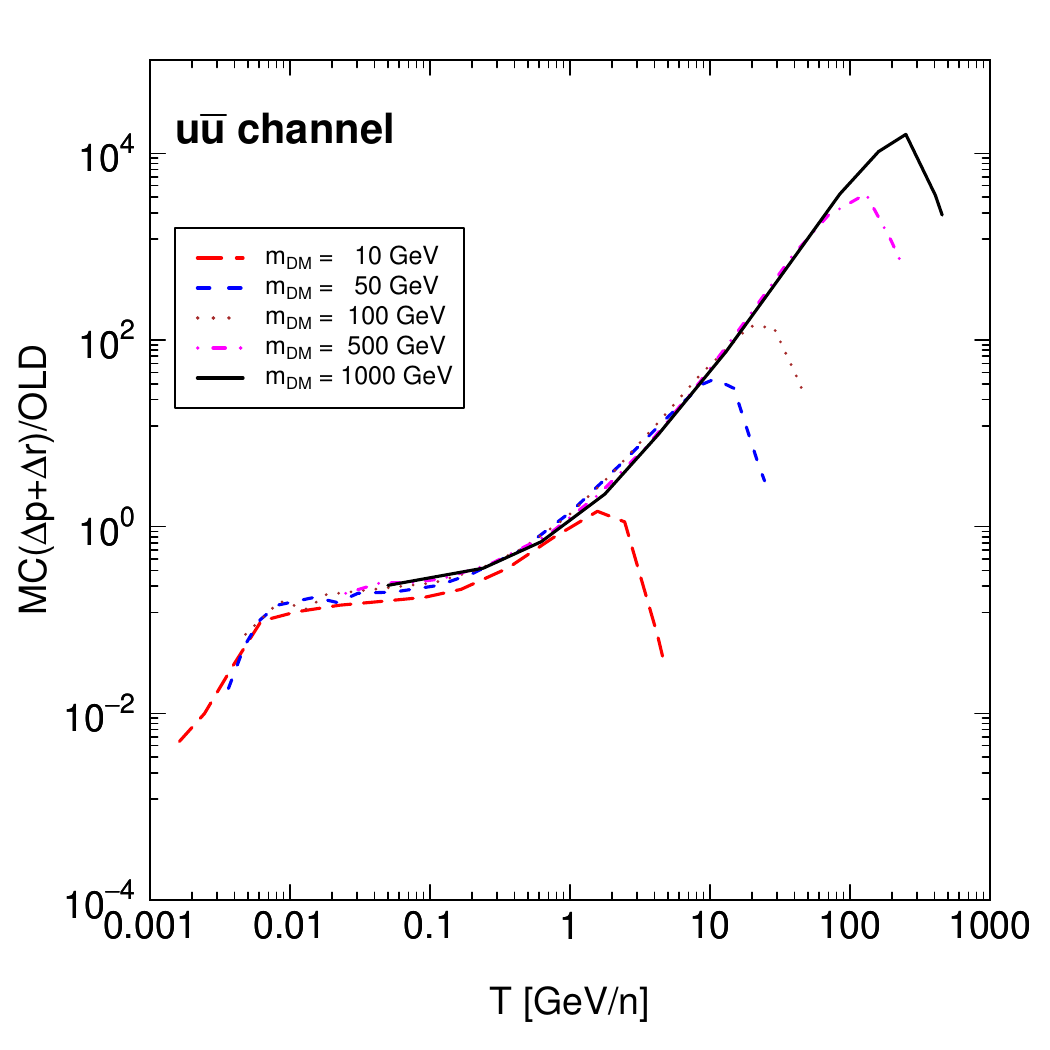}
\includegraphics[width=0.34\textwidth]{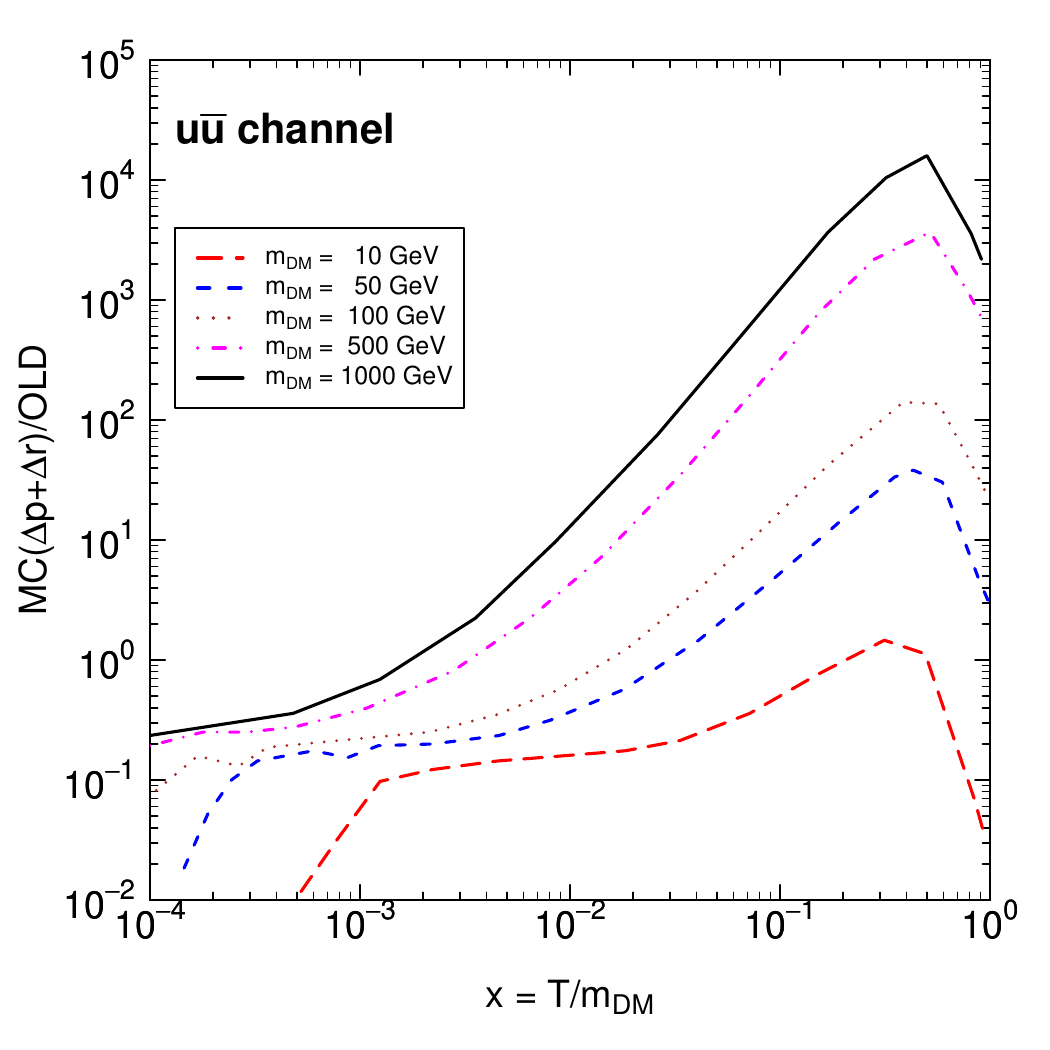}
\includegraphics[width=0.34\textwidth]{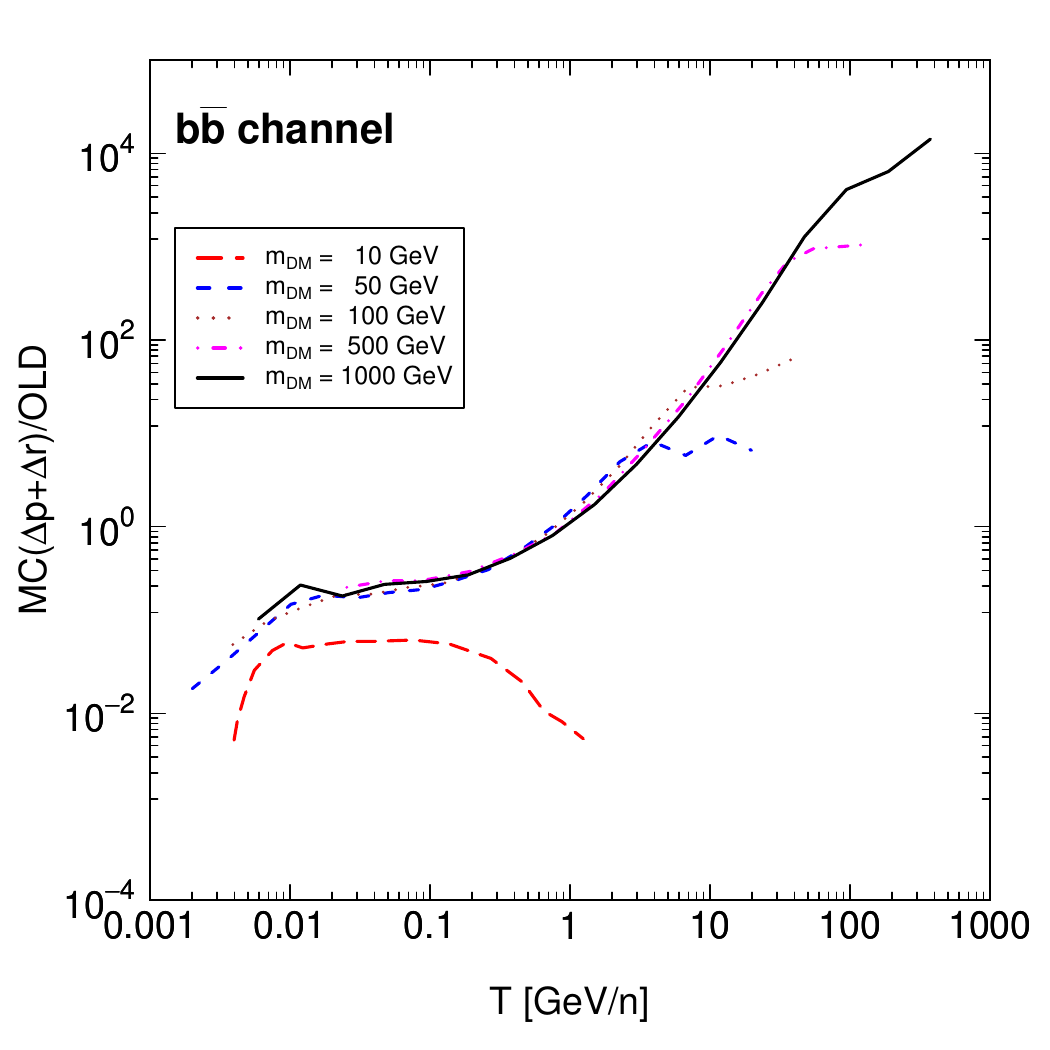}
\includegraphics[width=0.34\textwidth]{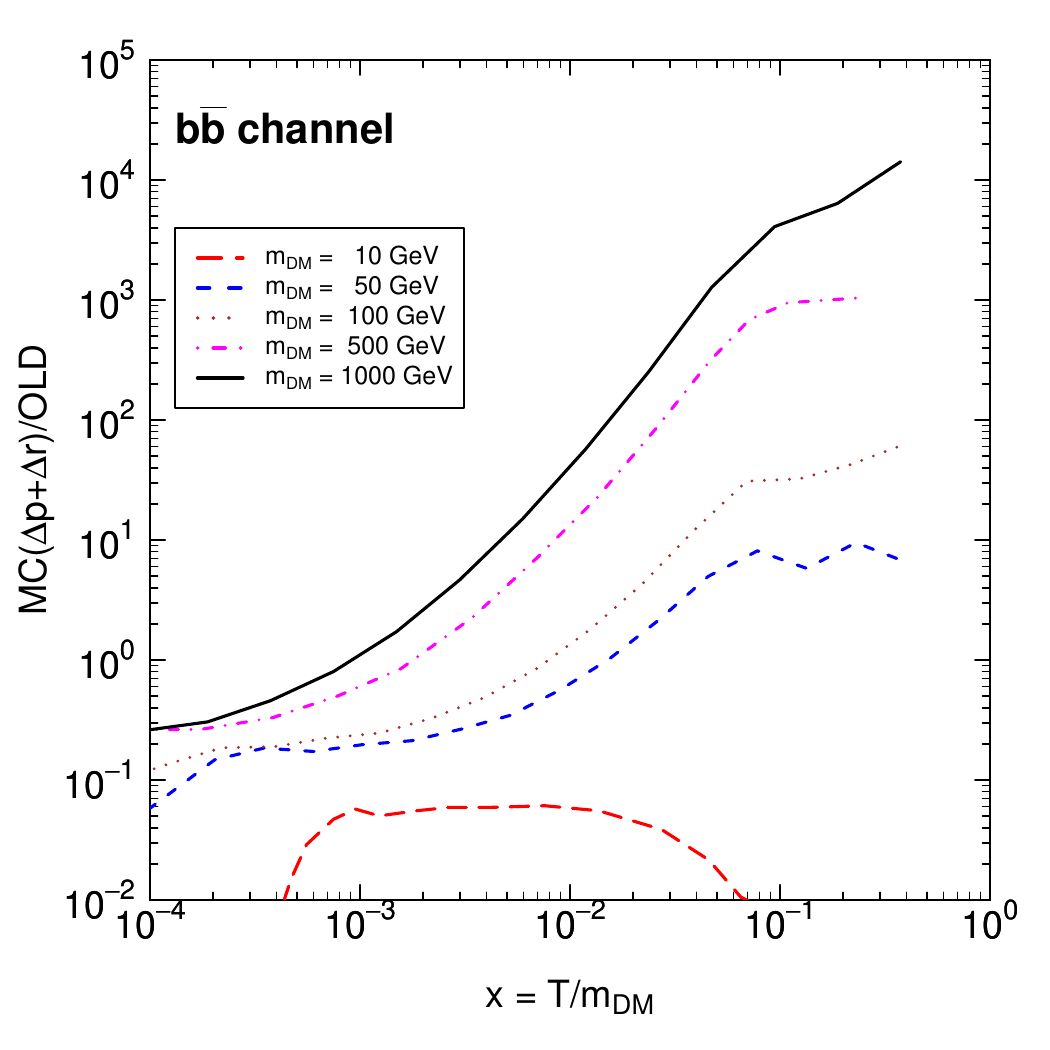}
\includegraphics[width=0.34\textwidth]{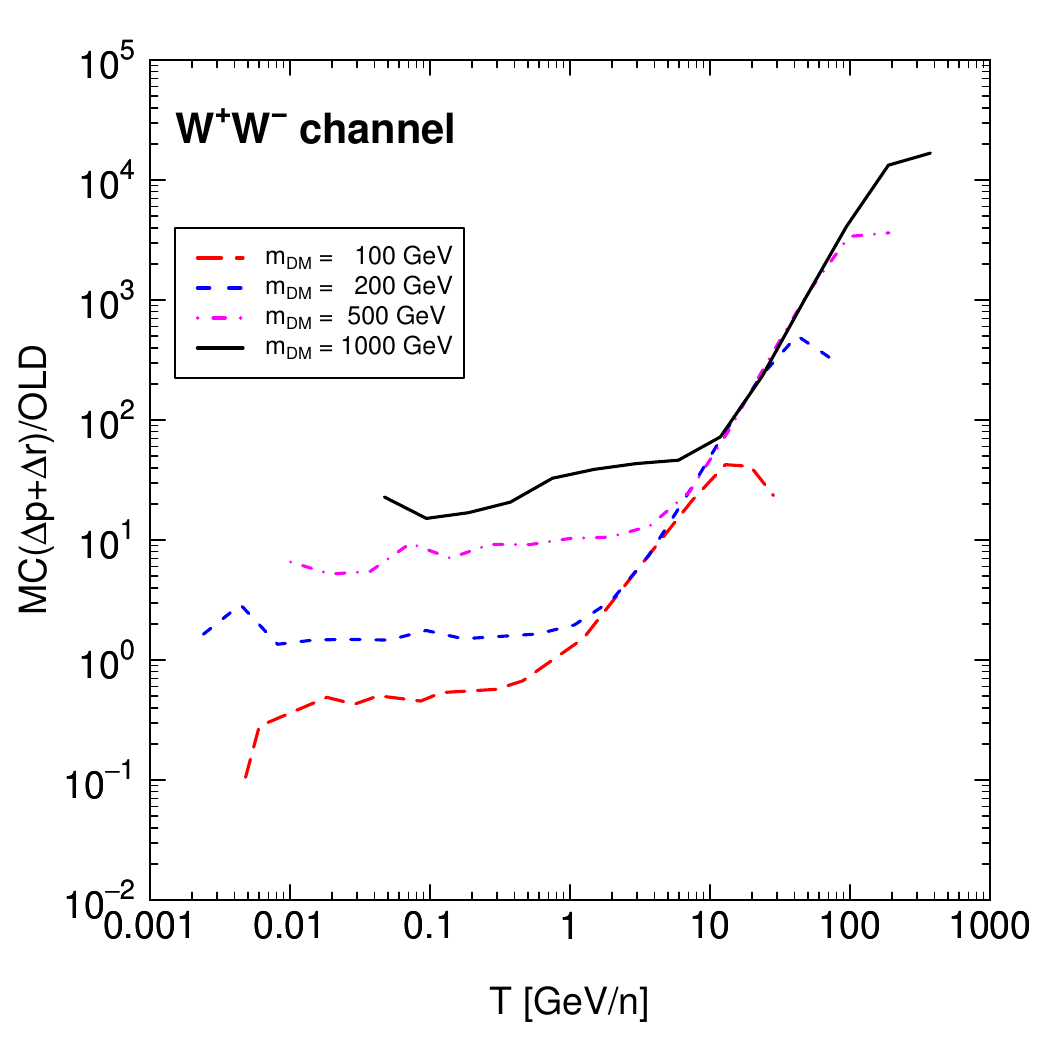}
\includegraphics[width=0.34\textwidth]{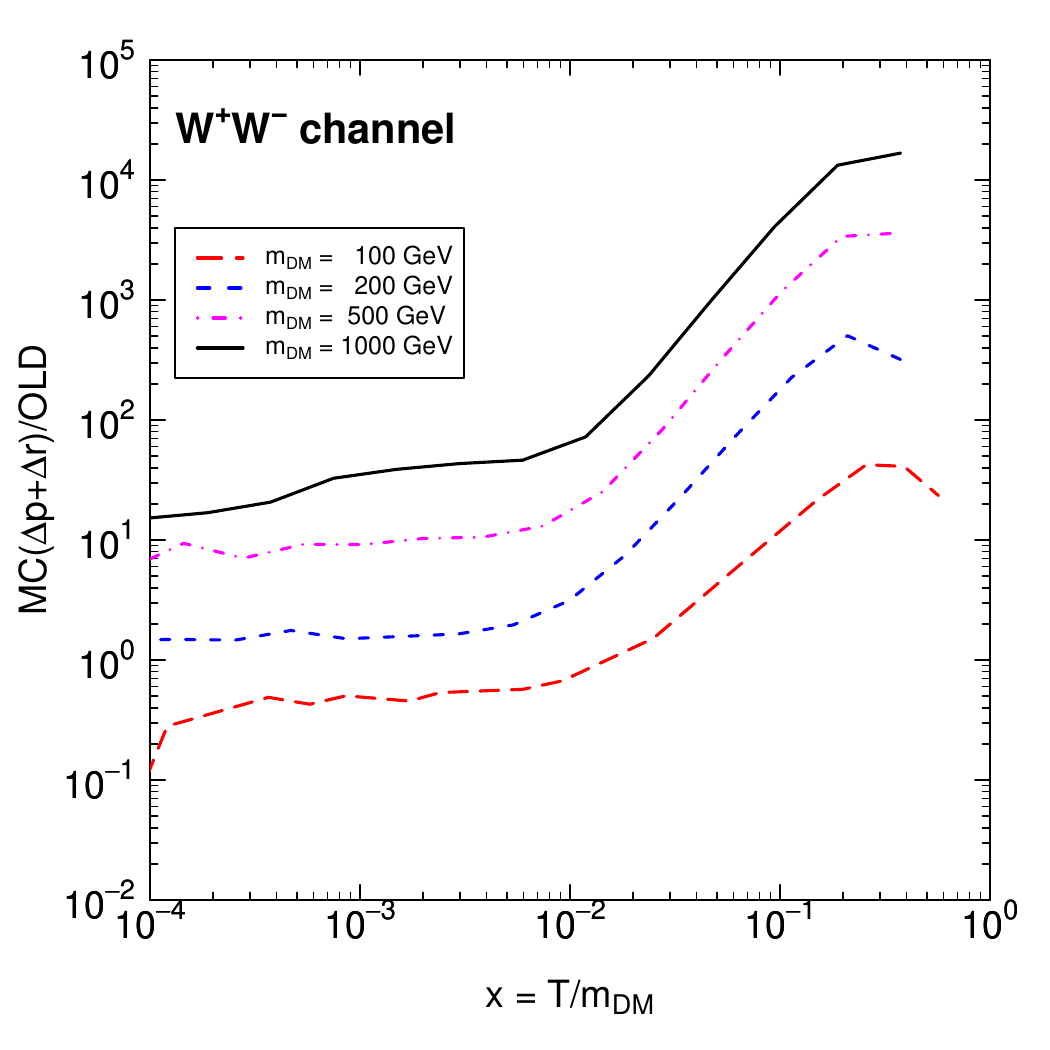}
\caption{Ratio between the $\dbar$ spectra obtained with the full MC($\Delta p + \Delta r$)
Monte Carlo coalescence modeling and the ``old" uncorrelated $\bar n \bar p$ production model.
The left column shows the ratio as a function of the $\dbar$ kinetic energy $T$, while the right
column as a function of the dimensionless variable $x=T/m_{DM}$.
The first, second, third rows refer to $\dbar$ production in the $u \bar u$, $b \bar b$ and $W^+W^-$
channels. Curves refer to different dark matter masses, as reported in the boxed inset.
\label{fig:MCratio1}}
\end{figure} 
%%%

%%%
\begin{figure}[t]
\centering
\includegraphics[width=0.34\textwidth]{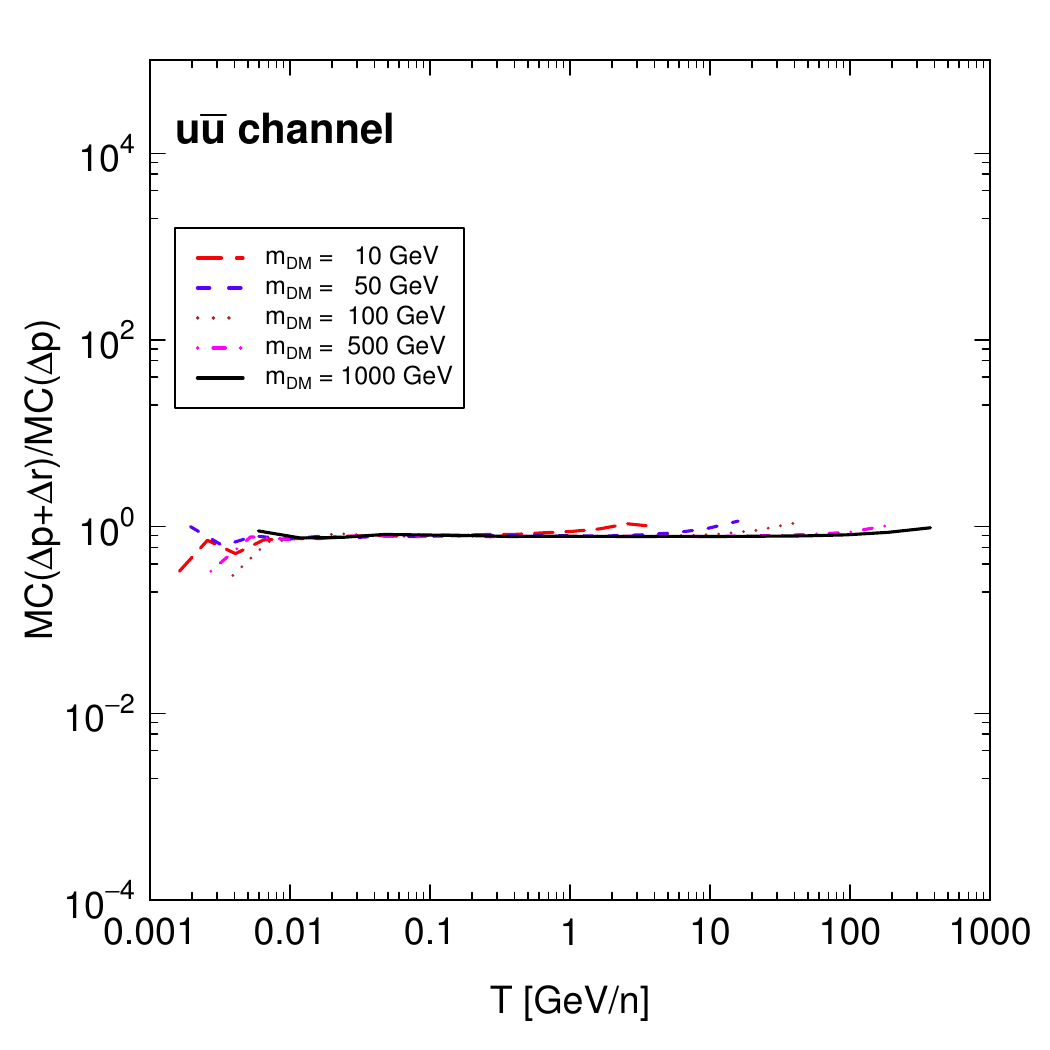}
\includegraphics[width=0.34\textwidth]{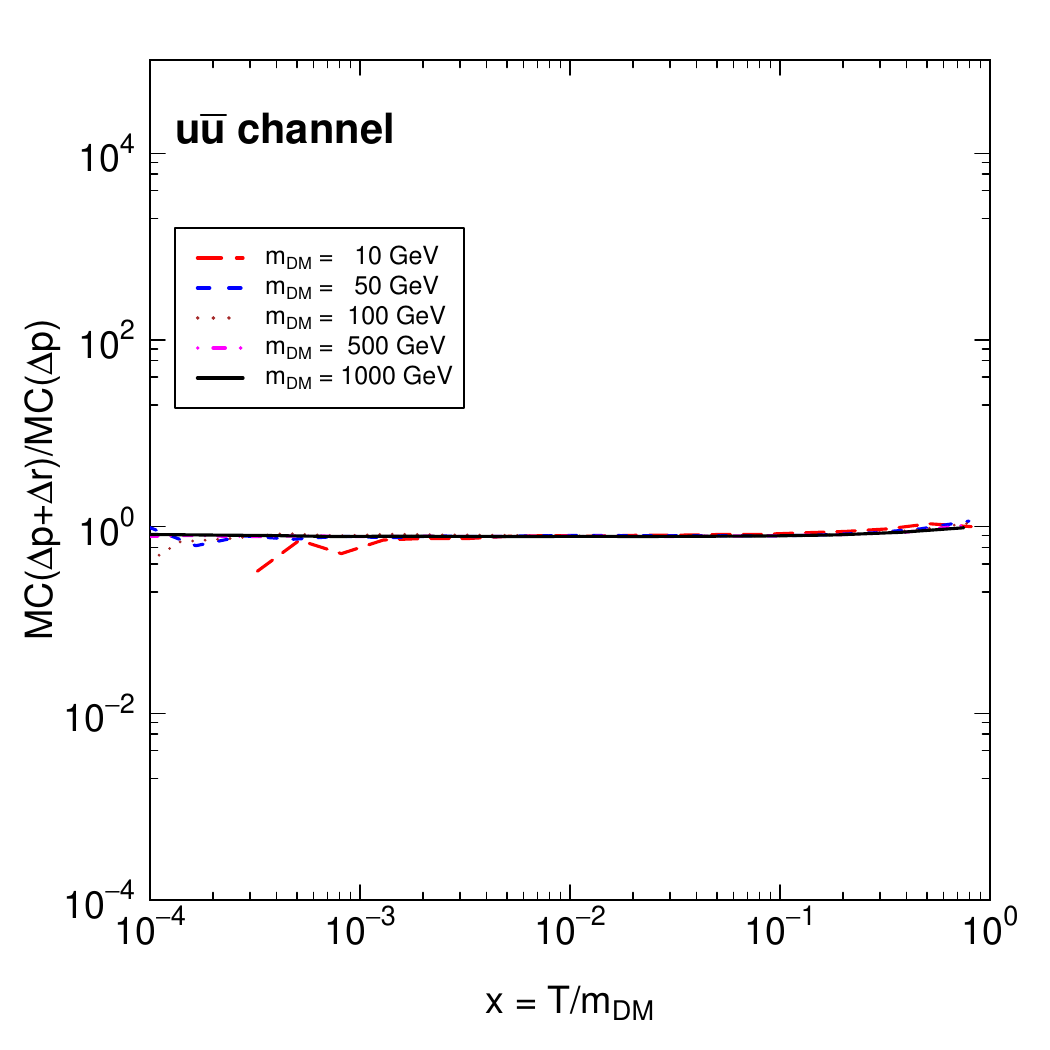}
\includegraphics[width=0.34\textwidth]{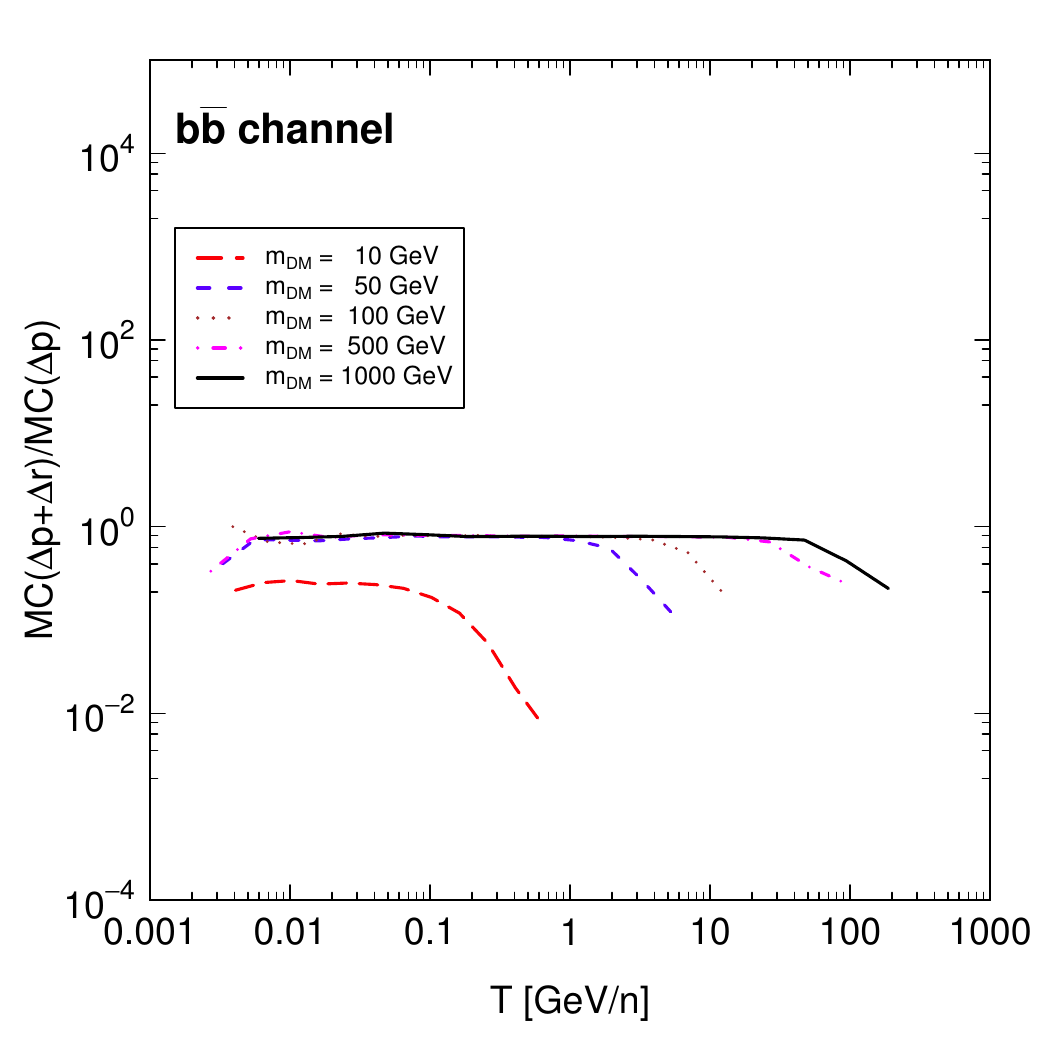}
\includegraphics[width=0.34\textwidth]{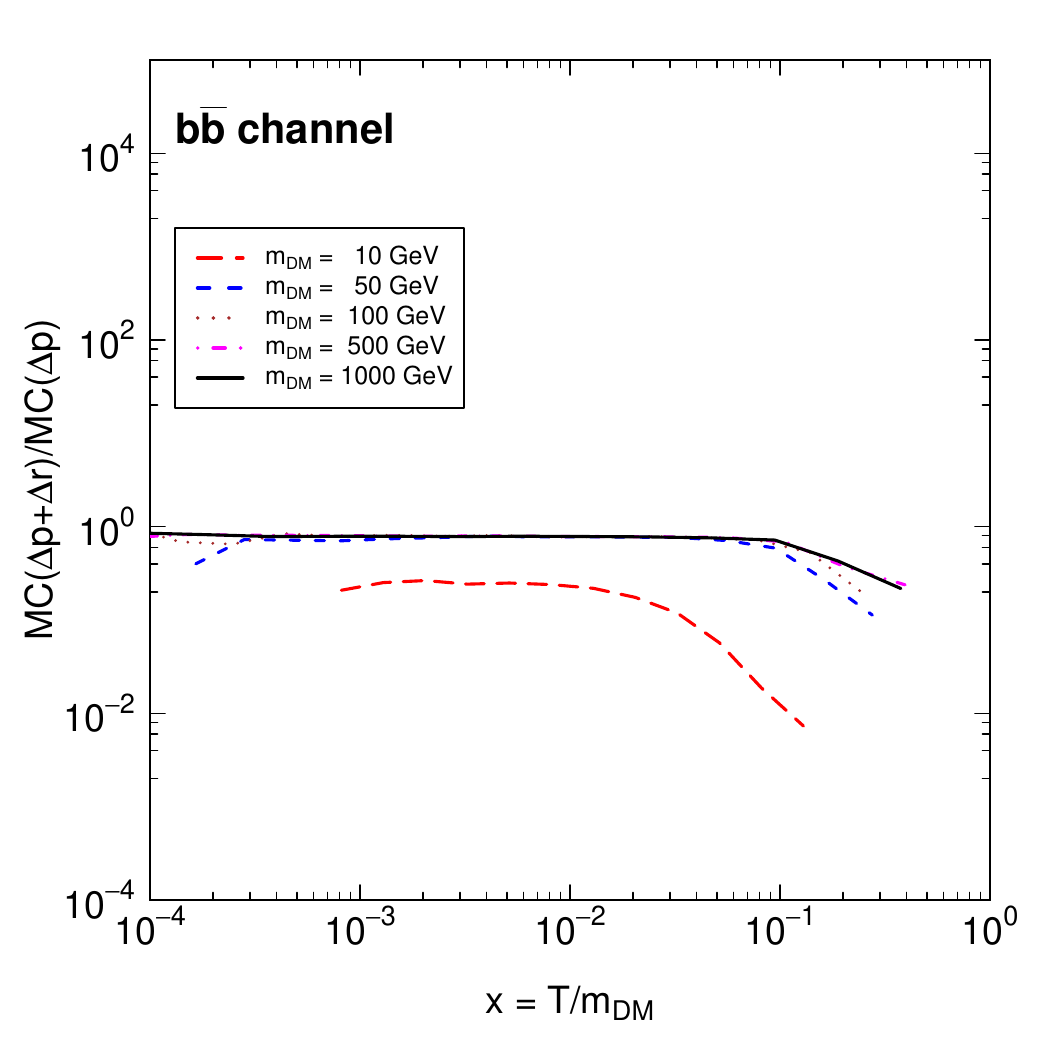}
\includegraphics[width=0.34\textwidth]{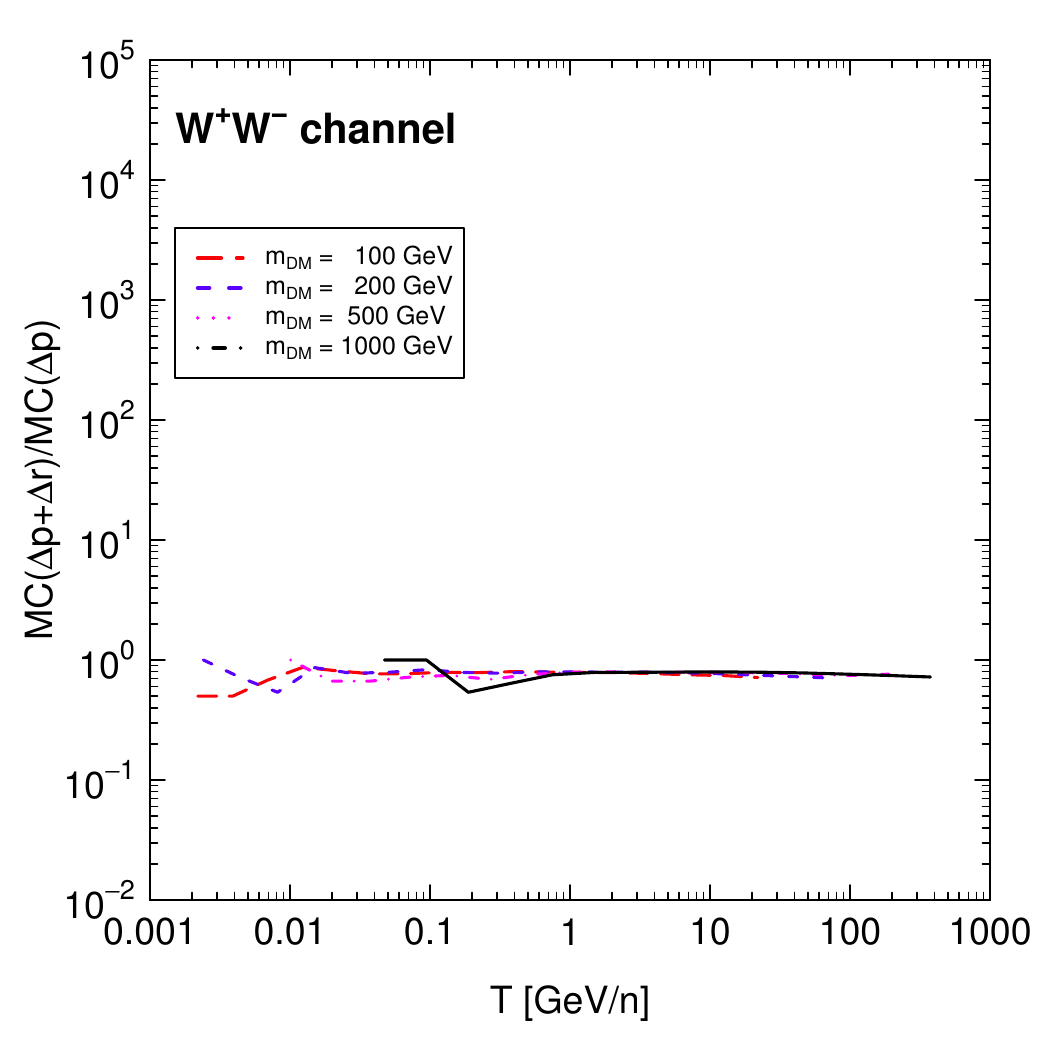}
\includegraphics[width=0.34\textwidth]{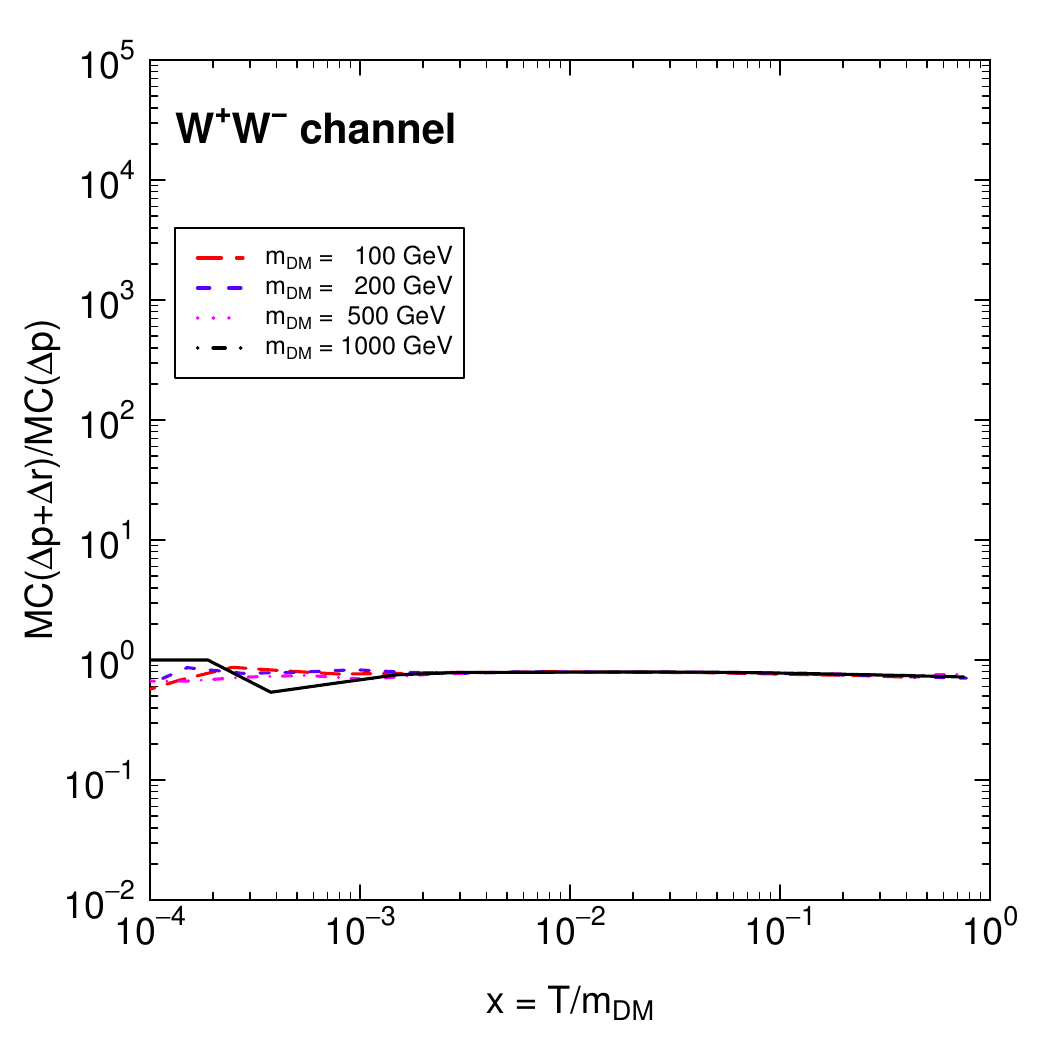}
\caption{Ratio between the $\dbar$ spectra obtained with the full MC($\Delta p + \Delta r$)
Monte Carlo coalescence modeling (cut-off condition imposed both on the relative momentum and on the physical distance of the $\bar n \bar p$ pair) and the MC($\Delta p$) model (cut-off condition imposed only on the relative momentum of the $\bar n \bar p$ pair). Labels and definitions are the same as in Fig. 
\ref{fig:MCratio1}.
\label{fig:MCratio2}}
\end{figure} 
%%%

The actual amount of reduction induced by the $\Delta r < 2$ fm condition is shown
in Fig.~\ref{fig:fractio2}, which displays the reduction of the number of $\pbar\nbar$ pairs (which then reflects in a  reduction in the $\dbar$ production rate) as due to the inclusion of the $\Delta r<2$ fm
condition. The reduction is shown as a function of the (absolute value of the)
relative momenta $\Delta p$ of the pair. We see that in the $\uubar$ channel 
60-70\% of the produced $\pbar\nbar$ pairs can actually coalesce, with no major 
differences in the result for different values of the DM mass, and with no relevant
dependence on $\Delta p$. On the contrary, for the $\bbbar$ channel a strong
evolution, both with the relative momentum and with the DM mass is present. We
notice that for very low relative momenta ($\Delta p< 50$~MeV) the suppression
is similar for the different center-of-mass energies (i.e.~for the different DM masses): about 60-70\% of the $\pbar\nbar$ pairs coalesce, similarly to the case of the
$\uubar$ channel. For larger relative momenta, the fraction of $\pbar\nbar$ pairs
able to merge is strongly reduced when $\mdm$ is small, while the effect is less
pronounced when $\mdm$ becomes larger than 40-50~GeV. The vertical
line denotes the coalescence momentum $p_0$ determined by adaptation
to the ALEPH data (and used in the \mcpr~analysis): we see that for light
DM a significant reduction is present in the \mcpr~vs.~the \mcp~model, 
as compared to the $\uubar$ case
and as compared to the same $\bbbar$ case at larger production energies.

To summarize, we find that for DM masses below 30-40 GeV, the $\dbar$ production
in the $\bbbar$ channel is suppressed, while this is not the case for the light quarks
channel. This may have relevant implications for $\dbar$ searches from DM annihilation
in the Galaxy. This discussion makes also clear that a proper implementation of the coalescence mechanism
in the event-by-event MC approach, must take into account both the momentum space
and the physical space conditions.

In the case of annihilation into gauge bosons, Fig.~\ref{fig:spectra2} and Fig.~\ref{fig:yield}
show that the most relevant effect is related to an incorrect determination of the $\dbar$
multiplicity which emerges with the technique of the old model \cite{strumia}. For
increasingly larger production energies, the multiplicity is suppressed by a factor
proportional to $1/\sqrts$ \cite{strumia}, while on the contrary a MC approach allows understanding the correct behavior of a constant multiplicity. This in fact is the expected trend, since the number of
$\dbar$ produced by a boosted gauge boson is the same as those arising from a gauge boson at rest (different is the case of the production of $q\bar q$ pair, where larger production energies yield more particles in the hadronization process). In addition to the multiplicity issue,
the MC approaches also predict a larger fraction of energetic $\dbar$.

Figures \ref{fig:MCratio1} and \ref{fig:MCratio2} show the spectral differences among the
various coalescence approaches. Figure \ref{fig:MCratio1} reports the ratio between
the $\dbar$ spectra obtained in the \mcpr~model and the old model. Figure \ref{fig:MCratio2} 
shows the ratio between the \mcpr~model and the \mcp~realization. The left columns
show the ratios as a function of the $\dbar$ kinetic energy (per nucleon) $T$, while the right
columns shows the same information as a function of the reduced variable $x=T/\mdm$.
We notice that in the case of the $\uubar$ channel a sort of  ``universal'' feature is
present, when the ratio is reported as a function of $T$. All the curves, referring to different
DM masses, show approximately a common trend, with a sharp cut-off when the
available phase space for $\dbar$ closes (at $T=\mdm$). This is approximately the case
also for the $\bbbar$ channel, when the heavy hadron production is not the dominant channel
(i.e.~for $\mdm \gtrsim$ 40-50~GeV, as discussed above), while at low DM masses there is significant separation of the curve from the ``universal'' trend, due to the larger impact on $\pbar\nbar$ production
from heavy baryons, with the ensuing reduction on the production of $\dbar$. 

In the case of
the $\ww$ channel, a common behavior is instead present when the ratio is shown as
a function of the $x$ variable: the curves at different $\mdm$ are almost parallel, rescaled
by the multiplicity factor shown in Fig.~\ref{fig:yield}. In general, Fig.~\ref{fig:MCratio1} shows that (anti)correlations in the production of $\pbar\nbar$,
which are not considered in the old model, produce a mild reduction for $\dbar$ kinetic
energies below 0.1~GeV (an energy range that is not relevant for $\dbar$ searches in space,
due to energy losses of $\dbar$ in the solar system, as it will be made clear later on), and strong enhancements at large kinetic energies (relevant mostly
for heavy DM matter). The largest enhancement occurs close to the maximal kinetic energy
available ($T \sim \mdm$). As a reference, at $T=1$~GeV, we find basically no enhancement
(except that in the $\bbbar$ channel for light DM matter),
while at 10 and 100~GeV we find about a factor of 100 and $10^4$ enhancement, respectively. 
The results reported in Fig.~\ref{fig:MCratio1} may be useful to infer an approximate $\dbar$ spectrum, that incorporates also the complex information on (anti)correlations (which requires time-consuming MC modeling), from the (simpler)
old uncorrelated model, corrected by exploiting the ``universal'' features discussed above.

Finally, Fig.~\ref{fig:MCratio2} shows that relevant corrections to the $\dbar$ spectra in the \mcp~modeling are necessary for the heavy quark case, both for low DM masses (below, as stated, 40-50~GeV) and for $\dbar$ kinetic energies close to their maximal value.

We wish to mention that details on the actual results would depend on the specific implementation of the hadronization process in the Monte Carlo (in our case PYTHIA), which is tuned on data coming
from observables (momentum spectra, multiplicities) different from those we are interested
in, in the present analysis. Dependences on the hadronization model and differences
between Monte Carlo packages  may also have impact on the reconstructed $\dbar$ spectra:
this issue has been investigated in \cite{Dal:2012my}.

\section{Propagation in the galactic environment}
\label{sec:galaxy}

%%%
\begin{table}[t]
\centering
\renewcommand{\arraystretch}{1.2}
    \begin{tabular}{ |l | c | c | c | c |}
\hline
     & $\delta$ & $ K_0$ (kpc$^2$/Myr) & $L$ (kpc) &  $V_c$ (km/s) \\
\hline
    MIN & 0.85  & 0.0016 & 1 & 13.5 \\  
    MED & 0.70  & 0.0112 & 4 & 12 \\
    MAX & 0.46  & 0.0765 & 15 & 5 \\
\hline
    \end{tabular}
\caption{Parameters of the galactic propagation models for charged cosmic rays \cite{diffusion1,minmedmax}.}
\label{tab:parameters}
\end{table}
%%%

After their injection in the Galaxy, antideuterons propagate through the galactic medium and their behavior can be expressed in terms of a transport equation. We closely adopt here 
the modeling and the solution discussed in Refs.  \cite{DFS,DFM}, to which we refer for additional
details. In the remainder we briefly recall the formalism and the relevant parameters, which will be
used in the analyses of the following Sections.

We in fact adopt the so-called two-zone diffusion model \cite{diffusion1,diffusion2,diffusion3} in which 
cosmic rays propagation is assumed to be confined in a cylinder of radius $R = 20 \: \mathrm{kpc}$ and vertical half-thickness $L$ out of the galactic plane. A thin disk coincident with the galactic plane and of vertical half-height $h = 100 \:\mathrm{pc}$ is the place where cosmic rays may interact with
 the ISM. In this setup, the transport equation can be cast in the form:
\begin{equation}
-\nabla [K(r,z,E) \nabla n_{\bar{d}}(r,z,E)] + \frac{\partial}{\partial{z}}\left[\mathrm{sign}(z)V_{c}n_{\bar{d}}(r,z,E)\right]+2h \delta (z)\Gamma_{\rm ann}^{\bar{d}}n_{\bar{d}}(r,z,E) = q_{\bar{d}}(r,z,E)
\label{eq:transport}
\end{equation}
where $n_{\bar{d}}(r,z,E)$ is the $\dbar$ number density at energy $E$ and position $\vec r = (r,z)$ in the galaxy ($r$ denotes the distance from the galactic center and along the galactic plane, while $z$ is the vertical distance away from the galactic plane), $q_{\bar{d}}(r,z,E)$ is the $\dbar$ source term, 
$K(r,z,E)$ denotes the diffusion coefficient, $V_{c}$ is a convection velocity (which may
have vertical dependence, although we will assume it to be constant),  $\Gamma_{\rm ann}^{\bar{d}}$ denotes the annihilation rate of
$\dbar$ due to interactions with the hydrogen and helium nuclei that populate the interstellar
medium (and the Dirac-delta function $\delta(z)$ expresses that these processes  occur in the
galactic plane).
In Eq. (\ref{eq:transport}) we have neglected both reacceleration and energy losses, since these terms have been shown to affect only sightly the final interstellar flux \cite{MaurinCombet}. Moreover, recent analyses of the proton LIS \cite{Loparco:2013pea}, of the positron absolute spectrum \cite{DiBernardo:2012zu} and of the galactic diffuse synchrotron emission \cite{Strong:2011wd,DiBernardo:2012zu}, have shown that strong reacceleration effects should not be expected in the propagated spectra. We assume a homogeneous diffusion coefficient $K(r,z,E)$, with the typical energy dependence:  
\begin{equation}
K(r,z,E) = \beta K_0 \left( \frac{\cal{R}}{1~\mbox{GV}}\right)^\delta
\end{equation}
where
$K_0$ and $\delta$ are constant parameters, $\beta = p/E$ is the particle velocity and ${\cal R} = pc/|e|Z$ is the rigidity of a particle with atomic number $Z$ ($Z=1$ in the case of $\dbar$), electric charge $e$ and momentum $p$. The galactic propagation model we are adopting depends on the four parameters $L$, $K_0$, $\delta$ and $V_c$: these parameters have been constrained mostly by
the study of B/C data \cite{diffusion1} (see also \cite{Putze:2010fr,donato_tomassetti,trotta}). We adopt here the reference sets of parameter typically
called MIN, MED and MAX \cite{minmedmax} which represent the best-fit (MED set) and
the uncertainty band (MIN and MAX) on the B/C data. The three sets of parameters are
listed in Table \ref{tab:parameters}. 

 We must warn the reader that subsequent multichannel analyses have further constrained the parameter space. For example, the analysis of the frequency spectrum and of the latitudinal profile of the diffuse galactic synchrotron emission \cite{DiBernardo:2012zu} yields a constraint on the halo height $L\gtrsim2$~kpc. This is also corroborated by the analysis of the positron absolute fluxes, which yields comparable constraints, albeit with larger uncertainties due to solar modulation effects \cite{DiBernardo:2012zu}. Nevertheless, it is instructive to consider this set of models, also because we can better compare our results with the ones in the literature.

Concerning the annihilation term in Eq.~(\ref{eq:transport}),  we model the $\dbar$ interactions with the Hydrogen and Helium nuclei that populate the ISM as:
\begin{equation}
\Gamma_{\rm ann}^{\bar{d}} = (n_{\rm H} + 4^{2/3}n_{\rm He})\sigma^{\bar{d}p}_{\rm ine}v_{\bar{d}}
\end {equation}
where $n_{H} = 1 \: \mathrm {g\;cm}^{-3}$ and $n_{He} = 0.1 \: \mathrm{g\;cm}^{-3}$ are the densities of Hydrogen and Helium nuclei, while $\sigma_{\rm ine}^{\bar{d}p}$ is the inelastic cross section for the process $\bar{d} + p \rightarrow \bar{d} + X$, expressed as:
\begin{equation}
\sigma^{\bar{d}p}_{\rm ine} = \sigma^{\bar{d}p}_{\rm tot} - \sigma^{\bar{d}p}_{\rm el}
\end{equation}
The cross section of this process has not been measured. We therefore adopt the technique of Ref.~\cite{Brauninger:2009pe}, by making the following assumptions:  $\sigma^{\bar{d}p}_{\rm tot} = \sigma^{\bar{p}d}_{\rm tot} $ and $\sigma^{\bar{d}p}_{\rm el} = 2 \sigma^{\bar{p}p}_{\rm el}$, where
both $\sigma^{\bar{p}d}_{\rm tot}$ and $ \sigma^{\bar{p}p}_{\rm el}$ are experimentally
available \cite{PDG}. Let us notice that we have used the total inelastic cross section instead of the annihilating one (the same approximation has been adopted in \cite{ibarra}): in other words, we have assumed that the ``tertiary'' antideuterons disappear rather than losing their energy and thus contributing to the low energy tail of the IS flux. This approximation is fully justified by the fact that the antideuteron binding energy is very small.

In the case of the $\dbar$ signal produced by DM annihilation, the source term on the right hand side of the transport equation is given by: 
\begin{equation}
 q_{\bar{d}}(r,z,E) = \frac{1}{2} \sigmav \frac {dN_{\bar{d}}}{dE}\left(\frac{\rho(r,z)}{m_{\chi}}\right)^2
\end{equation}
where $\sigmav$ is the thermally-averaged DM pair-annihilation cross section, ${dN_{\bar{d}}}/{dE}$ is the $\dbar$ injection spectrum, and  $\rho(r,z)$ denotes the galactic-halo DM density-distribution. For 
$\rho(r,z)$ we assume spherical symmetry and we adopt the profiles listed in Table \ref{tab:profiles}.
The galactocentric distance $r_{\odot}$ and the local DM density $\rho_{\odot}$ have been fixed
to the following values: $r_{\odot} = 8.5$~kpc, $\rho_{\odot}=0.39$~${\mathrm{GeV}\: \mathrm{cm}^{-3}}$.

%%%
\begin{table}[t]
\centering
\renewcommand{\arraystretch}{1.2}
  \begin{tabular}{ |l | c | c |}
\hline
    Profile & $\rho(r,z)/\rho_\odot$ & Parameters \\
\hline
    Isothermal  & $(1 + r_{\odot}^2/r_{s}^2 )/(1 + (r^2+z^2)/r_{s}^2)$ &  $r_s$ = 5~kpc\\  
    NFW          & $(r_{\odot}/\sqrt{r^2+z^2})(1 + r_{\odot}/r_{s})^2/ (1 + \sqrt{r^2+z^2}/r_{s})^2$  &  $r_s$ = 20~kpc\\  
    Einasto       & $\exp(-2[(\sqrt{r^2+z^2}/r_s)^{\alpha}-(r_{\odot}/r_s)^{\alpha}]/\alpha)$  &  $r_s$ = 20~kpc \; , \;$\alpha = 0.17$ \\  
\hline
    \end{tabular}
\caption{Dark matter density profiles $\rho(r,z)$ adopted in our analysis.}
\label{tab:profiles}
\end{table}
%%%

With the above definitions and assumptions, the transport equation may be solved analytically \cite{DFS,DFM}. We recall the main elements of the solutions (further details may be found  in Refs.~\cite{DFS,DFM}). The $\dbar$ density $n_{\bar{d}}(r,z,E)$ is expanded through a Bessel series:
\begin{equation}
n_{\bar{d}}(r,z,E) = \sum_{i} N^E_{i}(z)J_0\left(\frac{\zeta_i r}{R}\right)
\end{equation}
where $J_0$ is the zeroth-order Bessel function of the first kind and $\zeta_i$ are its zeros of index $i$. Due to the Bessel expansion, Eq. (\ref{eq:transport}) transforms into a set of ordinary differential equations for the functions $N_i^E(z)$ (with the energy $E$ merely playing the role of a label), each with the following source term:
\begin{equation}
Q_i(z) = \frac{2}{[J_1(\zeta_i)R]^2}\int_0^R dr\; r J_0\left(\frac{\zeta_i r}{R}\right) \left(\frac{\rho(r,z)}{\rho_\odot}\right)^2
\end{equation}
where $J_1$ is the first-order Bessel function of the first kind.
The solution at the Earth's position ($r=r_\odot$, $z=0$) is given by:
\begin{equation}
N^E_i(0) = \frac{e^{-aL}\;y^E_i(L)}{B_i \sinh(S_iL/2)}
\end{equation} 
where we have defined:
\begin{eqnarray}
&& a    = (V_c)/(2 K) \\
&& S_i = 2 \sqrt{a^2 + (\zeta_i/R)^2} \\
&& A_i = (V_c + 2h\Gamma_{\rm ann}^{\bar{d}})/(K S_i) \\
&& B_i = K \, S_i[A_i + \coth(S_i L/2)]
\end{eqnarray}
and where:
\begin{equation}
y^E_i(z) = 2 \int_0^z dz' \, {\rm e}^{a(z-z')}\, \sinh\left[\frac{S_i(z-z')}{2}\right]Q_i(z')\,.
\end{equation}

The interstellar flux can finally be expressed as:
\begin{equation}
\phi_{\bar{d}}(E) = \frac{v_{\bar{d}}}{4\pi}n_{\bar{d}}(r=r_{\odot},z=0,E) \;=\;\frac{v_{\bar{d}}}{4\pi}\left(\frac{\rho_{\odot}}{m_{\chi}}\right)^{2}R_{\bar{d}}(E)\;\frac{1}{2}\sigmav\frac{dN_{\bar{d}}}{dE}
\end{equation}
where the space dependence is contained in ``propagation function'' $R_{\bar{d}}(E)$ \cite{R_function}:
\begin{equation}
R_{\bar{d}}(E) = \sum_{i=1}^\infty J_0\left(\zeta_i\frac{r_\odot}{R}\right) 
\exp\left(-\frac{V_cL}{2K}\right)
\frac{y^E_i(L)}{B_i\sinh(S_iL/2)}\,.
\end{equation}

%
%%%
\begin{figure}[t]
\centering
\includegraphics[width=0.35\textwidth]{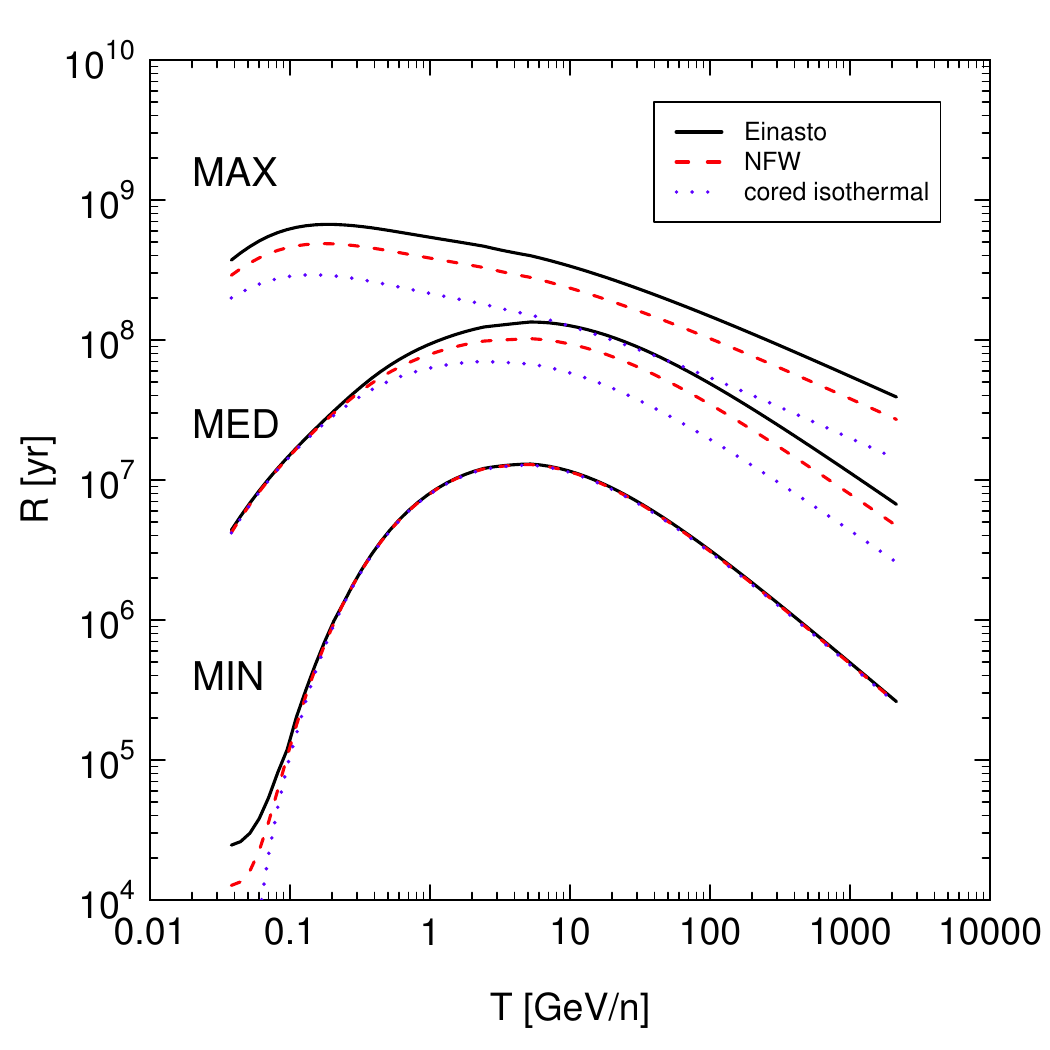}
\caption{Effective propagation function $R(T)$ as a function of the $\dbar$ kinetic energy, for
different galactic-transport parameter sets (MIN, MED, MAX, as defined in Table \ref{tab:parameters}
and Ref. \cite{minmedmax})
and for different galactic halo shapes (Einasto profile, NFW profile and a cored isothermal profile, as listed in Table \ref{tab:profiles}).
\label{fig:prop}}
\end{figure}
%%%

The ``propagation function'' $R_{\bar{d}}(E)$  is shown in Fig. \ref {fig:prop} for the three DM profiles in Table \ref{tab:profiles} and for the three propagation models listed in Table \ref{tab:parameters}: while the choice of the propagation set of parameters is extremely relevant to determine the interstellar flux (the difference between the MIN and MAX models is of almost four orders of magnitude), the choice of the DM profile affects only weakly the predicted $\dbar$ fluxes, and the impact is even less relevant in the case of the MIN model, where the three ``propagation functions'' coincide.  This is mainly due to the fact that CR escape in the $z$ direction is relatively fast for a small halo, hence the CR flux at Earth is mainly determined by the source density in the vicinity of the solar system \cite{Maurin:2002uc,Evoli:2011id}. For $L\simeq1$~kpc this region does not include the galactic center, where the differences among DM source density profiles are most relevant, thus making $R_{\dbar}$ insensitive to the actual profile model.   We will show and discuss predictions for the $\dbar$ fluxes in Section \ref{sec:signals}.
We first move to discuss the second transport phenomenon that affects cosmic rays fluxes at the
Earth, most notably at low energies, i.e.~the propagation inside the heliosphere.

\section{Propagation in the heliosphere: solar modulation}
\label{sec:solarmod}
%%%
\begin{table}[t]
\centering
    \begin{tabular}{ |l | c | c | c | c |}
\hline
     &  tilt angle $\alpha$ & mean-free-path $\lambda_{\|}$ & spectral index $\gamma$ & $\Phi$ \\
\hline
 Force Field                      & & & & 500 MeV \\
 CD\_60\_0.20\_1     & 60$^\circ$ & 0.20\; A.U. & 1.0 & \\
 CD\_60\_0.60\_1     & 60$^\circ$ & 0.60\; A.U. & 1.0  & \\
 CD\_20\_0.15\_0.5   & 20$^\circ$ & 0.15\; A.U. & 0.5 & \\
 CD\_20\_0.15\_1     & 20$^\circ$ & 0.15\; A.U. & 1.0 & \\
 CD\_60\_0.15\_1     & 60$^\circ$ & 0.15\; A.U. & 1.0 &\\
 CD\_60\_0.15\_0.5   & 60$^\circ$ & 0.15\; A.U. & 0.5 & \\
\hline
    \end{tabular}
\caption{Solar modulation models adopted in the analysis and their code name. The analytical force field model \cite{Gleeson_1968ApJ} is defined in terms of a single parameter (the modulation potential $\Phi$) which is not shared with the other solar modulation models, which instead refer to a full numerical 4D solution of the propagation equation
in the heliosphere \cite{Maccione:2012cu}.  Positive polarity (suitable for AMS and GAPS
operational periods) is always used for antideuterons calculations. Negative polarity (suitable for the PAMELA data-taking period) is always used for antiprotons calculations.
}
\label{tab:solarmod}
\end{table}
%%%

Before they are detected at Earth, CRs lose energy due to the solar wind while diffusing in the solar system \cite{Gleeson_1968ApJ}. This modulation effect depends, via drifts in the large scale gradients of the solar magnetic field (SMF), on the particle's charge including its sign \cite{1996ApJ...464..507C}. Therefore, it depends on the polarity of the SMF, which changes periodically every $\sim$11 years \cite{wilcox}. Besides the 11 year reversals, the SMF has also opposite polarities in the northern and southern hemispheres: at the interface between opposite polarity regions, where the intensity of the SMF is null, a heliospheric current sheet (HCS) is formed (see e.g.~\cite{1981JGR....86.8893B}). The HCS swings then in a region whose angular extension is described phenomenologically by the tilt angle $\alpha$. The magnitude of $\alpha$ depends on solar activity. Since particles crossing the HCS suffer from additional drifts because of the different orientation of the magnetic field lines, the intensity of the modulation depends on the extension of the HCS. This picture explains, at least qualitatively, the annual variability and the approximate periodicity of the fluctuations of CR spectra below a few GeV. 

The propagation of CRs in the heliosphere can be described by the following transport equation \cite{1965P&SS...13....9P}
\begin{equation}
\frac{\partial f}{\partial t} = -(\vec{V}_{\rm sw}+\vec{v}_d)\cdot \nabla f + \nabla\cdot (\bm{K}\cdot\nabla f) + \frac{P}{3}(\nabla\cdot\vec{V}_{\rm sw})\frac{\partial f}{\partial P}\;,
\label{eq:solartransport}
\end{equation}
where $f$ represents the CR phase space density, averaged over momentum directions, $\bm{K}$ represents the (symmetrized) diffusion tensor,  $\vec{V}_{\rm sw}$ the velocity of the solar wind, $\vec{v}_{d}$ the divergence-free velocity associated to drifts, $P$ the CR momentum. The transport equation is solved in a generic 3D geometry within the heliosphere, whose  boundary we place at 100~AU (see \cite{Bobik:2011ig} and Refs.~therein). The interstellar flux of CRs is given as a boundary condition and we assume that no sources are present within the solar system at the energies relevant to this work.

A model for solar propagation is specified by fixing the solar system geometry, the properties of diffusion and those of winds and drifts. 
We describe the solar system diffusion tensor by $\mathbf{K}(\rho) = {\rm diag}(K_{\|}, K_{\perp r}, K_{\perp \theta})(\rho)$, where $\|$ and $\perp$ are set with respect to the direction of the local magnetic field. We assume no diffusion in the $\perp\varphi$ direction and we describe as drifts the effect of possible antisymmetric components in $\mathbf{K}$. For the parallel CR mean-free-path we take $\lambda_{\|} = \lambda_{0}(\rho/1~\GeV)^{\gamma}(B/B_{\bigoplus})^{-1}$, with $B_{\bigoplus}=5~\nT$ the value of the magnetic field at Earth position, according to \cite{2011ApJ...735...83S,2012Ap&SS.339..223S}. We will consider $\lambda_{0}=0.2$ and $0.6~\AU$ in the following. For $\rho < 0.1~\GeV$, $\lambda_{\|}$ does not depend on rigidity. The values of $\lambda_{0}$ and the rigidity dependence of $\lambda_{\|}$ are roughly compatible both with the measured $e^{-}$ mean-free-path (see, e.g., \cite{Droge2005532}) and with the proton mean-free-path inferred from neutron monitor counts and the solar spot number \cite{Bobik:2011ig}. We then compute $K_{\|} = \lambda_{\|}v/3$. Perpendicular diffusion is assumed to be isotropic. According to numerical simulations, we assume $\lambda_{\perp r,\theta} = 0.02\lambda_{\|}$ \cite{1999ApJ...520..204G}. 

For the SMF, we assume a Parker spiral, although more complex geometries might be more appropriate for periods of intense activity
\begin{equation}
\vec{B} = AB_{0}\left(\frac{r}{r_{0}}\right)^{-2}\left(\hat{r} - \frac{\Omega r\sin\theta}{V_{\rm SW}}\hat{\varphi}\right)\;,
\end{equation}
where $\Omega$ is the solar differential rotation rate, $\theta$ is the colatitude, $B_{0}$ is a normalization constant such that $|B|(1~\AU)=5~\nT$ and $A=\pm H(\theta-\theta')$ determines the MF polarity through the $\pm$ sign. The presence of a HCS is taken into account in the Heaviside function $H(\theta-\theta')$. The HCS angular extent is described by the function $\theta' = \pi/2 + \sin^{-1}\left(\sin\alpha\sin(\varphi+\Omega r/V_{\rm SW})\right)$, where $0<\alpha<90^{\circ}$ is the tilt angle. The drift processes occurring due to magnetic irregularities and to the HCS are related to the antisymmetric part $K_{A}$ of the diffusion tensor as \cite{1977ApJ...213L..85J}
\begin{equation}
\vec{v}_{\rm drift} = \nabla\times(K_{A}\vec{B}/|B|) = {\rm sign}(q)v/3\vec{\nabla}\times\left(r_{L}\hat{B}\right)\;,
\label{eq:drifts}
\end{equation}
where $K_{A} = pv/3qB$, $r_{L}$ is the particle's Larmor radius and $q$ is its charge. We refer to \cite{2011ApJ...735...83S,2012Ap&SS.339..223S} for more details on the implementation of the HCS and of drifts. Adiabatic energy losses due to the solar wind expanding radially at $V_{\rm SW}\sim400~\km/\s$ are taken into account.

As clear from Eq.~\ref{eq:solartransport}, CRs lose energy adiabatically, due to the expansion of the solar wind, while propagating in the heliosphere.
 It is straightforward to notice that the larger their diffusion time (i.e.~the shorter their mean-free-path) the more energy they lose in propagation. 
 This fact is at the basis of the simplest modulation model used in the literature, the so called force-field model \cite{Gleeson_1968ApJ}. 
 In this picture, the heliospheric propagation is assumed to be spherically symmetric, and energy losses are described by 
 the modulation potential $\Phi \propto |\bm{K}|/V_{\rm sw}$ and $\Phi$ is to be fitted against data. However, this model 
 completely neglects the effects of $\vec{v}_{d}$, which may significantly alter the propagation path. $A$ and $\alpha$ are of 
 particular importance in this respect. If $q\cdot A<0$, drifts force CRs to diffuse in the region close to the HCS, which enhances their effective propagation 
 time and therefore energy losses, while if $q\cdot A>0$ drifts pull CRs outside the HCS, where they can diffuse faster \cite{2011ApJ...735...83S,2012Ap&SS.339..223S}. 
 %The relevance of this effect is further controlled by $\alpha$. %On the other hand, if $\alpha$ is small, the current sheet occupies only a small volume in the heliosphere, therefore the induced drift effects are reduced.
 As this is the only effect that depends on the charge-sign in this problem, and given that the force-field model does not account for it, the latter model cannot be used to 
 describe CR spectra below a few GeV, where charge-sign effects are demonstrated to be relevant \cite{1996ApJ...464..507C,GastSchael,Bobik:2011ig,DellaTorre:2012zz,Potgieter:2013cwj,Maccione:2012cu}.

\begin{figure}[t]
\centering
\includegraphics[width=0.6\textwidth]{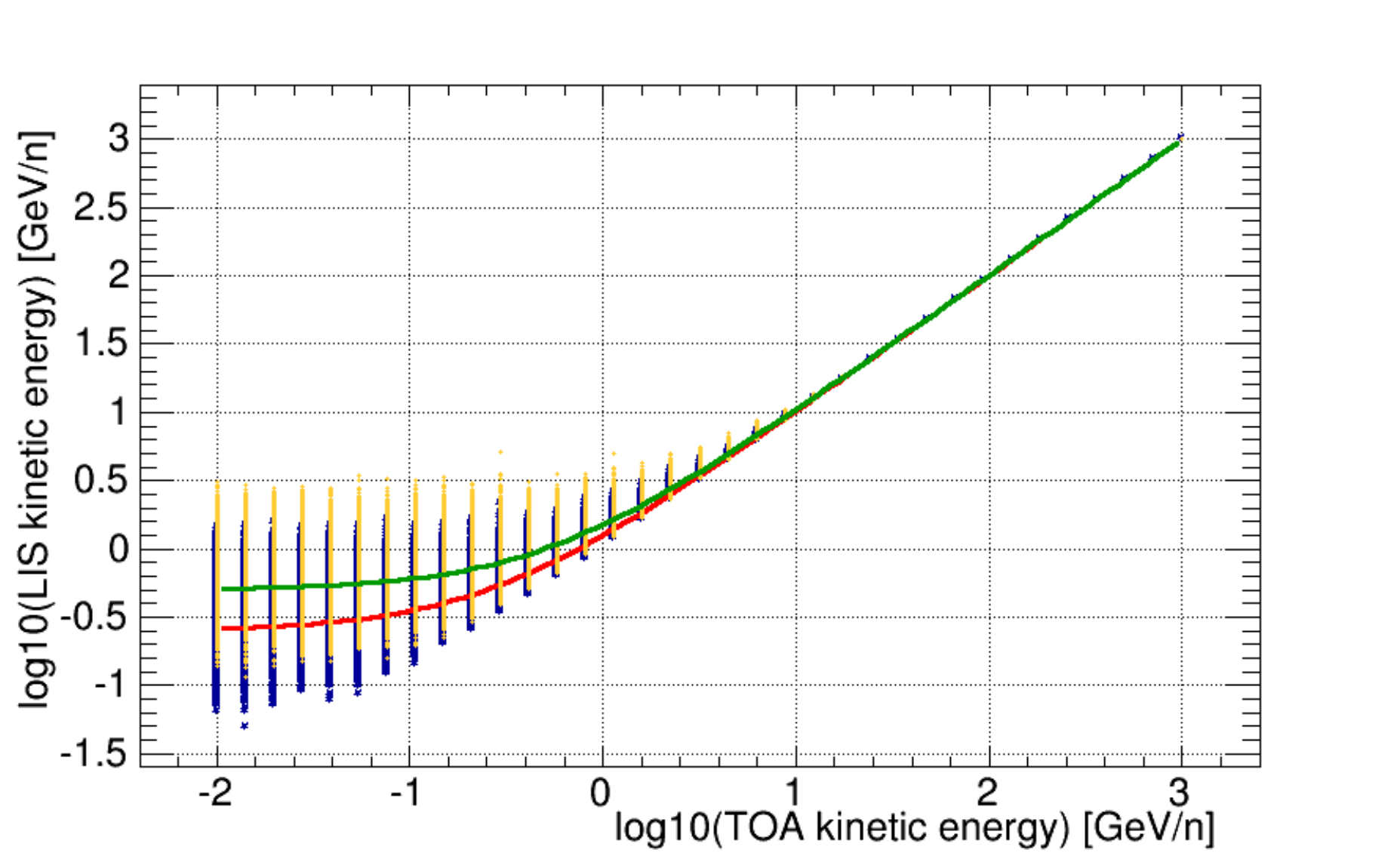}
\caption{LIS kinetic energy per nucleon corresponding to a given TOA kinetic energy per nucleon for antiprotons (orange) and antideuterons (blue) for a solar modulation model with a negative polarity of the solar magnetic field, a tilt angle $\alpha=20^\circ$, a mean free path $\lambda=0.15\:\mathrm{A.U.}$ and a spectral index $\gamma=1$. Energy losses are negligible above $\sim10~{\rm GeV/n}$, while below that energy they are larger for antiprotons than for antideuterons.  The upper (lower) solid line shows the relation between the LIS and TOA energies
for the force--field approximation for antiprotons (antideuterons), with $\Phi=500$ MV.}
\label{fig:sseloss}
\end{figure}

 We exploit then the recently developed numerical program \SolarProp\ \cite{Maccione:2012cu} for the 4D propagation of CRs in the solar system. 
The main effects of solar system propagation on antiprotons and antideuterons are demonstrated in Fig.~\ref{fig:sseloss}, where we show how the TOA 
energy of one of these particles corresponds to the LIS energy of the same particle, for a sample of $10^{4}$ particles generated at each $E_{\rm TOA}$ in \SolarProp. 
While at high energy $E_{\rm LIS}=E_{\rm TOA}$, because diffusion is so fast that no energy losses occur, at low energies, below a few GeV/n, $E_{\rm LIS}>E_{\rm TOA}$ 
and the actual energy lost can vary significantly from particle to particle in our sample,
 that shows the need for a MonteCarlo treatment
of heliospheric propagation. This is due to the fact that energy losses are a function of the actual path, and 
the path is determined by a combination of drifts and random walks. Operationally, the flux observed at Earth at $E_{\rm TOA}$ is determined as a proper weighted 
average of the LIS flux at the energies $E_{\rm LIS}$ corresponding to that $E_{\rm TOA}$, as in Fig.~\ref{fig:sseloss}. Interestingly, energy losses are larger for 
antiprotons than for antideuterons. This is due to the fact that the antideuteron rigidity is twice as large as the rigidity of antiprotons with the same kinetic energy 
per nucleon. Fig.~\ref{fig:sseloss} shows also the corresponding relation
between $E_{\rm TOA}$ and $E_{\rm LIS}$ for the force-field approximation (with $\Phi=500$ MV). The force-field model reproduces the energy-shift only on average. 

 %%%
%%%

Instead, the stochastic modeling
of the particle transport in the heliosphere accounts for fluctuations around the average behavior: these fluctuations are then convoluted with the LIS spectrum 
and the ensuing TOA spectrum depends not
only on the average energy-shift but also on the LIS spectral features.

The solar modulation models adopted in our analyses are listed in Table
\ref{tab:solarmod}. For all
the predictions for antideuterons, we will adopt a solar phase with
positive polarity, which is
appropriate for the expected operational phase of GAPS and which applies
to most of the data-taking
phase of AMS-02.  In the calculations of the antiproton fluxes (that we use
as a constraint), on the contrary, we
apply solar modulation models with a negative polarity, since we
confront our calculations with PAMELA data, which have been taken during a
negative polarity
phase of the solar cycle.

As for the choices of the numerical values of the tilt angle and of the mean free path, we use $\alpha=20^{\circ}$ and $\alpha=60^{\circ}$, that may roughly 
correspond to periods of solar minimum and maximum activity respectively, and we additionally consider $\lambda_{0}=0.15$ and $\lambda_{0}=0.6$~A.U.~to bracket situations in which diffusion is slow and fast respectively. It is understood that when the actual measurements will be available, the tilt angle $\alpha$ will become a fixed quantity, corresponding to the one measured in the data taking period, while $\lambda_{0}$ will still remain undetermined, but might be inferred from, e.g., proton data taken in the same solar phase. For the parameter $\gamma$, we consider the two cases $0.5$ and $1$, that have been discussed recently in the literature \cite{Potgieter:2013cwj,Loparco:2013pea}.

\section{Dark matter signals}
\label{sec:signals}

We now move to discuss the predictions for the $\dbar$ signals and analyze their dependence
on galactic transport and solar modulation modeling. The signals will be confronted with
 their
relevant astrophysical background, represented by secondary $\dbar$ production
in the Galaxy. For the background component we adopt our previous determination, reported in Ref.~\cite{DFM}(notice that a secondary $\bar{d}$ about a factor of two smaller than the one adopted here
has been discussed 
in Ref. \cite{Ibarra:2013qt}). For the $\dbar$ signal, we will adopt throughout the \mcpr~model for the coalescence process. 

%%%
\begin{figure}[t]
\centering
\includegraphics[width=0.30\textwidth]{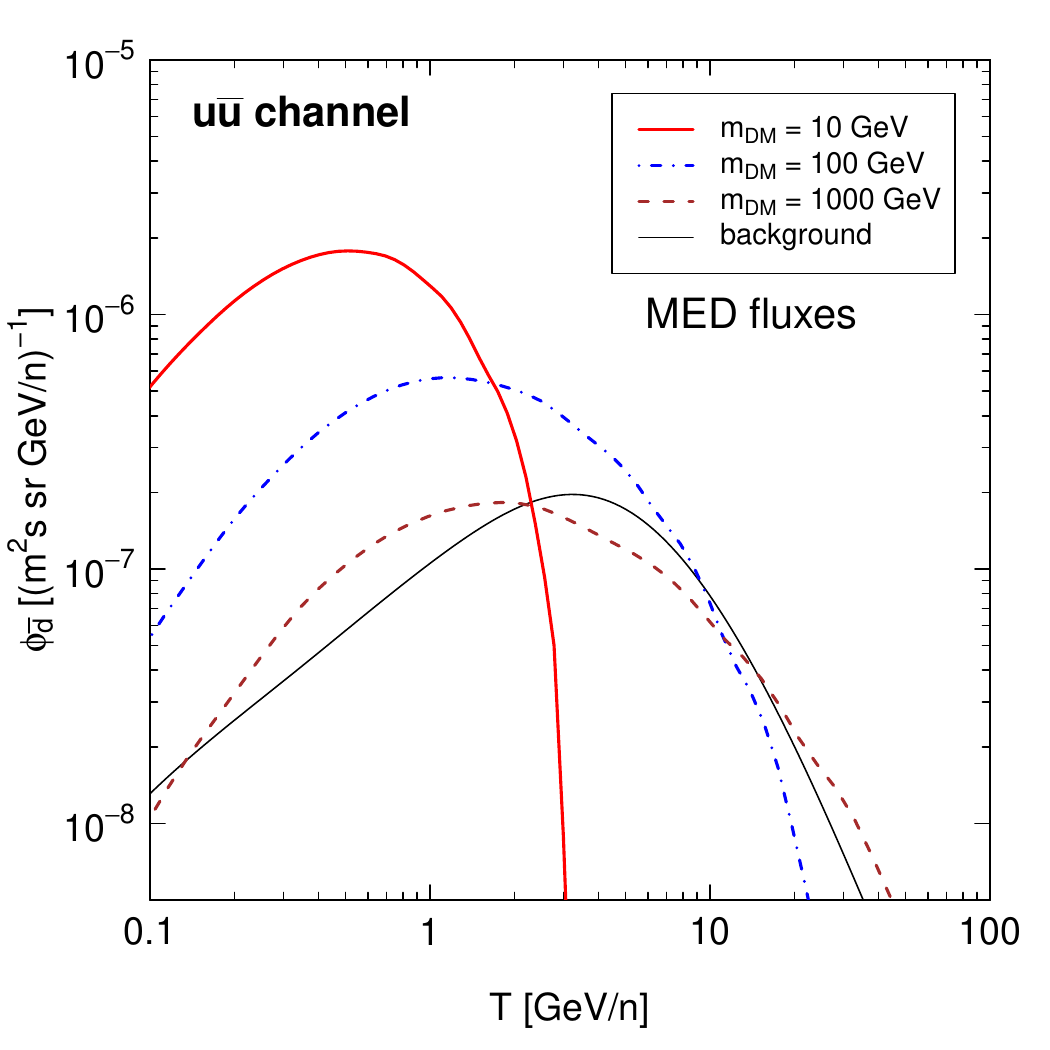}
\includegraphics[width=0.30\textwidth]{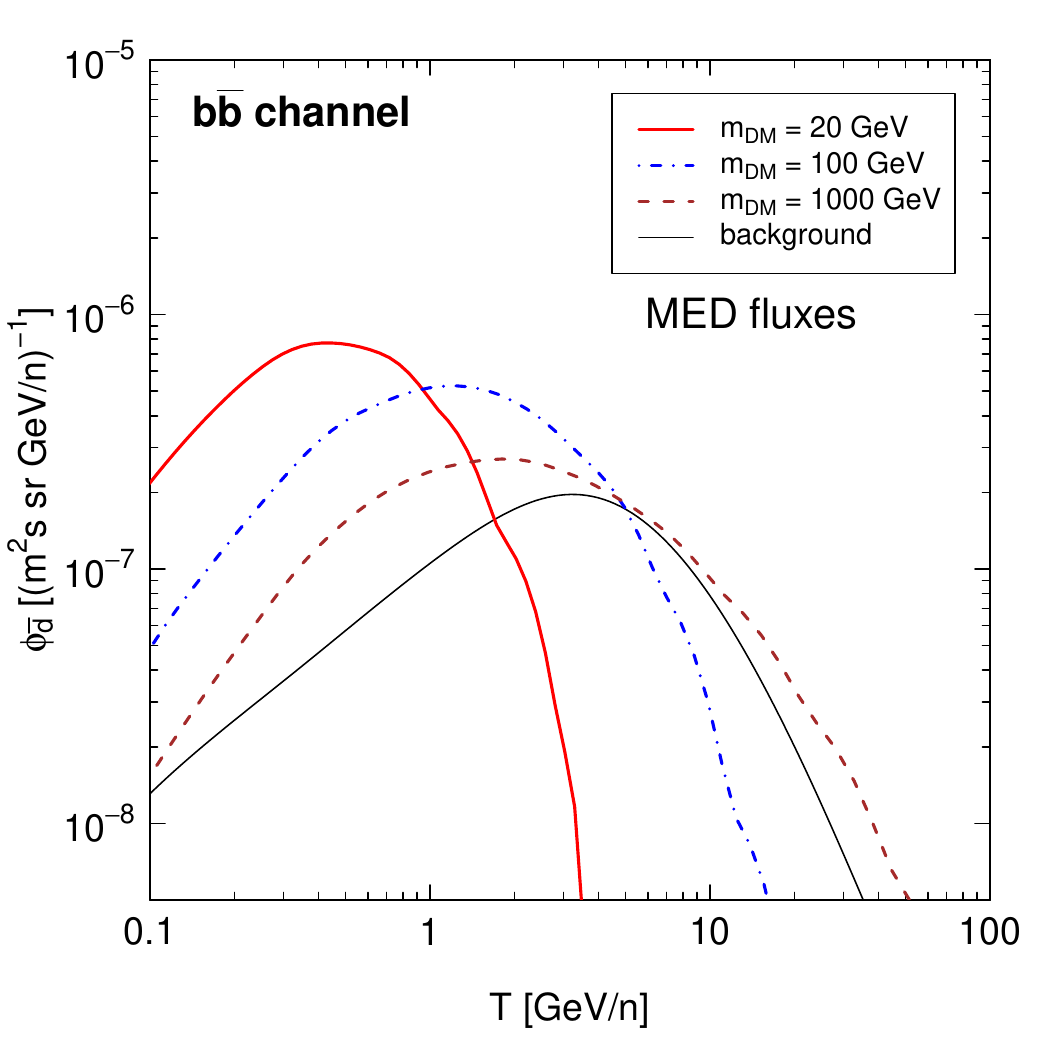}
\includegraphics[width=0.30\textwidth]{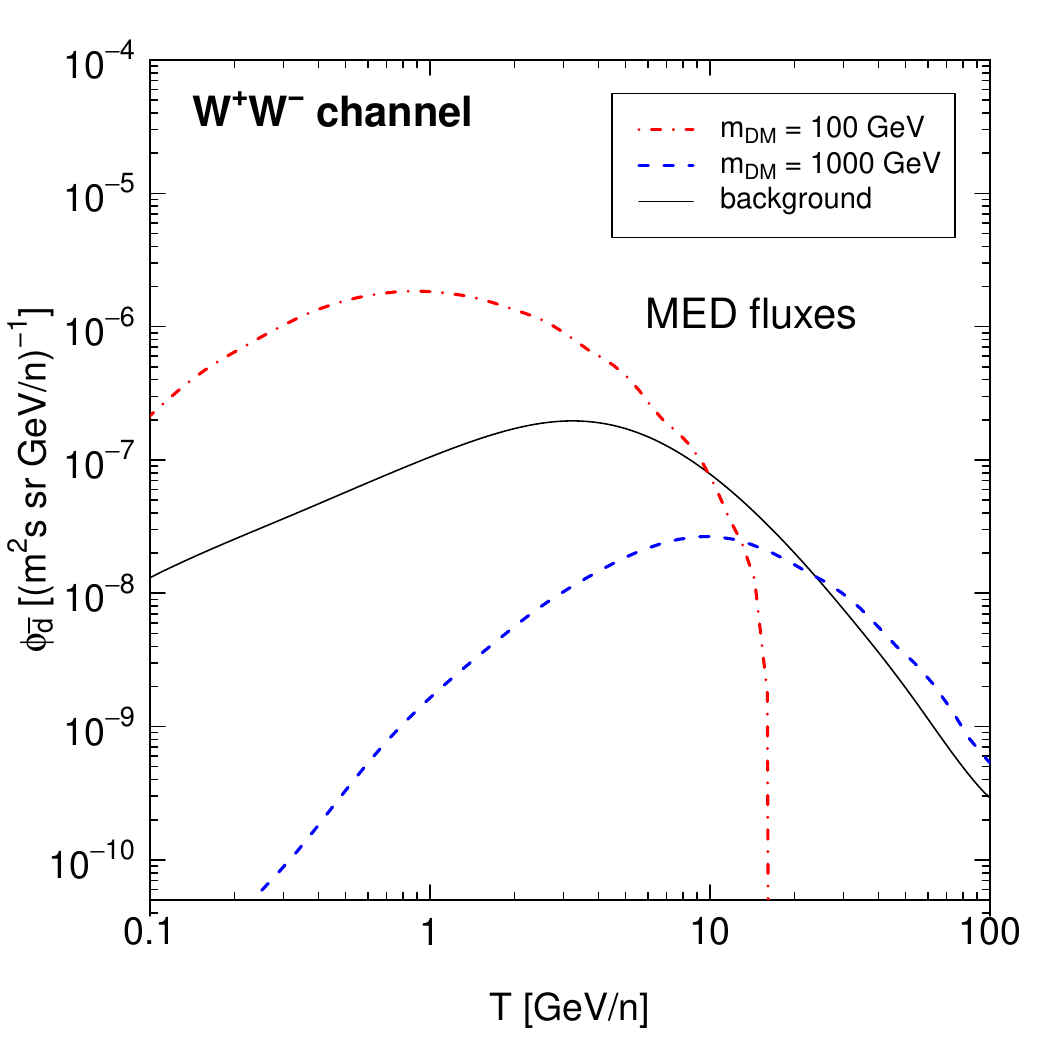}
\caption{Interstellar $\dbar$ spectra for dark matter annihilation in specific (and
representative) production channels: $u \bar u$ (left panel), $b \bar b$ (central panel)
and $W^+W^-$ (right panel). The solid (black) line shows the $\dbar$ secondary background,
the solid (red), dot-dashed and dashed lines refer to a signal produced by a dark matter of mass: 10 GeV (20 GeV for the $b \bar b$ channel), 100 GeV and 1 TeV, respectively.
The annihilation cross section has been fixed, for each curve,
at a value compatible with the antiproton bound coming from PAMELA \cite{Adriani:2010rc}. 
For the $\uubar$ channel: 
$\sigmav=2 \times 10^{-27}$ \cms~ for $\mdm=10$ GeV; 
$\sigmav=2 \times 10^{-26}$ \cms~ for $\mdm=100$ GeV; 
$\sigmav=4 \times 10^{-25}$ \cms~ for $\mdm=1000$ GeV.
For the $\bbbar$ channel: 
$\sigmav=1 \times 10^{-26}$ \cms~ for $\mdm=20$ GeV; 
$\sigmav=4 \times 10^{-26}$ \cms~ for $\mdm=100$ GeV; 
$\sigmav=6 \times 10^{-25}$ \cms~ for $\mdm=1000$ GeV.
For the $\ww$ channel: 
$\sigmav=6 \times 10^{-26}$ \cms~ for $\mdm=100$ GeV; 
$\sigmav=7 \times 10^{-25}$ \cms~ for $\mdm=1000$ GeV.
The galactic dark matter halo is described by an Einasto profile.
\label{fig:ISspectra}}
\end{figure} 
%%%

Figure \ref{fig:ISspectra} illustrates interstellar $\dbar$ fluxes for DM annihilation in specific (and
representative) production channels: $u \bar u$ (left panel), $b \bar b$ (central panel)
and $W^+W^-$ (right panel). The solid (black) line shows the $\dbar$ secondary background \cite{DFM}, the solid (red), dot-dashed and dashed lines refer to a signal produced by a DM of mass: 10 GeV (20 GeV for the $b\bar{b}$ channel), 100 GeV and 1 TeV, respectively. The fluxes have been obtained with the galactic propagation modeling discussed in Sec.~\ref{sec:galaxy}, for the MED set of propagation parameters
reported in Table \ref{tab:parameters}. Each curve for DM signals is compatible with the
the antiproton bound coming from PAMELA \cite{Adriani:2010rc}: cosmic antiprotons represent
a relevant constraint on $\dbar$ searches, since any process able to produce $\dbar$ clearly
produces also a (much larger) flux of antiprotons. We therefore adopt the PAMELA measurement
of the antiproton flux as a constraint on an exotic $\pbar$ component in cosmic rays. We  determine
the antiproton bound by means of a full spectral analysis on the PAMELA data (from 60 MeV to
180 GeV kinetic energy) \cite{Adriani:2010rc}. We use, in deriving the bounds, a 40\% uncertainty
on the theoretical secondary background estimate \cite{Donato:2008jk}.  All antiproton
limits on the DM signal reported in this paper refer to a 3$\sigma$ C.L. bound. 
The secondary $\pbar$ component is modeled as in Ref. \cite{Donato:2008jk}, while the $\pbar$ DM signal is calculated alongside the $\dbar$ signal, with the same
MC modeling of the injection spectra. For the $\pbar$ bound we adopt a set of solar modulation parameters compatible with the PAMELA data-taking period (July 2006 - December 2008): a negative polarity of the solar magnetic field, a tilt angle $\alpha=20^\circ$, a mean free path $\lambda=0.15\:\mathrm{A.U.}$ and a spectral index $\gamma=1$. More details
on the technique adopted to derive the $\pbar$ bounds
are presented elsewhere \cite{promise}. 

The antiproton bound translates into an upper limit on the DM annihilation cross section
$\sigmav$ as a function of the DM mass, once the annihilation channel is fixed. The $\dbar$ spectra reported in Fig.~\ref{fig:ISspectra}
refer to values of $\sigmav$ at the level of the antiproton bound. The specific values are reported
in the caption of the figure. Figure \ref{fig:ISspectra} shows that even with the current updated 
bounds on $\pbar$ from the PAMELA measurements, $\dbar$ fluxes largely in excess of the
secondary background are present: this occurs typically at low energies (below a few GeV
for the $\dbar$ kinetic energy) and for DM masses at least up to 100 GeV. On the
contrary, larger DM masses
(as can be seen in the right panel of Fig.~\ref{fig:ISspectra} for the $\mdm = 1$~TeV case)
predict $\dbar$ signals able to dominate the background at energies above a few tens of GeV
(an energy region which is currently not easily accessible experimentally).

 %%%
\begin{figure}[t]
\centering
\includegraphics[width=0.40\textwidth]{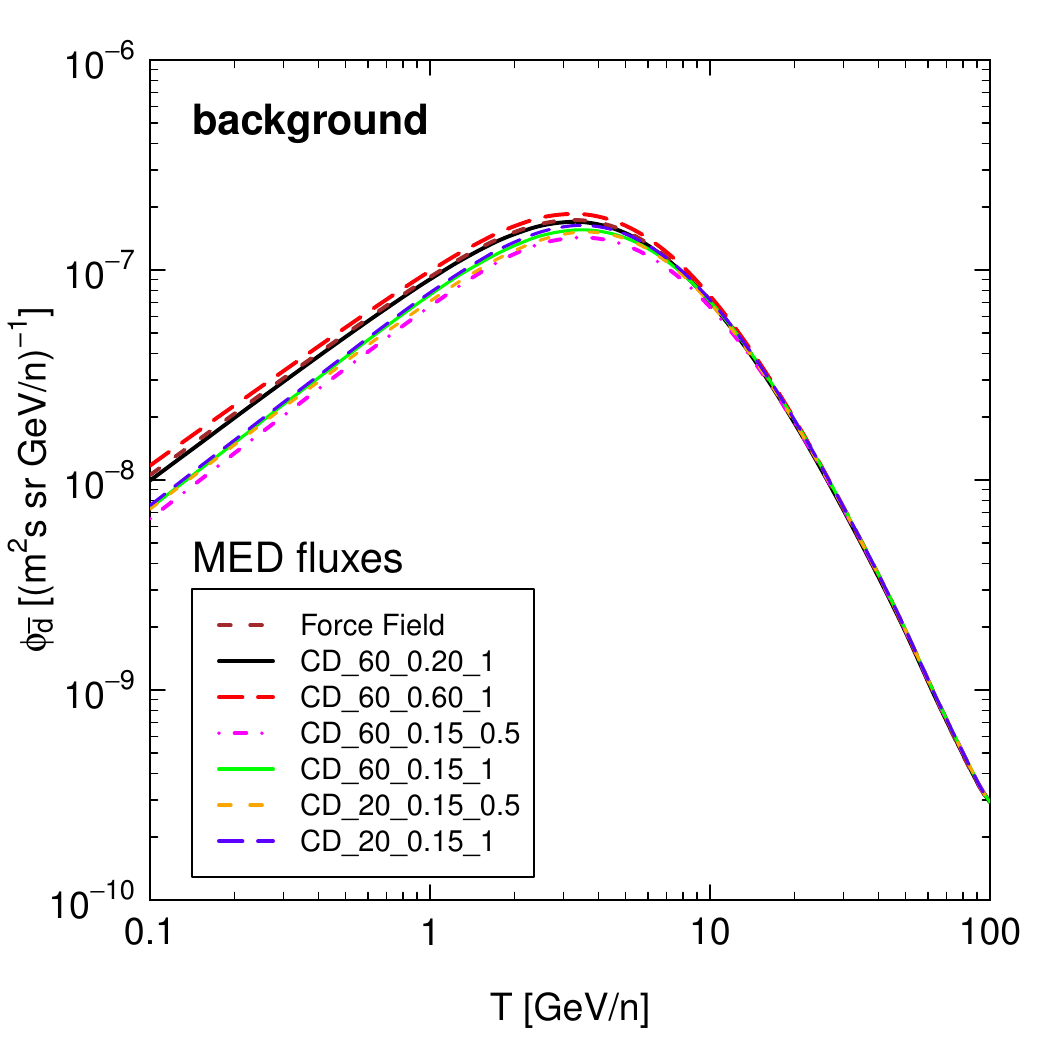}
\caption{Top-of-atmosphere $\dbar$ flux as a function of the $\dbar$ kinetic energy, for the
secondary background component. Galactic propagation adopts the MED parameters set of
Table \ref{tab:parameters}. The different curves refer to different solar modulation models
reported in Table \ref{tab:solarmod}, and outlined in the boxed inset.
\label{fig:TOAback}}
\end{figure}
%%%

\begin{figure}[t]
\centering
\includegraphics[width=0.35\textwidth]{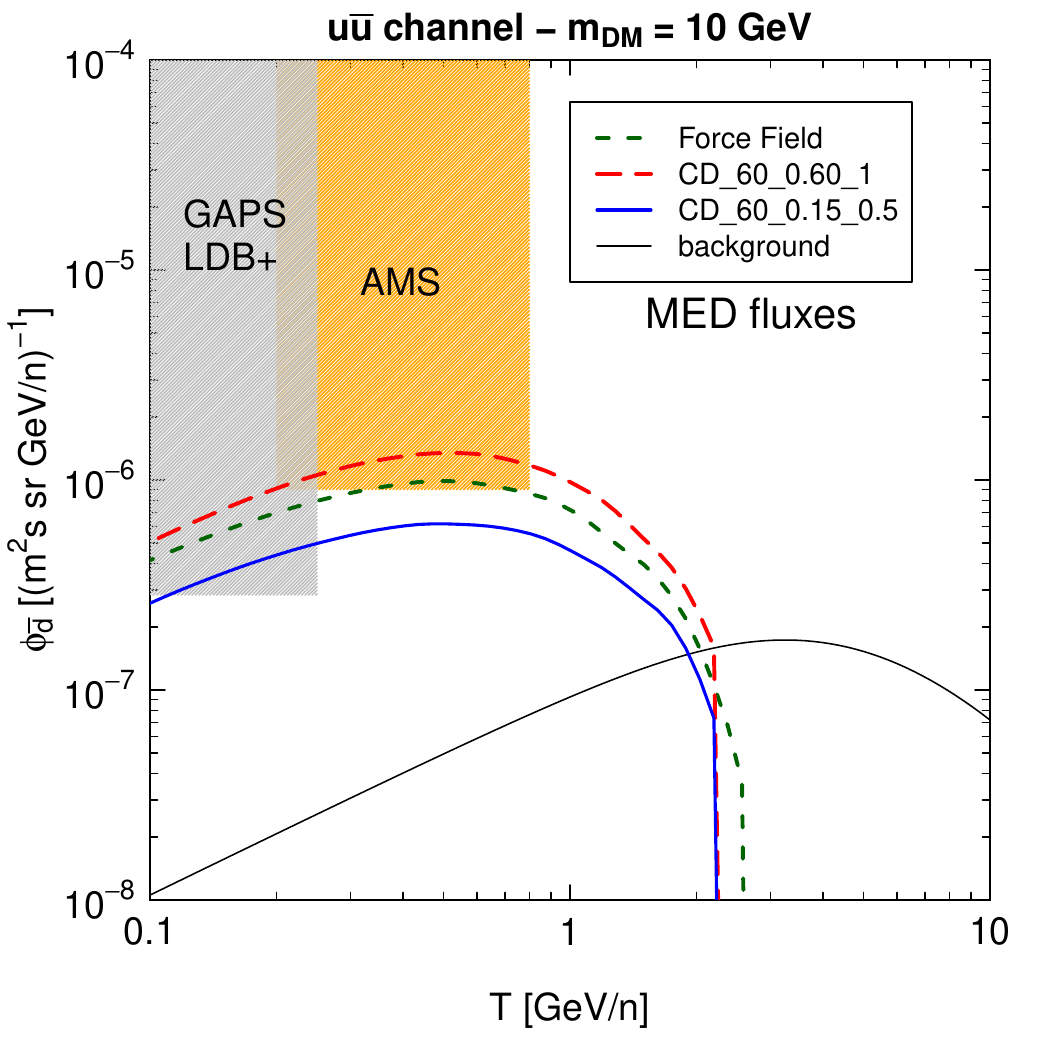}
\includegraphics[width=0.35\textwidth]{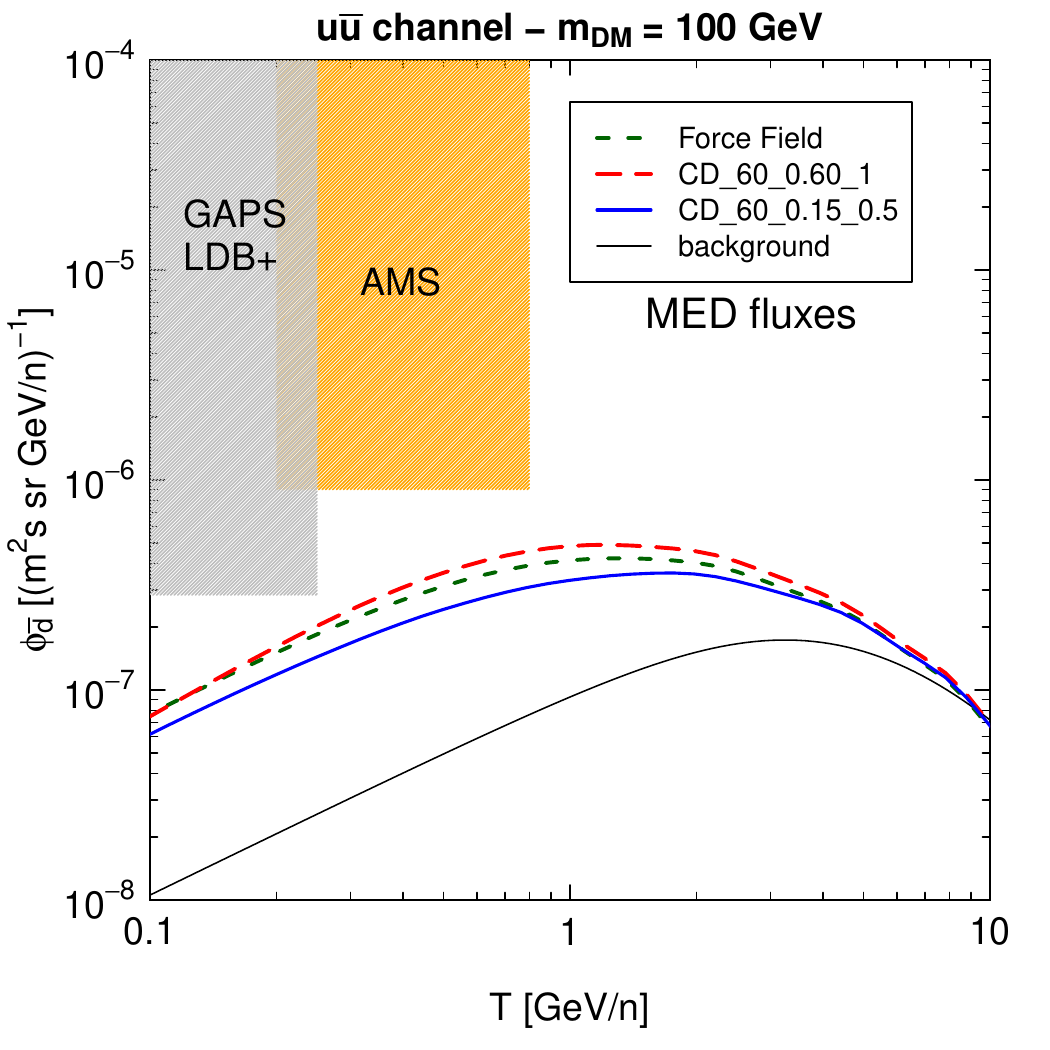}
\caption{Top-of-atmosphere $\dbar$ flux as a function of the $\dbar$ kinetic energy, for 
dark matter signal production in the $\uubar$ channel. The left panel refers to a dark matter
mass of 10 GeV, the right panel to 100 GeV. The solid (black, lower) curve stands for the secondary
background, calculated in the force-field solar modulation approach. In each panel, the other three curves report the signal calculated for three different
solar modulation models (defined in the boxed inset, code name refers to 
Table \ref{tab:solarmod}). Annihilation
cross sections are those reported in Fig. \ref{fig:ISspectra} for the $\uubar$ channel. 
The galactic dark matter halo is described by an Einasto profile. Galactic propagation adopts the MED parameters set of Table \ref{tab:parameters}. 
 Shaded regions refer to
the $3\sigma$ C.L. expected sensitivities of GAPS LDB+ (referring to 1 detected event) \cite{haileypriv} and AMS-02
(referring to 2 detected events, derived from the flux-limit of Refs. \cite{giovacchini,choutko})}.
\label{fig:TOAuu}
\end{figure}

\begin{figure}[t]
\centering
\includegraphics[width=0.35\textwidth]{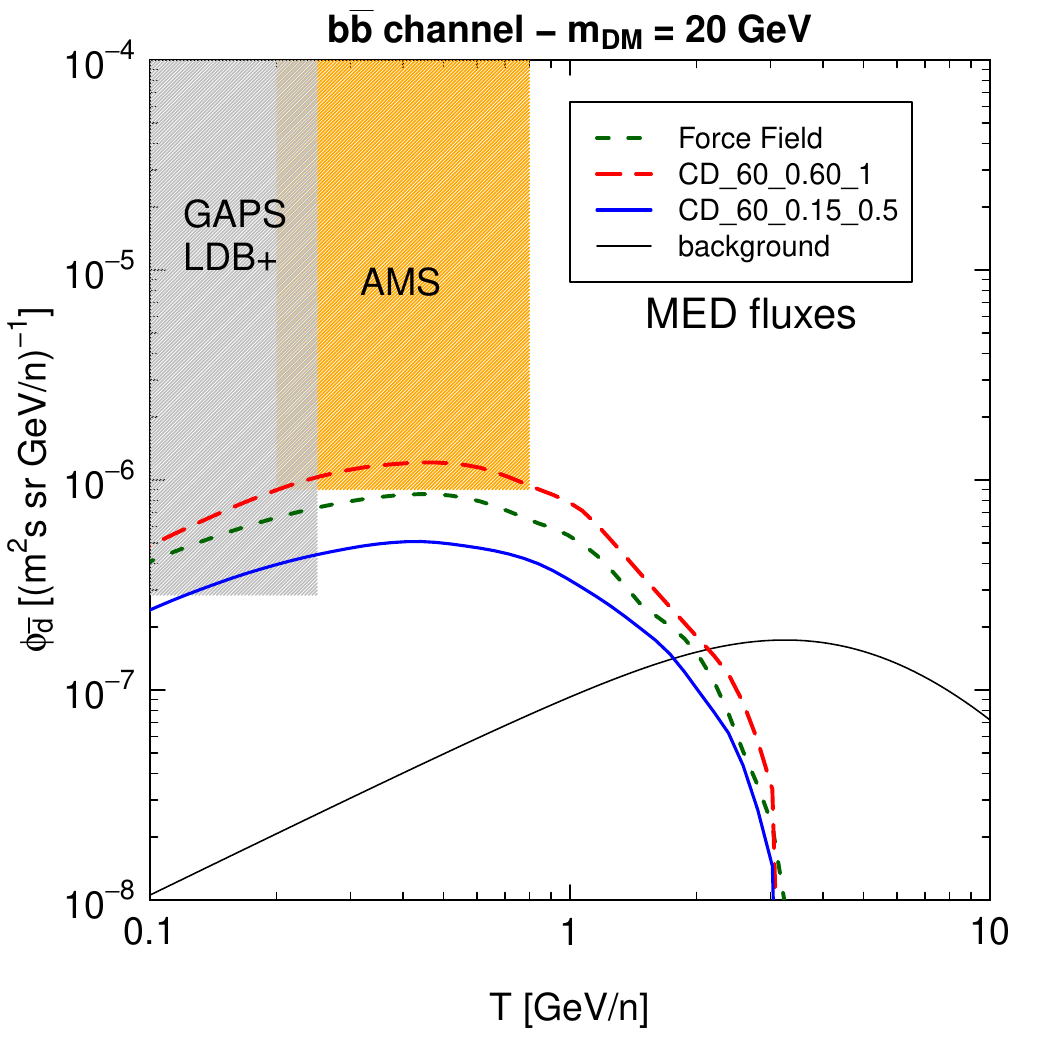}
\includegraphics[width=0.35\textwidth]{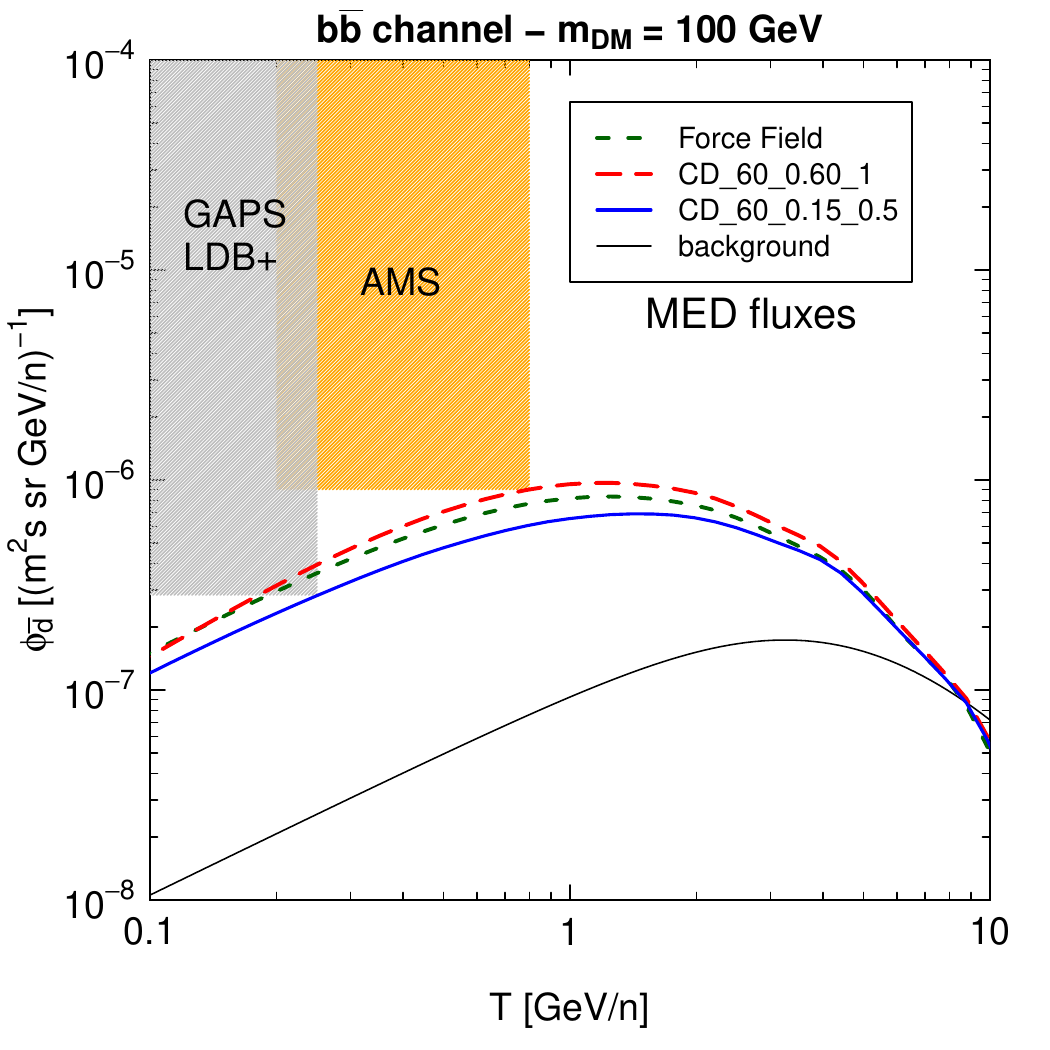}
\caption{Top-of-atmosphere $\dbar$ flux as a function of the $\dbar$ kinetic energy, for 
dark matter signal production in the $\bbbar$ channel. The left panel refers to a dark matter
mass of 20 GeV, the right panel to 100 GeV. Notations are as in Fig. \ref{fig:TOAuu}. Annihilation
cross sections are those reported in Fig. \ref{fig:ISspectra} for the $\bbbar$ channel. 
\label{fig:TOAbb}}
\end{figure}

\begin{figure}[t]
\centering
\includegraphics[width=0.35\textwidth]{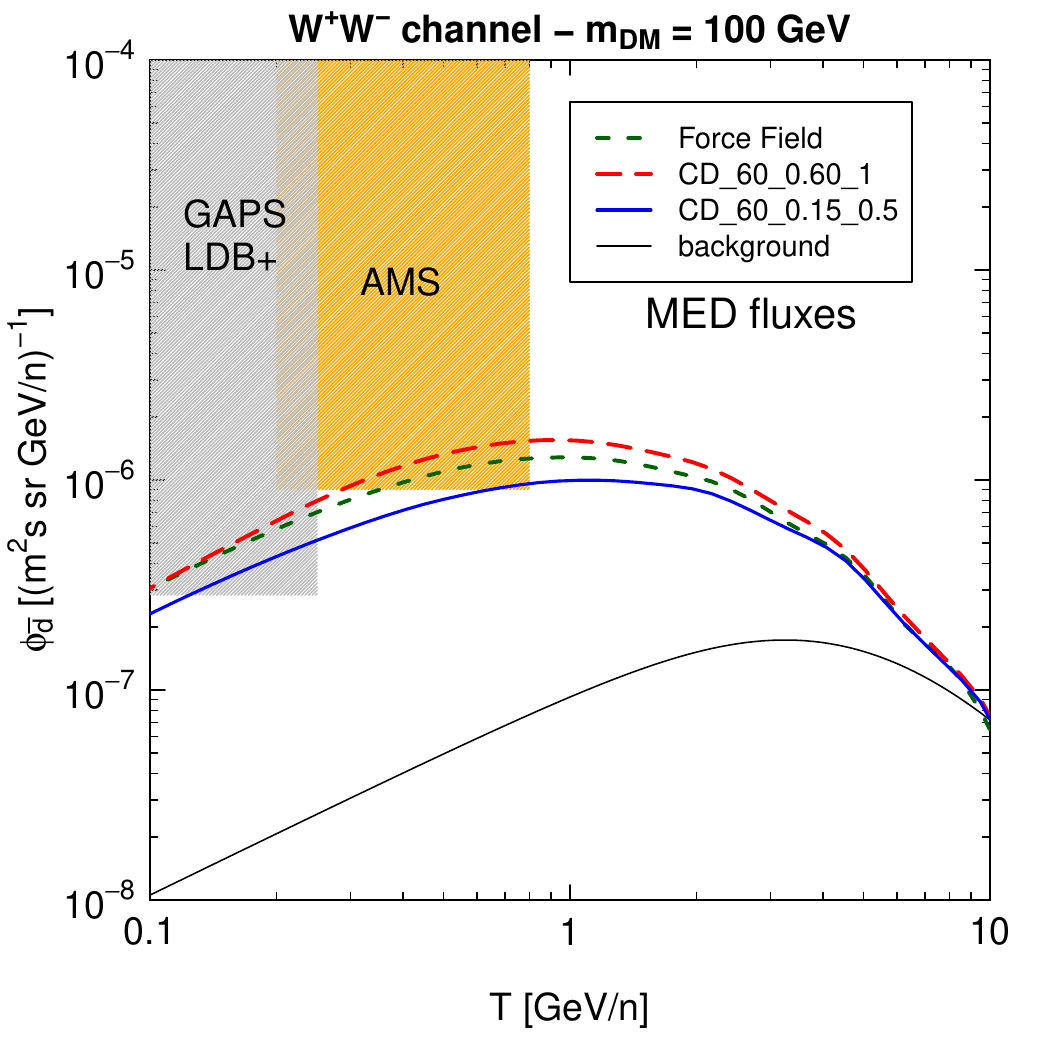}
\includegraphics[width=0.35\textwidth]{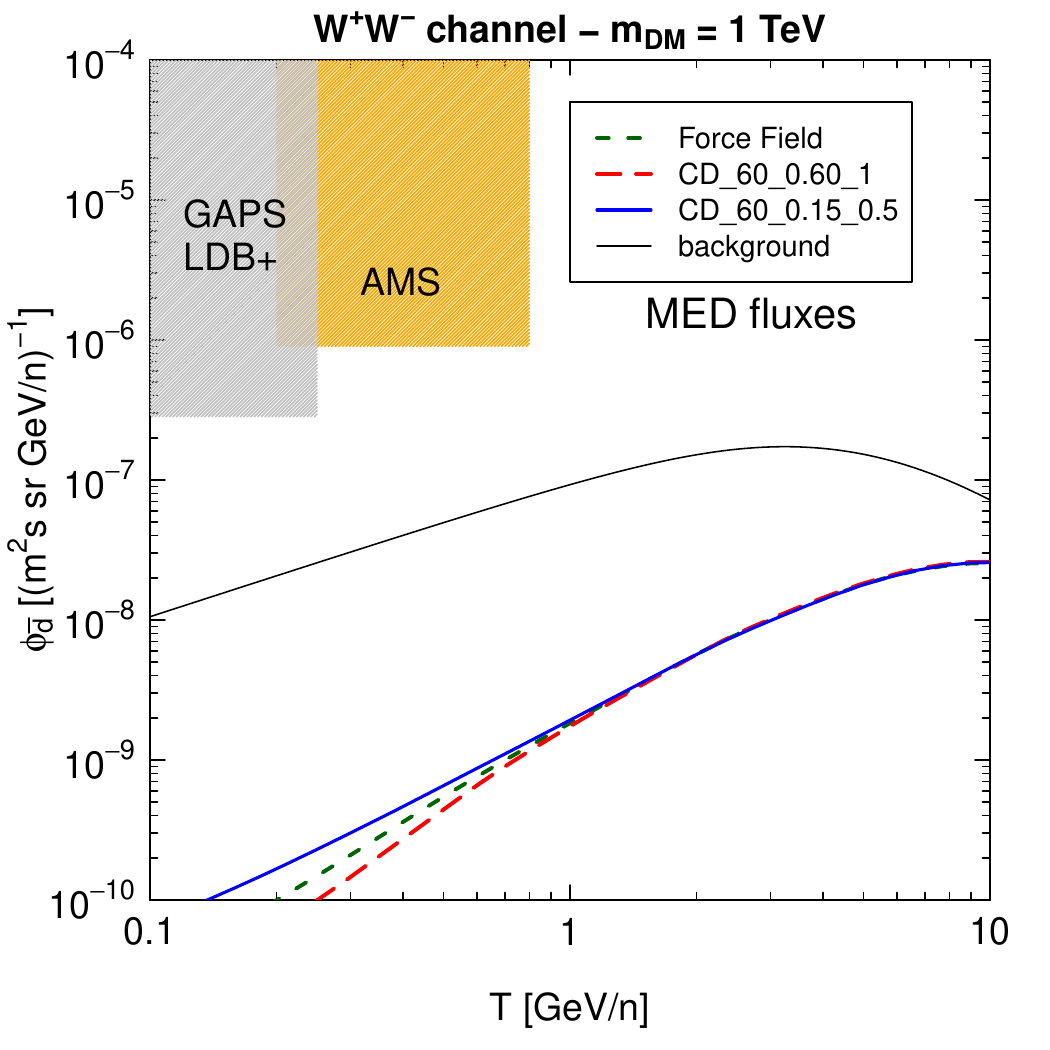}
\caption{
Top-of-atmosphere $\dbar$ flux as a function of the $\dbar$ kinetic energy, for 
dark matter signal production in the $\ww$ channel. 
The left panel refers to a dark matter
mass of 100 GeV, the right panel to 1 TeV. 
Notations are as in Fig. \ref{fig:TOAuu}. Annihilation
cross sections are those reported in Fig. \ref{fig:ISspectra} for the $\ww$ channel. 
\label{fig:TOAWW}}
\end{figure}

The actual relevance for cosmic $\dbar$ detection is nevertheless seen after solar modulation is applied.
Solar modulation affects mostly the low energy part of the cosmic rays spectra, which is
 the place where the $\dbar$ signal is expected to differ more significantly from the background,
 and the place where current experimental efforts concentrate. We therefore wish to understand
 how significant is the impact of modeling of  cosmic rays transport in the heliosphere. This
 effect is illustrated in Fig.~\ref{fig:TOAback} for the secondary background and in Fig.~\ref{fig:TOAuu}, \ref{fig:TOAbb} and \ref{fig:TOAWW} for the
 same representative DM models of Fig.~\ref{fig:ISspectra}. In each figure, the specific models of solar modulation
 which have been applied are reported in the boxed insets, where the labeling defined
 in Table \ref{tab:solarmod} has been adopted.
 
 %%%
%%%

The top-of-atmosphere (TOA) flux of the background component, reported in Fig.~\ref{fig:TOAback}, 
shows that solar modulation modeling affects the predicted fluxes by about 50\%
in all the energy range from $T=0.1$ GeV/n up to the peak of the $\dbar$ secondary component, 
around $T=$ 3--4 GeV/n. This variation is of the same order as the uncertainty on the
secondary $\dbar$ flux due to nuclear physics modeling \cite{DFM}. The impact of
solar modulation modeling on the $\dbar$ secondary flux is not dramatic, although a
change by about a factor of 1.5 from the lowest to the highest prediction can have impact
on the DM searches for a signal. When comparing solar modulation models with the
standard force-field approximation, a reduction of about 35\% of the background
component at low kinetic energies is predicted in solar modulation models with large 
tilt angle $\alpha$, small mean-free-path $\lambda_{\|}$ and small spectral index $\gamma$ of
the diffusion coefficient in the solar system (while an increase of a mere 10\% in the
other extremal case, with 
large $\alpha$, large $\lambda_{\|}$ and large $\gamma$).

Propagation inside the heliosphere has somewhat larger impact on the DM $\dbar$ fluxes, especially
for those fluxes produced by light DM, which are peaked at lower energies before
solar modulation is applied.
This is seen in Fig.~\ref{fig:TOAuu}, \ref{fig:TOAbb} and \ref{fig:TOAWW}, 
where also the
expected sensitivity for  GAPS LDB+  and  AMS-02
are reported. We define as ``expected sensitivity" the flux level that corresponds
to the detection of a $\dbar$ signal (on top of the secondary background) with a confidence level of $3\sigma$. The sensitivity of GAPS LDB+ is obtained from Ref. \cite{haileypriv}.
The sensitivity of AMS-02 is derived from the design flux limit of Ref. \cite{giovacchini,choutko,giovacchinipriv}, which corresponds to 1 detected event in the original
AMS-02 setup \cite{giovacchinipriv}. A further discussion on this point is reported in Sec. \ref{sec:prospects}.
As for AMS-02, we notice that the expected
sensitivity derived from Ref. \cite{giovacchini} should be updated to the new detector setup,
which has been changed from the original proposal to the actual flight configuration \cite{Battiston:2008zza,Incagli:2010zz,Bertucci:2011zz,Kounine:2012zz}.
While a reduction in the strength of the detector magnet may reduce performance, a much longer 
duration of the
flight onboard of the International Space Station can compensate in the final exposure. A dedicated
Monte Carlo modeling of the new setup is required to redefine the updated sensitivity.
For definiteness, we will adopt here the  AMS-02 design sensitivity \cite{giovacchini}
throughout the paper.

%%%
%%%

%%%
%%%

Signal fluxes produced by light (10--20~GeV) dark matter (left panels in Fig.~\ref{fig:TOAuu} and \ref{fig:TOAbb})
are peaked in $T$ around 400--500~MeV/n after solar modulation. This is the relevant range for
current operating and designed detectors. We see that the variation on the fluxes, due to solar
modulation modeling, reaches a maximum of a factor 2.5. For heavier DM (masses
around 100 GeV) the effect is reduced to the level of 1.5 from the lowest to the largest
predicted flux. The heliosphere models which predict larger DM fluxes are the same that also predict larger background
fluxes, as can be seen by comparing Fig.~\ref{fig:TOAuu}, \ref{fig:TOAbb}, \ref{fig:TOAWW}
with Fig.~\ref{fig:TOAback}. This behavior is reversed for DM spectra peaked at large energies
(above 10 GeV/n). This is manifest in the right panel of
Fig.~\ref{fig:TOAWW}, where the DM mass is set at 1 TeV. The solar model with parameters
$\alpha = 60^\circ, \lambda_{\|} = 0.60\;\mbox{A.U.}, \gamma=1.0$ (model CD\_60\_0.60\_1) now predicts the smallest
flux, contrary to the previous cases. This is due to the fact that in this case the peak of the $\dbar$ LIS happens to be at too large energies, at which diffusion in the solar system is so fast that energy losses are negligible. Therefore the modulated spectrum at low energy does not receive a relevant contribution from the peak, in contrast with the other cases, and its slope can be simply understood as the one of the LIS modified by the energy dependent diffusion time.

Figures \ref{fig:TOAuu}, \ref{fig:TOAbb} and \ref{fig:TOAWW}  show that, in the case of light DM annihilating dominantly into quarks, the expected sensitivities of GAPS and AMS-02 yield good prospects to see a DM signal, while for a DM mass greater than 100 GeV the sensitivity fades away. Even in the case of
a dominant $\ww$ annihilation, a DM with $\mdm = 100$ GeV has good chances to be detected.
%better chances to be detected than
%n the quark-dominated case. 
The signal from heavy DM becomes progressively depressed at low
kinetic energies when $\mdm$ increases: however, the signal starts exceeding the background for
$\dbar$ kinetic energies above a few GeV/n, as can be deduced for $T$ larger than 10 GeV
in the $\ww$ channel and  $\mdm=1$ TeV in Fig.~\ref{fig:TOAWW}. We will not specifically discuss
this energy range, since current designed and operating detectors will concentrate on $\dbar$ energies
below 1 GeV/n.

\begin{figure}[t]
\centering
\includegraphics[width=0.35\textwidth]{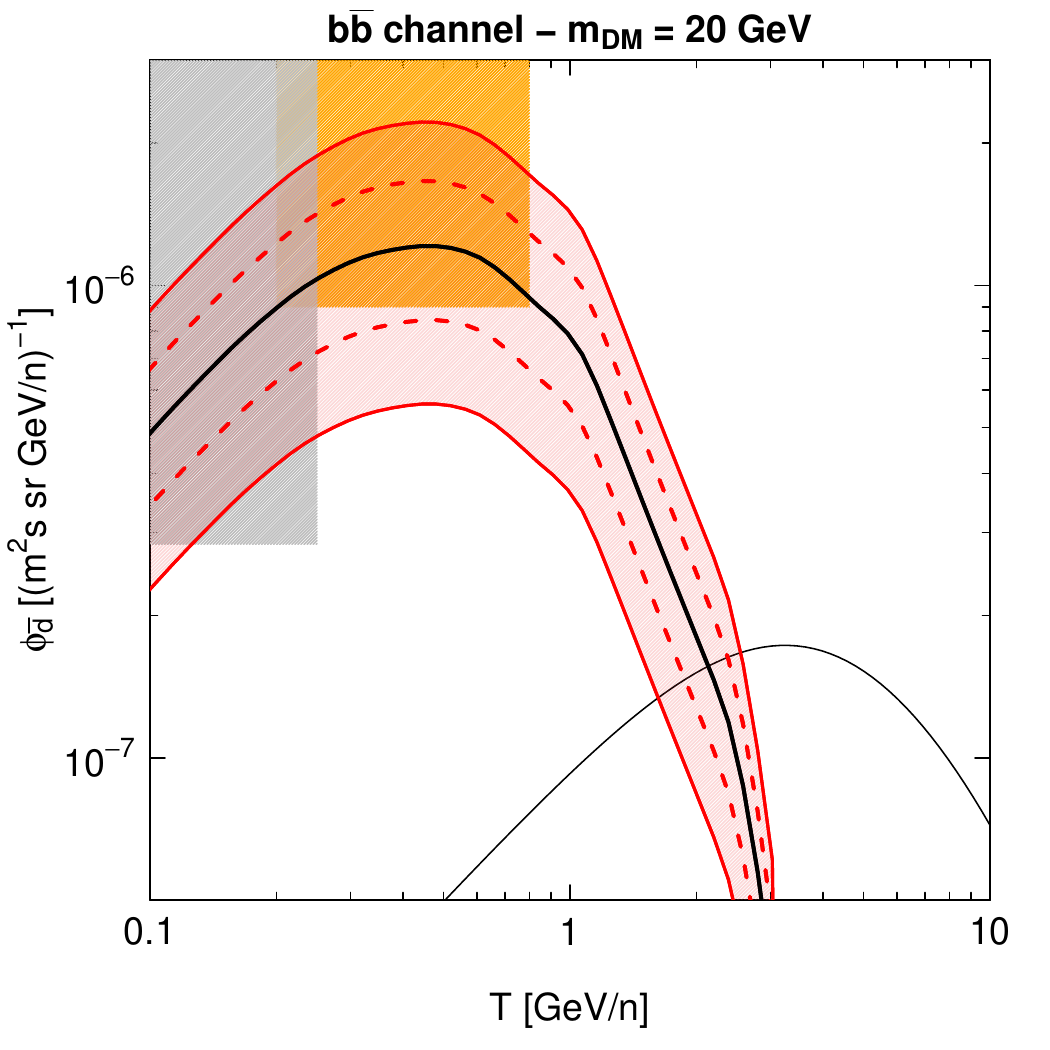}
\includegraphics[width=0.35\textwidth]{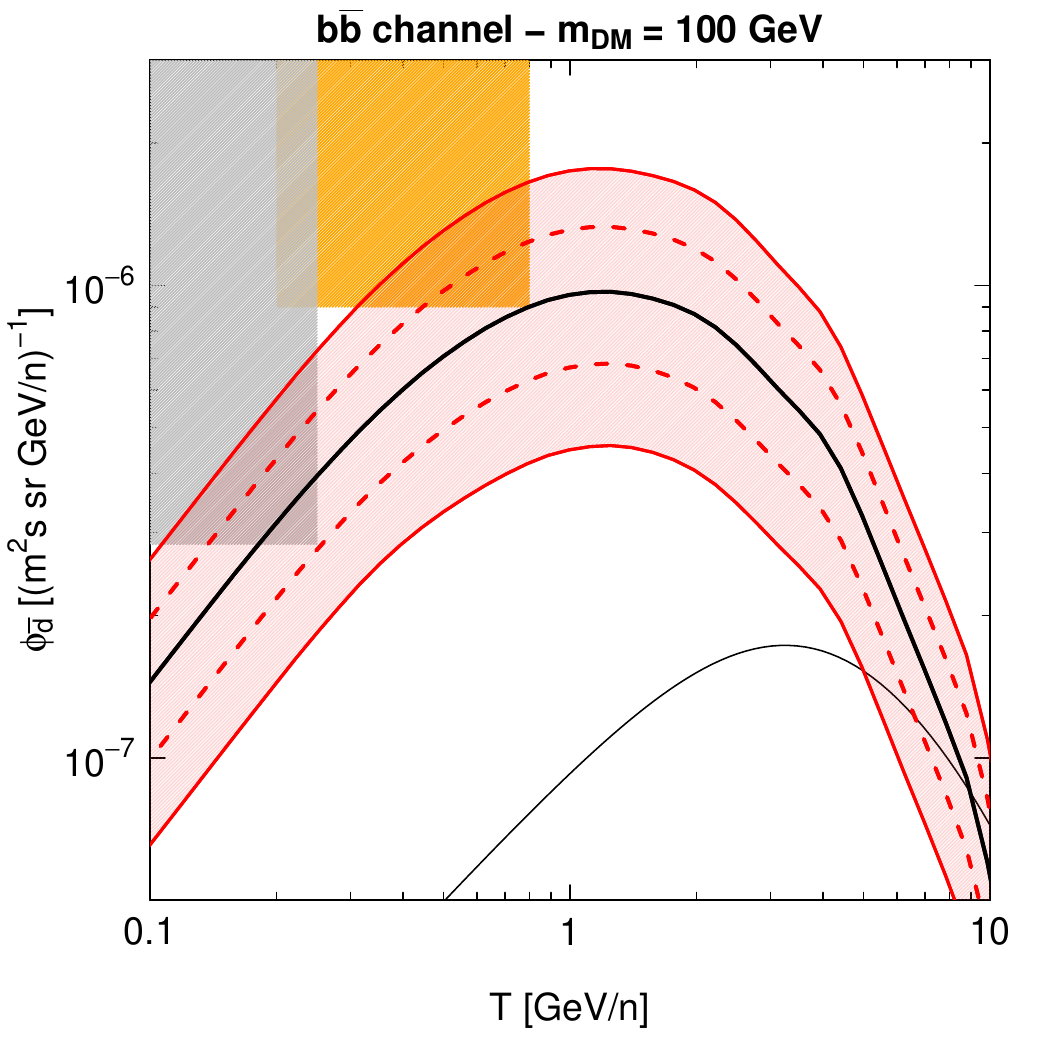}
\caption{The same as in Fig. \ref{fig:TOAbb}, for the solar modulation model with
 $\alpha = 60^\circ$, $\lambda_{\|} = 0.60\;\mbox{A.U.}$ and
 $\gamma=1.0$ (model CD\_60\_0.60\_1),
with a variation of the coalescence momentum $p_0$ inside its 1 and $2 \sigma$ allowed ranges.
\label{fig:TOAbb2}}
\end{figure}

The dependence of the above predictions on the coalescence momentum $p_0$ is 
analyzed in Fig.~\ref{fig:TOAbb2}, where we show the effect of varying $p_0$ inside its $2\sigma$ allowed interval, for the same configuration as Fig.~\ref{fig:TOAbb}
(i.e.~DM annihilating into $\bbbar$) and solar modulation model CD\_60\_0.60\_1.
 We notice that prospects of detection are now well within
reach also for $\mdm=100$~GeV. Similar extension of the uncertainty band arising from
the determination of $p_0$ applies also to the other cases shown in Figs.~\ref{fig:TOAuu}
and  \ref{fig:TOAWW}: approximately, the fluxes obtained with the best fit of $p_0$
reported in Table \ref{tab:p0} increase (decrease) by a factor of 1.8 (0.5) when  $p_0$ is
varied inside its $2\sigma$ allowed range. This implies that, considering theoretical uncertainties,
current and designed sensitivities may have the possibility to explore the $\dbar$ signal in the whole 
DM mass range up to few hundreds of GeV. We will come back to this point, with more quantitative statements, in the next Section.

\begin{figure}[t]
\centering
\includegraphics[width=0.35\textwidth]{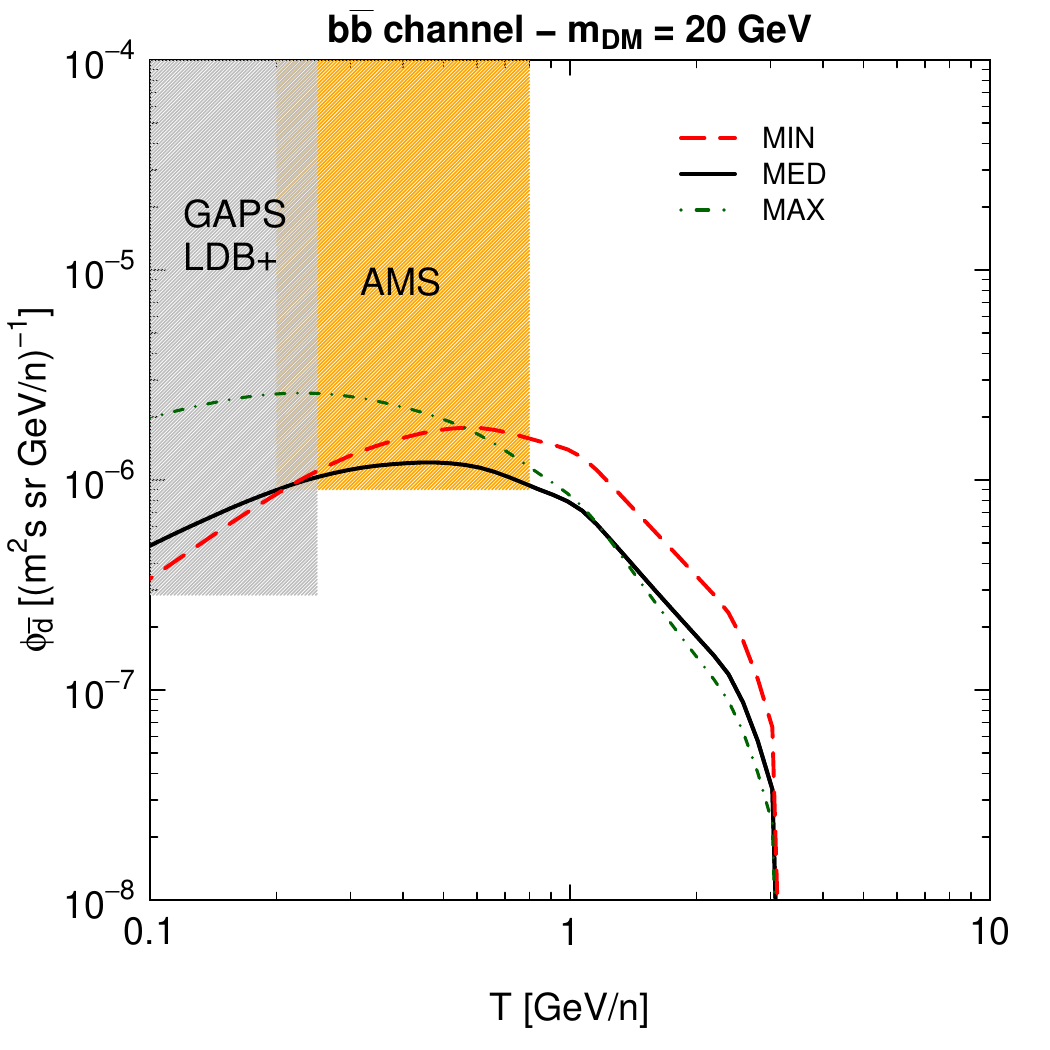}
\includegraphics[width=0.35\textwidth]{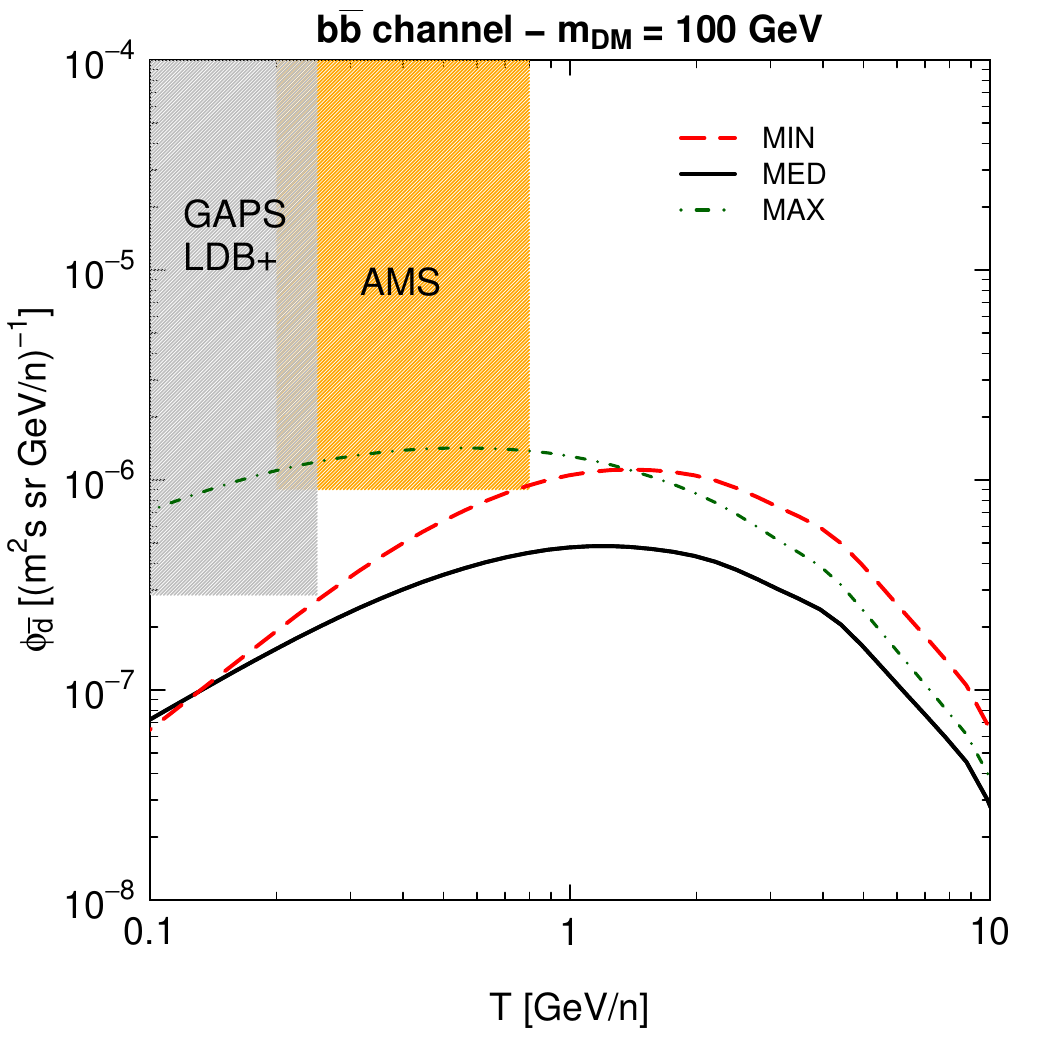}
\caption{The same as in Fig. \ref{fig:TOAbb}, for the different galactic propagation
models MIN/MED/MAX reported in Table \ref{tab:parameters}. The 
solar modulation model has been fixed to:
 $\alpha = 60^\circ$, $\lambda_{\|} = 0.60\;\mbox{A.U.}$ and
 $\gamma=1.0$ (model CD\_60\_0.60\_1). For each set of curves, the
 DM annihilation cross section has been fixed at its maximal value allowed by antiproton
 bounds. For $\mdm=20$ GeV: $\sigmav = 2\times10^{-25},1\times 10^{-26},2\times 10^{-27}$ cm$^3$ s$^{-1}$ for the MIN, MED, MAX model, respectively.
 For $\mdm=100$ GeV: $\sigmav = 5\times 10^{-25},4\times 10^{-26},1\times 10^{-26}$ cm$^3$ s$^{-1}$ for the MIN, MED MAX model, respectively.
\label{fig:TOAbb3}}
\end{figure}

Concerning the variation of the galactic propagation models, we show in Fig.~\ref{fig:TOAbb3}
the same models for $\bbbar$ annihilation of Fig.~\ref{fig:TOAbb}, with the variation
of the galactic propagation parameters among the MIN/MED/MAX models of Table \ref{tab:parameters}. For definiteness, the solar modulation model has been fixed to be
the one that provides the largest flux in Fig.~\ref{fig:TOAbb}, (i.e. the model CD\_60\_0.60\_1). Propagation in the MAX or MIN
 model deforms the spectral shape of the interstellar flux, and this reflects in modified 
 TOA fluxes, too.  In particular, the MAX model allows to replenish the low-energy tail of the $\dbar$ fluxes, and this implies increased prospects of detection at low kinetic energies, as can be seen in Fig.~\ref{fig:TOAbb3}, both for $\mdm=10$ GeV and, even more, for $\mdm=100$ GeV. The difference
 at 100 MeV/n between the MIN and MAX fluxes can reach one order of magnitude. We should
 mention here that in allowing the variation from the MED set  to e.g. the MAX set, we obtain both larger $\dbar$ fluxes
and larger antiproton fluxes, for a fixed annihilation cross section and DM mass. Since we are enforcing here the bound coming from the PAMELA $\pbar$ measurements, the MAX set imposes a stronger
bound on the  annihilation cross section $\sigmav$ to compensate the enhancement induced by the MAX propagation model. We have taken into account this feature, and in fact the annihilation cross sections
used in the calculations of the fluxes in Fig.~\ref{fig:TOAbb3} are different for the three
different galactic propagation models (as dictated by the $\pbar$ bounds). The specific values
are reported in the caption of Fig.~\ref{fig:TOAbb3}.

%%%
%%%

%%%
%%%

%%%
\begin{figure}[t]
\centering
\includegraphics[width=0.30\textwidth]{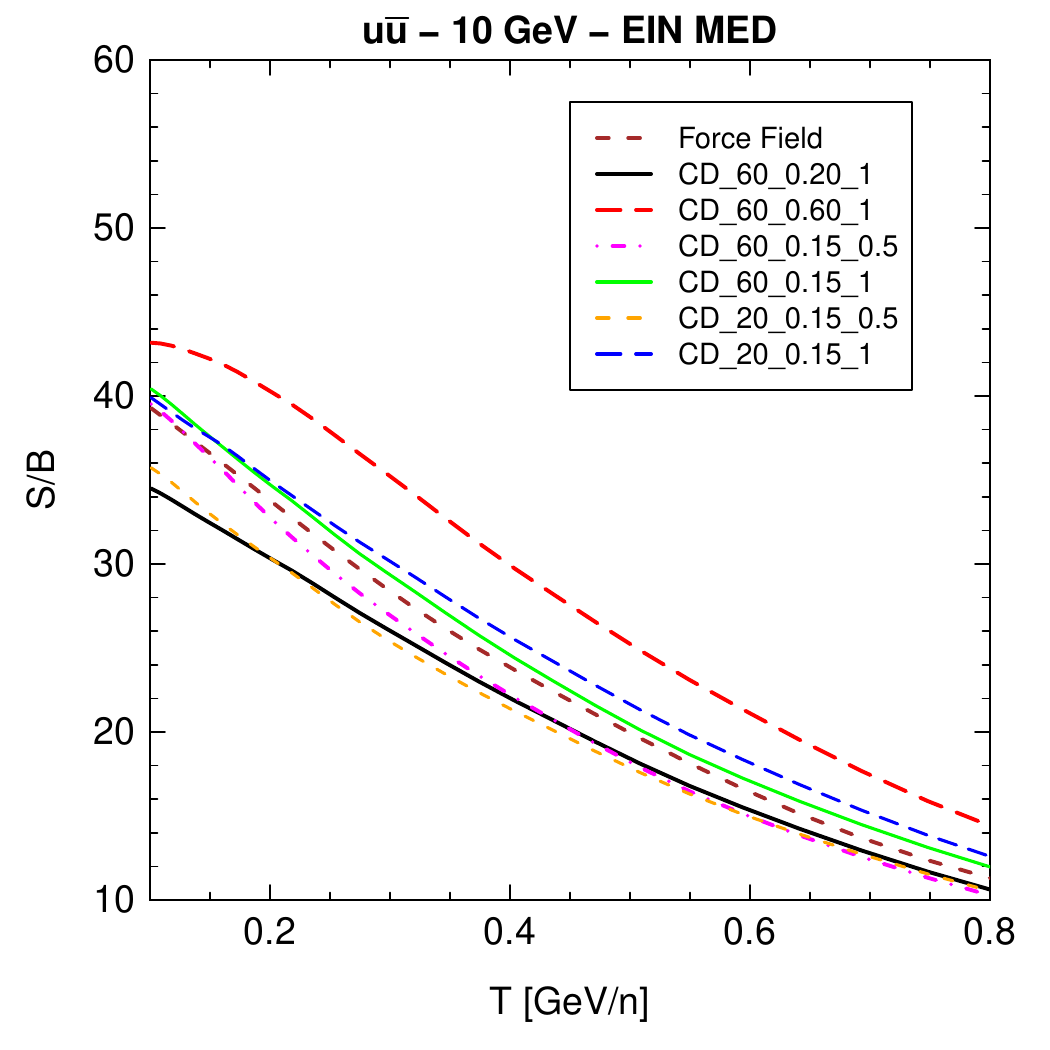}
\includegraphics[width=0.30\textwidth]{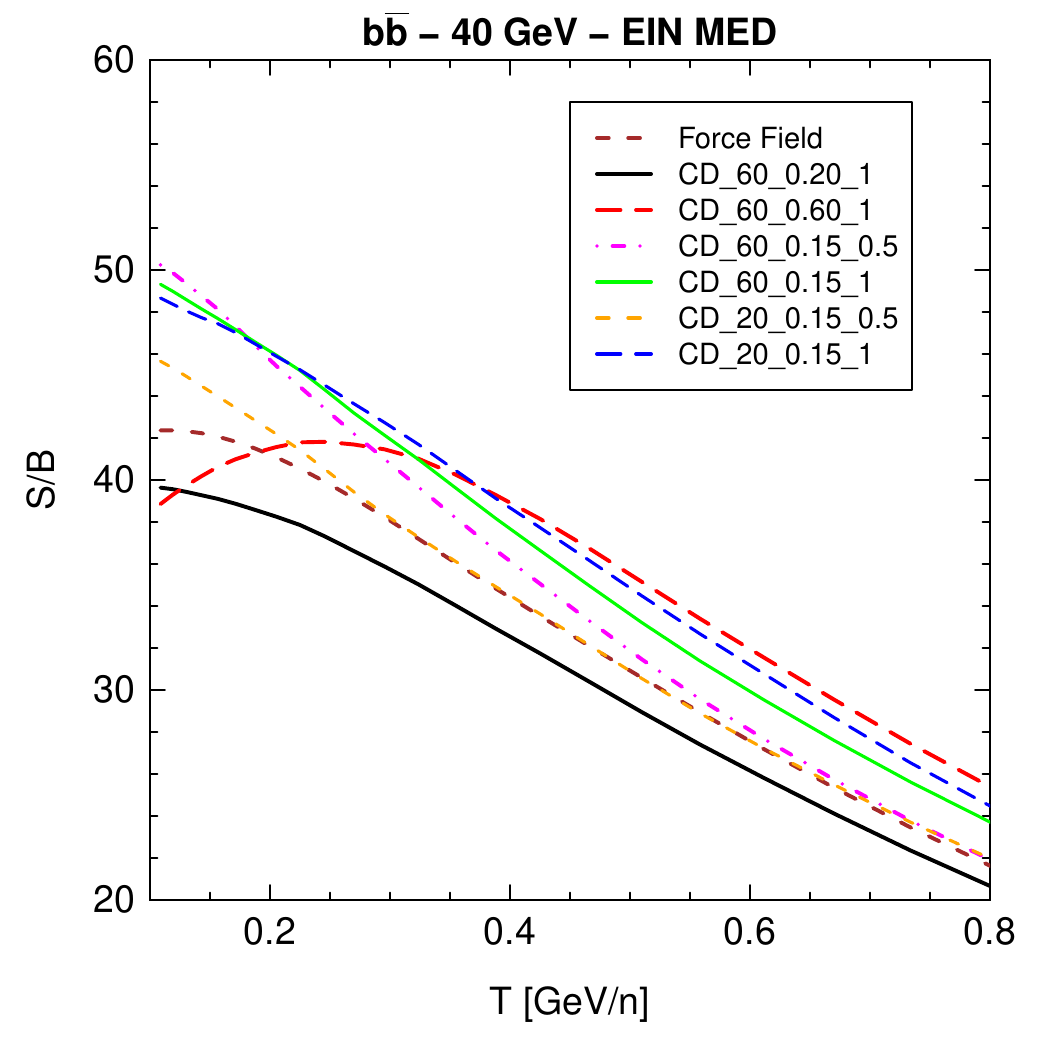}
\includegraphics[width=0.30\textwidth]{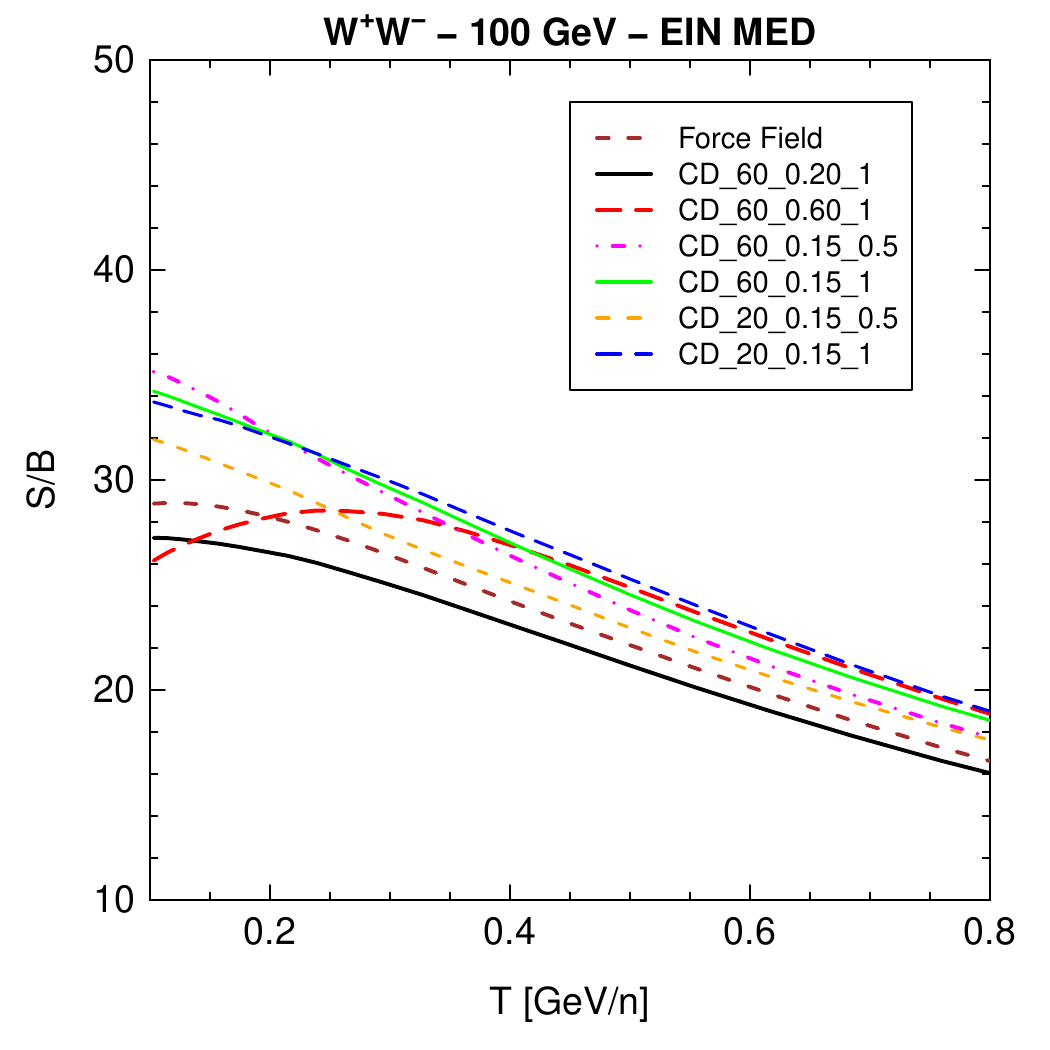}
\caption{Signal-to-background $S/B$ ratios as a function of the $\dbar$ kinetic energy, for three representative DM models
and for the different
solar modulation models listed in Table \ref{tab:solarmod} (the correspondence between the models
and the curves is reported in the boxed insets, code name refers to
Table \ref{tab:solarmod}). The left, central and right panels refer
to a dark matter with $\mdm=10$ GeV annihilating into $\uubar$, 
a dark matter with $\mdm=40$ GeV annihilating into $\bbbar$ (with a cross section $\sigmav=2 \times 10^{-26}$ \cms) and
a dark matter with $\mdm=100$ GeV annihilating into $\ww$, respectively.
For the $\uubar$ and the $\ww$ channels, the annihilation cross section is the same used in Fig. \ref{fig:ISspectra} and
Figs. \ref{fig:TOAuu}, \ref{fig:TOAbb} and \ref{fig:TOAWW} (i.e. allowed by the antiproton data measured by PAMELA). The dark matter halo is described by the Einasto profile and the
MED set is adopted for galactic cosmic-rays propagation.
\label{fig:sbratio}}
\end{figure}
%%%

Finally, we show in Fig.~\ref{fig:sbratio} the signal-to-background ($S/B$) ratios as a function of the $\dbar$ kinetic energy, for the three representative DM models 
and for the different
solar modulation models listed in Table \ref{tab:solarmod} (the correspondence between the model
and the curves is reported in the boxed insets, code name refers to
Table \ref{tab:solarmod}). The left, central and right panels refer
to a DM with $\mdm=10$ GeV annihilating into $\uubar$, 
a DM with $\mdm=100$ GeV annihilating into $\ww$ (with the same annihilation cross sections used in  Fig.~\ref{fig:ISspectra} and
Figs~ \ref{fig:TOAuu}, \ref{fig:TOAbb} and \ref{fig:TOAWW}),
and a DM with $\mdm=40$ GeV annihilating into $\bbbar$ (with an annihilation cross section $\sigmav=2 \times 10^{-26}$ \cms).
%a dark matter with $\mdm=100$ GeV annihilating into $\ww$, respectively.
%For each case, the annihilation cross section is the same used in Fig.~\ref{fig:ISspectra} and
%Figs~ \ref{fig:TOAuu}, \ref{fig:TOAbb} and \ref{fig:TOAWW} (i.e. allowed by the antiproton data measured by PAMELA). 
Again, the MED set is adopted for galactic cosmic-rays propagation.
This figure makes manifest that, especially at those low kinetic energies shown in the Figures,
the $\dbar$ signal can significantly exceed the expected secondary background, and that solar
modulation may have a non-trivial impact on predictions. Clearly, if we reduce the value
for the annihilation cross section $\sigmav$ we can make $S/B$ arbitrarily low: however,
this figure shows that the $\dbar$ signal has prospects of being clearly distinguishable
from the secondary background (i.e. $S/B \gg 1$) for a wide range of $\sigmav$, ranging
from the currently maximal value allowed by antiproton searches, down to about 40-50 times
smaller values (depending on the DM mass and annihilation channel). Even for a relatively
heavy DM annihilating to $\ww$ (like in the right panel of Fig. \ref{fig:sbratio}, where
$\mdm=100$ GeV) at low $\dbar$ kinetic energies the reaching capabilities of antideuteron
searches can explore cross sections a factor of 30 lower that the current bound from antiprotons.
This shows and confirms that antideuterons offer a remarkable prospects for DM indirect detection,
as was first pointed out in Ref. \cite{DFS},
for a wide range of the DM particle parameters.

Figure \ref{fig:sbratio} also puts into evidence the impact of solar modulation modeling. We notice
again that the largest fluxes for light DM occur for the solar modulation model CD\_60\_0.60\_1. The lowest fluxes instead occur typically
 when the mean-free-path is reduced to $\lambda_{\|} = 0.20\;\mbox{A.U.}$ 
 (model CD\_60\_0.20\_1).  While the last model typically predict the lowest $\dbar$ fluxes
 regardless of the DM annihilation channel and mass, the first one depletes the low-energy
 behavior of the $S/B$ when the DM is increased, i.e. for interstellar spectra which
are peaked to higher energies. Figure \ref{fig:sbratio} 
shows that the impact of solar modulation modeling has a non-trivial behavior, and cannot
be easily rescaled from the force-field approximation.

\section{Prospects for antideuterons observation}
\label{sec:prospects}

% GAPS
% energy: 0.1 - 0.25 (GeV/n)
% 3sigma sensitivity: 2.8e-07 (m^2 s sr GeV/n)^(-1)
% exposure: 2.36e07 (m^2 s sr)

% AMS
% energy: 0.2 - 0.8 (GeV/n) 
% flux limit for 1 event: 4.5e-07 (m^2 s sr GeV/n)^(-1)
% exposure: 3.70e06 (m^2 s sr)

\begin{figure}[t]
\centering
\includegraphics[width=0.35\textwidth]{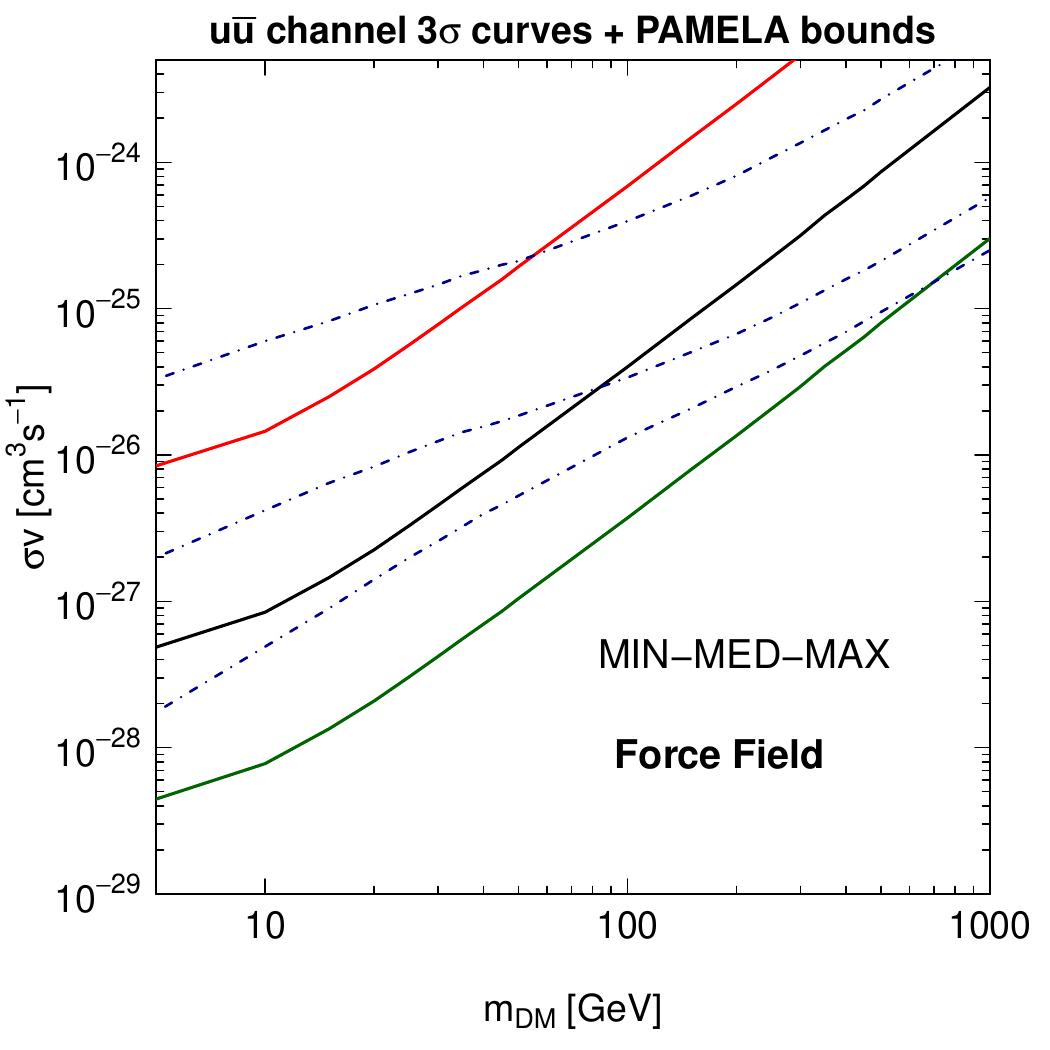}
\includegraphics[width=0.35\textwidth]{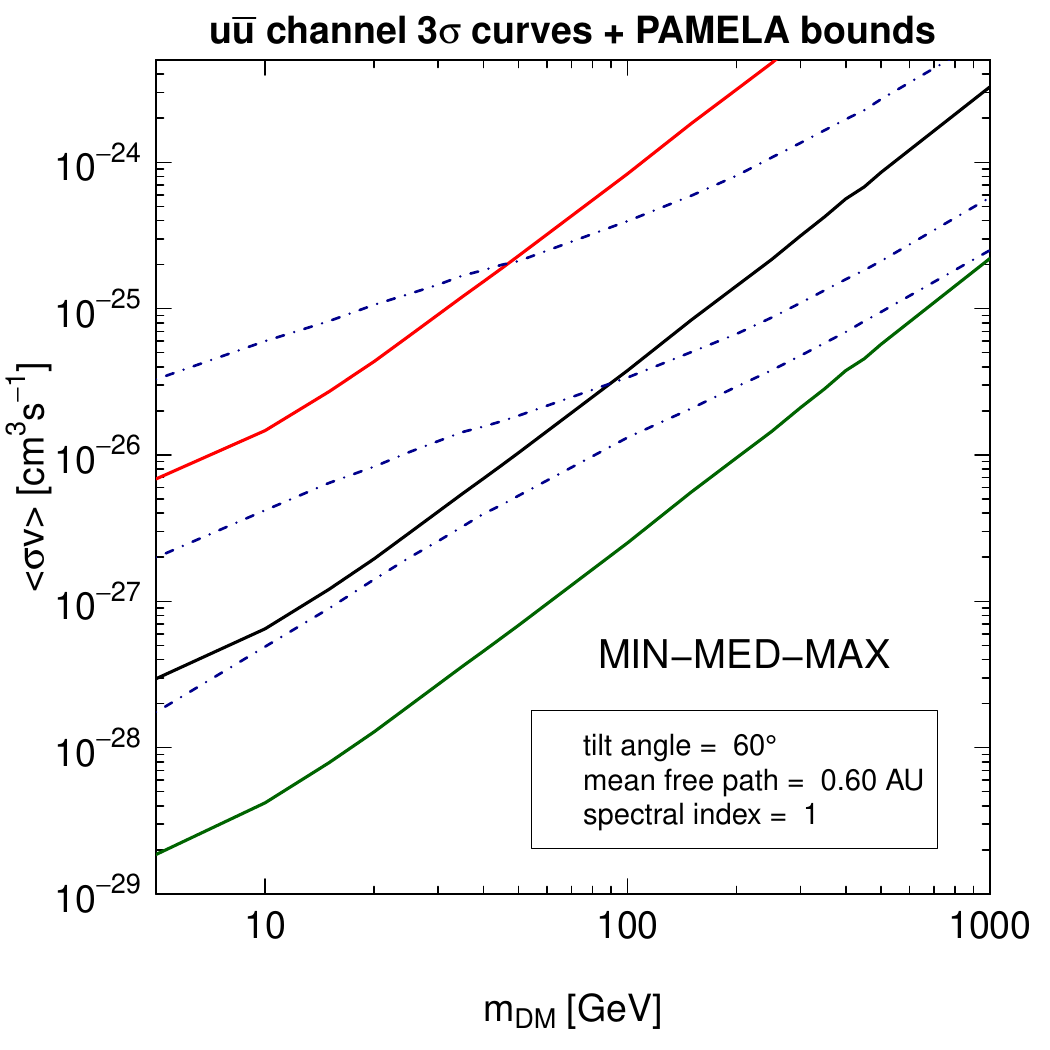}
\includegraphics[width=0.35\textwidth]{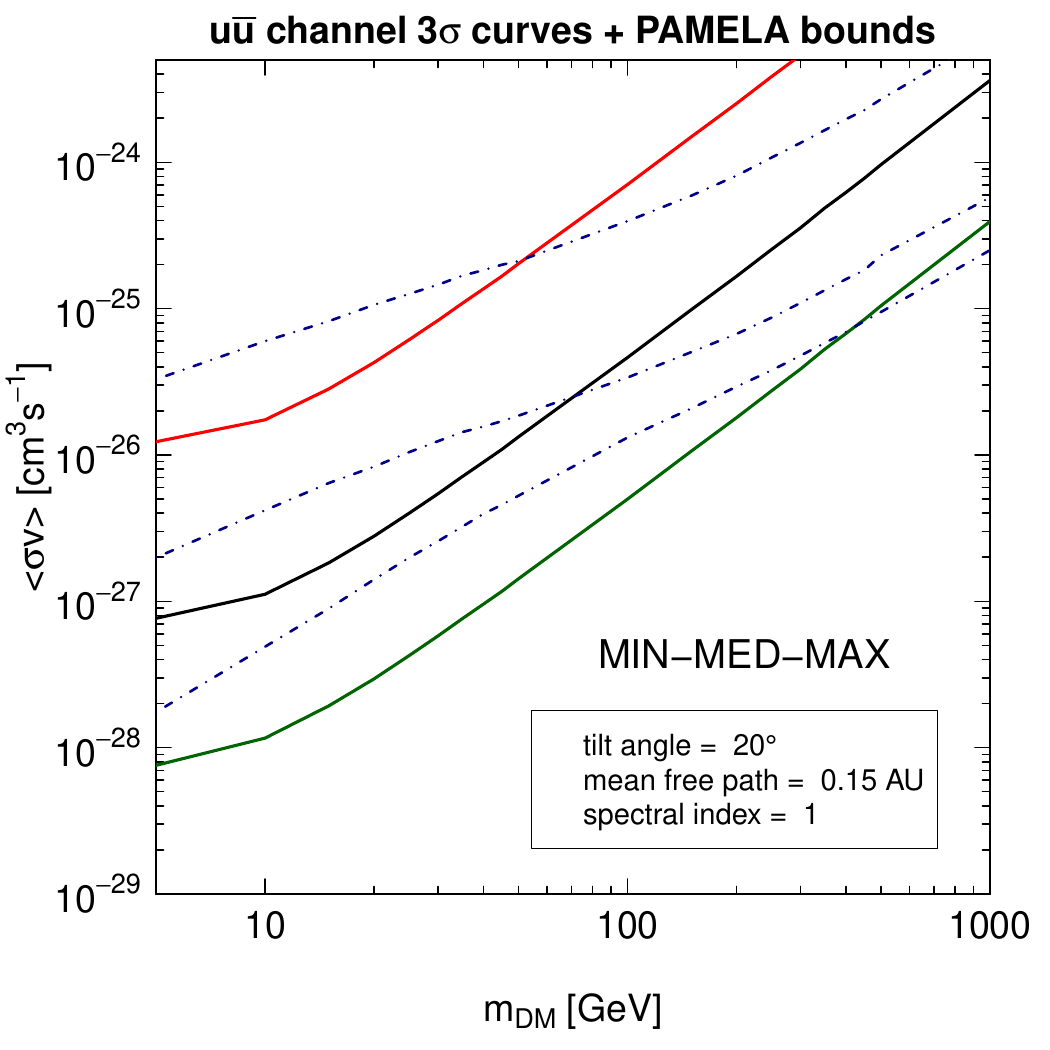}
\includegraphics[width=0.35\textwidth]{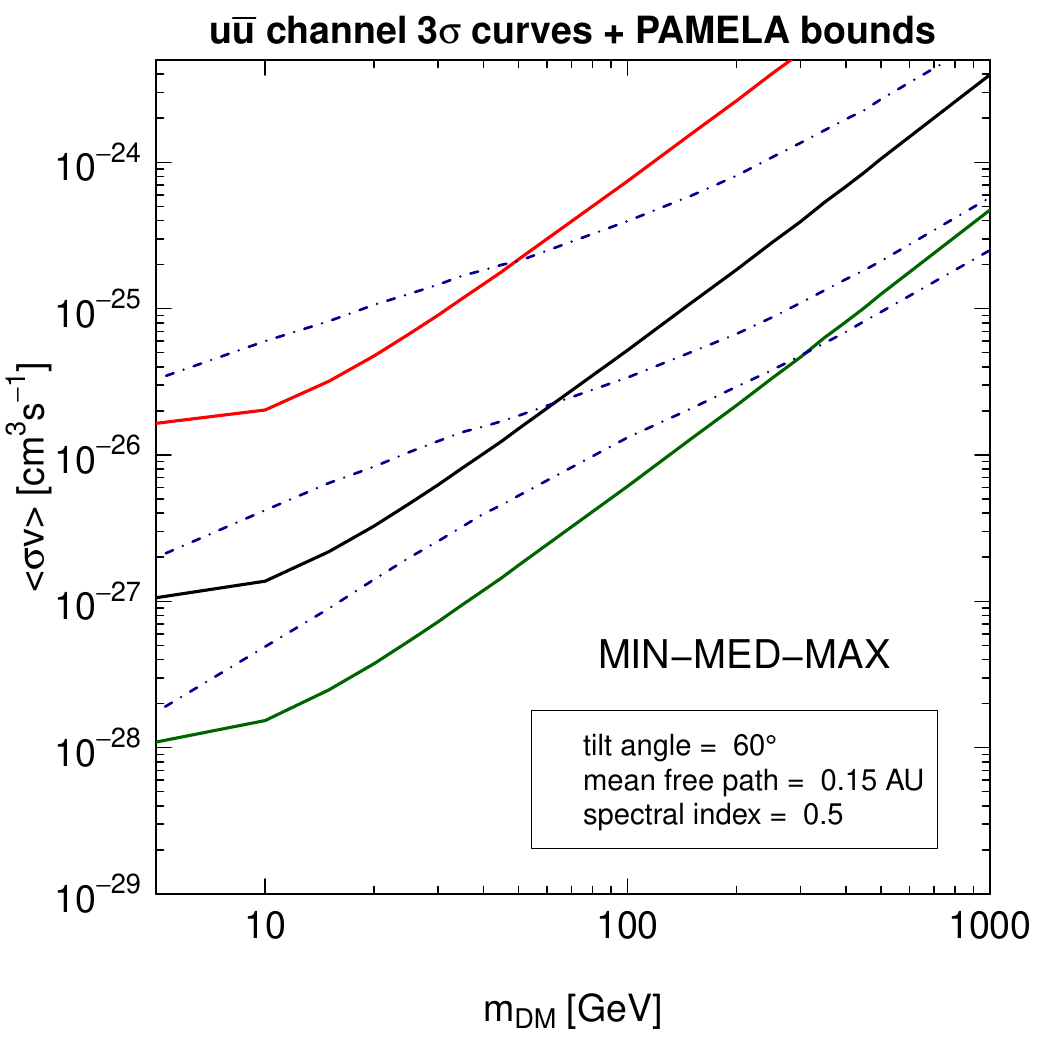}
\caption{Prospects for a $3\sigma$ detection of a $\dbar$ signal with GAPS, expressed in the plane
annihilation cross-section
$\sigmav$ vs the dark matter mass $\mdm$, for the $\uubar$ annihilation channel. The three upper/median/lower solid lines denote the detection limits (which correspond to detection of $N_{\rm crit} = 1$ event for GAPS with current design sensitivity \cite{haileypriv}) for the three galactic propagation models  MIN/MED/MAX of Table \ref{tab:parameters},
respectively. The three dot-dashed lines show the corresponding upper bounds  derived from the antiproton measurements of PAMELA \cite{Adriani:2010rc} (again for each of the three propagation models MIN/MED/MAX,
from top to bottom). 
The galactic dark matter halo is described by an Einasto profile.
Solar modulation has been modeled with the standard force-field approximation in the first panel, and
for solar modulation models as reported in the boxed insets, for the other panels.
\label{fig:GAPS1}}
\end{figure}%%%

%%%
\begin{figure}[t]
\centering
\includegraphics[width=0.35\textwidth]{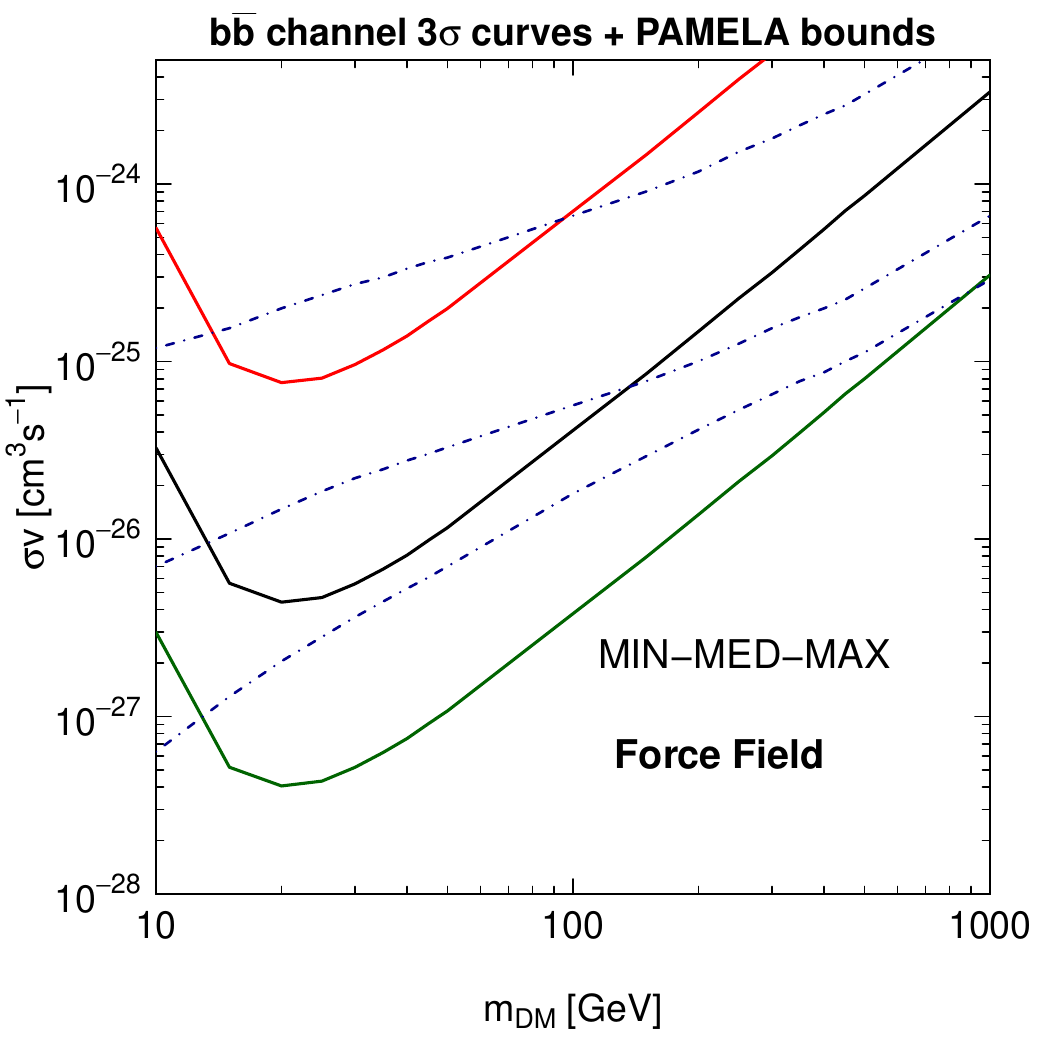}
\includegraphics[width=0.35\textwidth]{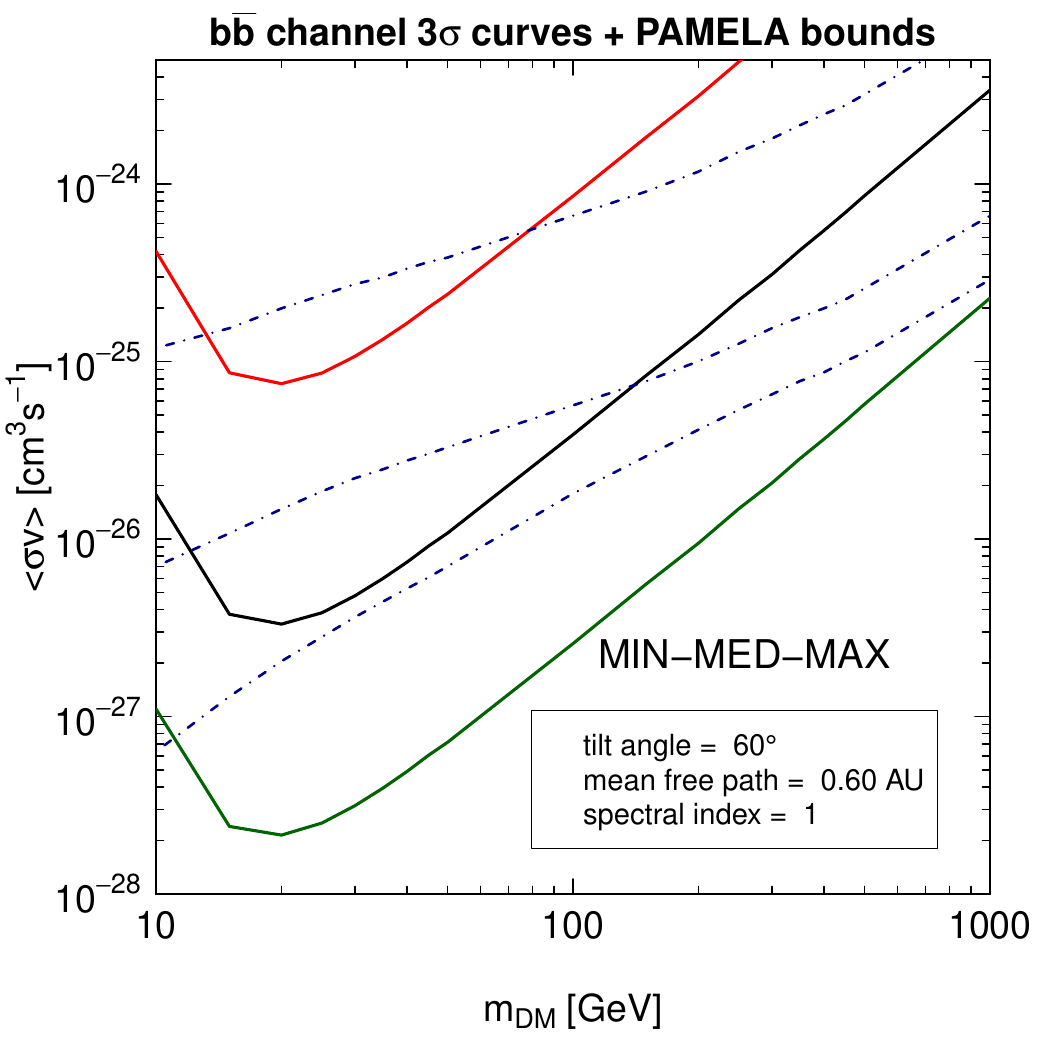}
\includegraphics[width=0.35\textwidth]{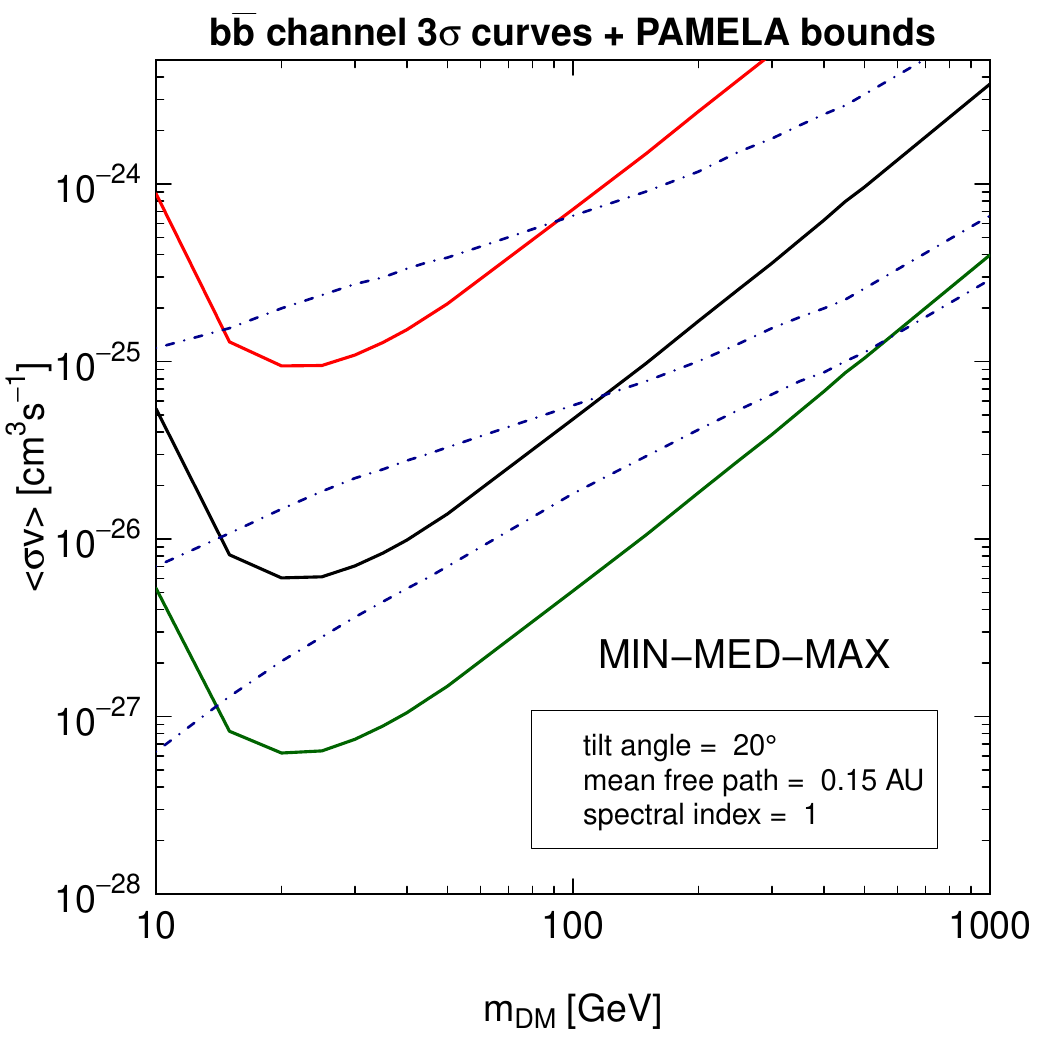}
\includegraphics[width=0.35\textwidth]{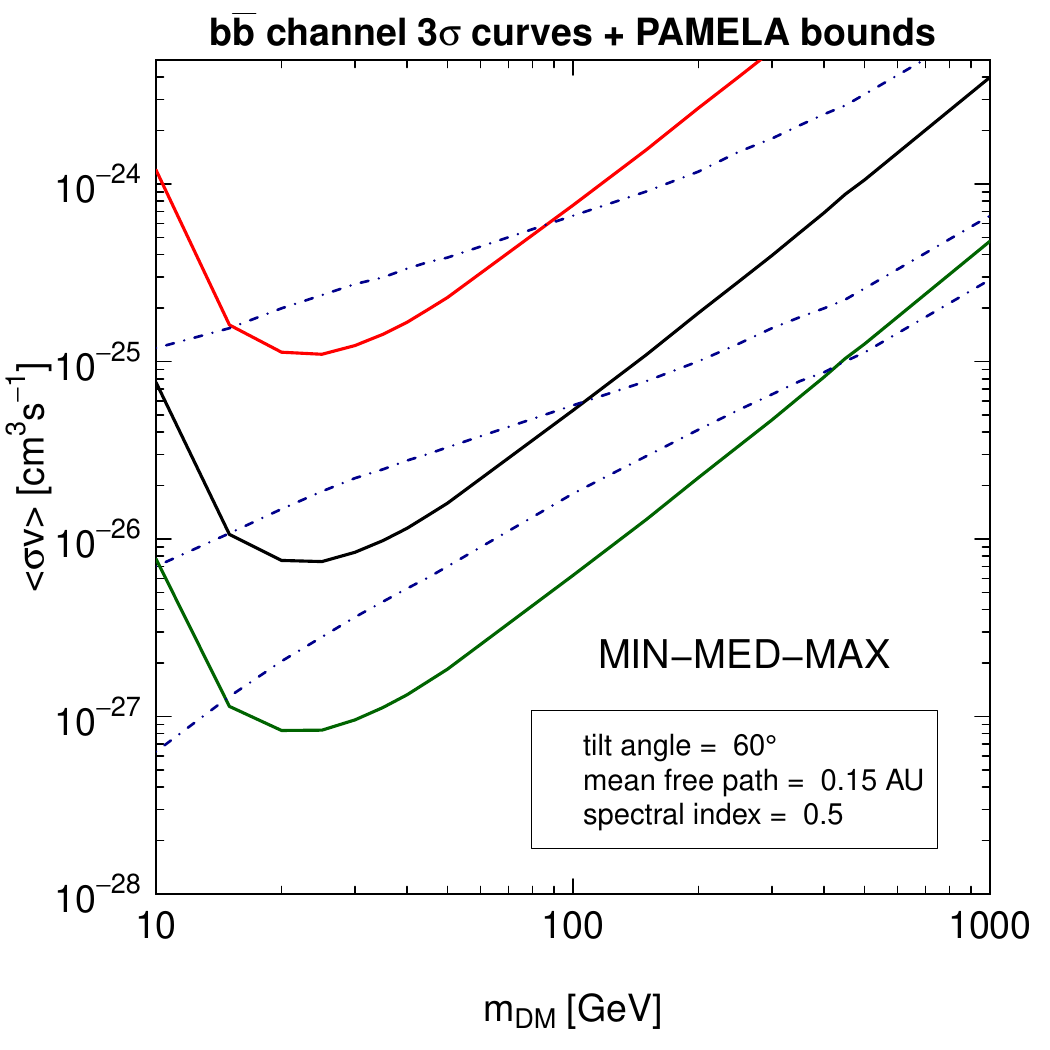}
\caption{Prospects for a $3\sigma$ detection of a $\dbar$ signal with GAPS, expressed in the plane
annihilation cross-section
$\sigmav$ vs the dark matter mass $\mdm$, for the $\bbbar$ annihilation channel. 
Notations are as in Fig. \ref{fig:GAPS1}. Solar modulation has been modeled with the standard force-field approximation in the first panel, and
for solar modulation models as reported in the boxed insets, for the other panels.
\label{fig:GAPS2}}
\end{figure}
%%%

%%%
\begin{figure}[t]
\centering
\includegraphics[width=0.35\textwidth]{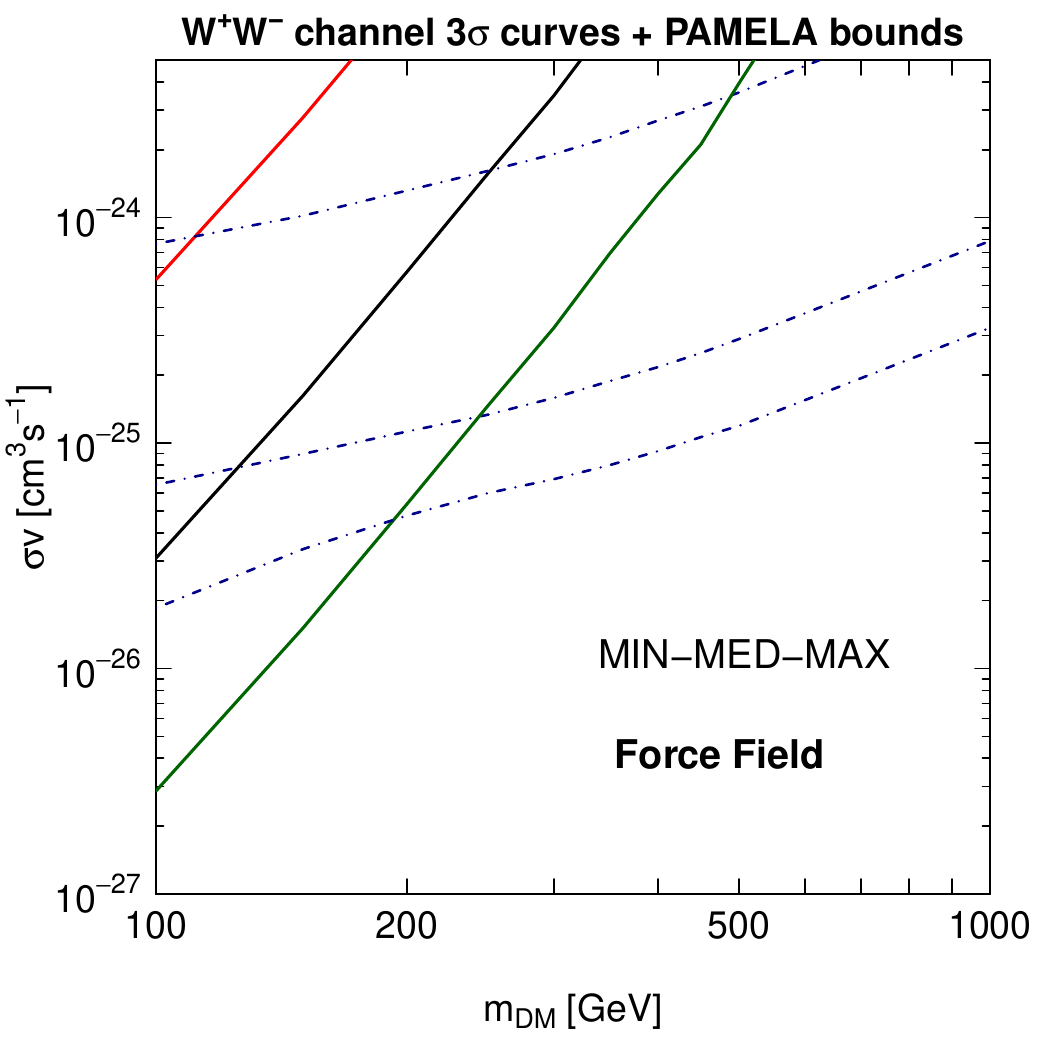}
\includegraphics[width=0.35\textwidth]{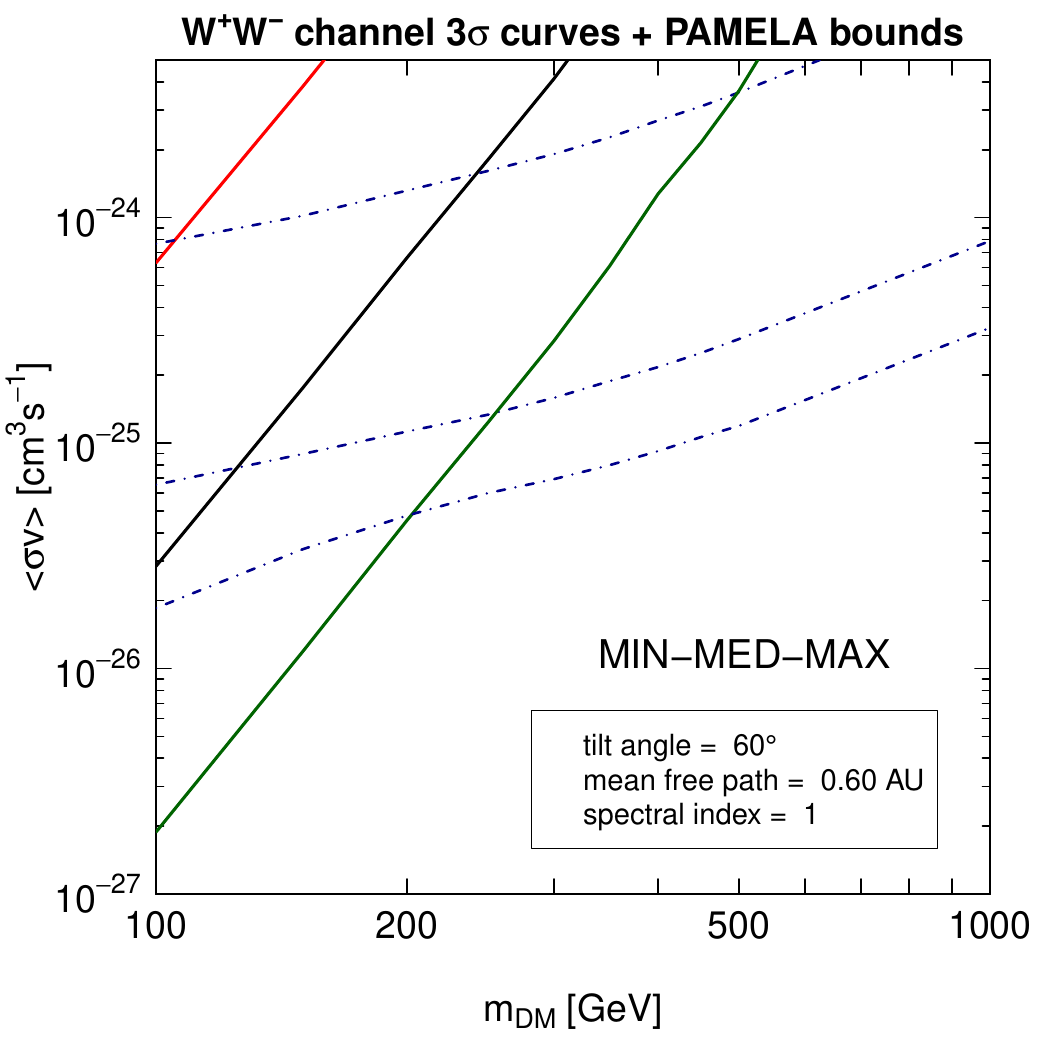}
\includegraphics[width=0.35\textwidth]{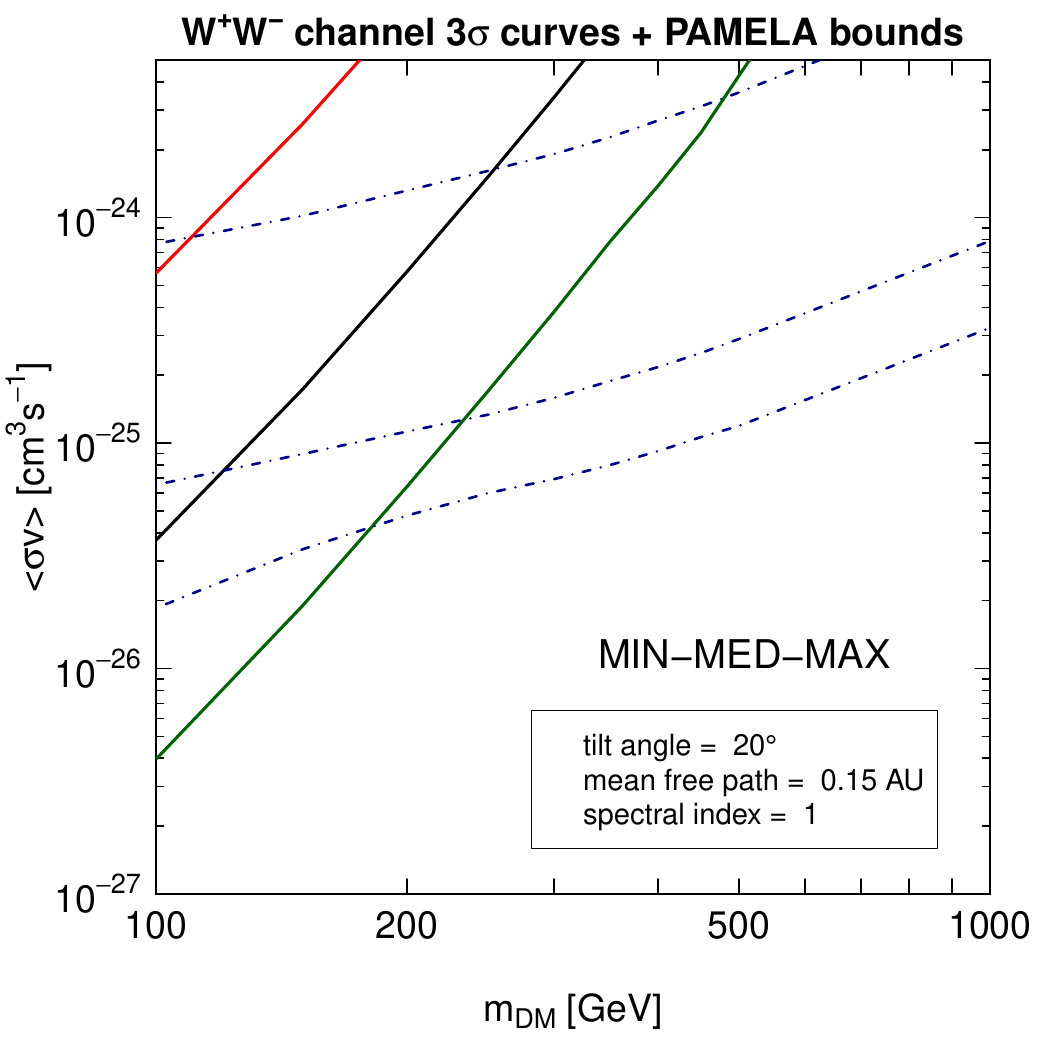}
\includegraphics[width=0.35\textwidth]{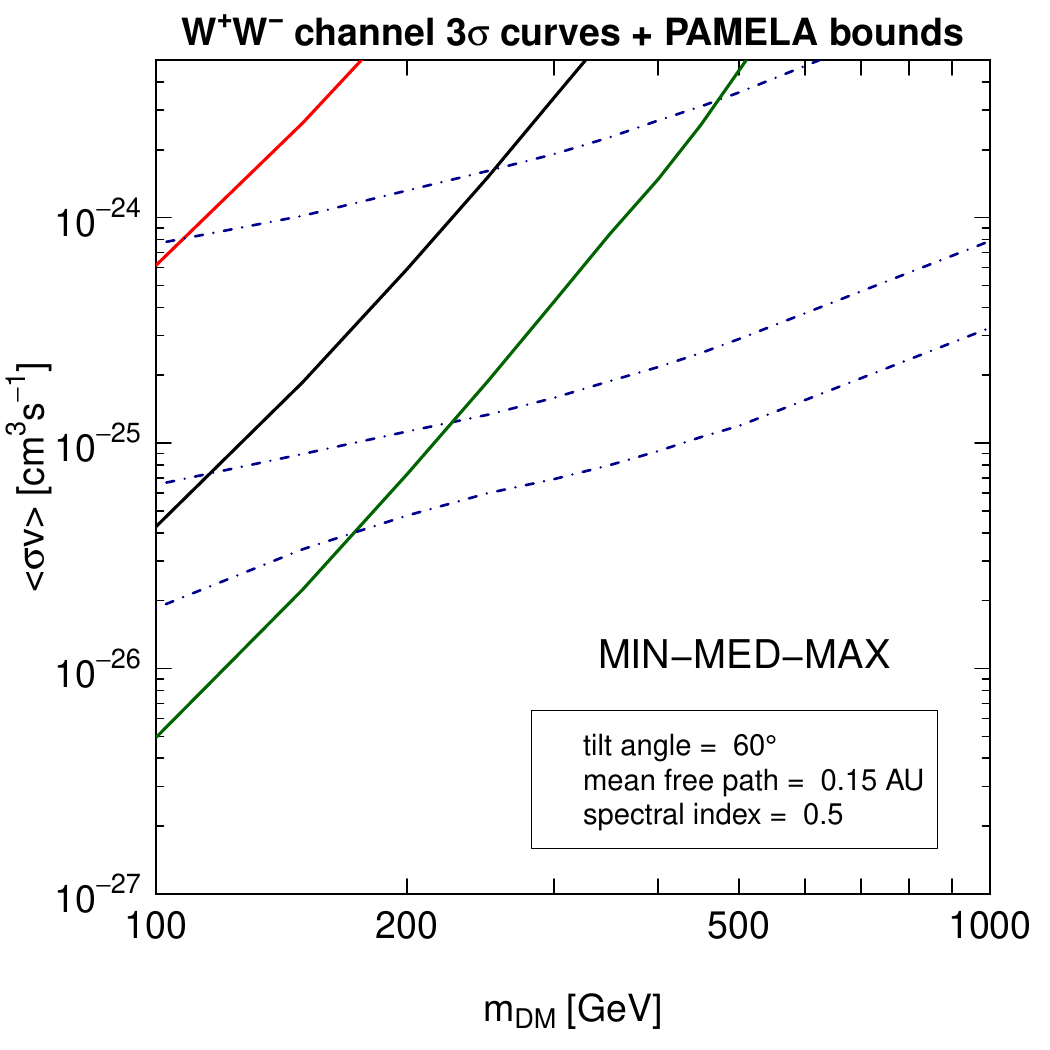}
\caption{Prospects for a $3\sigma$ detection of a $\dbar$ signal with GAPS, expressed in the plane
annihilation cross-section
$\sigmav$ vs the dark matter mass $\mdm$, for the $\ww$ annihilation channel. 
Notations are as in Fig. \ref{fig:GAPS1}. Solar modulation has been modeled with the standard force-field approximation in the first panel, and
for solar modulation models as reported in the boxed insets, for the other panels.
\label{fig:GAPS3}}
\end{figure}
%%%

%%%
\begin{figure}[t]
\centering
\includegraphics[width=0.35\textwidth]{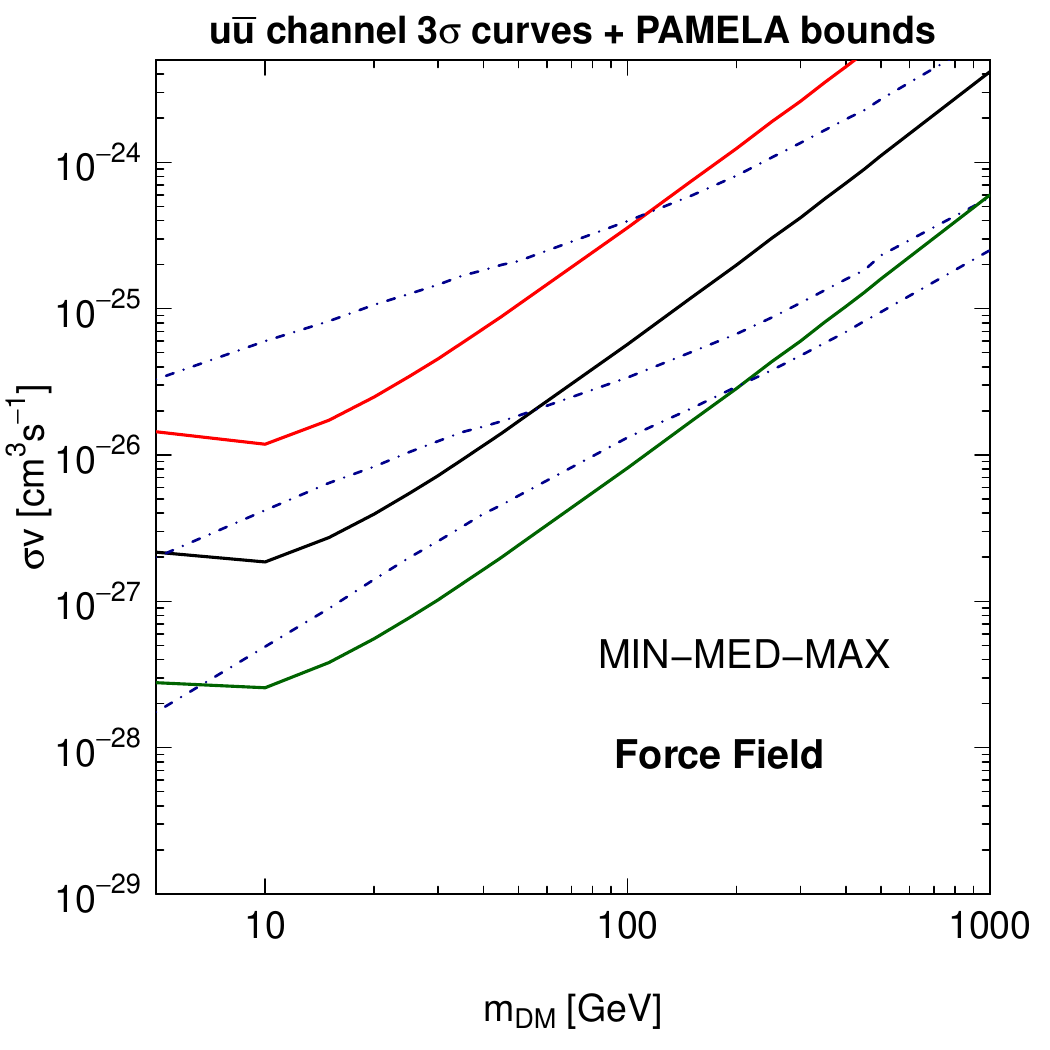}
\includegraphics[width=0.35\textwidth]{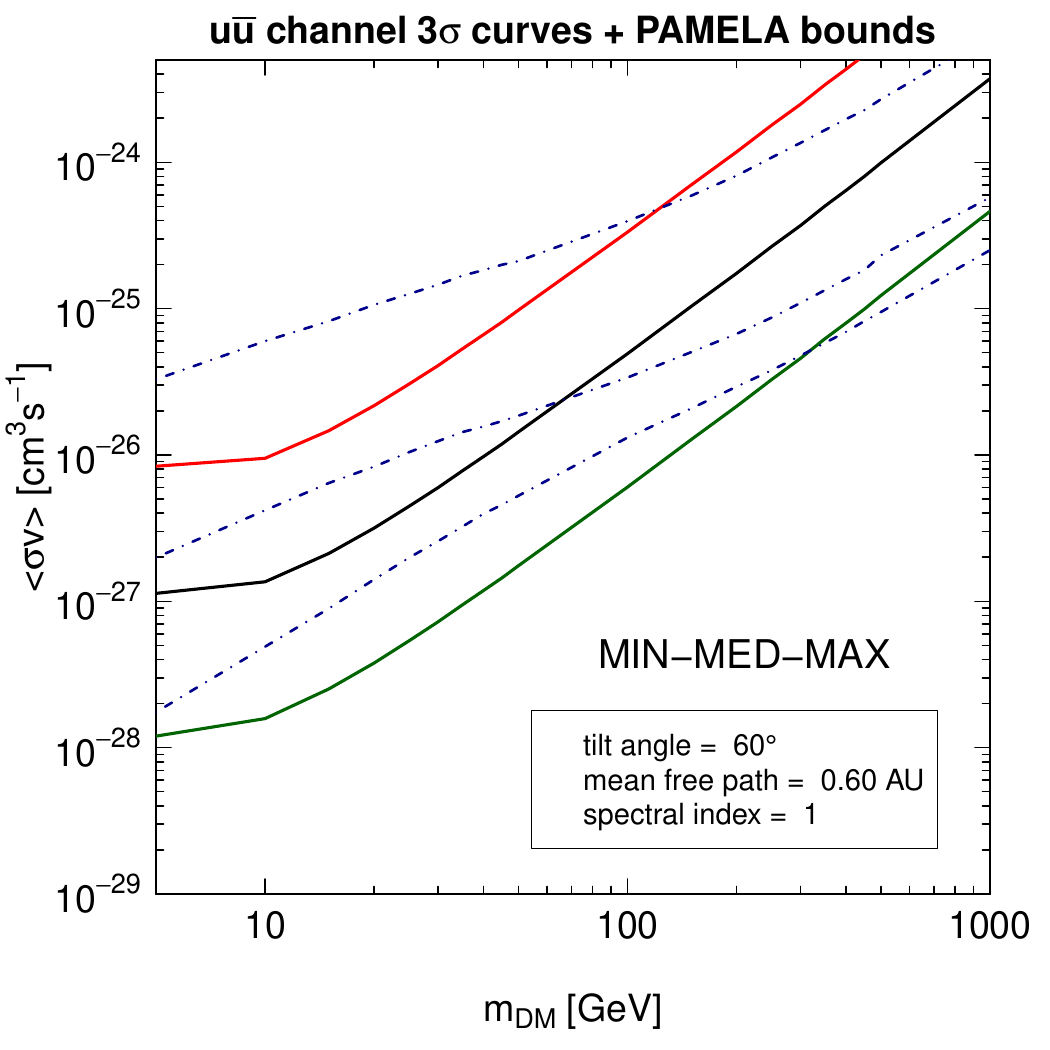}
\includegraphics[width=0.35\textwidth]{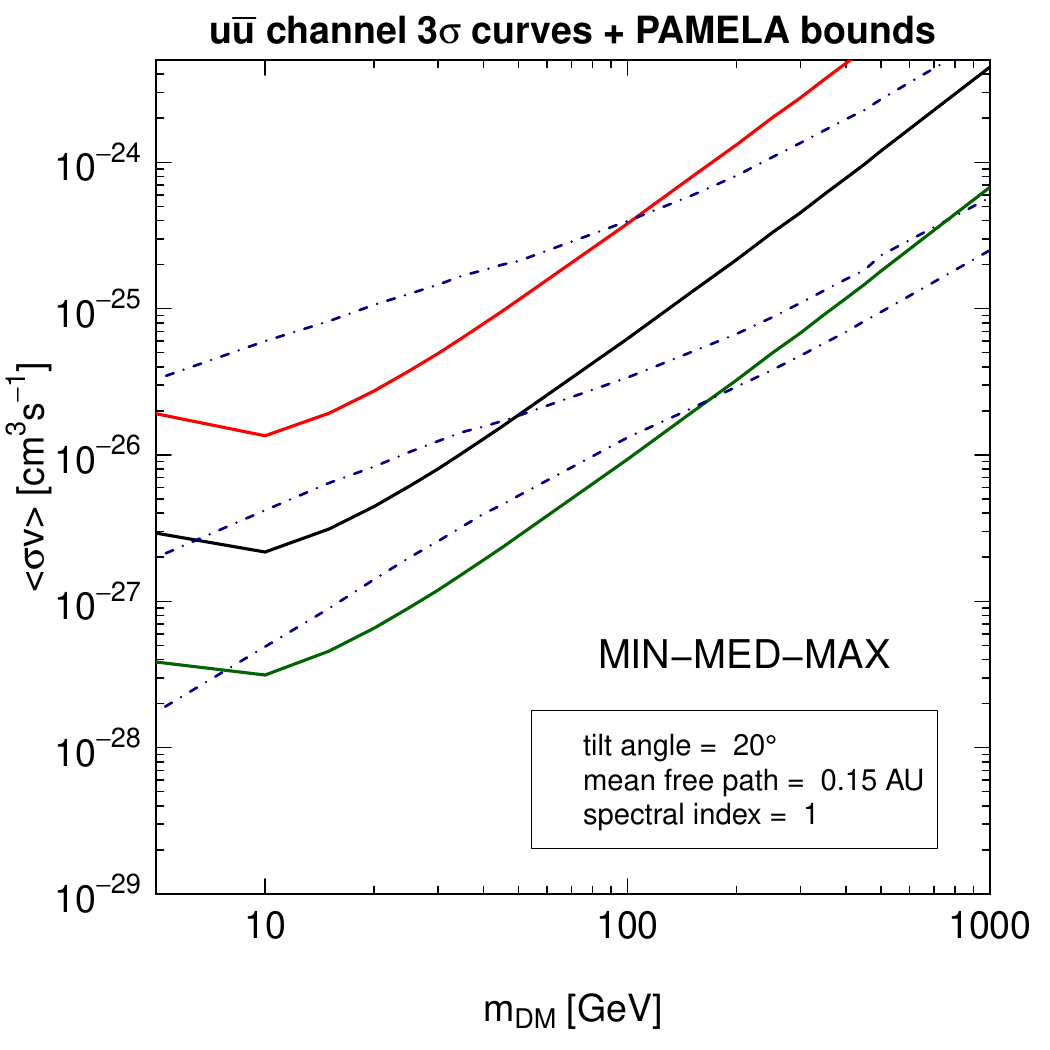}
\includegraphics[width=0.35\textwidth]{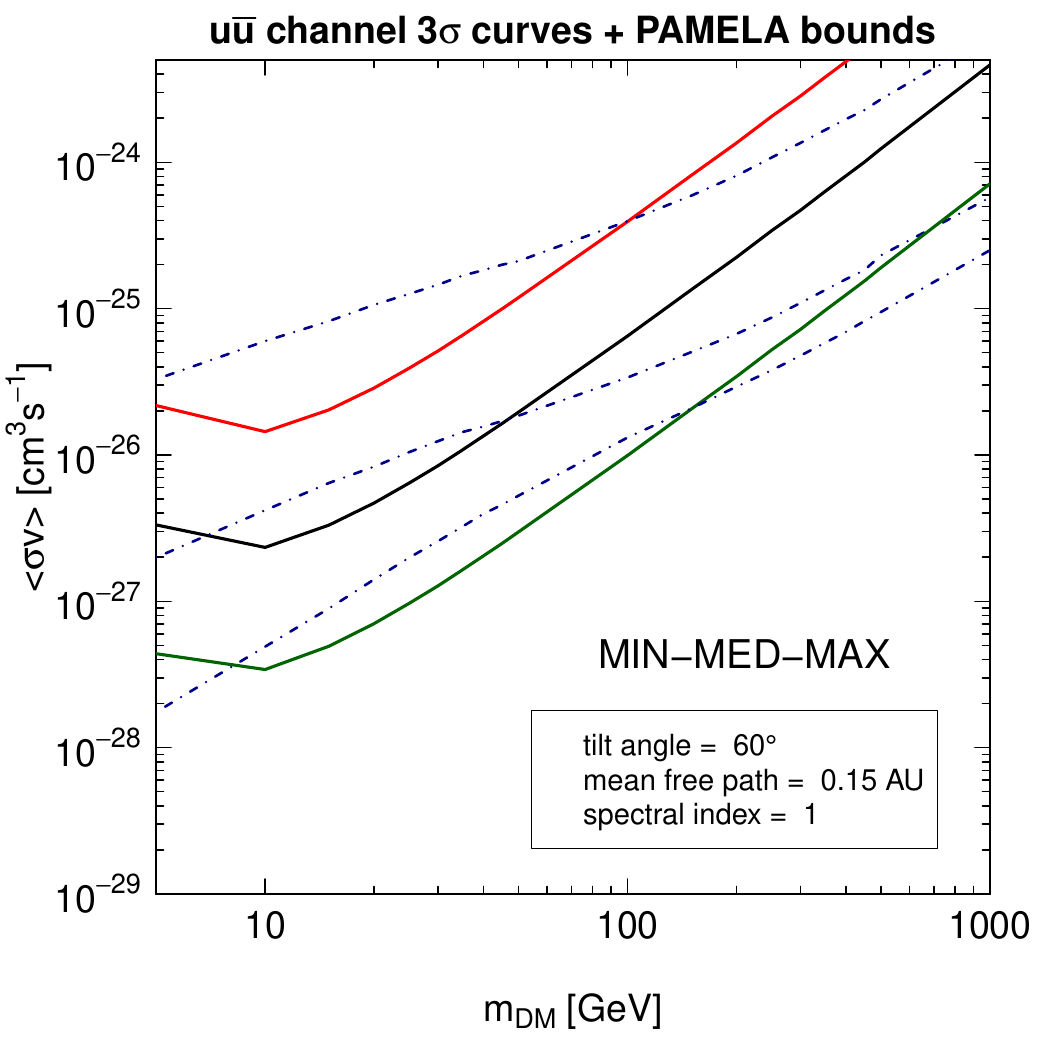}
\caption{Prospects for a $3\sigma$ detection of a $\dbar$ signal with AMS-02, expressed in the plane
annihilation cross-section
$\sigmav$ vs the dark matter mass $\mdm$, for the $\uubar$ annihilation channel. 
The three upper/median/lower solid lines denote the detection limits 
for the three galactic propagation models of Table \ref{tab:parameters} MIN/MED/MAX,
respectively ( $N_{\rm crit} = 2$ for the MED and MAX models and $N_{\rm crit} = 1$ for the  MIN model, according to the AMS-02 nominal sensitivity \cite{giovacchini}). The three dot-dashed lines show the corresponding upper bounds  derived from the antiproton measurements of PAMELA \cite{Adriani:2010rc} (again for each of the three propagation models MIN/MED/MAX,
from top to bottom). 
The galactic dark matter halo is described by an Einasto profile.
Solar modulation has been modeled with the standard force-field approximation in the first panel, and
for solar modulation models as reported in the boxed insets, for the other panels.
\label{fig:AMS1}}
\end{figure}
%%%

%%%
\begin{figure}[t]
\centering
\includegraphics[width=0.35\textwidth]{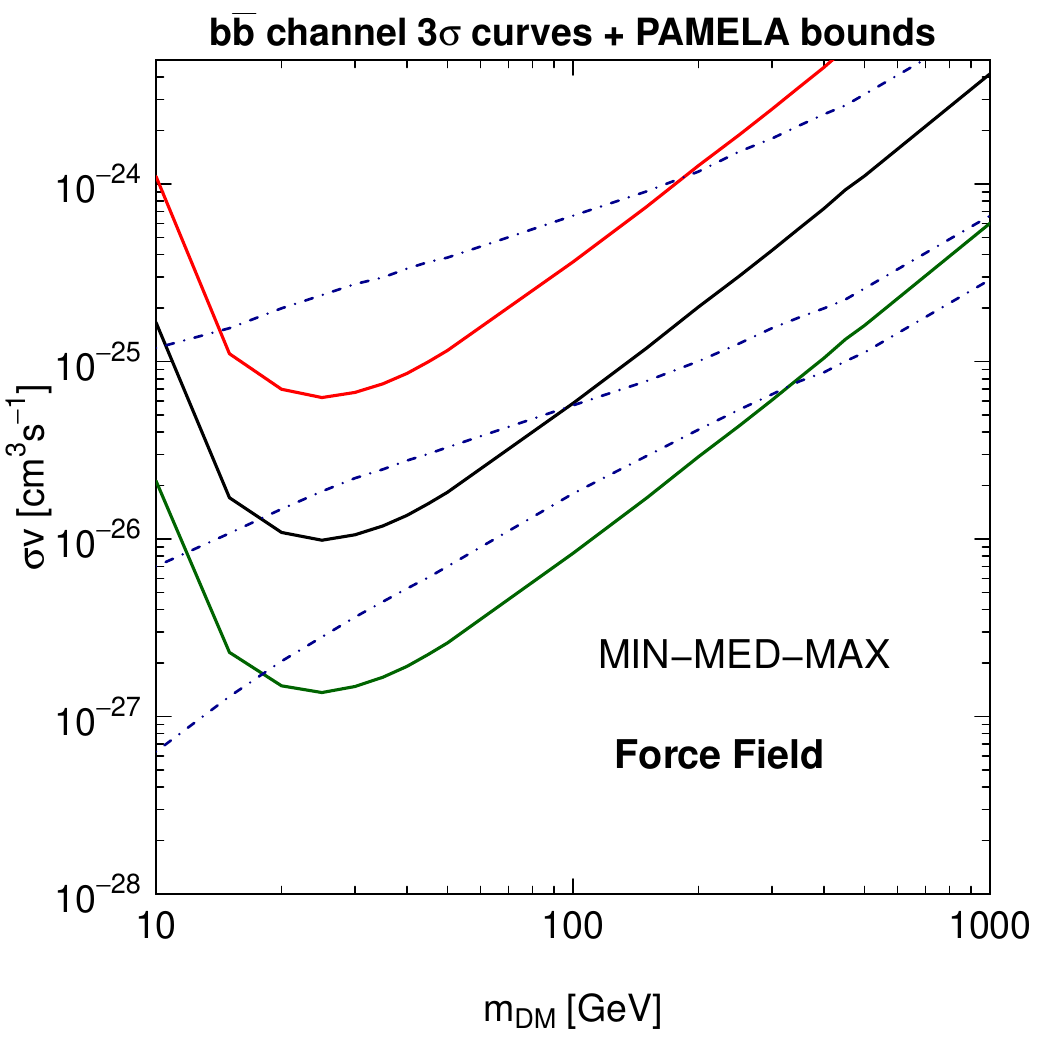}
\includegraphics[width=0.35\textwidth]{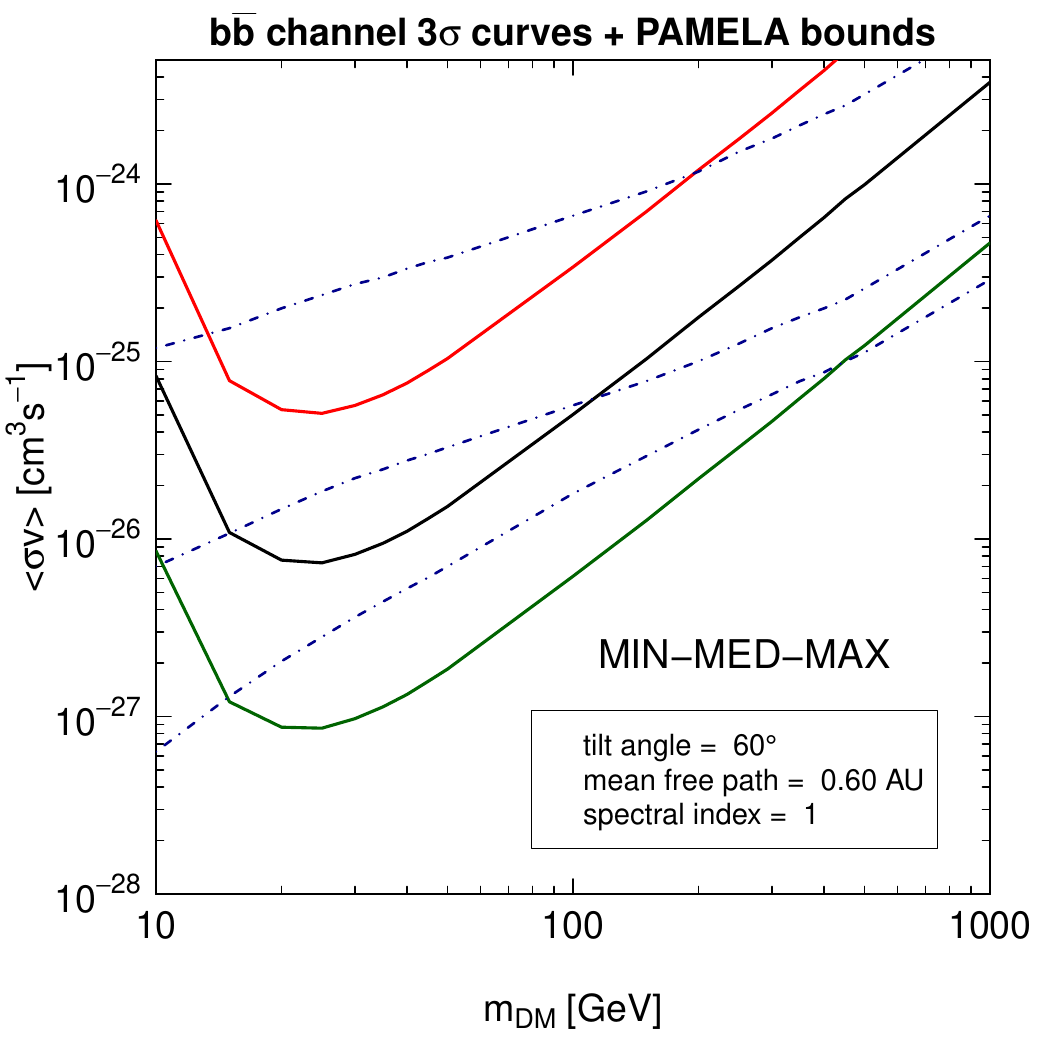}
\includegraphics[width=0.35\textwidth]{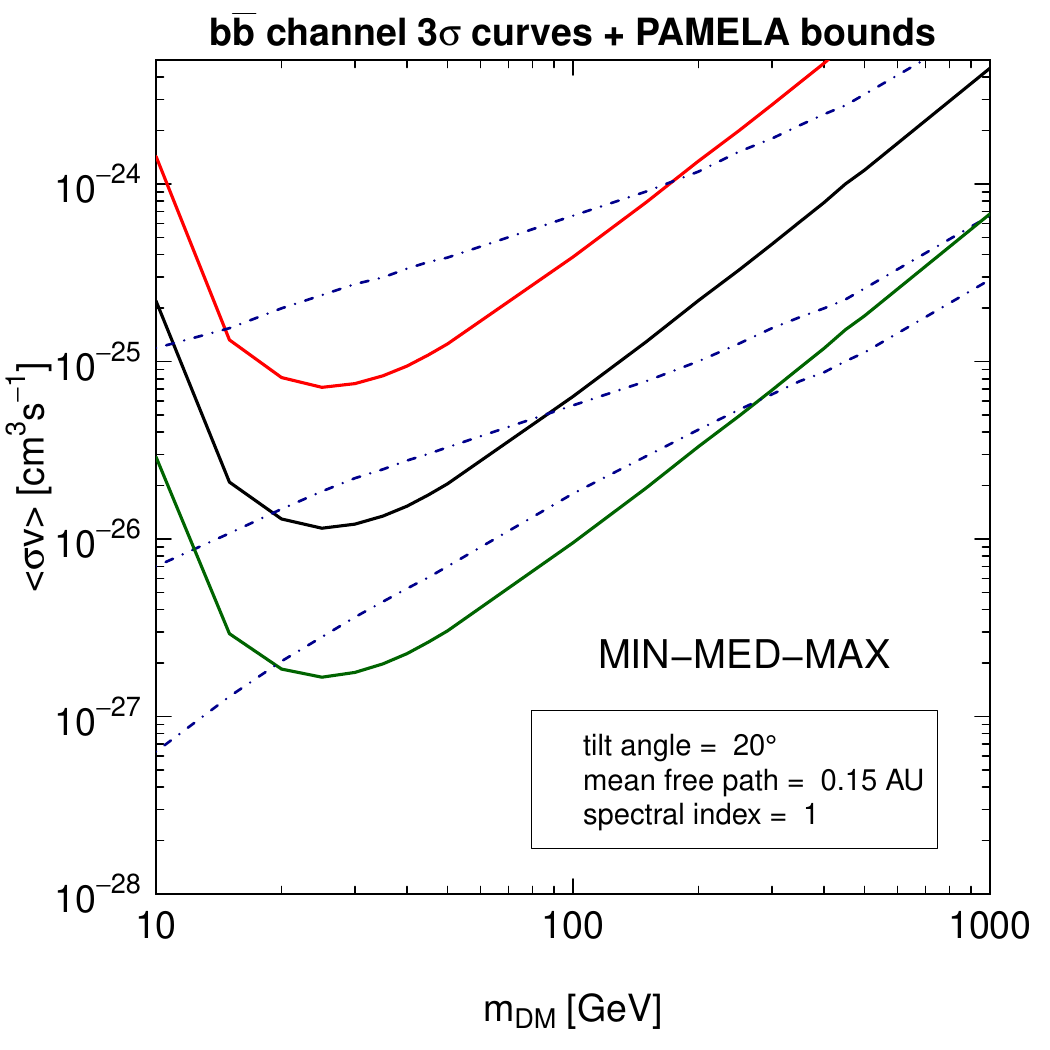}
\includegraphics[width=0.35\textwidth]{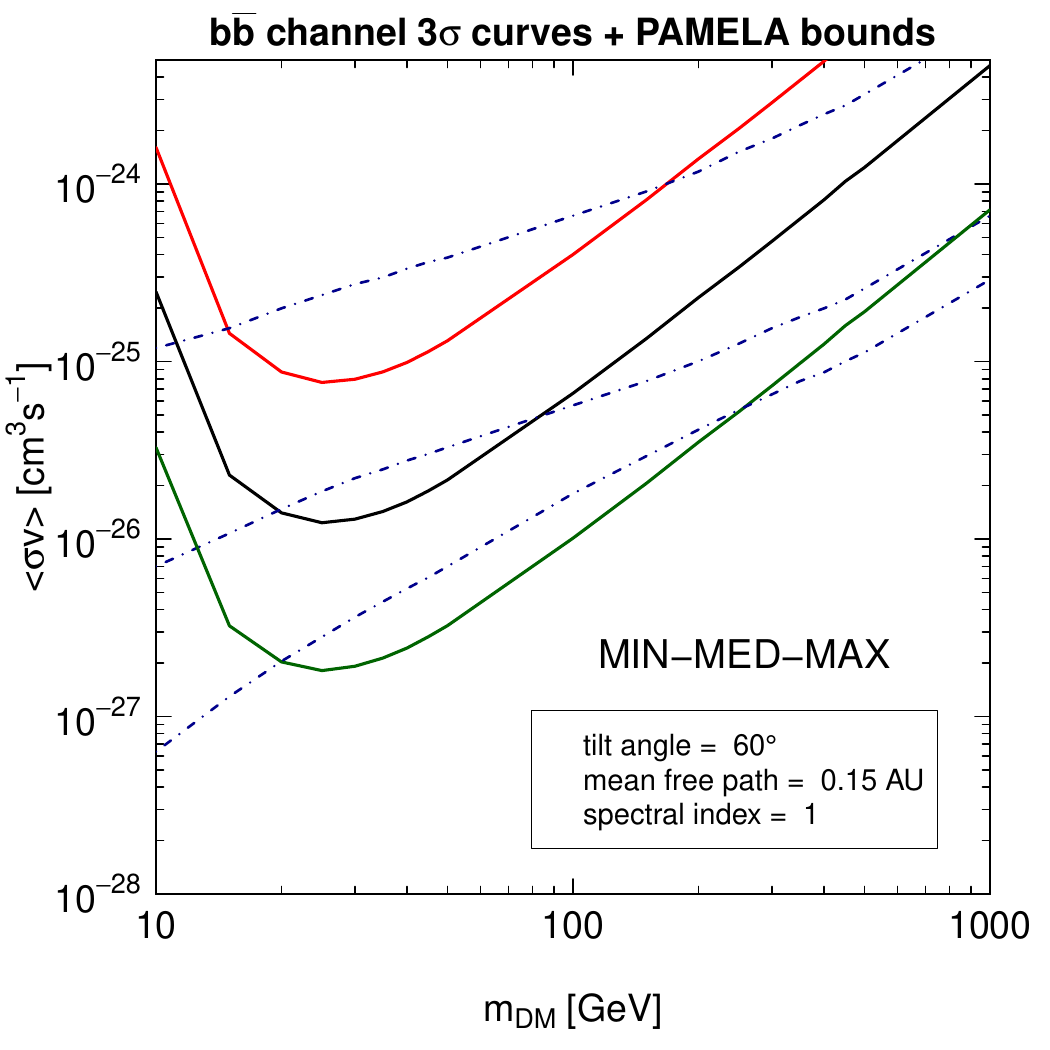}
\caption{Prospects for a $3\sigma$ detection of a $\dbar$ signal with AMS-02, expressed in the plane
annihilation cross-section
$\sigmav$ vs the dark matter mass $\mdm$, for the $\bbbar$ annihilation channel. 
Notations are as in Fig. \ref{fig:AMS1}. Solar modulation has been modeled with the standard force-field approximation in the first panel, and
for solar modulation models as reported in the boxed insets, for the other panels.
\label{fig:AMS2}}
\end{figure}
%%%

%%%
\begin{figure}[t]
\centering
\includegraphics[width=0.35\textwidth]{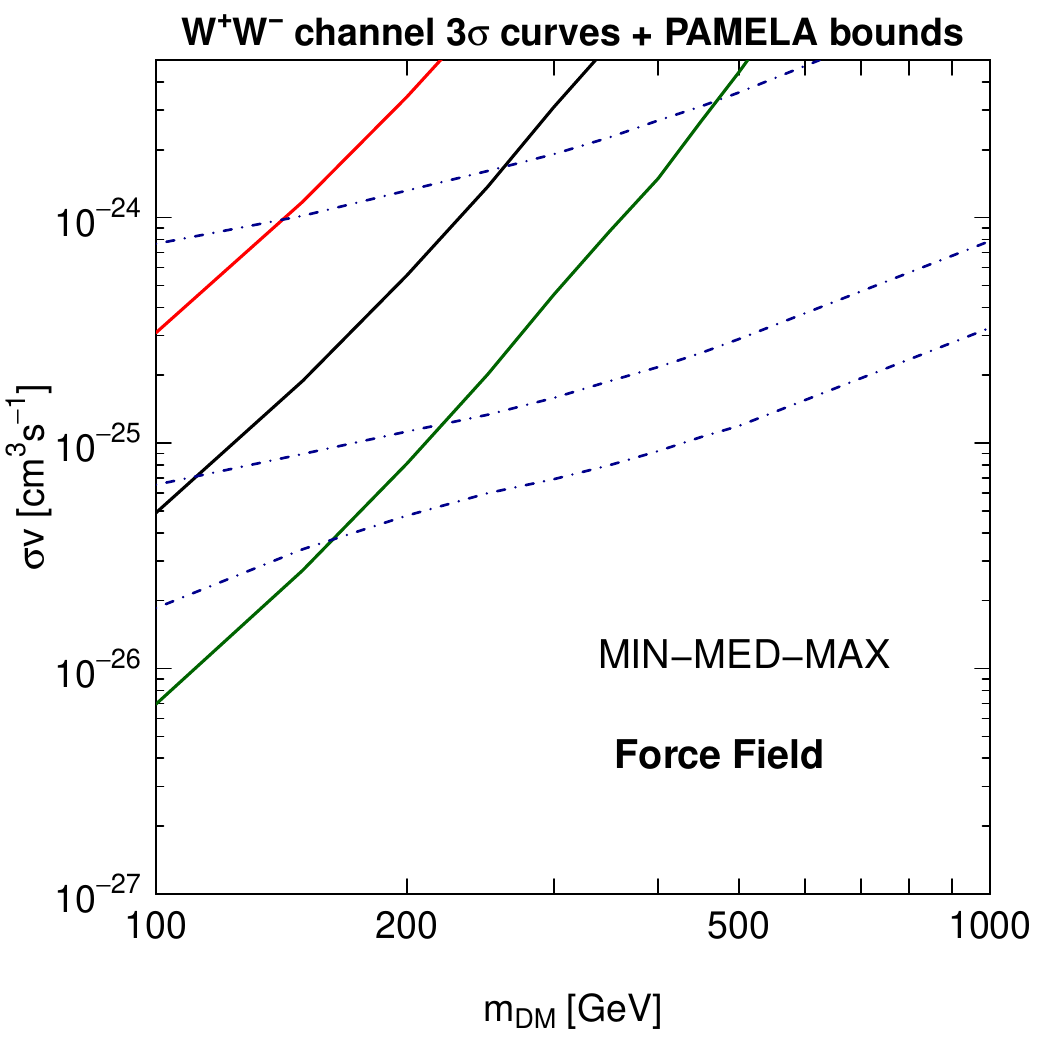}
\includegraphics[width=0.35\textwidth]{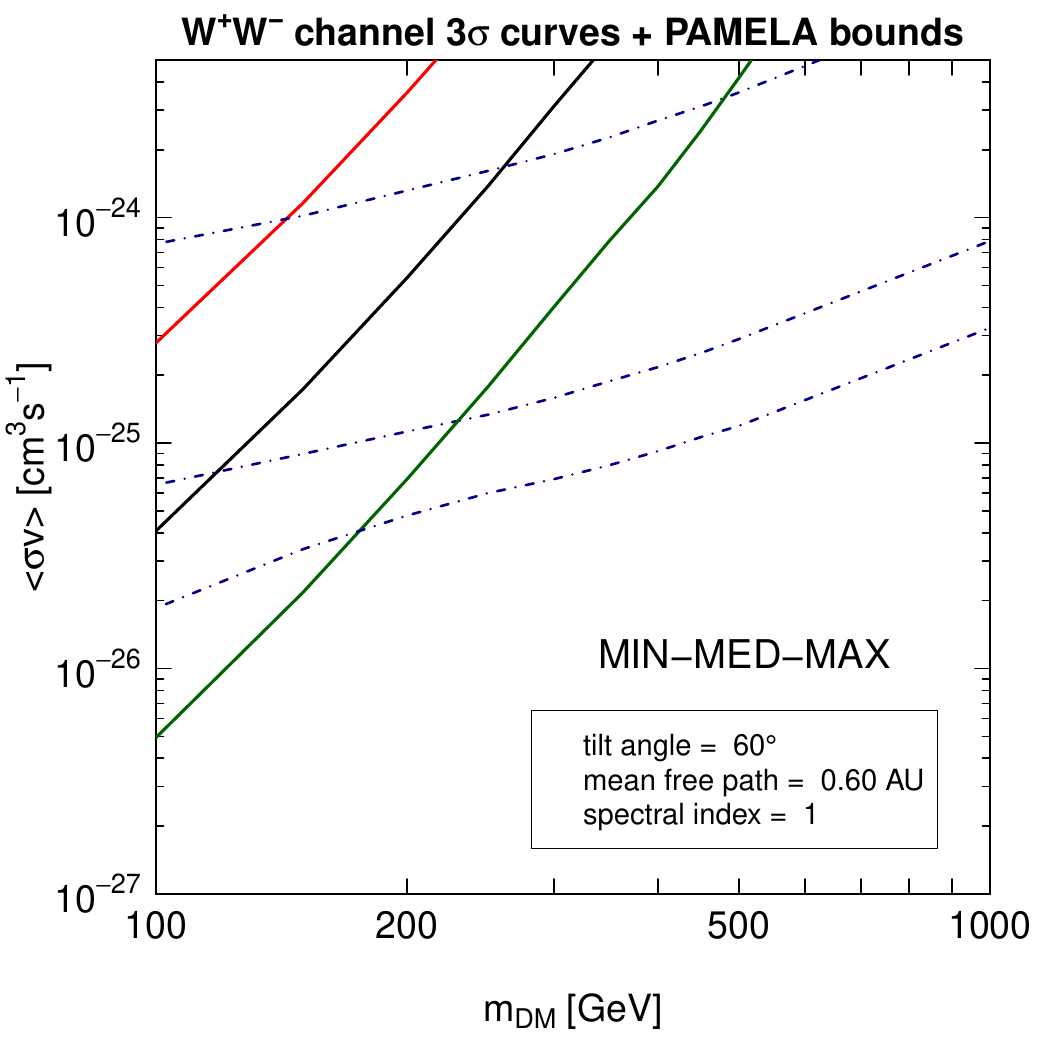}
\includegraphics[width=0.35\textwidth]{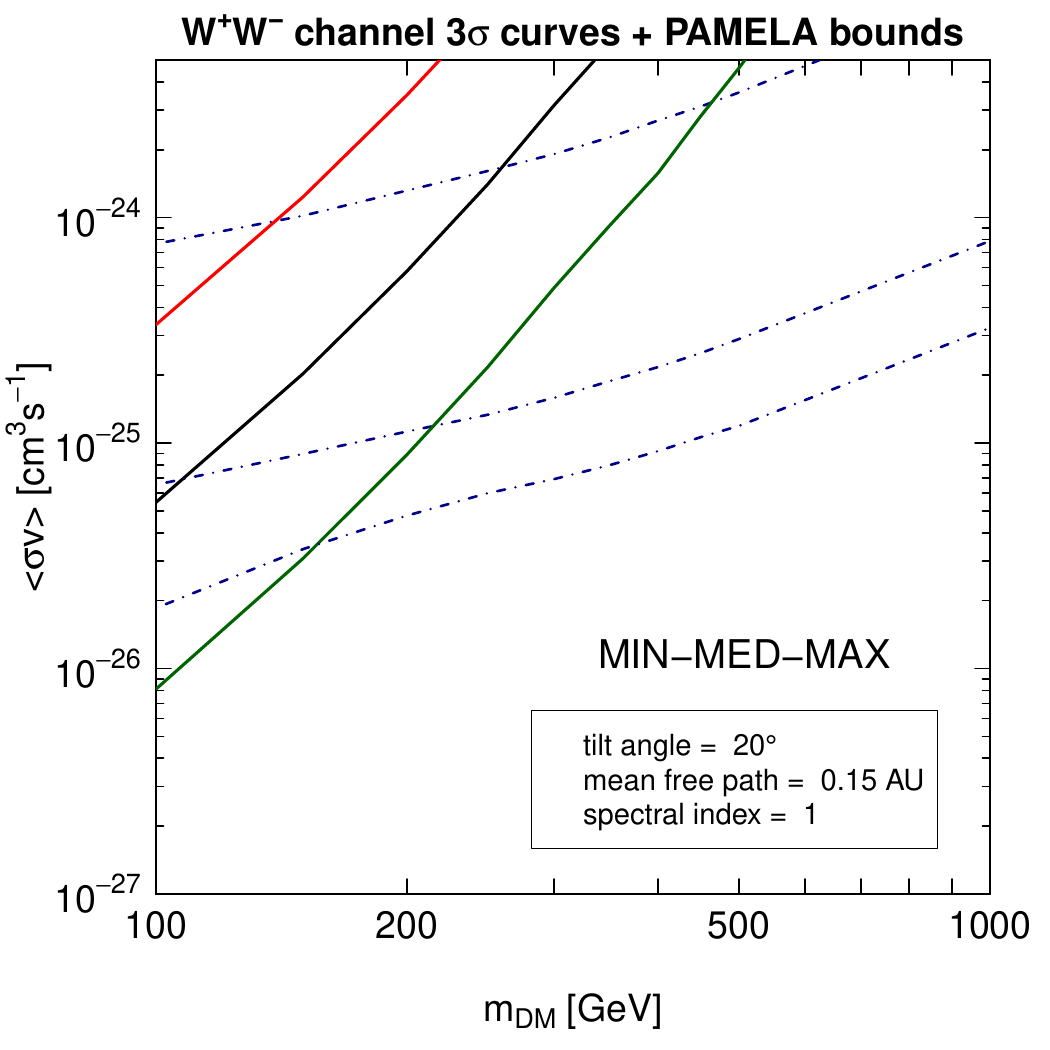}
\includegraphics[width=0.35\textwidth]{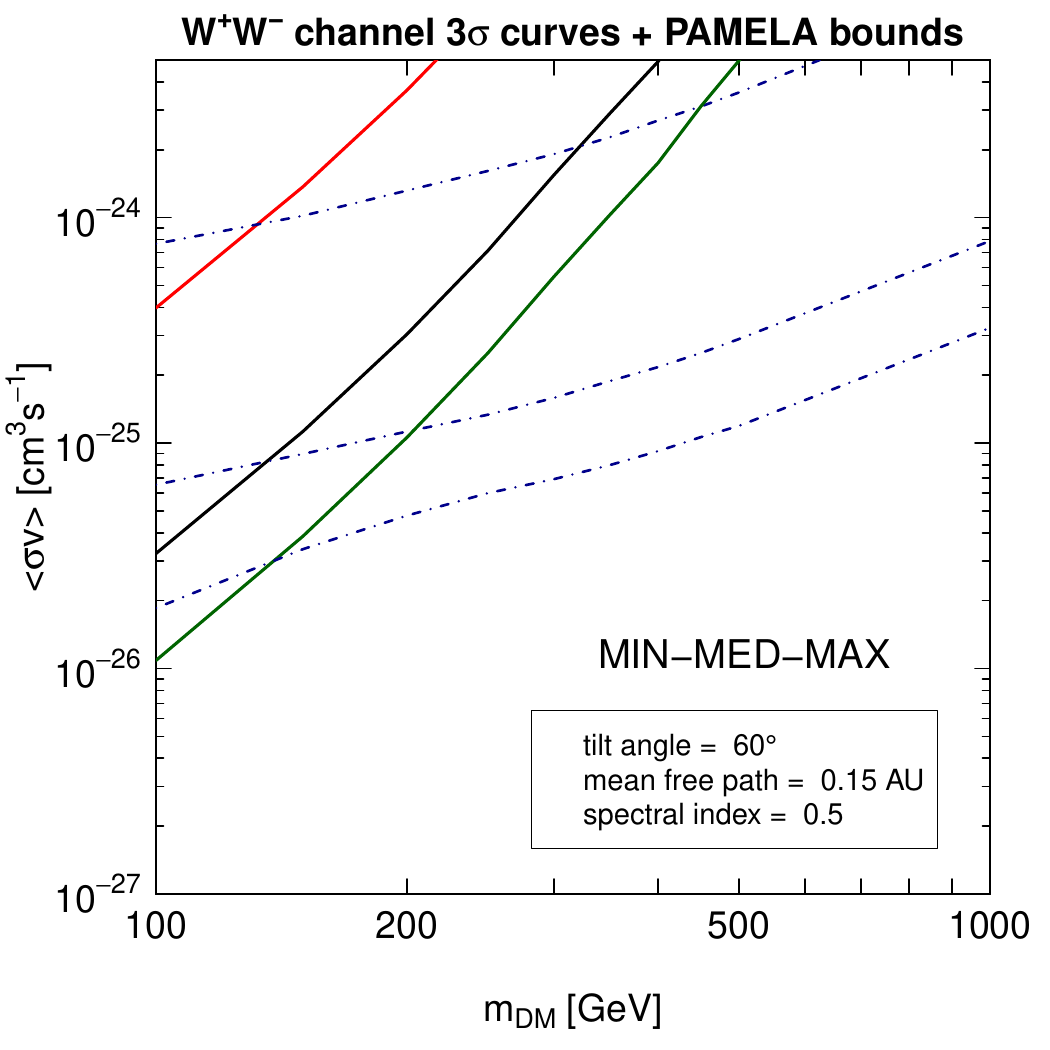}
\caption{Prospects for a $3\sigma$ detection of a $\dbar$ signal with AMS-02, expressed in the plane
annihilation cross-section
$\sigmav$ vs the dark matter mass $\mdm$, for the $\ww$ annihilation channel. 
Notations are as in Fig. \ref{fig:AMS1}. Solar modulation has been modeled with the standard force-field approximation in the first panel, and
for solar modulation models as reported in the boxed insets, for the other panels.
\label{fig:AMS3}}
\end{figure}
%%%

%%%
\begin{figure}[t]
\centering
\includegraphics[width=0.30\textwidth]{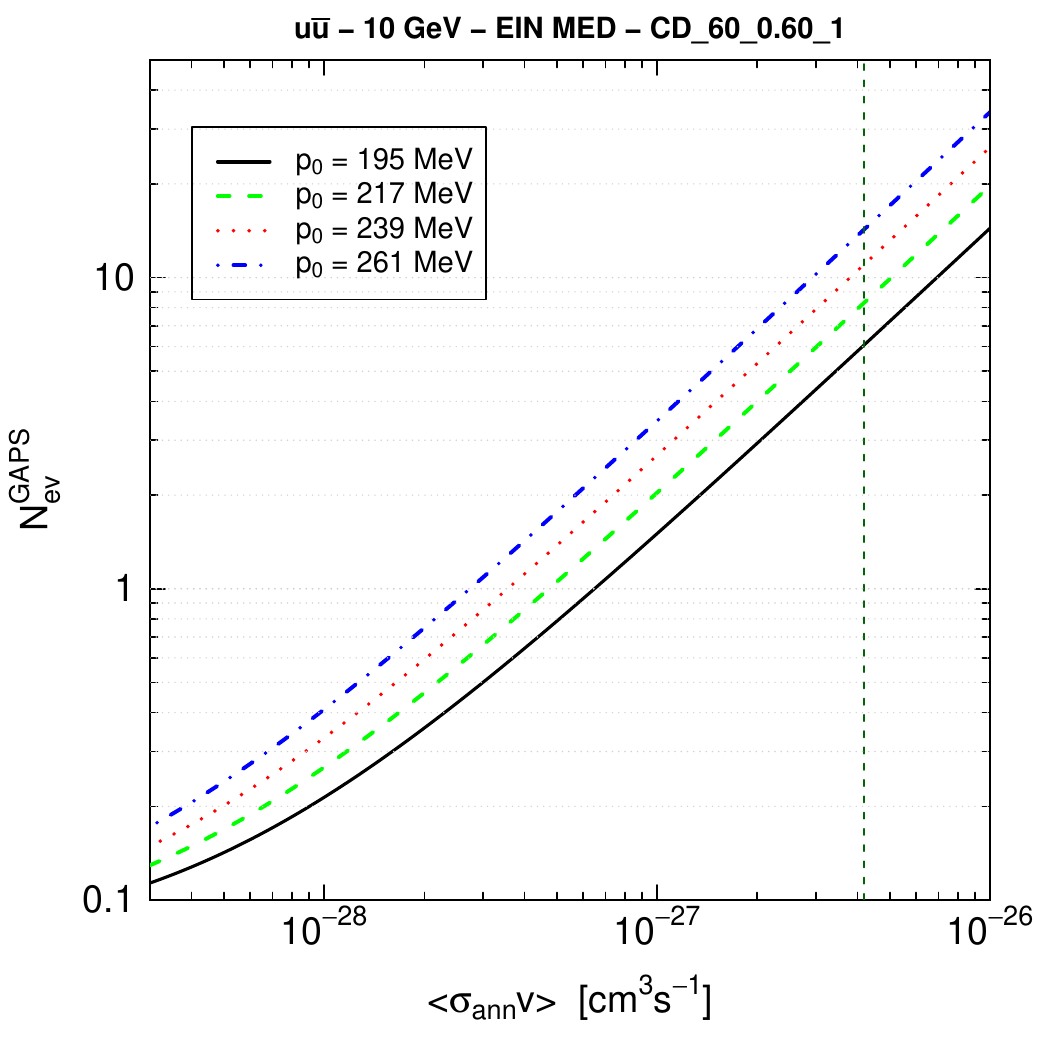}
\includegraphics[width=0.30\textwidth]{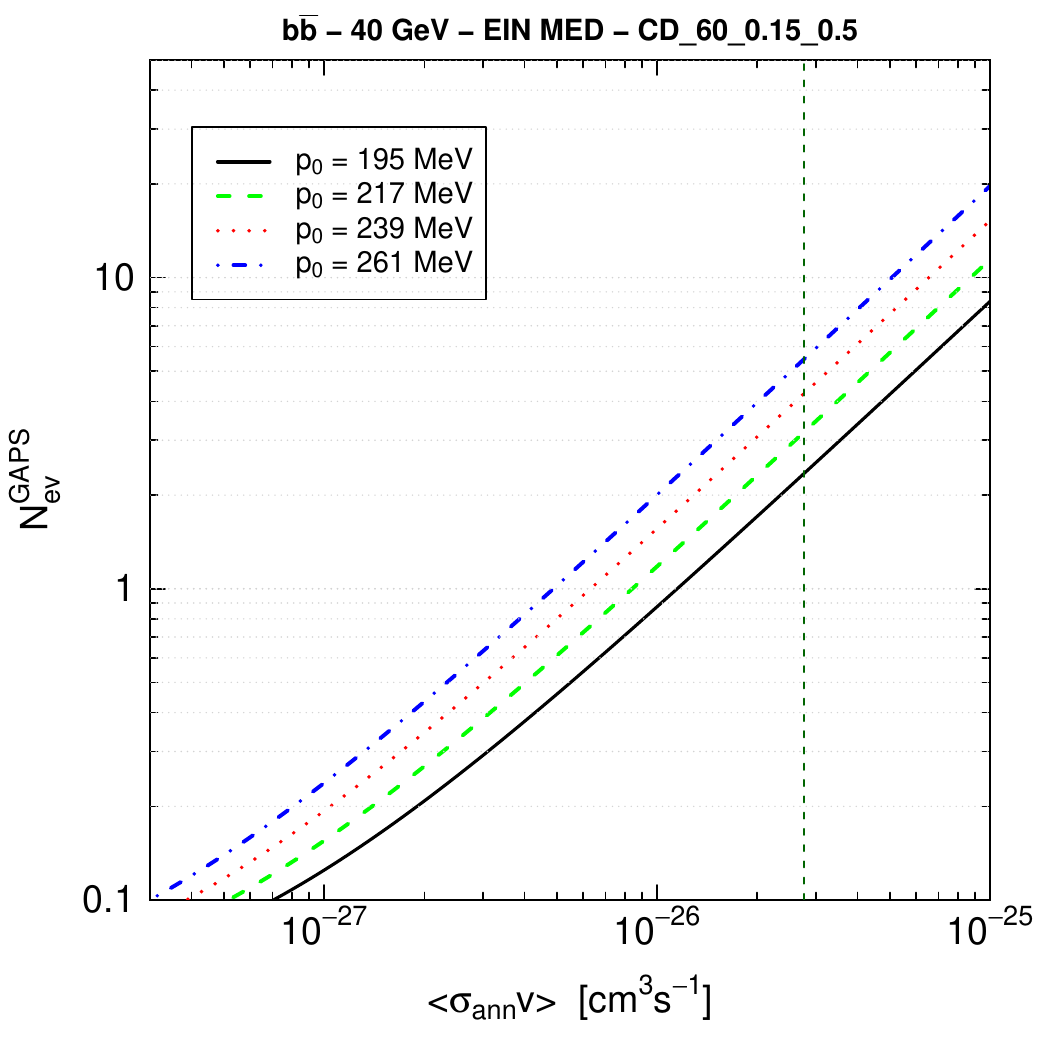}
\includegraphics[width=0.30\textwidth]{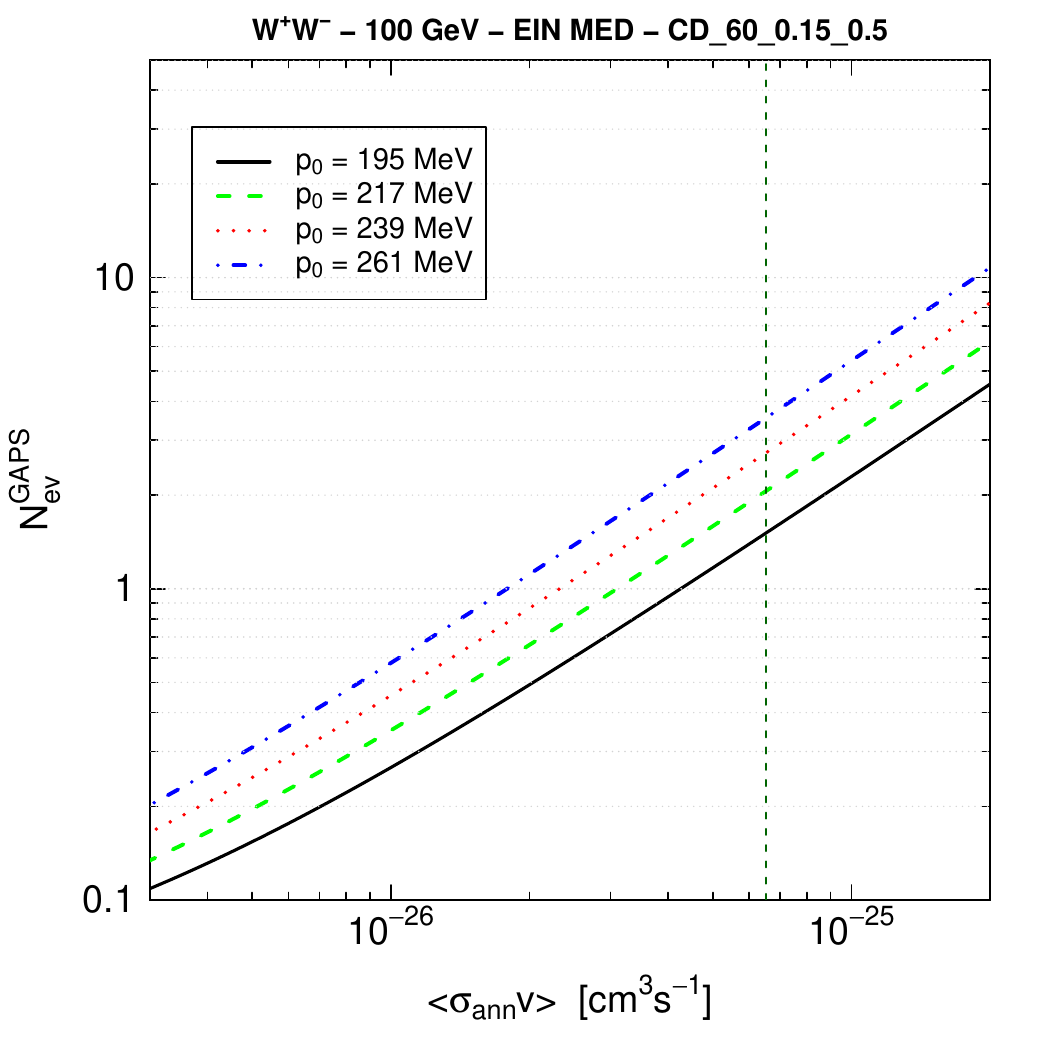}
\caption{Number of $\dbar$ expected for the GAPS experiment, with the LDB+ set up \cite{haileypriv} 
(shown in Fig. \ref{fig:TOAuu}) as a function of the annihilation cross section $\sigmav$. The left, central and right panels refer to a dark matter with $\mdm=10$ GeV annihilating into $\uubar$, 
a dark matter with $\mdm=40$ GeV annihilating into $\bbbar$ and
a dark matter with $\mdm=100$ GeV annihilating into $\ww$, respectively.
Each panel has been calculated for the solar modulation configuration which provides
the largest S/B ratio as shown in Fig. \ref{fig:sbratio} (for each panel, the code name of the adopted solar modulation model is reported in the upper label).
The vertical line represents the PAMELA \cite{Adriani:2010rc} (vertical line) bound on the dark matter annihilation cross section.
In each panel, the different curves show the effect induced by the uncertainty on the coalescence
momentum $p_0$: from bottom to top, the lines refer to $p_0 = 195, 217, 239, 261$ MeV
(as reported also in the boxed insets), values corresponding to the central value and the $1\sigma$, 
$2\sigma$ and $3\sigma$ positive deviations, as derived in this paper with the MC($\Delta p + \Delta r$) model. The background level from secondary $\dbar$ in GAPS LDB+ is below the horizontal axis reported in the figure.
\label{fig:GAPSevents1}
}
\end{figure}
%%%

%%%
\begin{figure}[t]
\centering
\includegraphics[width=0.30\textwidth]{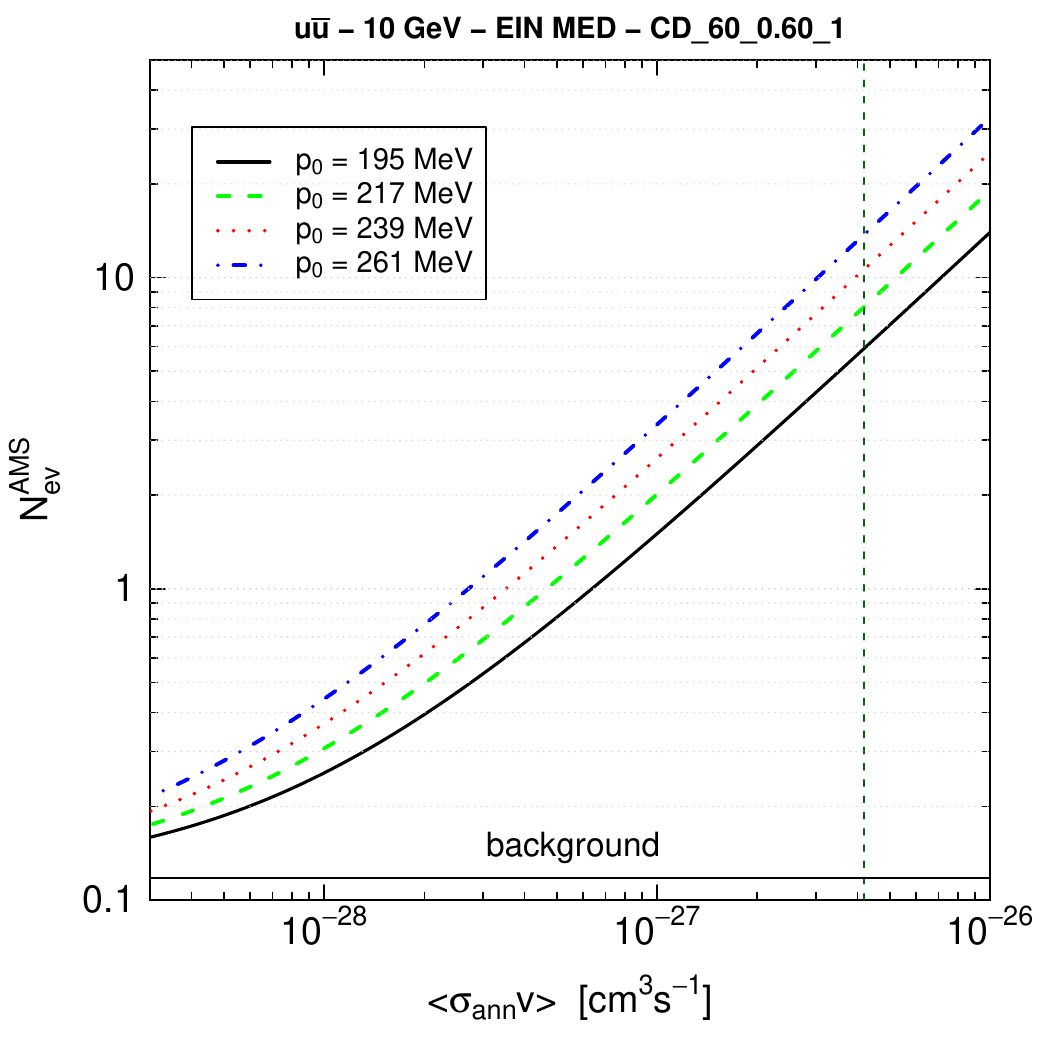}
\includegraphics[width=0.30\textwidth]{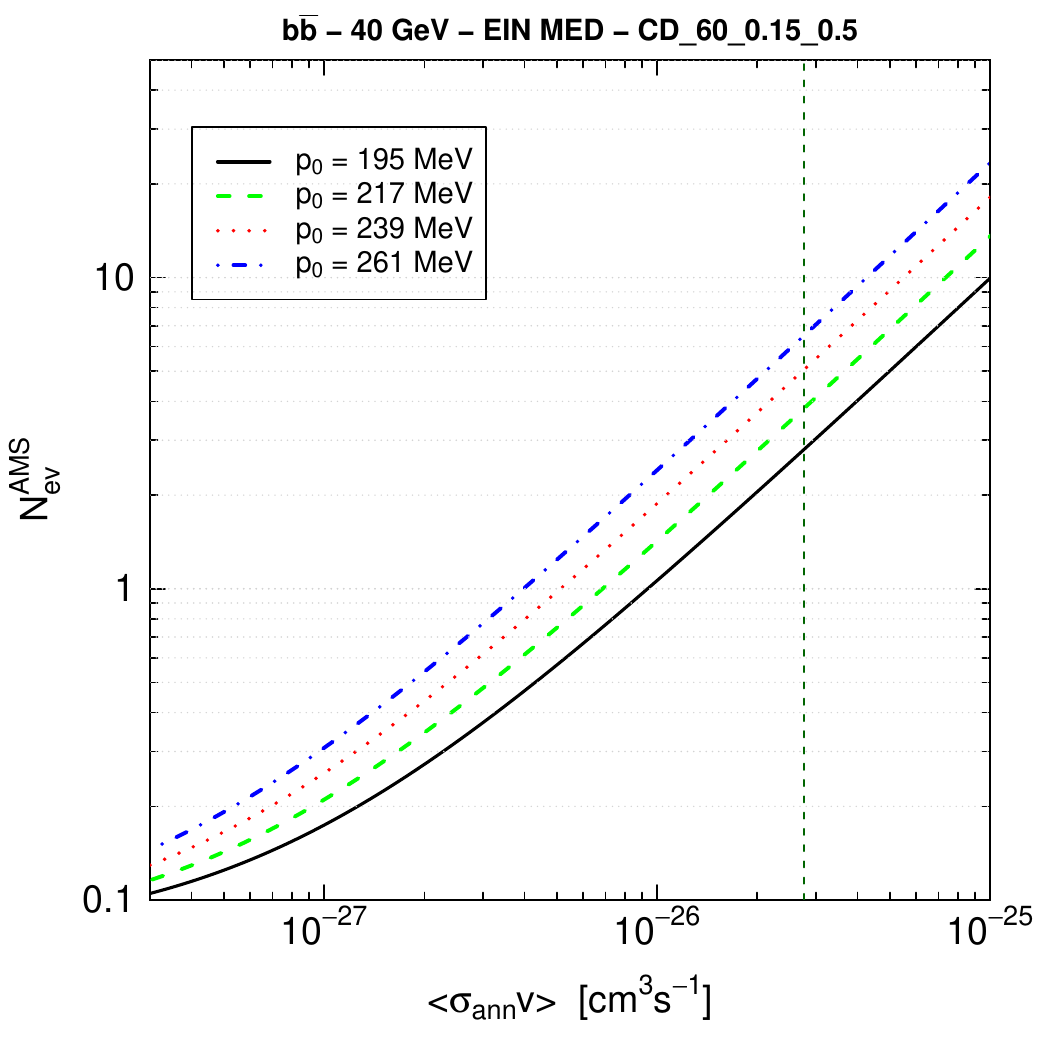}
\includegraphics[width=0.30\textwidth]{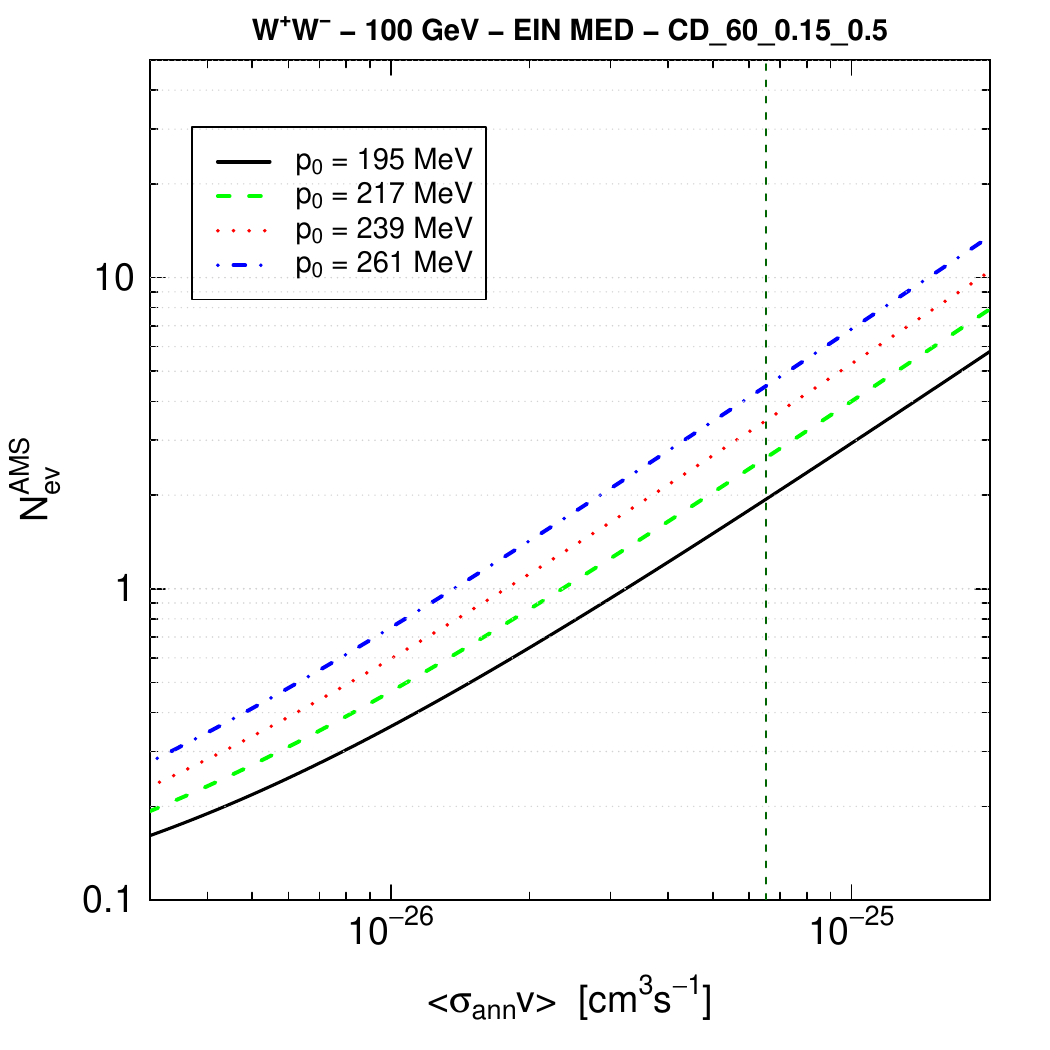}
\caption{The same as in Fig. \ref{fig:GAPSevents1}, for the AMS-02 nominal sensitivity \cite{giovacchini} (shown in Fig. \ref{fig:TOAuu}). The horizontal lines report the background
from secondary $\dbar$ in AMS-02 ( if it is above the horizontal axis)
%(they differ by changing panel, due to the different models
%adopted for solar modulation).
\label{fig:AMSevents1}
}
\end{figure}
%%%

%%%
\begin{figure}[t]
\centering
\includegraphics[width=0.30\textwidth]{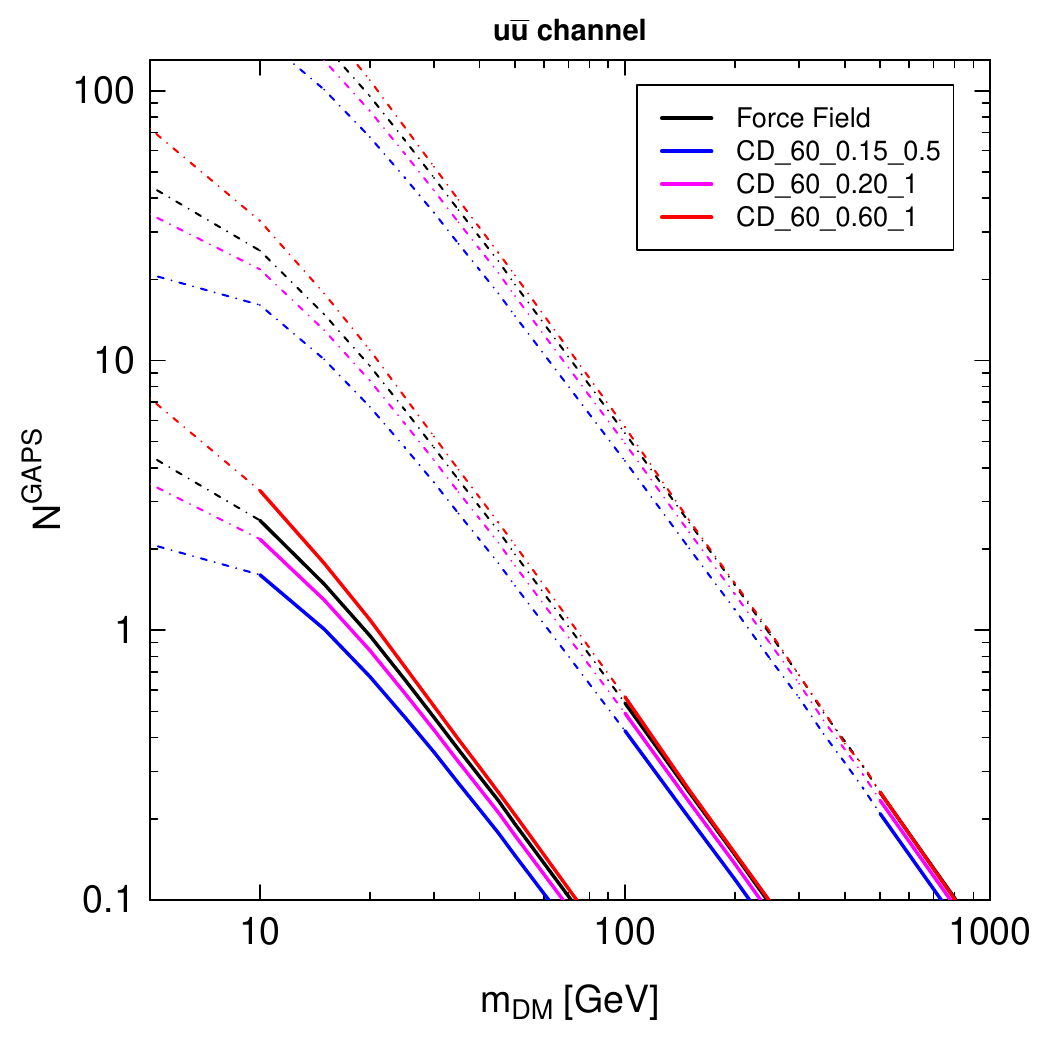}
\includegraphics[width=0.30\textwidth]{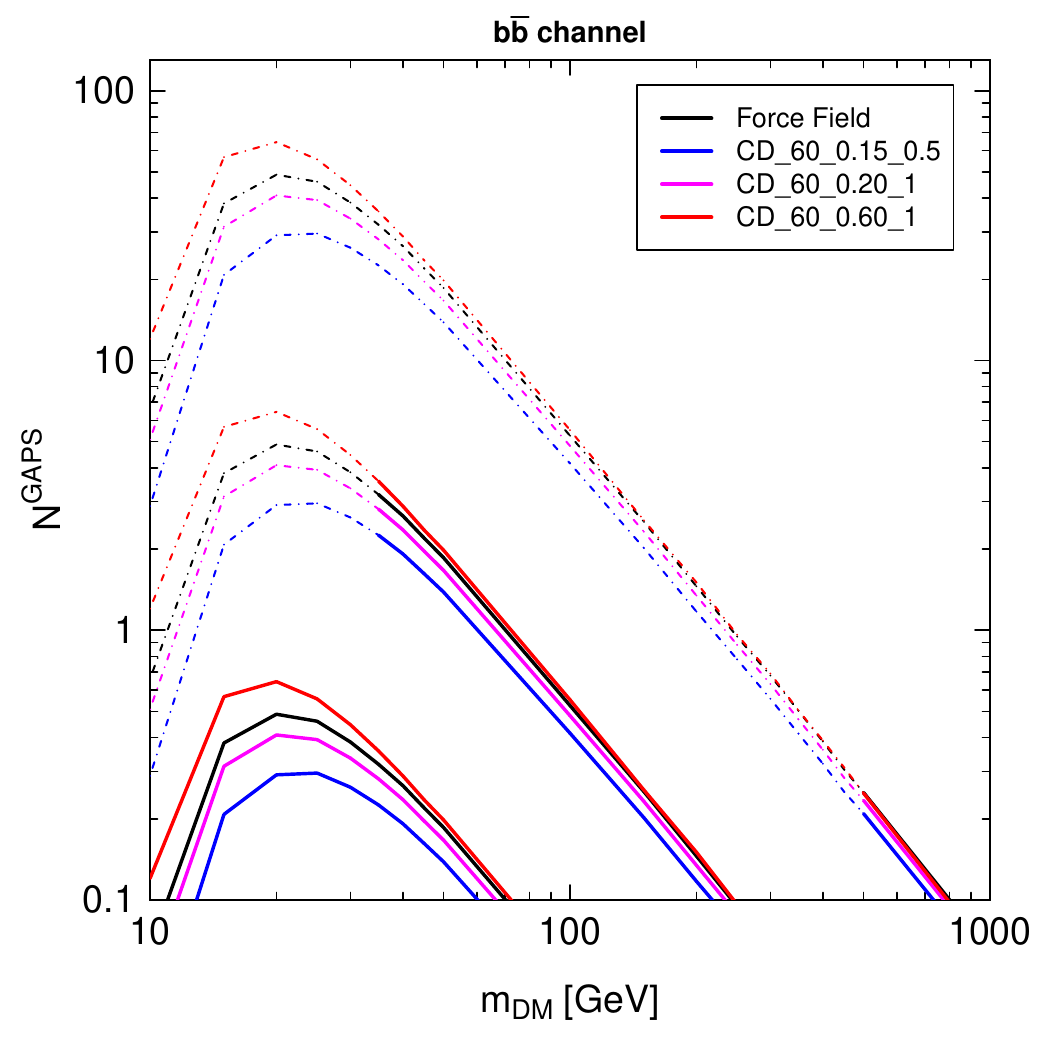}
\includegraphics[width=0.30\textwidth]{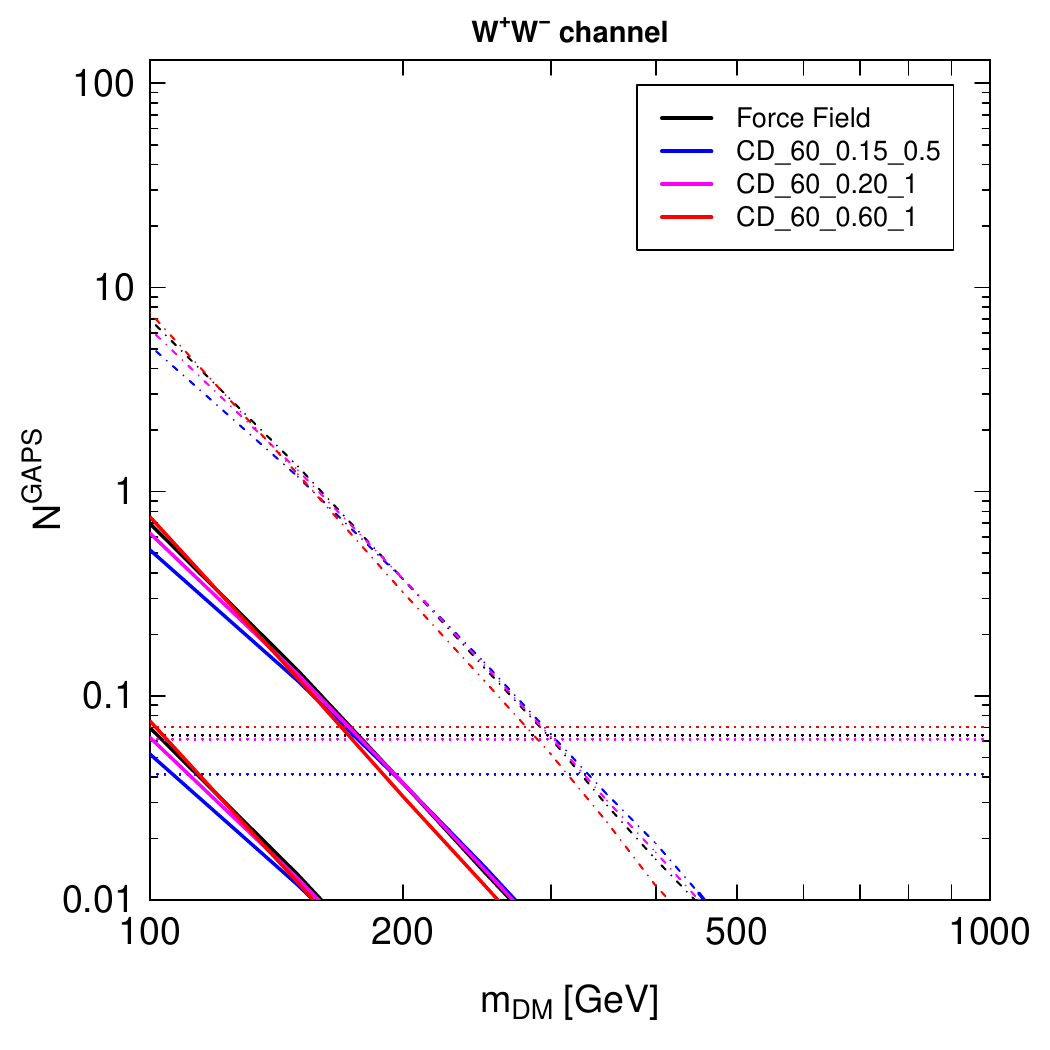}
\caption{Number of $\dbar$ expected for the GAPS experiment, with the LDB+ set up \cite{haileypriv} 
(shown in Fig. \ref{fig:TOAuu}) as a function of the dark matter mass $\mdm$.
The left, central and right panels refer to a dark matter particle
annihilating into $\uubar$,  into $\bbbar$ and into $\ww$, respectively. Each group of curves
is obtained for an annihilation cross section of 0.1, 1 and 10 times the thermal value
\thermal (from the lower set to the upper set). The different curves inside each group
are differentiated by the adopted solar modulation model (as reported in the boxed inset,  code name refers to Table \ref{tab:solarmod}).
For each line, the dashed part refers to configurations which are excluded by antiprotons searches
with the PAMELA detector \cite{Adriani:2010rc}, while the solid part is not in conflict with the PAMELA bound.
The horizontal lines report the level of the secondary $\dbar$ background (which changes according to
the different solar modulation models). The galactic dark matter halo is described by an Einasto profile, while the galactic cosmic-rays propagation model is MED.
\label{fig:GAPSevents2}
}
\end{figure}
%%%

%%%
\begin{figure}[t]
\centering
\includegraphics[width=0.30\textwidth]{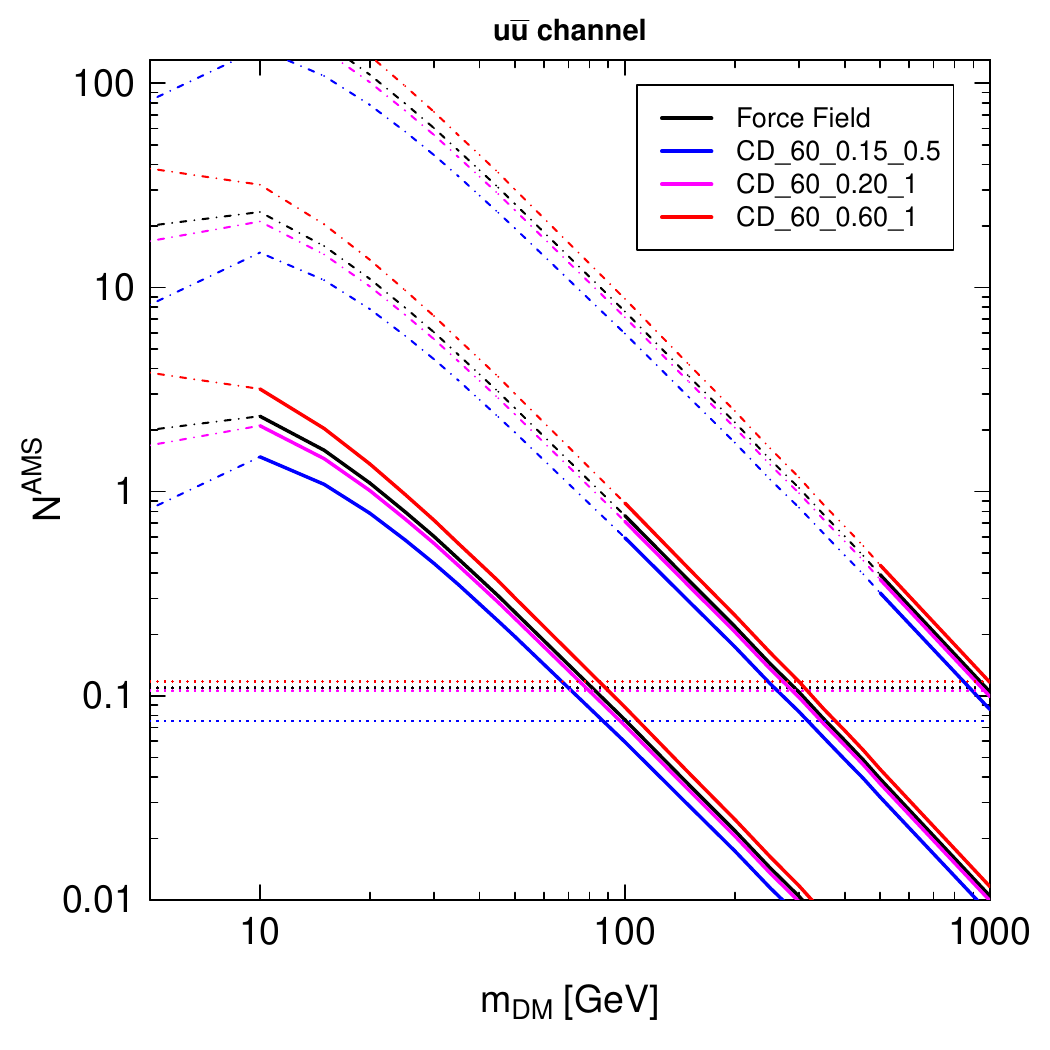}
\includegraphics[width=0.30\textwidth]{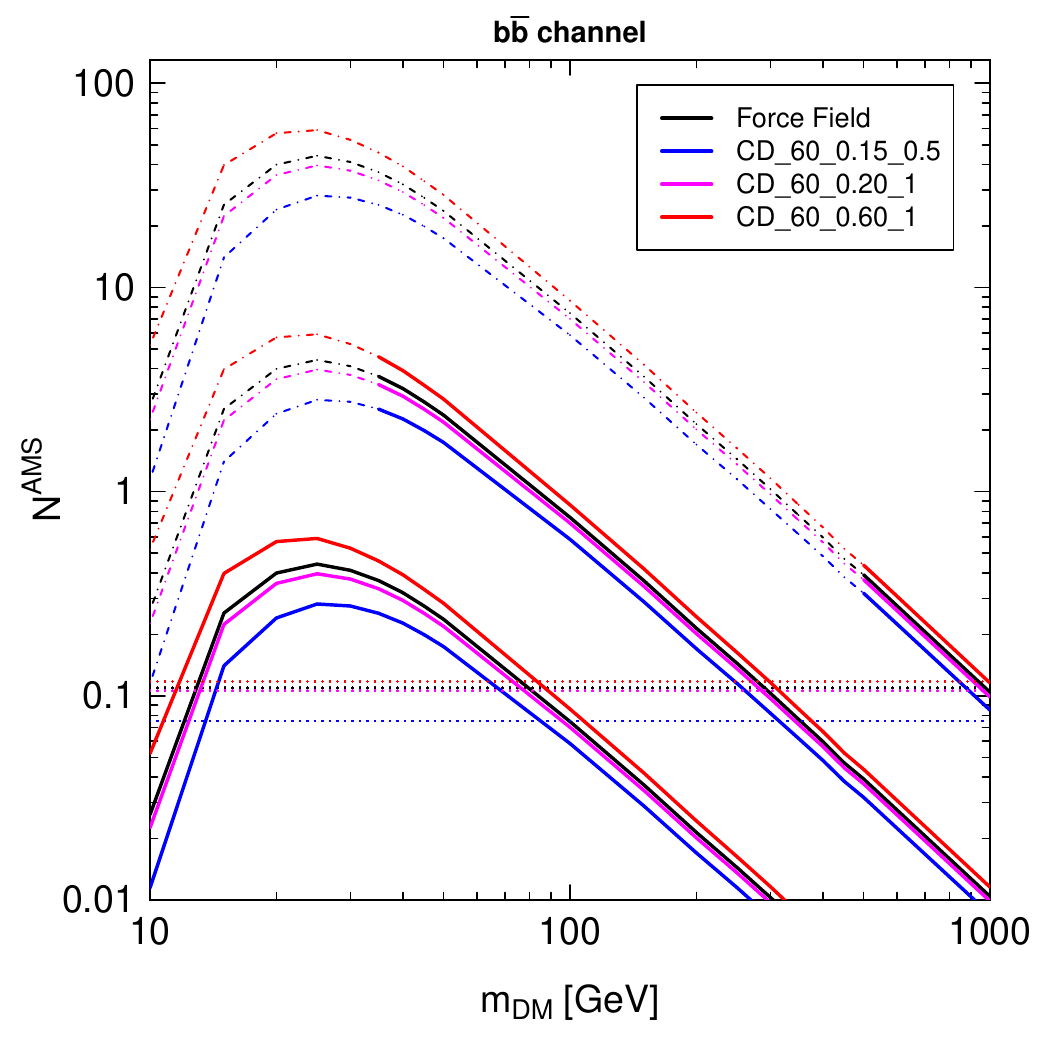}
\includegraphics[width=0.30\textwidth]{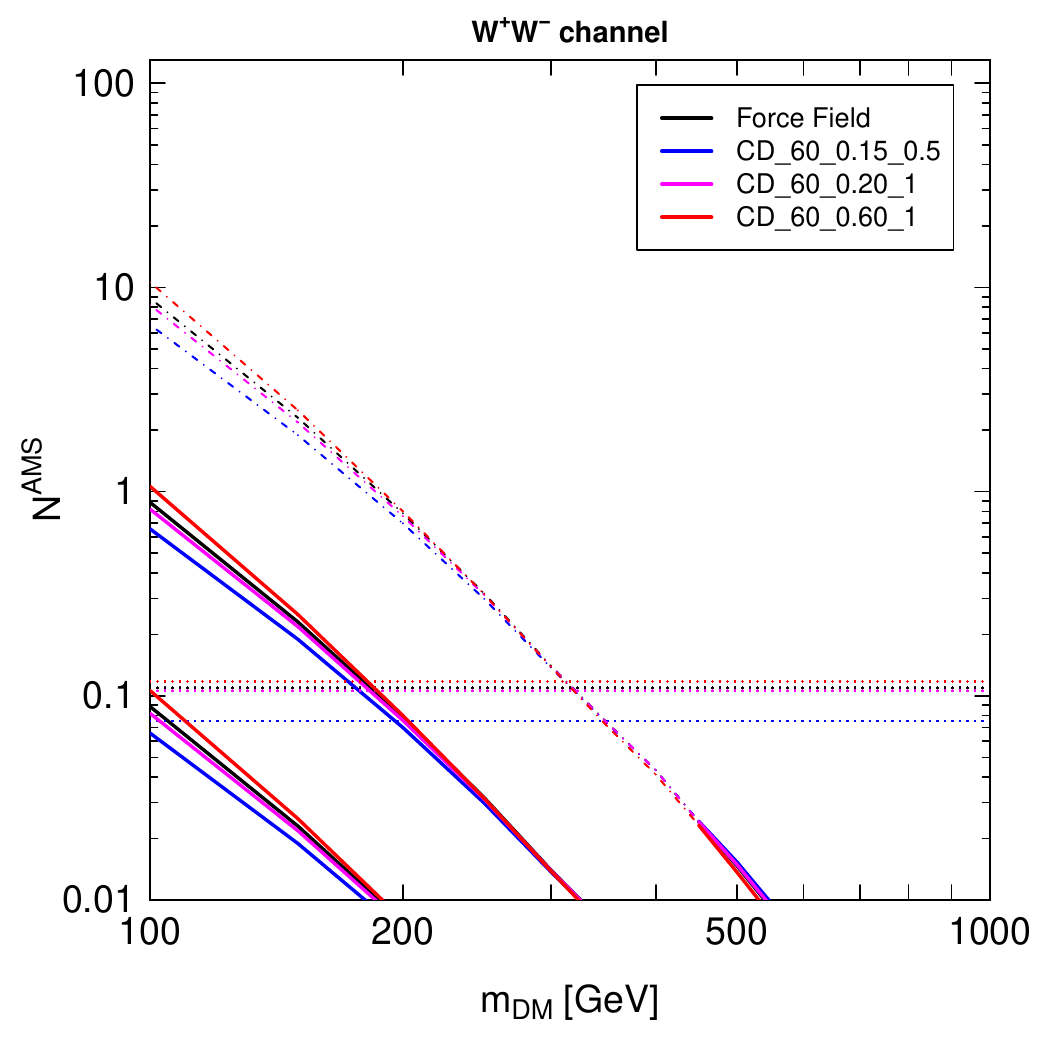}
\caption{The same as in Fig. \ref{fig:GAPSevents2}, for the AMS-02 nominal expected sensitivity \cite{giovacchini} (shown in Fig. \ref{fig:TOAuu}). 
\label{fig:AMSevents2}
}
\end{figure}
%%%

%%%
\begin{figure}[t]
\centering
\includegraphics[width=0.30\textwidth]{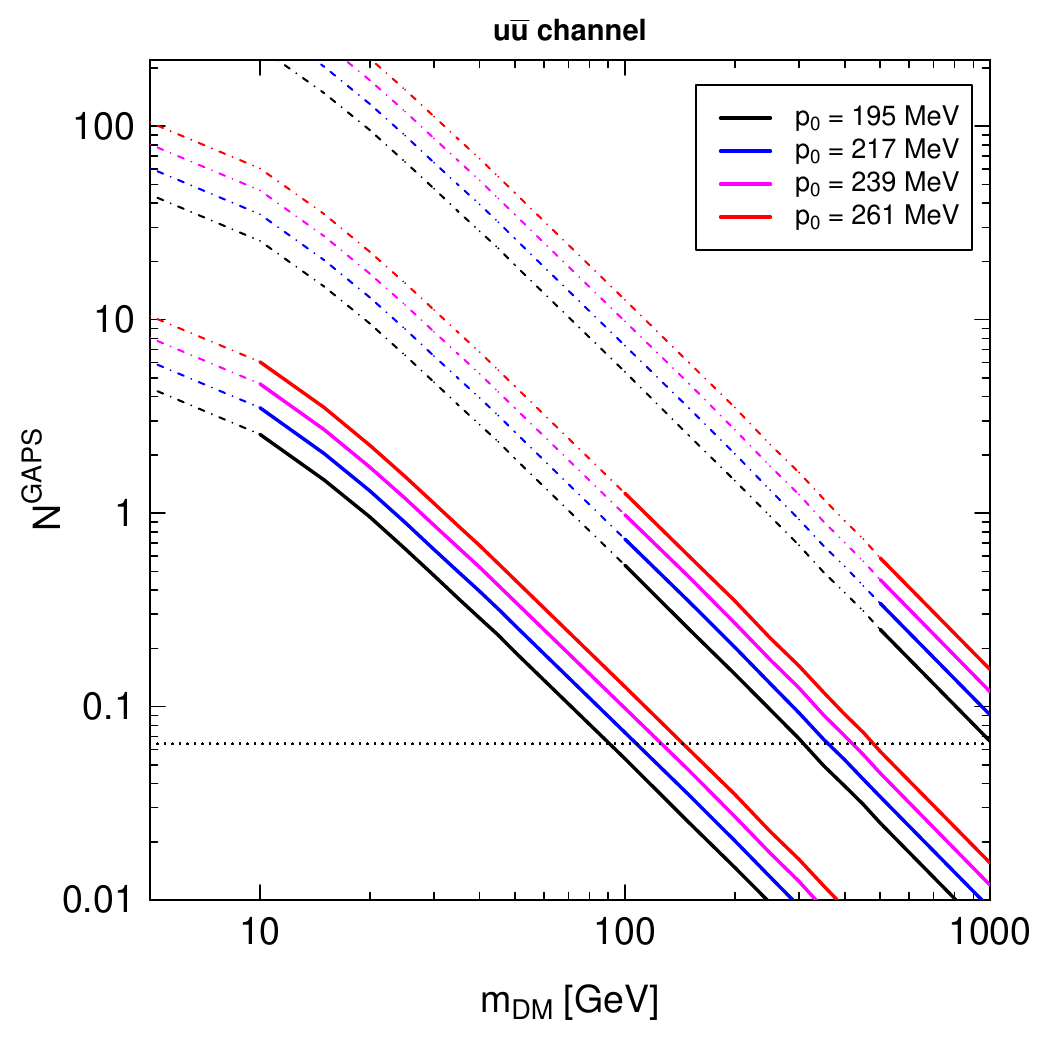}
\includegraphics[width=0.30\textwidth]{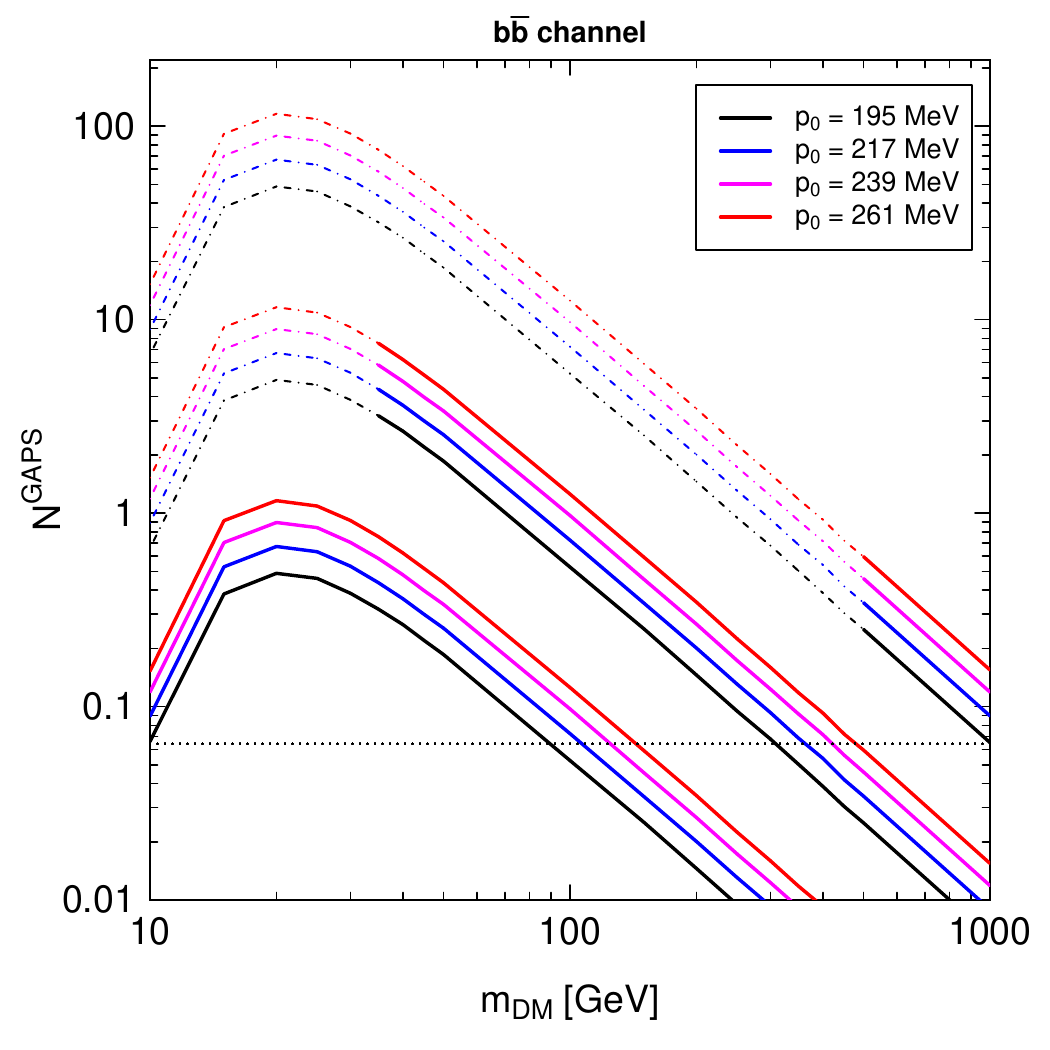}
\includegraphics[width=0.30\textwidth]{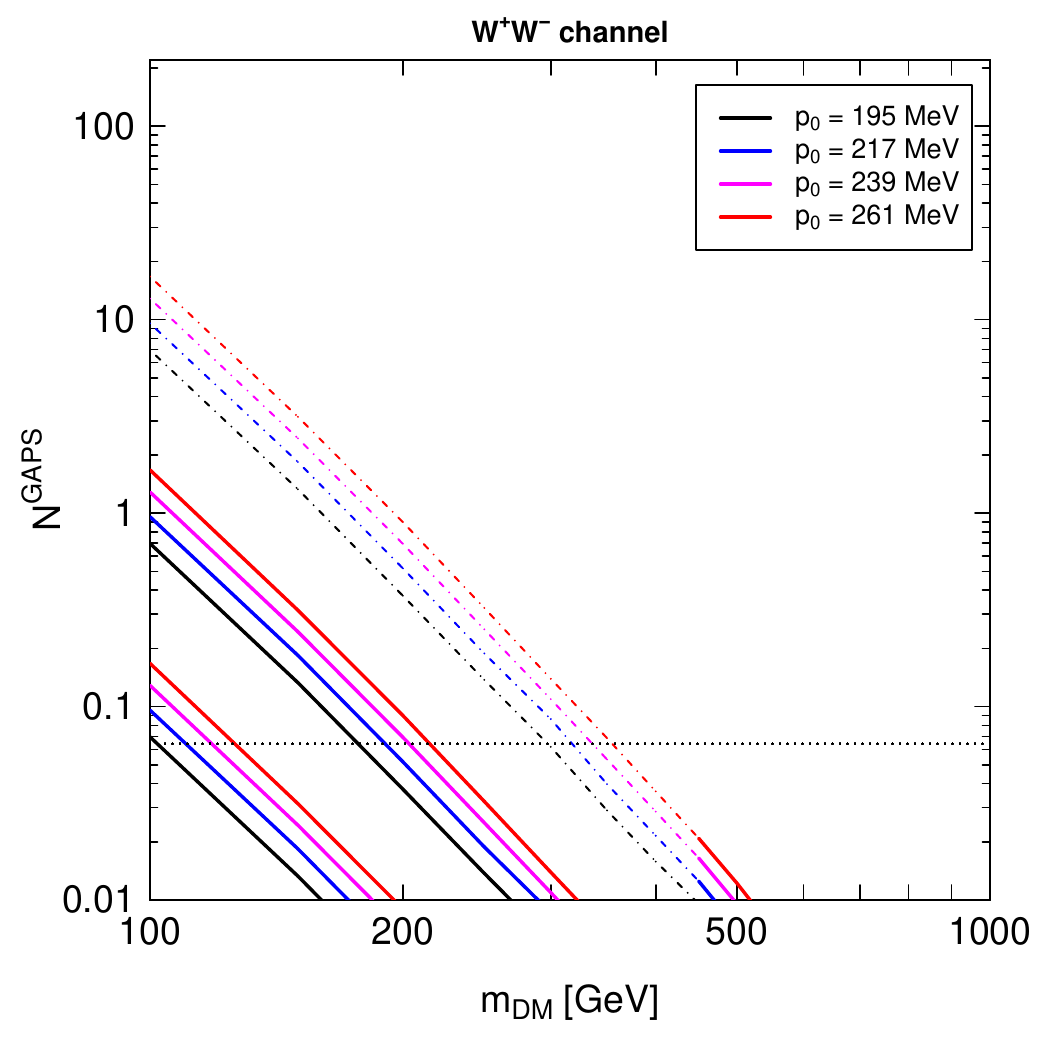}
\caption{The same as in Fig. \ref{fig:GAPSevents2}, except that here the solar modulation
is described by the standard force-field model, and the different curves inside each group
are differentiated by the adoption of different coalescence momenta.}
\label{fig:GAPSevents3}
\end{figure}
%%%

%%%
\begin{figure}[t]
\centering
\includegraphics[width=0.30\textwidth]{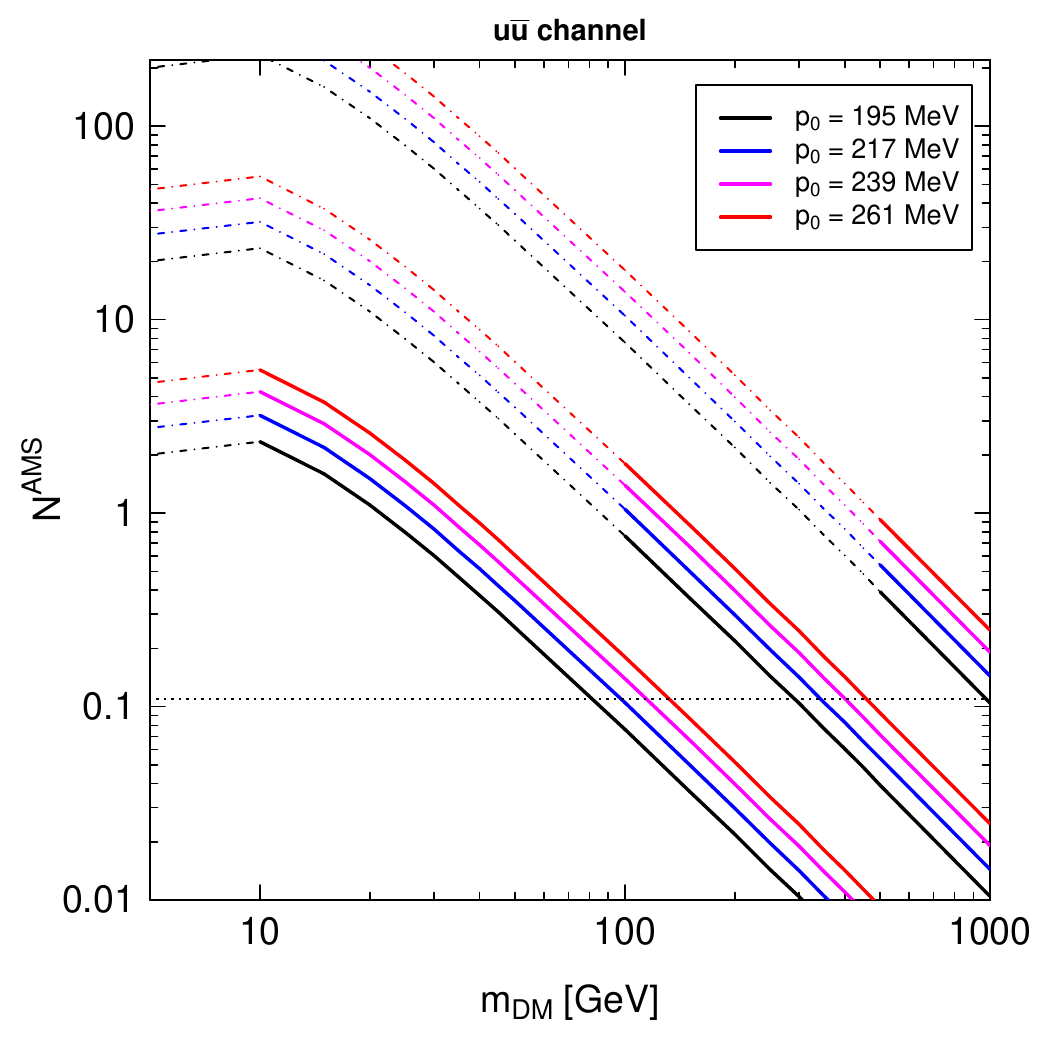}
\includegraphics[width=0.30\textwidth]{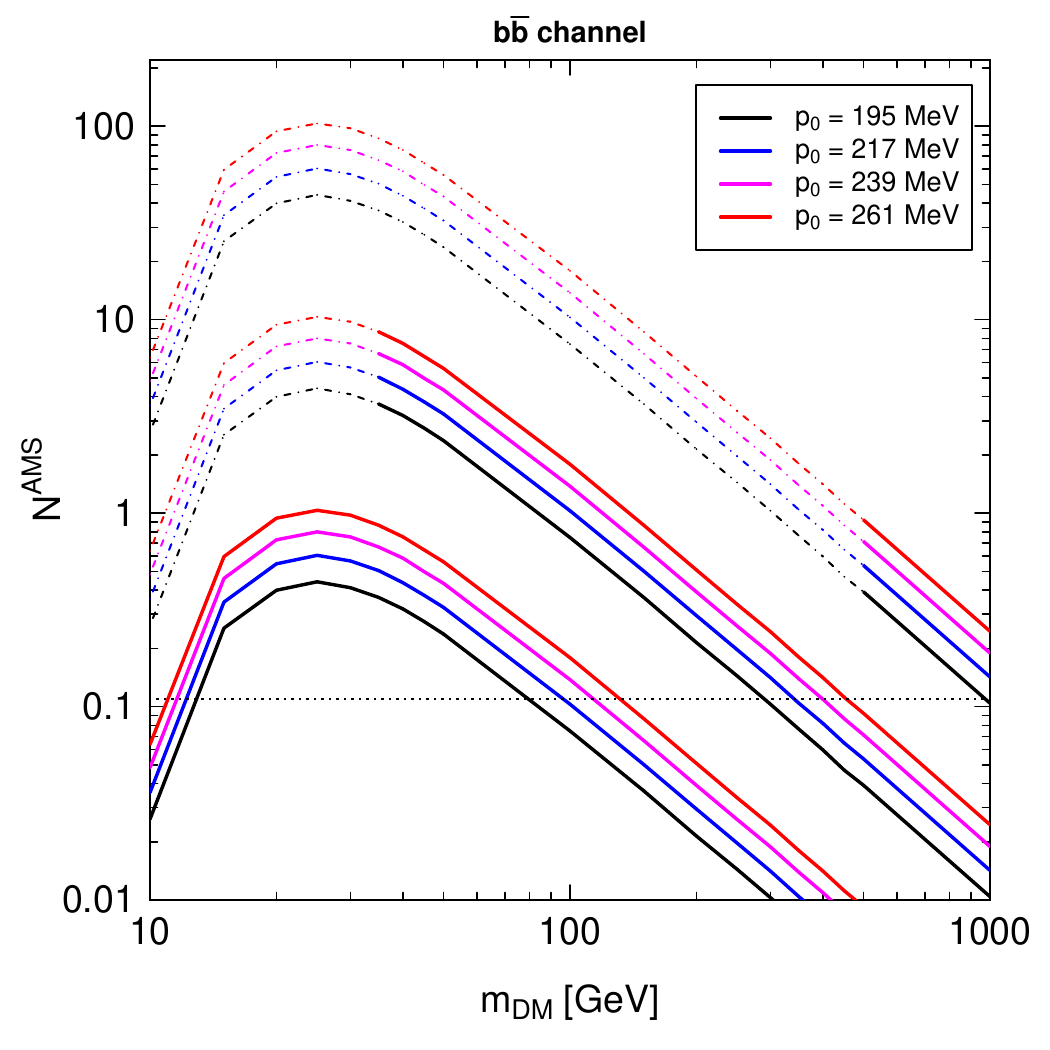}
\includegraphics[width=0.30\textwidth]{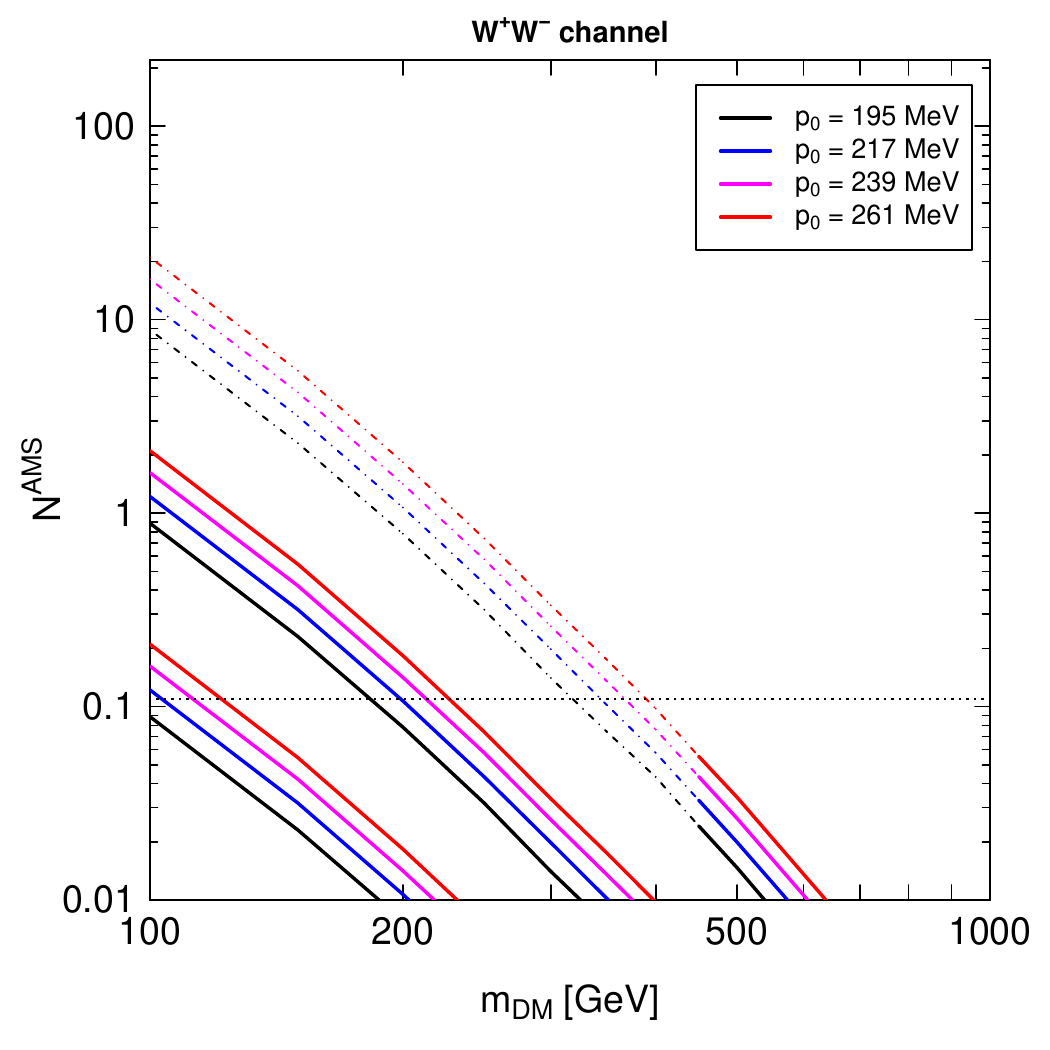}
\caption{The same as in Fig. \ref{fig:AMSevents2}, except that here the solar modulation
is described by the standard force-field model, and the different curves inside each group
are differentiated by the adoption of different coalescence momenta.}
\label{fig:AMSevents3}
\end{figure}
%%%

We now move to estimate the reaching capabilities of the GAPS and AMS-02 detector, to explore
what are the opportunities of $\dbar$ detection in the near future, for the different particle
physics, galactic propagation, solar modulation and coalescence models discussed in this analysis.
 We define the prospects of detection 
as the number $N_{\rm crit}$ of events a detector must observe in order to invoke a positive observation on the top of the background, at the $3\sigma$ level of confidence. $N_{\rm crit}$ is then transformed into a value for the annihilation cross section $\sigmav$ for any given DM mass $\mdm$ in the
specific propagation and coalescence model. The
number $N_{\rm crit}$ is the smallest integer $N$ for which the cumulative poissonian
distribution with mean equal to the expected number of background events $b$ is larger than the
chosen level of confidence, i.e. \cite{Profumo,Randall}:
\begin{equation}
\sum_{n=0}^{N-1}P(n,b)>0.997
\label{eq:detection}
\end{equation}
where $P(n,b)$ is the Poisson distribution for $n$ counts with mean $b$:
\begin{equation}
P(n,b)=\frac{b^ne^{-b}}{n!}
\end{equation}
$N_{\rm crit}$, as well as $b$,  depend on the detector exposure ${\cal E}(E)$ (which in
general is a function of the $\dbar$ energy and may also change in time) and on the energy window $\Delta E$ of the detector. These quantities are determined by design studies, calibration operations and
Monte Carlo modeling. 
Since we do not have the actual exposure functions for GAPS and
AMS-02, we infer effective exposures from the sensitivity fluxes quoted by the two collaborations: 
GAPS LDB+ has a sensitivity flux, corresponding to $D=1$ detected event, of
$\Phi^{\rm D}_{\Delta E} = 2.8 \times 10^{-7}$ m$^{-2}$ s$^{-1}$ sr$^{-1}$ (GeV/n)$^{-1}$ 
in the energy interval $(0.1 - 0.25)$ GeV/n \cite{haileypriv}; AMS-02 quotes a flux limit, corresponding to $D=1$ detected event, of
$\Phi^{\rm D}_{\Delta E} = 4.5\times 10^{-7}$ m$^{-2}$ s$^{-1}$ sr$^{-1}$ (GeV/n)$^{-1}$ 
in the energy interval $(0.2 - 0.8)$ GeV/n \cite{giovacchini,giovacchinipriv}. 
From this information we derive
an effective (energy and time independent) exposure $\langle {\cal E} \rangle$ as:
\begin{equation}
D = \int \Phi^{\rm D}(E) \; {\cal E}(E) \; dE \; = \; \Phi^{\rm D}_{\Delta E} \; \times \;
\langle {\cal E} \rangle \; \times \Delta E
\label{eq:sens}
\end{equation}
We therefore have: $\langle {\cal E} \rangle = 2.36 \times 10^{7}$ m$^2$ s sr for GAPS LBD+ 
and $\langle {\cal E} \rangle = 3.71 \times 10^{6}$ m$^2$ s sr for AMS-02, in their respective energy intervals.
These effective exposures will then be used to derive the prospects of detection as outlined
in connection to Eq. (\ref{eq:detection}). The number of background events is determined for
each specific (galactic and solar) propagation model by considering the actual theoretical estimate in the specific model. We notice that, in what follows, we always find $N_{\rm crit} = 1$ for GAPS LBD+,
while for AMS-02 $N_{\rm crit}$ varies according to the galactic propagation model adopted:  $N_{\rm crit} = 2$ for MED and MAX and $N_{\rm crit} = 1$ in the case of MIN. 
We stress that these numbers refer to
$\dbar$ signal detections with a confidence of $3\sigma$ C.L., since the number of expected events for
the background ranges from 0.04 to 0.07 events (depending, as stated, on
the specific propagations models) for GAPS LBD+ and from 0.06 to 0.15 for AMS-02.
Notice that the flux sensitivity lines reported in Figs. \ref{fig:TOAuu} -- \ref{fig:TOAWW}
have been determined in accordance with this discussion: they correspond to the flux level
which can be accessed by the detector (either GAPS LDB+ and AMS-02) in order to
observe a signal with a $3\sigma$ C.L. In Figs. \ref{fig:TOAuu} -- \ref{fig:TOAWW}, the
flux sensitivities correspond to 1 detected events for GAPS LDB+ and 2 events for AMS-02.

Fig. \ref{fig:GAPS1} shows the prospects for a $3\sigma$ detection of a $\dbar$ signal with GAPS
in the LDB+ setup, expressed in the plane
annihilation cross-section
$\sigmav$ vs the DM mass $\mdm$, for the $\uubar$ annihilation channel. The three upper/median/lower solid lines denote the detection limits (which correspond to detection of $N_{\rm crit} = 1$ event for GAPS with current design sensitivity \cite{haileypriv}) for the three galactic propagation models  MIN/MED/MAX of Table \ref{tab:parameters},
respectively. The three dot-dashed lines show the corresponding upper bounds  derived from the antiproton measurements of PAMELA \cite{Adriani:2010rc} in the same plane (again for each of the three propagation models MIN/MED/MAX, from top to bottom). 
The galactic DM halo is described by an Einasto profile.
Solar modulation has been modeled with the standard force-field approximation in the first panel, and
for solar modulation models as reported in the boxed insets, for the other panels. We notice that,
as already commented, the antiproton upper limit on $\sigmav$ is quite relevant in setting bounds
on the DM particle (for each set of lines, i.e. for each galactic propagation models, the portion of plane above the dot-dashed line is excluded at the 99\% C.L.). We then see that, regardless of the galactic
propagation model and of the solar modulation models, GAPS will have sensitivity to detect a
DM signal with a $3\sigma$ C.L. in large portions of the allowed parameter space. For instance, let's
concentrate on the MED galactic propagation model. In the case of a force-field solar-modulation,
prospects of a signal detection extend from very low masses up to
90 GeV, for annihilation cross sections ranging from $5\times 10^{-28}$ \cms\, up to 
to $2\times 10^{-26}$ \cms\,, depending on the DM mass. The reachability portion of the DM parameter
space is largely affected by galactic propagation: for the MAX set of propagation parameters, the region
moves to cover DM masses from very low values up to 700 GeV, for cross sections
from $4.5\times 10^{-29}$ \cms\, up to 
to $2\times 10^{-25}$ \cms\,, depending again on the DM mass. For the MAX set, the thermal annihilation cross section \thermal\, is within reach for masses around 300-400 GeV. Clearly,
the opposite situation, i.e. the exploration of larger annihilation cross sections, occurs for the MIN
set of galactic propagation parameters.

 Different solar modulation models shift the relevant region of exploration by some amount.
The heliosphere model that we already found to predict the largest fluxes for some specific 
DM masses, i.e. the model CD\_60\_0.60\_1, confirms this characteristic for DM masses below
approximately 100 GeV, while for large masses it does not deviate significantly from 
the force field-case. This can be appreciated in the upper-right panel in Fig. \ref{fig:GAPS1}, where
the reachability curves (and also the upper bounds from antiprotons) are shifted down by a factor
of 2--3 for light DM, while for heavy DM masses are similar to the force-field case.  

Figure \ref{fig:GAPS2} reports the prospects for a $3\sigma$ detection with GAPS in the case
of a DM annihilating into $\bbbar$. Notations are the same as in Fig.~\ref{fig:GAPS1}. Also
for this annihilation channel, GAPS has good prospects of DM detection, over a wide range of
DM masses and annihilation cross sections. The most noticeable difference from the $\uubar$
channel is the reduced reachability for very light DM, below about 15 GeV. This is expected, as
a consequence of the strong reduction in coalescence for light DM in the $\bbbar$ channel, due
to the higher production of nucleons from heavy baryon decays, as compared to the $\uubar$
channel (and to the same $\bbbar$ channel at higher values of $\mdm$). This phenomenon was discussed in Sec.~\ref{sec:spectra} and we see now that this fact limits DM searches in the
$\bbbar$  channel for light DM, since it brings the reachability curve above the antiproton bound.
On the contrary, for DM masses from 15 GeV up to 100 GeV (for the MED galactic propagation)
or up to 800--1000 GeV (for the MAX case), there are large portions of the DM parameter
space that can be explored (the actual portion, again, depends on the galactic propagation
and on solar modulation). The trend among the different solar modulation models is analogous
to what is found for the $\uubar$ channel.

The prospects for the $\ww$ annihilation channel are shown in Fig.~\ref{fig:GAPS3}. We notice that
for this channel the antiproton bounds leave very small space for reachability, expect in the MAX
case, where DM masses up to 200 GeV can be explored. 

Detection prospects for AMS-02 (nominal design sensitivity \cite{giovacchini}) are shown in Figs.~\ref{fig:AMS1},  \ref{fig:AMS2} and \ref{fig:AMS3}.  AMS-02 covers a different energy range,
as compared to GAPS, as can be seen in Fig.~\ref{fig:TOAuu}, extending to larger energies. 
The features of the reachability curves are quite similar to those discussed in connection to
GAPS, especially for the MIN propagation model where we have a similar coverage of the DM parameter space, while for the MED and MAX configurations the reachability appears somehow smaller but nevertheless a large range of DM masses is explorable: this is a welcome feature, since
it would be important to have redundancy in the signal searches for a such low signal as antideuterons are, with the ability to cover two different energy windows.

In Figs.~\ref{fig:GAPSevents1} and \ref{fig:AMSevents1} we instead show the actual number
of events we predict, under different assumption, for the GAPS LDB+ current design sensitivity
\cite{haileypriv} and for the AMS-02 nominal sensitivity \cite{giovacchini}. In both figures,
representative values of the DM mass for the three representative DM annihilation channels
are shown. Galactic propagation is fixed at the MED set of parameters.
Each panel has been calculated for the solar modulation configuration which provides
the largest S/B ratio as shown in Fig.~\ref{fig:sbratio} (for each panel, the code name of the adopted solar modulation model is reported in the upper label). The number of events are reported as a function of the DM annihilation cross section. The vertical dashed line shows the antiproton bound, while the
 horizontal line reports the background number of events, for the specific
 mass, galactic and solar modulation adopted in the plot.  The different lines show the increase
 in the predicted number of events when the coalescence momentum is increased from its central
 value to its $3\sigma$ upper allowed value. We see that both GAPS and AMS-02 can detect up to about
 15 events (in the most favorable situation). This is indeed a large number of events, considering that the secondary background is at the  level of 0.04 -- 0.07 for GAPS and 0.06 -- 0.15 in AMS-02 (depending on the specific
 galactic and solar modulation models; for AMS-02 the background levels are shown in the figure).

The number of expected events as a function of the DM mass is instead shown in Fig.~\ref{fig:GAPSevents2} for GAPS and Fig. \ref{fig:AMSevents2} for AMS-02.
The left, central and right panels refer to a dark matter particle
annihilating into $\uubar$,  into $\bbbar$ and into $\ww$, respectively. Each group of curves in each
panel, is obtained for an annihilation cross section of 0.1, 1 and 10 times the thermal value
\thermal (from the lower set to the upper set). The different curves inside each group
are differentiated by the adopted solar modulation model (as reported in the boxed inset,  where
the code name refers to Table \ref{tab:solarmod}).
For each line, the dashed part refers to configurations which are excluded by antiprotons searches
with the PAMELA detector \cite{Adriani:2010rc}, while the solid part is not in conflict with the PAMELA bound. Since we fix here the annihilation cross section and we vary the DM mass, not
all mass values are allowed for a given value of $\sigmav$, as discussed in connection to
Fig. \ref{fig:GAPS1} -- \ref{fig:AMS3}.
The horizontal lines report the level of the secondary $\dbar$ background (which changes according to
the different solar modulation models). The galactic dark matter halo is described by an Einasto profile, while the galactic cosmic-rays propagation model is MED. These figures show again the strong impact
of the antiproton bounds, which limits the maximal $\dbar$ expected signal. However, a few to almost
10 events (for this specific set of annihilation cross sections) are reachable. Solar modulation can change
the expected number of events by 1 or 2 units, depending on the annihilation channel and DM mass (this is not a small change, since this is
a rare-event signal). 

The effect induced by the uncertainty on the coalescence momentum is larger, and it is shown
in Fig.~\ref{fig:GAPSevents3} for GAPS and Fig.~\ref{fig:AMSevents3} for AMS-02, for the same
set of DM models of Fig.~\ref{fig:GAPSevents2} and \ref{fig:AMSevents2} for AMS-02. For reference,
solar modulation has been treated in the force field approximation. We see that a value
of $p_0$ larger than its reference value (although inside its $3\sigma$ allowed range) can boost 
the signal up to 10 events. 
%(or even slightly more). 
Considering theoretical uncertainties, prospects of exploration of DM in the $\dbar$ channel are therefore promising in a large fraction of the DM parameter space.

\section{Conclusions}
\label{sec:conclusions}

In this paper we have performed a detailed and complete re-analysis of the cosmic-ray antideuteron DM signal, proposed in Ref. \cite{DFS} as a promising channel for DM detection. We have here addressed
the main relevant issues related to antideuteron production and  propagation through the interstellar medium and the heliosphere. 

Specifically, we have first critically revisited the coalescence mechanism for antideuteron production in dark matter annihilation processes, by adopting a Monte Carlo modeling of the production of the
$\pbar\nbar$ pair in DM annihilation \cite{strumia,ibarra} (we focused the discussion on annihilation, but the results apply to DM decay as well), and by defining a coalescence mechanism that involves the vicinity in phase-space of the pair  (both in physical and momentum space, defined in the center-of-mass frame of the couple) in order for them to be merged to form a $\dbar$. Fine details
occurring mostly in the heavy-quark channels and for light DM mass have been discussed: basically,
for DM masses below 10-20 GeV, $\dbar$ production in the $\bar b b$ channel is suppressed, due
to a larger probability to produce $\pbar$ and $\nbar$  from heavy-baryon decays, a fact that
reduces the phase-space vicinity and consequently  inhibits the $\dbar$ production. A detailed
study of the dependence of the coalescence process on the $\dbar$ energy has then
be performed: while at large $\dbar$ kinetic energies the Monte Carlo approach predicts a
significant increase of the $\dbar$ rate \cite{strumia} as compared to the isotropic uncorrelated
model predictions, for kinetic energies below 1 GeV a small decrease is instead found. Some
regular trends in the Monte Carlo decrease/enhancement (as compared to the old model) are
found and discussed. Detailed modeling of the coalescence process, and its uncertainties, 
are found to have an impact that can be quantified as an increase (decrease) of a factor 1.8 (0.5) of the top-of-the-atmosphere flux.

Uncertainties arising from transport in the Galaxy remain the dominant source of variability,
 since they appear to affect the final flux up to one order of magnitude, as found also in Ref. \cite{DFM}. A refinement on this points requires
the upcoming data on cosmic rays nuclear species, expected in the near future from AMS-02, which
will potentially allow us to better shape the parameters of the propagation model. In particular, the increased accuracy in the determination of secondary/primary ratio (mainly the B/C) \cite{diffusion1,diffusion3,donato_tomassetti,Pato:2010ih}, together with the accumulation and ability to critically study multi-messenger data, may yield significant improvements on our understanding of galactic and solar transport, thereby reducing theoretical uncertainties.

The second relevant question addressed in the paper concerns the importance of the modeling
of the cosmic-rays transport in the heliosphere. 
Since antideuteron searches have their best prospects of detection at low kinetic energies, which is
where the effect of the solar wind and magnetic field are most relevant, it is important to assess
the relevance of solar modulation modeling beyond the standard and simple force-field approximation.
To this aim, we have used a full numerical 4D solution of cosmic rays transport in the heliosphere,
which offers the opportunity to take under consideration not only the average energy-shift due
to transport against the solar wind, but also stochastic fluctuations. We have quantified the effect, and found that 
uncertainties arising from solar modulation modeling are not very large, but they can nevertheless
reach a maximum of a factor 2.5 for the top-of-the-atmosphere fluxes of light DM particles.

Finally, we have used our improved predictions to provide updated estimates of the reaching capabilities for AMS-02 and GAPS, compatible with the current constraints imposed by the antiprotons measurements of PAMELA. After the
antiproton bound is applied, prospects of detection of up to about 15 events in GAPS LDB+
and AMS-02 missions are found, depending on the DM mass, annihilation rate and
production channel from one side, and on the coalescence process, galactic and solar
transport parameters on the other.

\acknowledgments

We wish to warmly thank R. Battiston, F. Donato, F.~Giovacchini, D.~Grasso, C. Hailey, D. Maurin, M. Regis and
P. Salati for insightful discussions. We thank M. Korsmeier for pointing out some typos and an inconsistency in the ``old'' coalescence model that have been fixed in this version.
N.F. and A.V. acknowledge the Research Grant funded by the Istituto Nazionale di Fisica Nucleare within the {\sl Astroparticle Physics Project}
(INFN grant code: FA51).  N.F. and A.V. acknowledge support of the {\sl Strategic Research Grant} jointly funded by the University of Torino and Compagnia di San Paolo 
(grant Unito-CSP 2010-2012 entitled {\sl Origin and Detection of Galactic and Extragalactic Cosmic Rays}). N.F. acknowledges support of the
spanish MICINN Consolider Ingenio 2010 Programme under grant MULTIDARK
CSD2009--00064. 
LM acknowledges support from the Alexander von Humboldt foundation and partial support from the European Union FP7 ITN INVISIBLES (Marie Curie Actions, PITN-GA-2011-289442).

% -------------------------------------------------------------------------------------------

% -------------------------------------------------------------------------------------------
\newpage

%%%

\end{document}